\documentclass[11pt, twoside]{Thesis} 

\graphicspath{{Graphics/}} 
\usepackage{graphicx}
\usepackage{pdflscape}
\usepackage{amsmath}
\usepackage[section]{placeins}
\usepackage{feynmf}
\usepackage{caption}
\usepackage[usenames,dvipsnames]{xcolor}
\usepackage[square, numbers, comma, sort&compress]{natbib} 
\hypersetup{urlcolor=blue, colorlinks=true} 
\title{\ttitle} 

\begin{document}
\setlength{\unitlength}{1mm}

\frontmatter 

\setstretch{1.3} 

\fancyhead{} 
\rhead{\thepage} 
\lhead{} 

\pagestyle{fancy} 

\newcommand{\HRule}{\rule{\linewidth}{0.5mm}} 

\hypersetup{pdftitle={\ttitle}}
\hypersetup{pdfsubject=\subjectname}
\hypersetup{pdfauthor=\authornames}
\hypersetup{pdfkeywords=\keywordnames}


\begin{titlepage}
\begin{center}

~\\\vspace{1.4cm}
{\Large \bfseries \ttitle}\\[0.4cm] 
\vspace{1.6cm}
{\large Donald Charles Jones\\
Charlottesville, VA
\vspace{1.5cm} \\
B.Sc.(Hon), Acadia University, 2008}
\vspace{2.6cm}
 
{\large A Dissertation presented to the Graduate Faculty\\
of the University of Virginia in Candidacy for the Degree of\\ Doctor of Philosophy\\ 
\vspace{1cm}
\large Department of Physics}\\[1cm]

{\large University of Virginia\\ 
October, 2015}\\[4cm] 
 
\vfill
\end{center}

\end{titlepage}


\addtotoc{Abstract} 

\abstract{\addtocontents{toc}{\vspace{1em}} 

The \Qs experiment which ran in Hall C at Jefferson Lab in Newport News, VA, and completed data taking in May 2012, measured the weak charge of the proton \qwps via elastic electron-proton scattering. Longitudinally polarized electrons were scattered from an unpolarized liquid hydrogen target. The helicity of the electron beam was flipped at approximately 1~kHz between left and right spin states. The \sms predicts a small parity-violating asymmetry of scattering rates between right and left helicity states due to the weak interaction. An initial result using 4\% of the data was published in October 2013 \cite{wien0} with a measured parity-violating asymmetry of $-279\pm 35(\text{stat})\pm 31$ (syst)~ppb. This asymmetry, along with other data from parity-violating electron scattering experiments, provided the world's first determination of the  weak charge of the proton. The weak charge of the proton was found to be $Q_W^p=0.064\pm0.012$, in good agreement with the Standard Model prediction of $Q_W^p(SM)=0.0708\pm0.0003$\cite{PDG2014}.

The results of the full dataset are expected to be published in early 2016 with an expected decrease in statistical error from the initial publication by a factor of 4-5. The level of precision of the final result makes it a useful test of Standard Model predictions and particularly of the ``running'' of $\sin^2\theta_W$ from the Z-mass to low energies. However, this level of statistical precision is not useful unless the systematic uncertainties also fall proportionately. This thesis focuses on reduction of error in two key systematics for the \Qs experiment. First, false asymmetries arising from helicity-correlated electron beam properties must be measured and removed. Techniques for determining these false asymmetries and removing them at the few ppb level are discussed. Second, as a parity-violating experiment, \Qs relies on accurate knowledge of electron beam polarimetry. To help address the requirement of accurate polarimetry, a Compton polarimeter built specifically for \Q. Compton polarimetry requires accurate knowledge of laser polarization inside a Fabry-Perot cavity enclosed in the electron beam pipe. A new technique was developed for \Qs that reduces this uncertainty to near zero.  
}

\clearpage 
\vspace{1cm}
\large We approve the dissertation of Donald C. Jones.\\
\vspace{1cm}~\\
\begin{minipage}{0.7\textwidth}
\rule[1em]{20em}{0.5pt}\\
Supervisor: Prof. Kent Paschke\\ 
~\\~\\~\\~\\
\rule[1em]{20em}{0.5pt}\\
Committee Member: Prof. Blaine Norum\\ 
~\\~\\~\\~\\
\rule[1em]{20em}{0.5pt}\\
Committee Member: Prof. Nilanga Liyanage\\ 
~\\~\\~\\~\\
\rule[1em]{20em}{0.5pt}\\
Committee Chair: Prof. James Fitz-Gerald\\ 
\end{minipage}
\begin{minipage}{0.3\textwidth}
\rule[1em]{10em}{0.5pt}\\
Date of Signature\\ 
~\\~\\~\\~\\
\rule[1em]{10em}{0.5pt}\\
Date of Signature\\ 
~\\~\\~\\~\\
\rule[1em]{10em}{0.5pt}\\
Date of Signature\\ 
~\\~\\~\\~\\
\rule[1em]{10em}{0.5pt}\\
Date of Signature\\ 
\end{minipage}

\clearpage 


\pagestyle{empty} 

\null\vfill 

\vspace{1in}
\textit{``Call unto me and I will answer you, and show you great and hidden things that you have not known."}
\begin{flushright}
Jeremiah 33:3
\end{flushright}

\vfill\vfill\vfill\vfill\vfill\vfill\null 

\clearpage 


\setstretch{1.3} 

\acknowledgements{\addtocontents{toc}{\vspace{1em}} 
The route from high school to Bachelors and on to PhD has been a rather circuitous track for me, and as I consider the key players who helped me reach this goal, I need to begin by acknowledging those who laid the foundations of education and character in my life. Without the encouragement and advice of others along the way, I might have been sidetracked numerous times perhaps permanently. Of course, I begin by acknowledging my parents, especially my mother. I think about my grade school teachers Ruth Morris, Shirley Harris and Murray Baker who encouraged me to go beyond high school and whose advice and confidence spurred me on. Or my older brothers Richard, Paul and Warren, who not only provided encouragement, but also helped materially when I would have otherwise had to quit for financial reasons. Richard has also been a key resource, helping me with difficult concepts all the way from undergraduate on through graduate studies. I want to acknowledge these as being key to my success.

I would next like to acknowledge my advisor, Kent Paschke, who I met as an undergraduate intern at Jefferson Lab. Although he demanded quality scientific output and I greatly benefited from his tutelage in this respect, perhaps greater was his example of integrity in research reporting. I learned to question my own conclusions and  be sensitive to self-introduced biases. I always felt under him that one gained the right to be vocal, confident or opinionated on an issue only after he had well-formulated ideas that could clearly presented and backed up with data. This was invaluable to a student who was quick to jump to conclusions, fast to verbalize them and then slower to test the hypotheses empirically. He not only taught me good research practices, but he also offered common sense advice for critical decisions relating to education and career and has always been supportive. I will continue to value his opinion  going forward.

Two others with whom I worked most closely and from whose expertise I most benefited were Dave Gaskell and Mark Dalton. They great resources, helping me to have confidence in tackling tasks on my own and willing to give credit, perhaps more than was due. These were a pleasure to work alongside.

I want to acknowledge Dave Armstrong, not only for the leadership role he played in the experiment, but also for taking the effort to break issues down into simple terms and for taking care not to assume that students were familiar with the science being discussed. I would like to thank Mark Pitt for his quick insight and leadership during the course of this difficult experiment. I want to also acknowledge the critical role Roger Carlini played, ensuring that the experiment ran smoothly, providing perspective, offering insights and taking, perhaps more than his due, of persecution from the rest, while maintaining a smile.

Finally, I want to thank my wife, Lydia, who has been a huge source of support and confidence through this whole process. She has been a tremendous encouragement, taking on more than her share of home duties to help me more rapidly reach the goal. 
}
\clearpage 


\pagestyle{fancy} 

\lhead{\emph{Contents}} 
\tableofcontents 

\lhead{\emph{List of Figures}} 
\listoffigures 

\lhead{\emph{List of Tables}} 
\listoftables 


\clearpage 

\setstretch{1.5} 

\lhead{\emph{Abbreviations}} 
\listofsymbols{ll} 
{
\textbf{IHWP} & \textbf{I}nsertable \textbf{H}alf-\textbf{W}ave \textbf{P}late \\
\textbf{PV} & \textbf{P}arity \textbf{V}iolation\\
\textbf{SM} & \textbf{S}tandard \textbf{M}odel of Particle Physics\\
\textbf{HC} & \textbf{H}elicity \textbf{C}orrelated\\
\textbf{HCBA} & \textbf{H}elicity \textbf{C}orrelated \textbf{B}eam \textbf{A}symmetry\\
\textbf{BPM} & \textbf{B}eam \textbf{P}osition \textbf{M}onitor\\
\textbf{BCM} & \textbf{B}eam \textbf{C}urrent \textbf{M}onitor\\
\textbf{ppm} & \textbf{p}arts \textbf{p}er \textbf{m}illion\\
\textbf{ppb} & \textbf{p}arts \textbf{p}er \textbf{b}illion\\

}


\clearpage 

\lhead{\emph{Physical Constants}} 

\listofconstants{lrl} 
{
Speed of Light & $c$ & $2.99792458\times10^{8}\ \mbox{ms}^{-\mbox{s}}$ (exact)\\
Planck constant & $h$ & $4.135667516(91)\times 10^{-15} eV\cdot s$\\
Mass of electron & $m_e$ & $0.510998928(11)~MeV/c^2$\\
Mass of protron & $m_p$ & $938.272046(21)MeV/c^2$\\
Fine structure constant & $\alpha=e^2/4\pi\epsilon_0\hbar c$ & $7.2973525698(24)\times 10^{-3}$\\
Fermi coupling constant & $G_F/(\hbar c)^3$ & $1.1663787(6)\times10^{-5}~GeV^{-2}$\\
$W^{\pm}$ boson mass & $ m_W$ & $80.385(15)~GeV/c^2$\\
$Z^0$ boson mass & $ m_Z$ & $91.1876(21)~GeV/c^2$\\
Strong coupling constant & $\alpha_s(m_Z)$ & $0.1185(6)$\\
Weak mixing angle & $\sin^2\theta_W(m_Z)_{\overline{MS}}$ & $0.23126(5)$\\
}


\mainmatter 

\pagestyle{fancy} 

\captionsetup{justification=justified,singlelinecheck=false}

\chapter{Introduction to the Standard Model}
\label{Chapter 1} 
\captionsetup{justification=justified,singlelinecheck=false}

\lhead{Chapter 1. \emph{The Standard Model}}

When great minds of science such as Galileo, Newton and Kepler turned their attention to understanding the world and the operations of nature, a whole new beauty unfolded -- a beauty of mathematical precision and simplicity. In a single concise statement Kepler accurately described not just the motion of a single planet but that of all planets. Newton with his Second Law now more tersely written as ${\bf{F}}=m{\bf{a}}$, captured the essential motion of all macroscopic bodies. Mankind is naturally attracted to simplicity, and concise statements explaining a large body of seemingly diverse phenomena are considered beautiful. Perhaps the more concise or fundamental a given statement is and the more diverse the phenomena that are encapsulated in that statement, the more beauty it holds and value it is given in the scientific community. 

A whole branch of physics developed in the 20th century called particle physics focused on understanding the fundamental particles which constitute the universe. Progress in this field advanced along with new particle discoveries, some predicted and some a surprise. After Planck had beautifully ``fixed'' the ultraviolet divergences of statistical mechanics analysis of blackbody radiation, Einstein took the concept a step further and proposed a fully quantized electromagnetic field, explaining the observed photo-electric in terms of a quantum particle called the ``photon''. Rutherford's work with alpha rays had proven that all elemental nuclei contained the hydrogen nucleus and by 1920 this particle was being termed the ``proton''. A decade later under the direction of Rutherford, James Chadwick discovered evidence of a second nucleon which had almost the same mass as the proton but which was electrically neutral. Around the same time, Anderson discovered evidence of a positive twin of the electron now called the positron, whose existence had been predicted by Dirac in 1928 \cite{Dirac}. 

By the mid-1930's scientists were having trouble explaining how energy was not conserved in beta-decay processes. Wolfgang Pauli proposed a new neutral invisible particle to carry away the missing energy and although this idea was initially met with skepticism, evidence began to accumulate for its existence. Firm evidence was not provided for another two decades when in 1956 Clyde Cowan and Frederick Reines published experimental confirmation for the existence of what is now called the neutrino. By this time it was becoming clear that older models of particle physics were insufficient. Unsatisfactory descriptions with seemingly arbitrary conservation laws were being used and modified to meet the latest data. An array of new particles being called ``baryons'' and ``mesons'' had been discovered and it was not altogether clear how to classify them. What were the best quantum numbers to describe these particles? The search for such a unifying principle was a priority. 

Such a model was on its way, and so successful was its description and predictions, that its formulation has remained largely unchanged since its development in the 1960's and 1970's. This model, now called the ``Standard Model of Particle Physics'' (SM) due to its broad acceptance in the scientific community, is a gauge quantum field theory and is based on a Lagrangian formulation that is invariant under continuous local transformations. The SM is said to be  an $SU(3)\times SU(2)_L\times U(1)_Y$ theory where the language of group theory is used to describe the internal symmetries of the SM Lagrangian.  $SU(3)$ describes the color symmetry of the strong force and $SU(2)_L\times U(L)_Y$ describes the symmetries of the electroweak interaction. Thus the SM accounts for the strong, weak and electromagnetic forces.

\begin{figure}[ht]
\centering
\framebox{\includegraphics[width=5.2in]{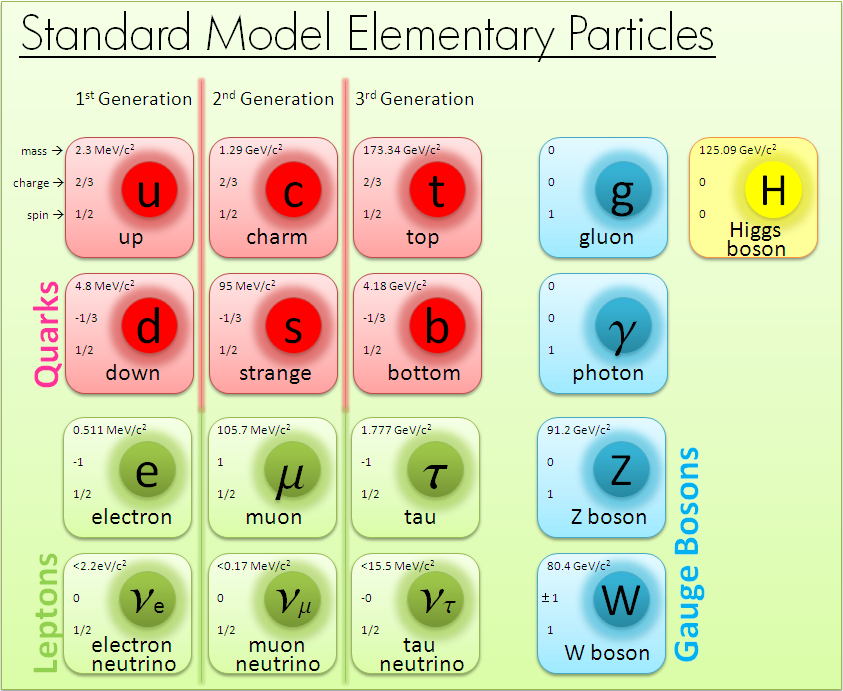}}
\caption {Chart of fundamental particles in the Standard Model of Particle Physics. There are three generations of leptons and three generations of quarks. Interactions are mediated by gauge bosons. The scalar Higgs field with its accompanying interaction carrier, the Higgs boson, gives mass to the leptons via spontaneous symmetry breaking in the Standard Model formulation.}
\label{fig:SM_particles}
\end{figure}

There are four known fundamental forces: the electromagnetic, weak, strong and gravitational forces. Although the relative strengths of these forces depends upon the energy scale at which they are evaluated, Table \ref{tab:fundamental_forces} shows the relative strengths at energy scales where the strong coupling constant is unity. In the SM, three of these forces (electric, weak and strong) are mediated by gauge boson force carriers. All fundamental particles are categorized as either fermions, the basic building blocks of the universe which carry half integer spin, or bosons, which have integer spin. The SM includes 12 fundamental fermions divided into three generations and categorized as either quarks or leptons. With the possible exception of neutrinos, the higher the generation, the higher the mass of the quarks and leptons. The SM also includes 5 fundamental bosons\footnote{Technically there are 6 if you count the $W^+$ and $W^-$ separately.}. The chart in figure \ref{fig:SM_particles} summarizes the elementary particles of the SM. These particles interact with each other by exchange of gauge bosons. Therefore, force, which in the context of classical physics is continuous, is quantized in the SM and mediated by gauge bosons. Gauge bosons that carry charge do not conserve particle identity  or ``flavor''. Thus, the $W^{\pm}$ bosons change the charge of a given particle by $\pm1$, whereas the photon or Z boson can only change its spin configuration or its energy state. Gluons, the mediators of the strong force, are the only bosons that carry color charge and only interact with quarks and with themselves. Leptons interact with quarks and other leptons via the electric and weak forces mediated by the photon and the W and Z bosons respectively. Quarks interact via the strong, weak and electric forces. The strong force is also responsible for holding the nucleus together against the repulsion of the Coulomb forces. The Higgs boson is not a gauge boson because it does not arise as a necessity of gauge symmetry as do the others, and thus the ``Higgs'' force is not considered a fifth fundamental force.\footnote{Although the Higgs field interacts with all massive particles and gives them mass via the Higgs mechanism, this is not accomplished by the exchange of Higgs bosons. Rather, massive particles interact with the ubiquitous Higgs field whose presence has been confirmed by the discovery of the Higgs boson, which can be considered a region of high density in the Higgs field. Although Higgs boson exchange creates a force, the Yukawa coupling is very small and important only for very massive particles like the top quark.}

Although the SM has been very successful in its description of fundamental particle interactions, it is not considered satisfactory as a complete or fundamental theory. The SM description is given in terms of many free parameters which must be determined experimentally. These free parameters include all the fermion masses, the mass of the Higgs boson and the strength of the gauge couplings to mention a few. A great mystery of modern physics, the presence of dark matter and energy, is not accounted for in the SM. Perhaps its most glaring deficiency, however, is the total absence of gravity in the model. Because the strength of the gravitational force is many orders of magnitude lower than even the weak force (see Table \ref{tab:fundamental_forces}) this is not a problem for experimentalists; however, the complete absence of a force so fundamental to the existence of the universe as we know it and of life itself is a gaping hole in this otherwise beautiful model. Furthermore, in the SM neutrinos do not acquire mass via the Higgs mechanism and when it was found that they are not massless, their masses had to be inserted {\it ad hoc}.
\begin{table}[h]
\caption{Four fundamental forces of nature. Relative strengths are normalized to the energy regime where the strong force is close to unity. The values in this table were taken from {\it An Introduction to the Physics of Nuclei and Particles} by Richard Dunlop. Other resources quote slightly different relative strengths which can be attributed to the running of the coupling constants and the freedom to choose any energy scale for comparison.}
\begin{center}
\begin{tabular}{l|c|c|c|c}\hline
Force & Mediator & Charge & Strength & Range (fm)\\ \hline
Strong & gluon & color & 1 & 1\\
Electromagnetic & photon & electric & $10^{-2}$ & $\infty~(\frac{1}{r^2})$\\
Weak & $W^{\pm},~Z$ bosons & weak & $10^{-5}$ & $10^{-3}$\\
Gravity & graviton(undiscovered) & mass & $10^{-39}$ & $\infty~(\frac{1}{r^2})$\\
\hline
\end{tabular}
\end{center}
\label{tab:fundamental_forces}
\end{table}

Since the research in this thesis is mainly limited to electroweak physics and constitutes a test of electroweak predictions of the SM, the remainder of any theoretical discussions will be focused on the electroweak sector. 

\section{Electroweak Physics}
In 1961, Sheldon Glashow discovered a way to naturally combine the electric and weak interactions into a unified ``electroweak'' theory using an $SU(2)\times U(1)$ gauge group with four massless gauge bosons. Peter Higgs and others showed how fields can be made to acquire mass terms in a Lagrangian description by introducing a scalar field with a non-zero vacuum expectation value. Steven Weinberg and Abdus Salam independently condensed these ideas into a single model showing that through spontaneous symmetry breaking in non-Abelian gauge theories both massive and massless bosons are produced and that through the Higgs mechanism fermions naturally acquire mass. These seminal publications by Weinberg and Salam laid the foundations of the Standard Model\cite{Weinberg}\cite{Salam}.

One of the most important features of the SM relevant to the \Qs experiment is the phenomenon of parity violation, a phenomenon that was perhaps unsatisfactorily written into the model as opposed to arising naturally from it. Parity violation, the lack of symmetry in a physical process under spatial inversion, was known to be a signature of the charged weak interaction years before Weinberg and Salam published their early formulation of what is now the bedrock of the SM in 1967.  Ten years earlier Wu and collaborators had measured parity violation in weakly-mediated beta decays\cite{Wu} by observing that the measured decay rate of $^{60}$Co atoms was not symmetric with respect to the direction of a magnetic holding field. This was interpreted as a preference in nature towards left-handed neutrinos. Further experiments examining the decay of pions, $\pi^-\longrightarrow \mu^-+\bar{\nu_{mu}}$, were found to be consistent with a maximal violation of parity, that is, with no right-handed neutrino production\cite{Backenstoss}. Therefore, when Weinberg published his $SU(2)_L \times U(1)_Y$  formulation of electroweak theory, parity violation was simply included in the model by writing the left-handed electron and neutrino as a weak isodoublet and the right handed electron as an isosinglet\cite{Weinberg} as follows
\[
\left(\begin{array}{c}e^-\\\nu \end{array}\right)_L,~~~~~\left(e^-\right)_R.
\]
This model, although not widely accepted at first, solved a number of outstanding problems. First, it created mass for the fermions without imposing them {\it ad hoc} on a massless gauge theory, a process which was known to introduce non-renormalizable infinities. Second, by choosing a suitable gauge transformation, the unobserved massless Goldstone bosons were replaced by four bosonic fields: the massive charged boson field ($W^{\pm}$), a massive neutral boson field (Z), a massless neutral field ($\gamma$), and a ubiquitous, massive Higgs field. Two of these were interpreted as mediating known interactions: $\gamma$, the electromagnetic interaction, and $W^{\pm}$, the observed weak charged decays. The bold prediction of an unobserved massive neutral boson made this theory imminently testable. Six years later in 1973, neutral current processes were observed at CERN in the Gargamelle bubble chamber, validating a key claim of the Weinberg-Salam (W-S) electroweak model.  

Parity violation, although maximal for both charged and neutral weak currents in the W-S model, had not been confirmed in weak neutral current interactions, and many symmetric models predicted equal left and right-handed neutral weak currents\cite{Staley}. In 1977, two experiments attempting to measure parity violation in atomic transitions in bismuth published null results creating a dismal outlook for the W-S model\cite{Lewis}\cite{Baird}. Although the Weinberg model was considered the simplest, other ``hybrid'' models were created to account for parity violation including models with right-handed electron doublets paired with left-handed quark isospin doublets and right-handed quark isospin singlets. In 1978, Charles Prescott and collaborators published the results of electron-deuteron inelastic scattering in which they observed parity violation (non-zero at 10$\sigma$!) in the neutral-weak current, consistent with the simple Weinberg-Salam model \cite{Prescott}. The results of this experiment (E122) at SLAC, so established the W-S model that the following year Steven Weinberg, Abdus Salam and Sheldon Glashow shared the Nobel Prize in Physics ``for their contributions to the theory of the unified weak and electromagnetic interaction between elementary particles, including, inter alia, the prediction of the weak neutral current''\footnote{http://www.nobelprize.org/nobel\_prizes/physics/laureates/1979/}.

By the early 1970's non-Abelian gauge theories had been extended to include the strong interaction and the SM was established in its modern form and has remained nearly unchanged over the past four decades. Although known to be incomplete, the SM has stood up to a barrage of experimental tests over this period and, in almost every case, its predictions have proven to be correct. Parity violation was established as a key tool for probing weak physics processes and more recently as a precision test of key SM predictions.

One such experiment which utilized parity violation as a precision probe of electroweak physics is the recently completed \Qs experiment which ran in Hall C at Jefferson Lab in Newport News, VA, USA, from the Fall of 2010 to the Spring of 2012. This experiment ran with the intention of providing a first determination of the weak charge of the proton. The precision of this experiment also allows it to solidly test the ``running'' of the weak mixing angle, $\theta_W$, which is accurately predicted by the SM. Deviation from the predicted value can be interpreted as physics beyond the SM, whereas agreement with the predicted value would place constraints on some models of physics beyond the SM. 

The author was deeply involved in the setup, commissioning and running of the \Qs experiment and has been participating in data analysis over the 3+ years since the completion of the experimental data-taking phase in May 2012. The author's greatest contributions to the experiment have been in two main areas. First, building, commissioning and analyzing data for the laser system of the Compton polarimeter built specifically for \Qs and key to its small systematic error. Second, removing false asymmetries arising from helicity-correlated electron beam properties from the main data set. These topics will constitute the majority of this thesis.

The format of the thesis is as follows: 1). an introduction to the theory underlying the \Qs experiment and the motivation for the experiment. 2). an overview of the design of the experiment both from hardware and operational perspectives.
3). a detailed analysis of the authors contributions to Compton polarimetry and removal of false asymmetries and 4). a summary of the analysis steps to get from the raw measurement of a parity-violating asymmetry to the final result of the proton's weak charge. 


\chapter{Theoretical Basis for the \Qs Experiment} 
\captionsetup{justification=justified,singlelinecheck=false}

\label{Chapter2} 

\lhead{Chapter 2. \emph{Theoretical Basis}} 


The \Qs experiment which ran in Hall C at Jefferson Lab in Newport News, VA, began taking data in the Fall of 2010 and completed production running in May 2012. The experiment measured the parity-violating asymmetry of elastic electron-proton (ep) scattering of longitudinally polarized electrons from unpolarized protons. In the SM, parity is maximally violated  in weak interactions and it is precisely this broken symmetry that provides unique access to weak sector physics such as that being measured by \Q. It would not be fitting to proceed with discussion of the experimental analysis and results without first introducing the basic underlying theory. This chapter gives a limited overview of electroweak theory necessary for the interpretation of the \Qs experimental results. It further includes the basic theory governing the use of electron scattering as a tool for probing nuclear and nucleon structure and concludes with a section on the use of parity violation as a gateway to weak sector physics and as a precision tool for testing the limits of the SM.

\section{A Peek at Electroweak Theory}
In the SM formulation of electroweak unification, the electric and weak fields arise from a set of four massless gauge fields: $W^-, W^0, W^+$ form an isotriplet and $B^0$ an isosinglet. These acquire mass through the Higgs mechanism which spontaneously breaks the apparent gauge symmetry of the SM Lagrangian.  The $W^{\pm}$ bosons are mass eigenstates of the charged weak interactions. The remaining two neutral gauge fields, $B^0$ and $W^0$, are not observable fields with associated definite mass eigenstates. However, in this beautiful formulation, the mass matrix that arises from neutral field interaction terms in the SM Lagrangian has two eigenstates of definite mass -- one massless and one massive.  These mass eigenstates are formed by rotating the neutral gauge fields through an angle called ``Weinberg angle'' or ``weak mixing angle'' as follows: 
\begin{equation}
\left(\begin{array}{c}\gamma\\Z^0\end{array}\right)=\left(\begin{array}{cc}\cos\theta_W&\sin\theta_W\\-\sin\theta_W&\cos\theta_W\end{array}\right)\left(\begin{array}{c}B^0\\W^0\end{array}\right),
\label{eq:weaking_mixing}
\end{equation}
where is $\gamma$ is the massless photon of electromagnetic interactions and $Z^0$ is the massive neutral weak boson. 

The SM interaction of fermions with $W^{\pm}$ bosons is given by
\begin{equation}
  \frac{ig_w}{2}\bar{\psi}^f\gamma^{\mu}(1-\gamma^5)\psi^f,
\label{eq:charged_weak_vertex}
\end{equation} 
where $\psi^f$ is the fermion wave-function, $g_w$ is the weak coupling constant and the $\gamma^i$'s are the Dirac matrices for spin-$\frac{1}{2}$ particles. 

\begin{figure}[t]
\centering
\begin{fmffile}{feynfile}

\begin{fmfgraph*}(35,30)
\fmfleft{i1,i2}\fmfright{Z}
\fmflabel{$f$}{i1}
\fmflabel{$f$}{i2}
\fmf{boson,label=$\gamma,,Z$}{v,Z}
\fmf{fermion}{i1,v,i2}
\fmfdot{v}
\end{fmfgraph*}
\end{fmffile}
\caption{Feynman diagram of neutral current interaction (Equations \ref{eq:neutral_weak_vertex} and \ref{eq:neutral_em_vertex}) which describes electroweak elastic scattering of a fermion.}
\label{fig:feynman_neutral_current}
\end{figure}
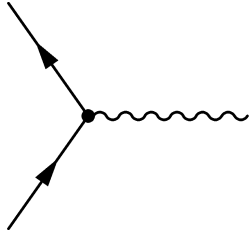

 The component of the interaction involving $\bar{\psi}\gamma^{\mu}\psi$ is familiar from electromagnetism and is called a ``vector'' interaction since it transforms like a vector in Euclidean space. In particular, this term changes sign under a parity transformation. The component involving $\bar{\psi}\gamma^{\mu}\gamma^5\psi$ is called the pseudovector or ``axial vector'' interaction since it transforms like a pseudovector and does not change sign under a parity transformation. Thus, charged weak interactions combine terms that transform with opposite signs under parity. These interactions are said to violate parity and because the terms come with equal strength, the parity violation is said to be ``maximal''.  The interaction of fermions with the $Z^0$ is given by
\begin{equation}
  \frac{ig_wM_Z}{4M_W}\bar{\psi}^f\gamma^{\mu}(g_V^f-g_A^f\gamma^5)\psi^f,
\label{eq:neutral_weak_vertex}
\end{equation} 
where $M_W$ and $M_Z$ are the masses of the W and Z bosons respectively\footnote{Here we make use of the fact that $g_e=g_w\sin\theta_W$ and $\cos_{\theta_W}=\frac{M_W}{M_Z}$ to express electromagnetic and hypercharge currents in terms of $g_w$ the weak coupling constant.}, and where $g_v^f$  and $g_A^f$ are the so-called fermion weak vector and axial-vector charges and are defined as 
\begin{equation}
g_V^f = 2T_3^f-4q^f\sin^2\theta_W, ~~~~~ g_A^f=-2T_3^f,
\label{eq:fermion_Qw}
\end{equation}
where $T_3^f$ is the third component of the fermion weak isospin and $q^f$ is the fermion electric charge\cite{Musolf1994}.  The $T_3^f$ operator acts on the weak isodoublets of the SM\footnote{Lepton isodoublets of the SM are given by $\left(\begin{array}{c}\nu_e\\e^-\end{array}\right)$, $\left(\begin{array}{c}\nu_{\mu}\\\mu^-\end{array}\right)$, $\left(\begin{array}{c}\nu_{\tau}\\\tau^-\end{array}\right)$ and quark isodoublets as $\left(\begin{array}{c}u\\d\end{array}\right)$, $\left(\begin{array}{c}c\\s\end{array}\right)$ and  $\left(\begin{array}{c}t\\b\end{array}\right)$. $T_3^f$ acting on these gives $+\frac{1}{2}$ for the upper components and $-\frac{1}{2}$ for the lower.}. In the form of Equation \ref{eq:neutral_weak_vertex} it is easy to compare with the well established neutral interaction vertex from quantum electrodynamics (QED) given by 
\begin{equation}
 \frac{ig_eq^f}{2}\bar{\psi}^f\gamma^{\mu}\psi^f,
\label{eq:neutral_em_vertex}
\end{equation}
where $g_e$ is the electric coupling constant, $q^f$ is the fermion charge. The basic neutral electroweak fermion vertices given by equations \ref{eq:neutral_weak_vertex} and \ref{eq:neutral_em_vertex} are shown in Figure \ref{fig:feynman_neutral_current}. Comparison of equations \ref{eq:neutral_weak_vertex} and \ref{eq:neutral_em_vertex} shows that the isovector term (involving $g_V^f$) in the neutral weak interaction plays a similar role to the photon in neutral electromagnetic interactions. The \Qs experiment measured the weak charge of the proton at tree level which is precisely this weak vector charge summed over the constituent quarks and is given by 
\begin{equation}
Q_W^p=g_V^f=1-4\sin^2\theta_W.
\label{eq:Qwp}
\end{equation}
Table \ref{tab:fermion_charges} gives the fermion charges at tree level calculated from Equation \ref{eq:fermion_Qw}.
\begin{table}[h]
\caption{Tree level charges of fundamental particles in the Standard Model. This convention is the one used in the review article by Musolf {\it et al.}\cite{Musolf1994} and is used in the \Qs analysis but differs by a factor of 2 from the convention used, for example, in {\it Quarks and Leptons} by Halzen and Martin \cite{Halzen}.}
\begin{center}
\begin{tabular}{l|c|c|c|c}
Fermion&$T_3^f$&$q^f$&$g_v^f$&$g_A^f$\\\hline
~&~&~&~&~\\
$\nu_e,\nu_{\mu},\nu_{\tau}$&$+\frac{1}{2}$&0&1&-1\\
~&~&~&~&~\\
$e^-,\mu^-,\tau^-$&$-\frac{1}{2}$&-1&$-1+4\sin^2\theta_W$&+1\\
~&~&~&~&~\\
$u, c, t$&$+\frac{1}{2}$&$\frac{2}{3}$&$1-\frac{8}{3}\sin^2\theta_W$&-1\\
~&~&~&~&~\\
$d, s, b$&$-\frac{1}{2}$&$-\frac{1}{3}$&$-1+\frac{4}{3}\sin^2\theta_W$&+1\\
~&~&~&~&~\\
\hline
\end{tabular}
\end{center}
\label{tab:fermion_charges}
\end{table}
\\

\section{Fundamentals of Electron Scattering}
In 1924, a French physics student, Louis de Broglie, made a rather stunning proposal during his PhD defense, suggesting that all particles had both wave-like and particle-like properties. He went on to predict that the wavelength $\lambda$ of a particle was related to its momentum $p$ by the following relationship: $p=h/\lambda$, where $h$ is Planck's constant. This idea opened up a whole new world of physics. Visible light waves cease to be useful for viewing objects whose size is of order their own wavelength due to diffraction effects. The energy of a massive particle like an electron is inversely proportional to its de Broglie wavelength, which means that the resolution of a beam of electrons increases with the electron energy. 

Jefferson Lab (JLab) in Newport News, VA, is an electron accelerator facility whose beam energy ranges from $<$1~GeV up to 11~GeV. This range of energies was chosen for examining the structure of neutrons and protons, and from this perspective the JLab facility is sometimes referred to as a giant microscope. The pictures of nucleonic structure typically come in the form of differential scattering cross sections and asymmetries which are then interpreted in the light of the models that best fit the data.

The elastic scattering cross section of a spin-$\frac{1}{2}$ electron from a point-like spin-$\frac{1}{2}$ proton is called the Mott scattering cross section and is given in the lab frame as
\begin{equation}
\left(\frac{d\sigma}{d\Omega}\right)_{Mott}=\frac{\alpha^2}{4E^2\sin^4\frac{\theta}{2}}\left(\frac{E^{\prime}}{E}\right)\left(\cos^2\frac{\theta}{2}+\frac{Q^2}{2m_p^2}\sin^2\frac{\theta}{2}\right),
\label{eq:mott}
\end{equation} 
where $\alpha$ is the fine structure constant, $E(E^{\prime})$ is the energy of the electron before(after) scattering, $Q^2$ is the four-momentum transfer squared\footnote{The four-momentum transfer squared $Q^2$, the energy transferred from the electron to the proton, is sometimes written equivalently as $-q^2$, where $q^2$ is negative and can be thought of as the negative energy of the virtual photon mediating the scattering.}, $m_p$ is the proton mass and $\theta$ is the scattering angle of the electron in the lab frame\cite{Halzen}\cite{Bernauer}. The factor of $(E^{\prime}/E)$ comes from target recoil. $E^{\prime}$ and $q^2$ are explicitly defined as 
\[
E^{\prime}=\frac{E}{1+\frac{2E}{m_p}\sin^2\frac{\theta}{2}},
\]
\[
q^2=-4EE^{\prime}\sin^2\frac{\theta}{2}.
\]

To account for a proton with structure, the simple Mott cross section must be modified to include electric and magnetic form factors. The elastic scattering cross section for electrons on protons with structure is 
\begin{equation}
\left(\frac{d\sigma}{d\Omega}\right)_{lab}=\frac{\alpha^2}{4E^2\sin^4\frac{\theta}{2}}\left(\frac{E^{\prime}}{E}\right)\left(\frac{G_E^2+\tau G_M^2}{1+\tau}\cos^2\frac{\theta}{2}+2\tau G_M^2\sin^2\frac{\theta}{2}\right),
\label{eq:proton_cx}
\end{equation} 
where $\tau=\frac{Q^2}{4m_p^2}$. $G_E$ and $G_M$ are the Sachs electric and magnetic proton form factors respectively. Although it is tempting to think of these as the Fourier transforms of the magnetic and charge distributions of the proton, the recoil of the proton renders this interpretation imperfect. However, in the Breit frame in which the norm of the electron's spatial momentum remains constant but the direction is exactly reversed $({\bf p}=-{\bf p}^{\prime})$ or when $|{\bf q}|^2<<m_p^2$, this interpretation is approximately valid\cite{Halzen}. In this limit, the form factors then become
\begin{equation}
G_{(E,M)}(q^2) = \int_V e^{i{\bf q\cdot r}}\rho_{(E,M)}({\bf x})d^3{\bf x}= \int_0^R\int_{-1}^{+1}2\pi e^{i{\bf qr\cos\phi}}\rho_{(E,M)}({\bf x})r^2drd(\cos\phi),
\label{eq:fourier_ff}
\end{equation}
where $\rho({\bf x})$ is the charge or magnetic distribution of the nucleon.  Assuming that $\rho({\bf x})$ is spherically symmetric gives
\[
G_{(E,M)}(q^2) =\frac{4\pi}{q}\int_0^{\infty}\sin(qr)\rho_{(E,M)}(r)rdr.
\]

Expanding in terms of $qr$,  valid for small $|{\bf q}|$, gives the following expansion:
\[
 \begin{array}{ll}G_{(E,M)}(q^2) &=\frac{4\pi}{q}\int_0^{\infty} \left(qr-\frac{1}{6}(qr)^3+\frac{1}{120}(qr)^5-...\right)\rho_{(E,M)}(r)rdr\\~&=4\pi\int_0^{\infty} \left(1-\frac{1}{6}(qr)^2+\frac{1}{120}(qr)^4-...\right)\rho_{(E,M)}(r)r^2dr\\~&=1-\frac{1}{6}q^2\langle r^2\rangle +\frac{1}{120}q^4\langle r^4\rangle-... ,\end{array}
\]
where $\langle r^n \rangle$ is the nth moment of the charge or magnetization distribution $\rho({\bf x})$, explicitly defined as $\langle r^n \rangle=\int r^n\rho({\bf x})d^3{\bf x}=\int 4\pi r^n\rho(r)r^2dr$, and the charge/current normalizations have been set equal to unity: $\int_0^{\infty} \rho_{E,M}(r)d^3{\bf x}=1$. Thus, the mean square charge(magnetic) radius of the nucleon is given by the slope of $G_{E}^p(G_{M}^p)$ evaluated at $Q^2=0$:
\begin{equation}
\langle r^2 \rangle=-6\left.\frac{dG(Q^2)}{dQ^2}\right|_{Q^2=0}.
\label{eq:proton_radius}
\end{equation}
For the proton, the charge and magnetic radius are approximately the same at 0.88~fm\footnote{2010 CODATA recommended value}.
At $Q^2=0$, called the ``static limit'' because there is no target recoil, the electric form factor $G_E$ evaluates to the charge of the nucleon in units of the elementary charge, and the magnetic form factor goes to the magnetic moment of the nucleon in units of the nuclear magneton $\mu_N=\frac{e\hbar}{2m_p}$. Thus for the proton and neutron at $Q^2=0$ these become
\[
\begin{array}{cc}G_E^p(Q^2=0)=1, ~~G_E^n(Q^2=0)=0\\G_M^n(Q^2=0)=\mu_p\approx2.793, ~~G_M^n(Q^2=0)=\mu_n\approx-1.913.\end{array}
\]
At energies high enough to probe nucleonic structure ($>200$~ MeV), target recoil renders the Fourier interpretation of the charge and magnetization distributions invalid for the lab frame. 

\begin{figure}[ht]
\centering
\begin{minipage}{0.5\textwidth}
\centering
\includegraphics[width=0.9\linewidth]{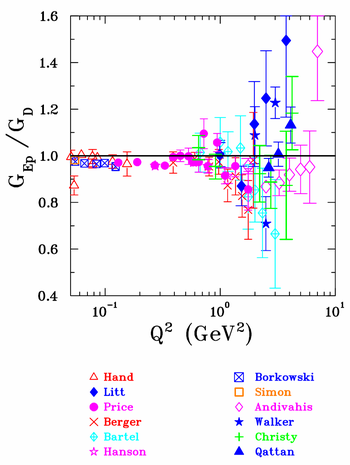}
\caption{Proton electric form factor data from electron scattering. Plot taken from \cite{Perdrisat}.}
\label{fig:Gep}
\end{minipage}%
\begin{minipage}{0.5\textwidth}
\centering
\includegraphics[width=0.9\linewidth]{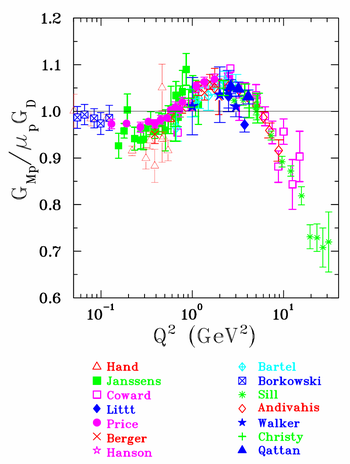}
\caption{Proton magnetic form factor data from electron scattering. Plot taken from \cite{Perdrisat}.}
\label{fig:Gmp}
\end{minipage}

\end{figure}

Many parametrizations and approximations have been developed to characterize the electric and magnetic form factors of the proton and neutron. One particularly useful approximation of the charge and magnetic distributions is the exponential distribution $\rho({\bf x})=e^{-\alpha r}$ first proposed by Hofstadter\cite{Hofstadter}. Using Equation \ref{eq:fourier_ff}, it can be shown that this distribution produces what is called the dipole form factor, $G_D$, where
\[
G_D\propto\frac{1}{1+\frac{Q^2}{\alpha^2}}.
\]
Electron scattering data shows that for the proton $\alpha^2\approx 0.71$. In 1971 Galster {\it et al.} published a parametrization of the proton and neutron form factors using dipole form factors given as \cite{Galster}
\begin{equation}
\begin{array}{c}
G_D=\left(1+\frac{Q^2}{0.711}\right)^{-2}\\
G_E^p(Q^2)=G_D(Q^2)\\
G_M^p(Q^2)=\mu_M^pG_D(Q^2)\\
G_M^n(Q^2)=\mu_M^nG_D(Q^2)\\
G_E^n(Q^2)=-\left(\frac{\mu_M^n\tau}{1+5.6\tau}\right)G_D(Q^2),\\
\end{array}
\label{eq:galster_ff}
\end{equation}
where $\mu_M^{p(n)}$ is the magnetic moment of the proton (neutron) and $\tau=Q^2/4M^2$. It has become common to show form factor measurements normalized to this parametrization for $G_E^p, G_M^p$ and $G_M^n$ as seen in a compilation of electron scattering data showing the proton form factors in figures \ref{fig:Gep} and  \ref{fig:Gmp}.

In the limit of no target recoil, the cross section for electron-proton scattering is proportional to the square of the sum of the electric and weak scattering amplitudes $\sigma^{r,l}\propto\left|\mathcal{M}\right|^2=\left|\mathcal{M}_{\gamma}+\mathcal{M}_{Z}^{(r,l)}\right|^2$, where r(ight) and l(eft) signify that there is a helicity dependence in the weak neutral amplitudes. The elastic scattering amplitude of electrons from an unpolarized proton target is proportional to the electric and weak currents \cite{Perdrisat}\cite{Goldhaber}
\begin{equation}
\mathcal{M_{\gamma}}=\left(-\frac{1}{q^2}\right)(J_{\gamma}^p)_{\mu}(J_{\gamma}^e)^{\mu},
\label{eq:gamma_scattering_amplitude}
\end{equation}
\begin{equation}
\mathcal{M}_{Z^0}=\left(-\frac{1}{q^2-M_{Z^0}^2}\right)(J_{Z^0}^p)_{\mu}(J_{Z^0}^e)^{\mu}\approx\left(\frac{1}{M_{Z^0}^2}\right)(J_{Z^0}^p)_{\mu}(J_{Z^0}^e)^{\mu},
\label{eq:z_scattering_amplitude}
\end{equation}
where $J^{p(e)}$ are the proton (electron) currents and $q^2$ is the square of the four-momentum transferred from the electron to the proton and $M_{Z^0}$ is the mass of the $Z_0$ boson. The approximation $-q^2+M_{Z^0}^2\rightarrow M_{Z^0}^2$ is valid for low energy scattering. For the electron vertex, the electric and weak neutral currents are given by
\begin{equation}
(J_{\gamma}^{e})_{\mu}=-e\bar u_e(p)\gamma_{\mu}u_e(p^{\prime})
\label{eq:electron_gamma_current}
\end{equation}
\begin{equation}
(J_{Z^0}^{e})_{\mu}=g_z\bar u_e^L(p)\gamma_{\mu}u_e^L(p^{\prime})=\frac{g_z}{2}\bar u_e(p)\gamma_{\mu}(g_V-g_A\gamma^5)u_e(p^{\prime}),
\label{eq:electron_Z_current}
\end{equation}
where $u_e$ are electron spinors, $g_z=g_e/(\sin\theta_W\cos\theta_W)$ is the neutral weak coupling constant, $p~(p^{\prime})$ is the incoming (outgoing) electron four-momentum and $g_V$ and $g_A$ are the vector and axial vector weak charges respectively (see Table \ref{tab:fermion_charges} for weak vector and axial charges of fermions).
The electric and weak neutral proton currents expressed using form factors to account for the proton structure are given by
\begin{equation}
J_{\gamma}^{\mu}=e\bar u(p^{\prime})\left(\gamma^{\mu} F_1^{\gamma}(Q^2) + \frac{i\gamma^{\mu}}{2m_p}\sigma^{\mu\nu}q_{\nu} F_2^{\gamma}(Q^2)\right)u(p),
\label{eq:proton_gamma_current}
\end{equation} 
\begin{equation}
J_{Z}^{\mu}=g_z\bar u(p^{\prime})\gamma^{\mu}\left( F_1^Z(Q^2) + \frac{i\kappa}{2m_p}\sigma^{\mu\nu}q_{\nu} F_2^Z(Q^2) + \gamma^{\mu}\gamma^5 G_A^Z + \gamma^5 G_p^Z\right)u(p),
\label{eq:proton_Z_current}
\end{equation} 
where $F_1^{\gamma}$ and $F_2{\gamma}$ are the the electromagnetic Dirac and Pauli form factors, $F_1^{Z}$ and $F_2^{Z}$ are the weak analogs of the electromagnetic form factors, and $\kappa$ is the anomalous magnetic moment of the proton. The Sachs form factors, most often used for their interpretability, are linear combinations of the Dirac and Pauli form factors and are chosen such that all cross terms cancel in the cross section calculation. In particular $G_E=F_1-\tau\kappa F_2$ and $G_M=F_1+\kappa F_2$. Notice that the weak neutral current requires two further form factors to account for axial terms (terms with a $\gamma^5$) in the cross section: $G_A^{Z}$, the axial-vector proton form factor and $G_P^{Z}$, the pseudoscalar form factor both of which are associated with the axial-vector component of the $Z^0$ interaction \cite{Arrington2007}. These terms violate parity and on this basis are excluded from the electromagnetic current. Neither of these contribute greatly to the ep elastic cross section at the kinematics of the \Qs experiment and can be constrained by other electron scattering data. 

\section{Accessing the Weak Sector via Parity Violation}
Theoretically, the cleanest access to weak sector physics is through neutrinos which, for purposes of experimental physics, only interact weakly\footnote{Since neutrinos have mass they also interact with the Higgs field and are affected by gravity, but these effects are many orders of magnitude weaker than even the weak force.};  however, this also makes neutrino experiments notoriously difficult. In the SM, parity violation is unique to the weak force, making it another gateway to the weak sector. As previously mentioned, the parity operator reverses the sign of all spatial components of a system, that is $P|x,y,z\rangle=|-x,-y,-z\rangle$. The parity operation can be decomposed into a mirror inversion plus a 180 degree rotation around the axis of inversion. Although a mirror only inverts one direction (normal to the mirror surface), yet because the laws of physics are invariant under rotations,\footnote{This implies conservation of angular momentum as shown by Emmy Noether in her famous theorem where she proved that any symmetry in physics implies a conserved quantity.} the parity inversion of a physical process can be accomplished by a mirror inversion. This implies that any difference in measured results between an experiment and its exact mirror experiment must be due to the weak interaction.

\begin{figure}[ht]
\centering
\includegraphics[width=2.5in]{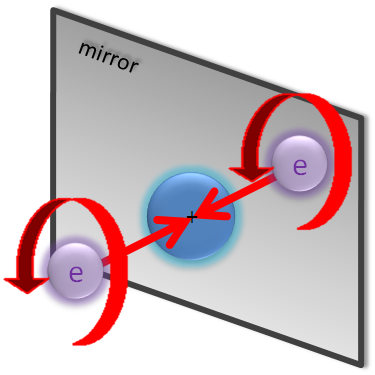}
\caption{Illustration of parity inversion for \Qs experiment which was accomplished by flipping electron helicity.}
\label{fig:mirror_parity_inversion}
\end{figure}

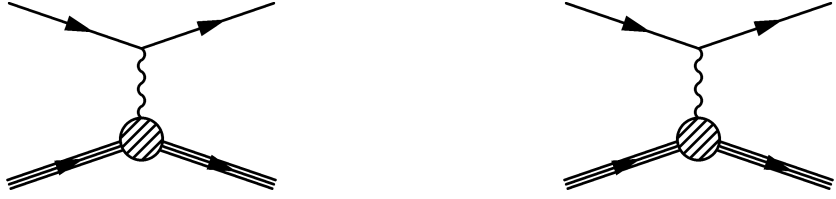
\begin{figure}[t]
\centering

\begin{minipage}{0.5\textwidth}
\centering

\begin{fmffile}{epgammascat}
\begin{fmfgraph*}(35,30)
\fmfbottom{pin,pout}\fmftop{ein,eout}
\fmflabel{$e^-$}{ein}
\fmflabel{$e^-$}{eout}
\fmflabel{$p$}{pin}
\fmflabel{$p$}{pout}
\fmf{boson,label=$\gamma$}{v1,v2}
\fmf{fermion}{pin,v1,pout}
\fmfblob{0.16w}{v1}
\fmf{fermion}{ein,v2,eout}
\fmffreeze
\renewcommand{\P}[3]{\fmfi{plain}{vpath(__#1,__#2) shifted (thick*(#3))}}
\P{pin}{v1}{0.3,-0.8}
\P{pin}{v1}{-0.3,0.8}
\P{v1}{pout}{0.3,0.8}
\P{v1}{pout}{-0.3,-0.8}
\end{fmfgraph*}
\end{fmffile}

\end{minipage}%
\begin{minipage}{0.5\textwidth}
\centering
\begin{fmffile}{epZscat}
\begin{fmfgraph*}(35,30)
\fmfbottom{pin,pout}\fmftop{ein,eout}
\fmflabel{$e^-$}{ein}
\fmflabel{$e^-$}{eout}
\fmflabel{$p$}{pin}
\fmflabel{$p$}{pout}
\fmf{boson,label=$Z$}{v1,v2}
\fmf{fermion}{pin,v1,pout}
\fmfblob{0.16w}{v1}
\fmf{fermion}{ein,v2,eout}
\fmffreeze
\renewcommand{\P}[3]{\fmfi{plain}{vpath(__#1,__#2) shifted (thick*(#3))}}
\P{pin}{v1}{0.3,-0.8}
\P{pin}{v1}{-0.3,0.8}
\P{v1}{pout}{0.3,0.8}
\P{v1}{pout}{-0.3,-0.8}
\end{fmfgraph*}
\end{fmffile}
\end{minipage}
\caption{Tree level Feynman diagrams of neutral electric ($\gamma$-mediated) and neutral weak (Z-mediated) electron-proton scattering.}
\label{fig:feynman_ep}
\end{figure}

The \Qs experiment utilized parity violation to determine the weak charge of the proton. This was accomplished by scattering longitudinally polarized electrons off an unpolarized liquid hydrogen target and measuring the difference in elastic scattering rates between electrons of positive and negative helicity. Figure \ref{fig:mirror_parity_inversion} demonstrates that flipping the helicity of longitudinally polarized electrons is equivalent to creating a mirror image experiment. The two processes that contribute to elastic scattering at tree level are shown in Figure \ref{fig:feynman_ep}.

The Feynman diagram of elastic ep scattering in Figure \ref{fig:feynman_ep} showing a clean leptonic vertex and a blob at the proton vertex is intended to demonstrate the uncertainty of hadronic interactions that come into play when scattering from a composite object like the proton. In the quark model, the proton is made up of three valence quarks, two up quarks and one down quark, bound together by the strong interaction. As shown in the previous section, the clean interaction equations for fundamental particles (equations \ref{eq:neutral_weak_vertex} and \ref{eq:neutral_em_vertex}) must be modified to account for the internal structure of the proton.

\Qs measured the parity-violating elastic e-p scattering asymmetry by rapidly flipping the helicity of the electron beam. The asymmetry in terms of cross sections of positive and negative helicity electrons on an unpolarized proton target is given by
\begin{equation}
A_{PV}=\frac{\sigma_+-\sigma_-}{\sigma_++\sigma_-},
\label{eq:qw_asymmetry_cx}
\end{equation}
where the cross sections $\sigma$ are averaged over the detector acceptance. At tree level this asymmetry takes the form
\begin{equation}
A_{PV}=\frac{-G_FQ^2}{4\pi \alpha\sqrt{2}}\left[\frac{{\epsilon}G^{\gamma,p}_EG^{Z,p}_E+{\tau}G^{\gamma,p}_MG^{Z,p}_M-\left(1-4sin^2{\theta}_W\right){\epsilon}^{\prime}G^{\gamma,p}_MG^{Z,p}_A}{{\epsilon}(G^{\gamma,p}_E)^2+{\tau}(G^{\gamma,p}_M)^2} \right],
\label{eq:qw_asymmetry}
\end{equation} 
where $G_F$ is the Fermi coupling constant, $\alpha$ is the electromagnetic coupling constant, $Q$ is the four-momentum transfer, $G_E^{\gamma,p}$ and $G_M^{\gamma,p}$ are the Sachs electric and magnetic form factors of the proton and $G_E^{Z,p}$ and $G_M^{Z,p}$ and the analogous form factors for the neutral weak current. $G_A^Z$ is the isovector axial form factor. The kinematic factors $\tau$, $\epsilon$ and $\epsilon^{\prime}$ are defined as follows:
\begin{equation}
\tau=\frac{Q^2}{4m_p^2},~~\epsilon=\frac{1}{1+\left(2+2\tau\tan^2\frac{\theta}{2}\right)},~~\epsilon^{\prime}=\sqrt{\tau(1+\tau)(1-\epsilon^2)},
\label{eq:kinematic_factors}
\end{equation} 
where $m_p$ is the proton rest mass and $\theta$ is the electron scattering angle in the lab frame. All kinematic quantities are acceptance-averaged and the form factors are evaluated at the acceptance-averaged $Q^2$ of \Q.

This asymmetry can be recast as a reduced asymmetry where the leading $Q^2$ dependence has been divided out giving an equation with the proton weak charge $Q_W^p$ as a constant plus terms with higher order dependence on $Q^2$\cite{Erler2005}:
\begin{equation}
\begin{array}{c}
\bar A=\frac{A_{PV}}{A_0}=Q_W^p+Q^2B(Q^2,\theta),\\A_0=\frac{-G_FQ^2}{4\pi\alpha\sqrt{2}}.
\end{array}
\label{eq:reduced_asym}
\end{equation}
$B(Q^2,\theta)$ contains terms with the proton electromagnetic, axial vector and strange quark form factors. In this form the proton weak charge is found by evaluating the reduced asymmetry $\bar A$ at $Q^2=0$.

\section{Radiative Corrections for the \Qs Experiment}
Two types of radiative corrections must be applied to the \Qs data. First, electromagnetic corrections must be applied  for the accurate calculation of experimental kinematics. The extraction of the proton weak charge $Q_W^p$ from the measured asymmetry is accomplished using Equation \ref{eq:reduced_asym}. Although $Q_W^p$ is the value of the reduced asymmetry at $Q^2=0$, all the experimental data were taken at $Q^2>0$. Since all the data are in the second term of the RHS of Equation \ref{eq:reduced_asym} and $Q_W^p$ is given by fitting the data and extrapolating to $Q^2=0$, radiative corrections are vital to proper evaluation of $Q^2$ over the detector acceptance. A small correction must also be applied for depolarization of the electron beam in the target. Second, due to the running of the electric and weak coupling constants, electroweak corrections must be calculated to compare the measurements of Qweak with those of experiments at different kinematics or with different interactions. 

\subsection{Electromagnetic Corrections}
The electron as a light fundamental particle provides a clean probe for particle physics; however, electrons tend to radiate in the presence of target nuclei complicating the interpretation of scattering measurements. The process of correcting electron scattering data (in which only the scattered electron is detected) to account for ``radiative effects'' of the electron beam is a well-established procedure documented by Mo and Tsai\cite{MoAndTsai} in a paper written in 1969 and updated by Tsai\cite{Tsai1971} in 1971.\footnote{Appendix A of Collen Ellis's PhD Thesis {it Measurement of the Strange Quark Contribution to Nucleon Structure through Parity-Violating Electron Scattering}\cite{ColleenEllis} outlines the procedure for electromagnetic radiative corrections for low energy parity-violating electron scattering experiments.} The \Qs experiment measured an asymmetry of scattering rates (cross section difference) between left and right longitudinally polarized electrons on unpolarized protons. Although the detector acceptance allowed a range of kinematic variables, the asymmetry is reported at a set scattering angle and $Q^2$. The data must be corrected for radiative processes in order to accurately report the acceptance-averaged $Q^2$ and scattering angle. For the \Qs experiment there are three basic first-order electromagnetic processes whose contributions must be calculated to accurately report the average energy of the electron at the scattering vertex. These corrections are given the names ``vertex'', ``self-energy'', and ``bremsstrahlung''. A fourth next-to-leading order correction is the vacuum polarization correction which gives the running of the electromagnetic coupling constant\footnote{Although the theory of renormalization is beyond the scope of this thesis, loop diagrams such as the vacuum polarization introduce infinities in calculating scattering amplitudes. These are removed by redefining the coupling constants to depend on the energy scale. Thus, the coupling constants are said to ``run'' with $Q^2$.}. Although a detailed look at radiative corrections is beyond the scope of this thesis, a basic explanation for the four processes can be found in Table \ref{tab:em_radiative_corr}.
\begin{table}[ht]
\caption{Four first order electromagnetic radiative processes contributing to correction for electron-proton scattering.}
\begin{center}
\begin{tabular}{c|p{5cm}|c}\hline
\begin{minipage}{2.0cm}\centering Vertex\\correction\end{minipage}&\begin{minipage}{4.7cm}\vspace{0.25in}Photon emitted before the electron scatters and reabsorbed after scattering\vspace{0.25in}\end{minipage}&
\begin{minipage}{0.25\textwidth}
\centering
\begin{fmffile}{vertex_correction}
\setlength{\unitlength}{0.75cm}
\begin{fmfgraph*}(3,3)
\fmfright{b}\fmfleft{ein,eout}
\fmf{fermion}{ein,v2,v3,v4,eout}
\fmf{photon}{v3,b} 
\fmf{photon,left=0.5,tension=0.2}{v2,v4} 
\end{fmfgraph*}
\end{fmffile}
\end{minipage}
\\\hline
\begin{minipage}{2.0cm}\centering Vacuum\\ polarization\end{minipage}&\begin{minipage}{4.7cm}\vspace{0.15in}Virtual photon propagator emits an electron-positron pair which recombine into a virtual photon\end{minipage}\vspace{0.15in} &
\begin{minipage}{0.25\textwidth}
\centering
\begin{fmffile}{vacuum_polarization}
\setlength{\unitlength}{0.75cm}
\begin{fmfgraph*}(3.5,3.2)
\fmfright{b}\fmfleft{ein,eout}
\fmf{phantom,tension=5}{v1,v2}
\fmf{phantom,tension=5}{v3,b}
\fmf{fermion,tension=1.5}{ein,v1,eout}
\fmf{photon}{v1,v2}
\fmf{fermion,left,tension=1.6}{v2,v3,v2}
\fmf{photon}{v3,b}
\end{fmfgraph*}
\end{fmffile}
\end{minipage}
\\\hline

Self energy&\begin{minipage}{4.7cm}\vspace{0.2in}Incoming electron emits and re-absorbs a photon before scattering or scattered electron emits and re-absorbs a photon \end{minipage}\vspace{0.25in}&\begin{minipage}{0.17\textwidth}
\centering
\begin{fmffile}{self_energy1}
\setlength{\unitlength}{0.75cm}
\begin{fmfgraph*}(2.2,3)
\fmfbottom{in,out}\fmftop{t}
\fmf{plain}{in,v1}
\fmf{fermion}{v1,v2}
\fmf{photon,left,tension=0.0}{v1,v2} 
\fmf{plain}{v2,v3}
\fmf{fermion, tension=0.33}{v3,out}
\fmf{photon}{v3,t} 
\end{fmfgraph*}
\end{fmffile}
\end{minipage}
\vspace{-0.1in}
\begin{minipage}{0.17\textwidth}
\centering
\begin{fmffile}{self_energy2}
\setlength{\unitlength}{0.75cm}
\begin{fmfgraph*}(2.2,3)
\fmfbottom{in,out}\fmftop{t}
\fmf{fermion,tension=0.33}{in,v1}
\fmf{plain}{v1,v2}
\fmf{fermion}{v2,v3}
\fmf{photon,left,tension=0.0}{v2,v3} 
\fmf{plain}{v3,out}
\fmf{photon}{v1,t} 
\end{fmfgraph*}
\end{fmffile}
\end{minipage}
\vspace{-0.1in}
\\\hline
Bremsstrahlung& \begin{minipage}{4.7cm}Incoming electron or outgoing electron emits a real photon losing energy\end{minipage}\vspace{0.2in}&\begin{minipage}{0.17\textwidth}
\vspace{0.1in}
\centering
\begin{fmffile}{brem1}
\setlength{\unitlength}{0.75cm}
\begin{fmfgraph*}(2.2,3)

\fmfbottom{in,out}\fmftop{t}
\fmfleftn{l}{5}
\fmf{plain}{in,v1}
\fmf{fermion}{v1,v2}
\fmf{plain}{v2,v3}
\fmf{fermion, tension=0.33}{v3,out}
\fmf{photon}{v3,t} 
\fmf{photon, tension=0}{v2,l4} 
\end{fmfgraph*}
\end{fmffile}
\end{minipage}
\begin{minipage}{0.17\textwidth}
\vspace{0.1in}
\centering
\begin{fmffile}{brem2}
\setlength{\unitlength}{0.75cm}
\begin{fmfgraph*}(2.2,3)

\fmfbottom{in,out}\fmftop{t}
\fmfrightn{r}{5}
\fmf{fermion, tension=0.33}{in,v1}
\fmf{plain}{v1,v2}
\fmf{fermion}{v2,v3}
\fmf{photon}{v1,t} 
\fmf{plain}{v3,out}
\fmf{photon, tension=0}{v2,r4} 
\end{fmfgraph*}
\end{fmffile}
\end{minipage}
\\\hline
\end{tabular} 
\end{center}
\label{tab:em_radiative_corr}
\end{table}

\subsection{\label{Sctn:EWcorr}Electroweak Corrections}

Although physical measurements of cross sections involve processes of all orders, results are typically reported at tree level for ease of interpretability. Tree level reporting requires that higher order processes be subtracted off the final result. Thus before the measured Qweak parity-violating asymmetry in Equation \ref{eq:qw_asymmetry_cx} can be interpreted as a tree level process as in Equation \ref{eq:qw_asymmetry}, contributions from higher orders must be calculated and removed.

The weak charge of the proton including one loop radiative corrections can be written as\cite{Erler2003}\cite{Carlson2013}
\begin{equation}
Q_W^p=(1+\Delta\rho+\Delta_e)(1-4\sin^2\hat{\theta}_W+\Delta_e^{\prime})+\Box_{WW}+\Box_{ZZ}+Re\Box_{\gamma Z},
\label{eq:electroweak_rad_corr}
\end{equation}
where $\Delta\rho$ renormalizes the ratio of neutral to charged current interaction strength from the $Z^0$ mass to low energy, $\Delta_e=-\alpha/2\pi$ is an electron vertex correction correction to the axial vector $Zee$ coupling and $\Delta_e^{\prime}$ is a correction to the $\gamma ee$ coupling corresponding to the anapole moment of the electron. The three box corrections are generated by diagrams shown in Figure \ref{fig:box_diagrams}. The box corrections involving only weak interactions, $\Box_{WW}$ and $\Box_{ZZ}$, can be perturbatively calculated due to the large $W$ and $Z^0$ masses\cite{Erler2003}. These two corrections create a cumulative correction of nearly 30\% to $Q_W^p$ but have small associated errors. However, the $\Box_{\gamma Z}$ contribution has a larger uncertainty due to the low $Q^2$ contributions allowed by the photon propagator. The correction $\mathcal{R}e\Box_{\gamma Z}$ has an axial-electron, vector-proton component $\mathcal{R}e\Box_{\gamma Z}^V$ which vanishes as electron energy goes to 0, and a non-vanishing vector-electron, axial-proton component $\mathcal{R}e\Box_{\gamma Z}^A$. Four groups have published calculations of the $\Box_{\gamma Z}^V$ correction at $Q^2=1.165~GeV$, the momentum transfer of the \Qs experiment and obtained similar results (see Table \ref{tab:gamma_Z_correction}). However, the uncertainty on the corrections varies widely with the smallest adding a systematic uncertainty to $Q_W^p$ of $\pm0.00036$. Two published calculations of $\mathcal{R}e\Box_{\gamma Z}^A$ agree within uncertainty on this correction at $E=1.165~GeV$. The correction for $\Box_{\gamma Z}$ and the total correction $\Box_{\gamma Z}^V+\Box_{\gamma Z}^A$ are plotted versus electron energy in Figure \ref{fig:gamma_Z}. 

Table \ref{tab:ew_corrections} shows the size of each correction term in Equation \ref{eq:electroweak_rad_corr} and the error associated with the $\Box_{\gamma Z}$ which dominates the uncertainty of the electroweak radiative corrections.

\begin{figure}[ht]
\centering
\begin{tabular}{ccc}
\begin{fmffile}{wwbox}
\begin{fmfgraph*}(35,25)
  \fmfbottom{i1,d1,o1}
  \fmftop{i2,d2,o2}
    \fmf{fermion,label=$q$,l.side=right}{i1,v1}
    \fmf{plain,tension=.9}{v1,v2}
    \fmf{fermion,label=$q$,l.side=right}{v2,o1}
    
    \fmf{fermion,label=$e^-$,l.side=left}{i2,v3}
    \fmf{plain,tension=.9}{v3,v4}
    \fmf{fermion,label=$e^-$,l.side=left}{v4,o2}
    \fmf{dashes,label=$W$,tension=0.4,l.side=left}{v1,v3}
    \fmf{dashes,label=$W$,tension=0.4,l.side=right}{v2,v4}
\end{fmfgraph*}

\end{fmffile}&
\begin{fmffile}{zzbox}
\begin{fmfgraph*}(35,25)
  \fmfbottom{i1,d1,o1}
  \fmftop{i2,d2,o2}
    \fmf{fermion,label=$q$,l.side=right}{i1,v1}
    \fmf{plain,tension=.9}{v1,v2}
    \fmf{fermion,label=$q$,l.side=right}{v2,o1}
    
    \fmf{fermion,label=$e^-$,l.side=left}{i2,v3}
    \fmf{plain,tension=.9}{v3,v4}
    \fmf{fermion,label=$e^-$,l.side=left}{v4,o2}
    \fmf{dashes,label=$Z^0$,tension=0.4,l.side=left}{v1,v3}
    \fmf{dashes,label=$Z^0$,tension=0.4,l.side=right}{v2,v4}
\end{fmfgraph*}

\end{fmffile}&
\begin{fmffile}{gammazbox}
\begin{fmfgraph*}(35,25)
  \fmfbottom{i1,d1,o1}
  \fmftop{i2,d2,o2}
    \fmf{fermion,label=$q$,l.side=right}{i1,v1}
    \fmf{plain,tension=.9}{v1,v2}
    \fmf{fermion,label=$q$,l.side=right}{v2,o1}
    
    \fmf{fermion,label=$e^-$,l.side=left}{i2,v3}
    \fmf{plain,tension=.9}{v3,v4}
    \fmf{fermion,label=$e^-$,l.side=left}{v4,o2}
    \fmf{photon,label=$\gamma$,tension=0.4,l.side=left}{v1,v3}
    \fmf{dashes,label=$Z^0$,tension=0.4,l.side=right}{v2,v4}
\end{fmfgraph*}
\end{fmffile}
\\
\end{tabular}
\caption{Three box diagrams corresponding to the last three terms in Equation \ref{eq:electroweak_rad_corr}. Crossing diagrams also contribute\cite{Hall2013}.}
\label{fig:box_diagrams} 
\end{figure}
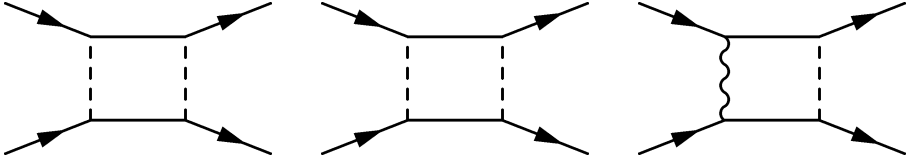
\begin{figure}[ht]
\centering
\includegraphics[width=3in]{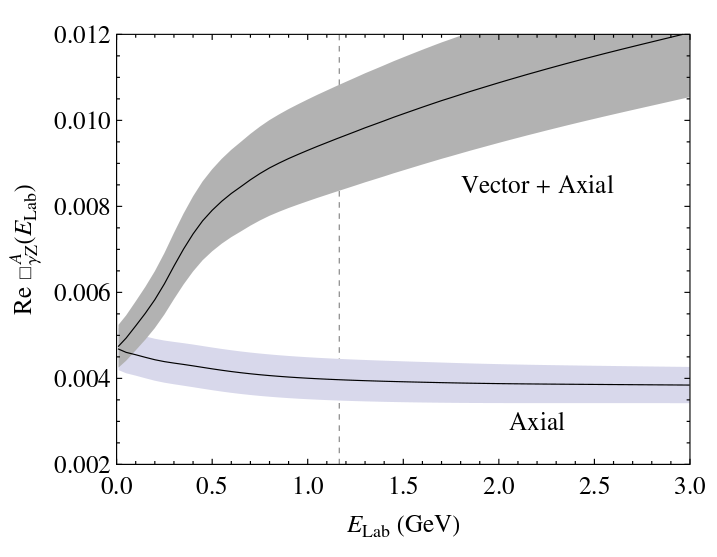}
\caption{Correction to $Q_W^p$ from $\Box_{\gamma Z}$ process shown versus electron energy. The axial proton contribution $\mathcal{R}e\Box_{\gamma Z}^A$ is nearly flat at the \Qs kinematics (shown as vertical dashed line), whereas the vector proton contribution $\mathcal{R}e\Box_{\gamma Z}^V$ is energy dependent. Plot taken from Rislow {\it et al.} (2013) \cite{Carlson2013}.}
\label{fig:gamma_Z}
\end{figure}

\begin{table}[ht]
\begin{center}
\caption{Correction to Standard Model $Q_W^p=0.0713(8)$ from axial proton ($\mathcal{R}e\Box_{\gamma Z}^A$) and vector proton ($\mathcal{R}e\Box_{\gamma Z}^V$) components of $\mathcal{R}e\Box_{\gamma Z}$ evaluated at \Qs kinematics $Q^2=1.165~GeV$.}
\label{tab:gamma_Z_correction}
\begin{tabular}{l|c|c}\hline
$\mathcal{R}e\Box_{\gamma Z}^V$&~&~\\
\hline
~&Sibirtsev {\it et al.} (2010) \cite{Sibirtsev2010}&$0.0047_{-0.0004}^{+0.0011}$\\
~&Carlson and Rislow (2011) \cite{Carlson2011}&$0.0057\pm 0.0009$\\
~&Gorchtein {\it et al.} (2011) \cite{Gorchtein2011}&$0.0054\pm 0.0020$\\
~&Hall {\it et al.} (2013) \cite{Hall2013}&$0.00557\pm0.00036$\\
\hline
$\mathcal{R}e\Box_{\gamma Z}^A$&~&~\\
\hline
~&Blunden {\it et al.} (2011) \cite{Blunden2011} &$0.0037_{-0.0004}^{+0.0011}$\\
~&Carlson and Rislow (2013) \cite{Carlson2013}&$0.0040\pm 0.0005$\\
\hline

\end{tabular}
\end{center}
\end{table}

\begin{table}
\begin{center}
\caption{Values for radiative correction terms for $Q_W^p$ found in Equation \ref{eq:electroweak_rad_corr}. Errors are shown for the $\Box_{\gamma Z}$ correction only since its uncertainty dominates the correction. The 2014 Particle Data Group (PDG) values used in the calculations below are as follows: $\alpha = 7.2973525698(24)\times 10^{-3}$, $\hat{\alpha}=\alpha(M_Z)_{\overline{MS}}\approx\frac{1}{128}$,  $\alpha_s(M_Z)=0.1185(6)$ and $\hat s^2=1-\hat c^2=\sin^2\hat{\theta}_W(M_Z)_{(\overline{MS})}=0.23126(5)$.}
\label{tab:ew_corrections}
\begin{tabular}{c|c|c}\hline
Term & Value & Reference\\\hline
$\Delta\rho$ & $0.00833$ & Chetyrkin {\it et al.} (1995) \cite{Chetyrkin1995}\\~&~&~\\
$\Delta_e=\frac{-\alpha}{2\pi}$ & -0.001161 & Erler {\it et al.} (2003) \cite{Erler2003}\\~&~&~\\
$\Delta_e^{\prime}=\frac{-\alpha}{3\pi}(1-4\hat{s}^2)\left(\ln{\frac{M_Z^2}{m_e^2}}+\frac{1}{6}\right)$ & -0.00142 & Erler {\it et al.} (2003) \cite{Erler2003}\\~&~&~\\
$\Box_{WW}=\frac{\hat{\alpha}}{4\pi\hat{s}^2}\left[2+5\left(1-\frac{\alpha_s(M_W^2)}{\pi}\right)\right]$ & $0.01832$ & Erler {\it et al.} (2003) \cite{Erler2003}\\~&~&~\\
$\Box_{ZZ}=\frac{\hat{\alpha}}{4\pi\hat{s}^2\hat{c}^2}\left(\frac{9}{4}-5\hat s^2\right)(1-4\hat s^2 +8\hat s^4)$ & $0.001923$ & Erler {\it et al.} (2003) \cite{Erler2003}\\~&~&~\\
$\Box_{\gamma Z}=\frac{5\hat{\alpha}}{2\pi}(1-4\hat{s}^2)\left(\ln{\frac{M_Z^2}{\Lambda^2}}+C_{\gamma Z}(\Lambda)\right)$ & $0.0093_{-0.0008}^{+0.0015}$ & Hall {\it et al.} (2013) \cite{Hall2013}\\
~& $0.0097\pm0.0014$ & Carlson {\it et al.} (2013) \cite{Carlson2013}\\
\hline
\end{tabular}
\end{center}
\end{table}

\begin{figure}[ht]
\centering
\includegraphics[width=3.0in]{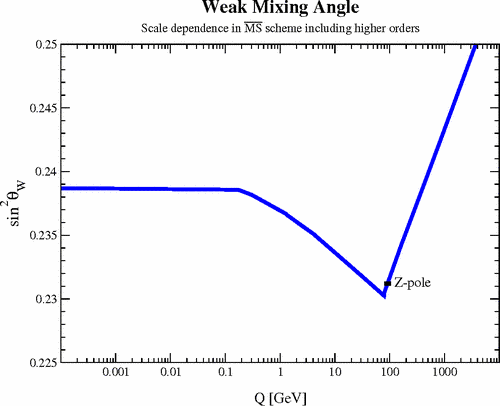}
\caption{Standard Model predicted running of $\sin^2\theta_W$ from the Z-pole to low energies evaluated in the $\overline{MS}$ renormalization scheme. The width of the line gives the error in the prediction. Plot taken from Erler {\it et al.} (2013) \cite{Musolf2005}.}
\label{fig:running_sine_squared_thetaW}
\end{figure}

Equation \ref{eq:electroweak_rad_corr} shows that $\sin^2\theta_W$ is a function of $Q^2$ which is a key prediction of the SM. As already mentioned, this ``running'' comes from the necessity of including higher order diagrams as $Q^2$ increases. Although $\sin^2\theta_W$ is not an observable and its value depends upon the renormalization scheme chosen, in a given scheme the running of $\sin^2\theta_W$ can be precisely calculated and proves to be a convenient tool for comparing experimental results at different energy scales. In particular, one can compare the SM prediction of the running of $\sin^2\theta_W$ from its precisely determined values near the $Z^0$ resonance. Figure \ref{fig:running_sine_squared_thetaW} shows the predicted running of $\sin^2\theta_W$ from the Z-pole to low energies in the  Modified Minimal Subtraction ($\overline{MS}$) renormalization scheme. A $\pm 4\%$ measurement of $Q_W^p$ at the kinematics of the \Qs experiment would translate to a $\pm 0.3\%$ measurement of $\sin^2\theta_W$ giving a $10\sigma$ test of the SM predicted running from the Z-pole. Deviations from the prediction may be interpreted as signatures of physics beyond the SM, whereas agreement puts significant bounds on certain models of new physics. 
 
\chapter{Instrumentation and Operation of the \Q~Experiment} 
\captionsetup{justification=justified,singlelinecheck=false}

\label{Ch:instrumentation}

\lhead{Chapter 3. \emph{Instrumentation and Operation}}

The \Qs experiment measured the weak charge of the proton \qwps via elastic electron-proton scattering. This experiment began commissioning in July 2011 and began production running in February 2011. Data from the first few days of production running, referred to after this as ``Run 0'', comprising $\sim 4\%$ of the full data set, were published in October 2013 providing the world's first determination of the proton's weak charge \cite{wien0}. The production running after Run 0 was divided into two periods termed ``Run 1'' and ``Run 2''. Run 1 extended from February 2011 to May 2011 after which there was a scheduled accelerator down time. During this time many hardware and configuration changes took place to improve the data quality. From November 2011 to May 2012 a period of efficient data taking took place (Run 2) in which over 60\% of the data of the experiment was collected.

Longitudinally polarized electrons were scattered from an unpolarized liquid hydrogen target. The helicity of the electron beam was flipped at approximately 1~kHz between left and right spin states. The \sms predicts a small parity-violating asymmetry of scattering rates between right and left helicity states due to the weak interaction. The \sm, along with information from measured nucleon form-factors, provides a prediction for this asymmetry of approximately $2.2\times 10^{-7}$, making it the smallest ep scattering asymmetry measured at the date of this writing. Furthermore, the proposed error on the asymmetry measurement was 2.5\% (see Table 4 in the 2007 proposal \cite{Jeopardy}), which would be, if achieved, the smallest absolute uncertainty in a parity-violating electron scattering experiment by a factor of a few. 

The raw asymmetry measured by the \Qs experiment can be written as \cite{QweakNIM}
\begin{equation}
\begin{array}{rl}A_{raw}=&\frac{Y^+-Y^-}{Y^++Y^-}\\~&~\\=&P\left[\frac{(1-\sum f_b)A_{PV}}{R}+\sum f_bA_b\right]+A_{beam}+A_{bb}+A_T+A_{\epsilon},\end{array}
\label{eq:raw_asymmetry}
\end{equation} 
where $Y^{+(-)}$ are the integrated detector yields over a window with $+(-)$ electron beam helicity, $A_{PV}$ is the desired parity-violating ep scattering asymmetry, $P$ is the electron beam polarization, $A_b$ are asymmetric polarization-dependent backgrounds and $f_b=\langle Y_b \rangle/\langle Y \rangle$ is the fraction of the total signal coming from that background. $A_{beam}$ is the total asymmetry arising from helicity correlated changes to the beam energy, position and angle. $A_{bb}$ is a helicity-correlated neutral background associated with scattering in the beamline downstream of the target. $A_T$ is the QED asymmetry arising from residual transverse asymmetry on the electron beam and $A_{\epsilon}$ is a false asymmetry arising in the electronics of the data acquisition system (DAQ) from the presence of the helicity signal. Finally, $R$ is a correction factor that accounts for the difference between $\langle A(Q^2) \rangle$ and $A(\langle Q^2 \rangle)$, non-uniformity in the $Q^2$ distribution across the detector bars and electron radiative energy losses in the target. With the information from this equation together with Equation \ref{eq:reduced_asym} it becomes apparent that \Qs requires accurate polarimetry and determination of average $Q^2$, precision measurement of backgrounds, an accurate measurement of helicity correlated beam properties, a means to minimize and remove residual transverse asymmetry on the beam and a clean detector signal chain without contamination from helicity-correlated signals. 

During the design and operation of \Q, particular attention was paid to each term involved in the extraction of $A_{PV}$ from $A_{raw}$ with the intention of minimizing error. Terms such as $P$, $R$ and $f_b$ require accurate measurement, whereas the false asymmetry terms need to be minimized and removed.  This chapter 
provides an overview of the experiment highlighting the focus on error reduction from both operational and instrumentation perspectives.
 
\section{Experimental Overview}
The \Qs experiment was designed specifically to be a high luminosity experiment in order to meet its statistical goals. \Qs took the highest continuous current($180~\mu A$) ever delivered to an experiment at Jefferson Lab. As a result of the high current heat load, a liquid hydrogen (\LH) target was designed that at the time of this writing is the world's most powerful \LHs target. In order to accommodate the required rates (850~MHz per detector at $180~\mu A$), the experiment was designed to be an integrating experiment with only the total flux in a given detector being stored for each window. Typical operating parameters for \Qs are given in Table \ref{tab:kinematics}.

\begin{table}[hbtbp]
\begin{center}
\caption{Typical operating parameters for Run 2 of the \Qs experiment. Table taken from Qweak instrumentation publication. \cite{QweakNIM}}
\label{tab:kinematics}
\begin{tabular}{ll}
Quantity & Value   \\ \hline
Beam energy & 1.16 GeV \\
Beam polarization & 89\% \\
Target length & 34.4 cm \\
Beam current & 180 $\mu$A \\
Luminosity & 1.7x10$^{39}$ cm$^{-2}$ s$^{-1}$ \\
Beam power in target & 2.1 kW \\
$\theta$ acceptance& $5.8^\circ - 11.6^\circ$ \\
$ \phi$ acceptance & $49$\% of $2\pi$  \\
$Q^2$ & 0.025 GeV$^2$ \\
$\Delta \Omega_{\rm elastic}$ & 43 msr  \\
$\int |\vec B| dl$ & 0.9 T $\cdot$ m  \\
Total detector rate  & 7 GHz \\
\end{tabular}
\end{center}
\end{table}

\Qs measured the parity-violating asymmetry of longitudinally polarized electrons on unpolarized protons by rapidly flipping the helicity of the electron beam in a quartet pattern of $+--+$ or $-++-$, with the sign of the first macro-pulse (MPS) in each quartet chosen by a pseudo-random number generator. The MPS repetition rate was $960.15~s^{-1}$ making each quartet approximately $240~\mu s$ long. The asymmetry of scattering rates between states of positive and negative helicity was measured for each quartet as given in Equation \ref{eq:raw_asymmetry} where $Y^{+(-)}$ is the integrated flux over the +(-) states of a pattern. The pattern order was design to remove slow linear drifts in the detector rates. 

\begin{figure}[!t]
\begin{center}
\includegraphics[width=\textwidth]{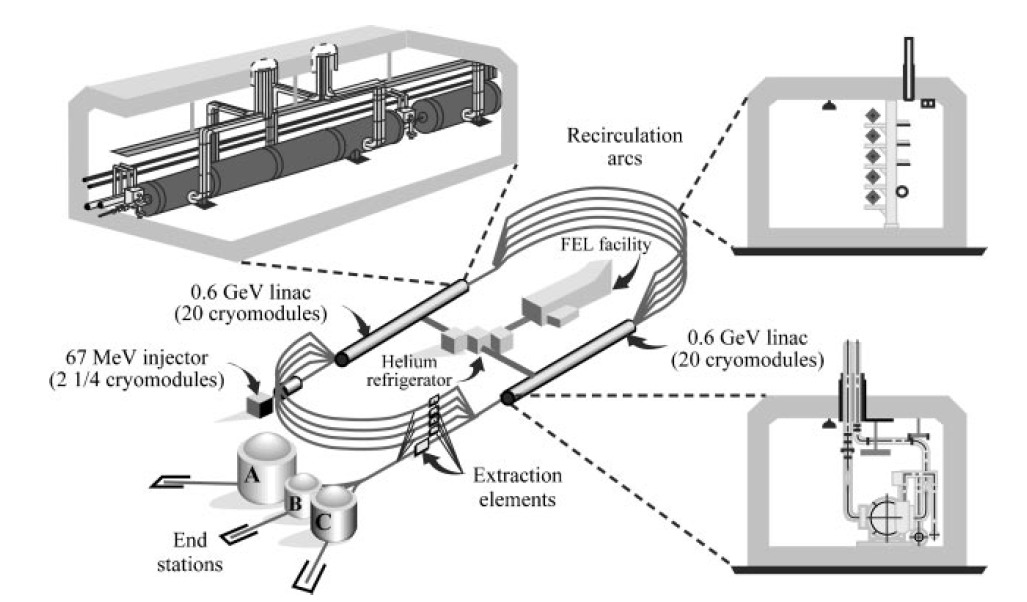}
\end{center}
\caption{\label{fig:CEBAF}
Schematic of the accelerator at Jefferson Lab showing the electron source, linear accelerating regions (linacs) and the recirculating arcs. At the time of the \Qs experiment electrons could pass around the accelerator from 1-5 times depending upon the desired energy. Vertical dipole magnets momentum analyze the electrons distributing them in various recirculation arcs at different heights according to the number of passes through the linacs with higher arcs corresponding to lower energy electrons. The three experimental halls A, B and C are also shown.\cite{Leeman}}
\end{figure}

The electron beam at Jefferson Lab is created by shining circularly polarized laser light on a super-strained GaAs photo-cathode. Superconducting radio frequency (RF) cavities in the injector region accelerate the electrons to highly relativistic energies before entering the accelerator proper\cite{Leeman}. The accelerator consists of two parallel arrays of superconducting RF cavities connected by recirculating arcs such that the shape of the whole accelerator resembles a racetrack. Figure \ref{fig:CEBAF} gives a schematic of the accelerator showing the source, linear accelerators and recirculating arcs. The accelerator was designed to give 1096~MeV per round trip with the ability to tune the energy by adjusting accelerating cavity electric field gradients. For most of \Qs the beam energy of 1160~MeV was obtained by a single pass around the accelerator. However, for the first two months during Run 2 due to an operating accident that took some RF cavities offline, the same energy had to be obtained by two full passes with lower cavity gradients.

Electron beam polarimetry for \Qs was accomplished using an existing M\o ller polarimeter in Hall C as well as a new Compton polarimeter built specifically for \Q. Polarization values from these were also compared to results from a less accurate Mott polarimeter in the injector region that measures the electron beam polarization at low energies where the Mott scattering cross section is large.

\begin{figure}[!hhhbb]
\begin{center}
\includegraphics[width=\textwidth]{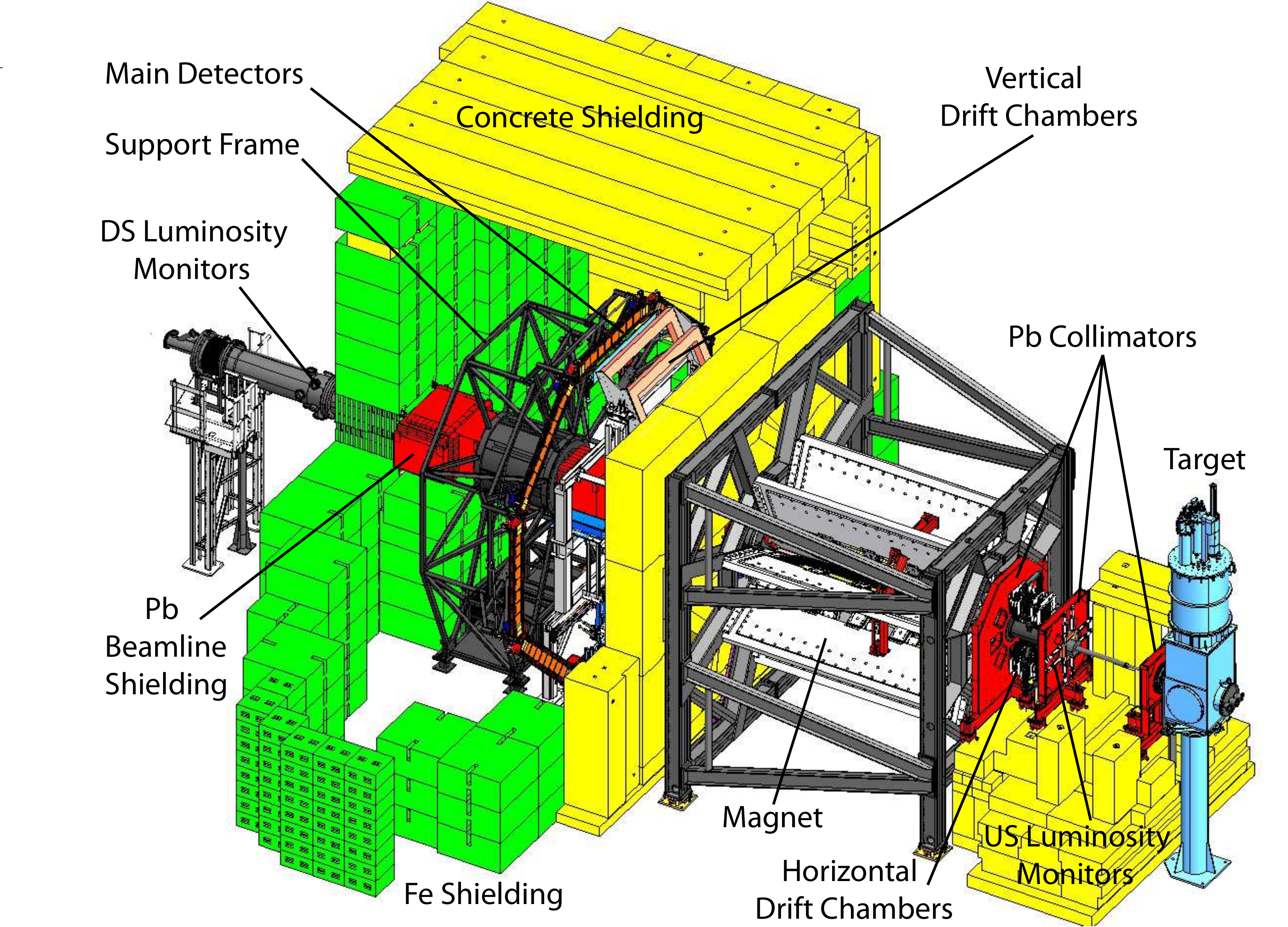}
\end{center}
\caption{\label{fig:QweakApparatus}
  CAD drawing of the main \Qs experimental apparatus. The electron beam is incident from the right. The main high current elements shown include the target, torriodal spectrometer, the octogonal array of quartz \v{C}erenkov detectors and the collimator and shielding systems.  Vertical drift chambers directly in front of the detector bars and horizontal drift chambers between the target and spectrometer were used to reconstruct individual tracks during low current running. 
}
\end{figure}

A CAD drawing of the experimental apparatus specific to \Qs installed in Hall C is shown in Figure \ref{fig:QweakApparatus}. This section of the beamline was completely reconstructed to accommodate the \Qs installation including the target, collimators and detector system. The detector system was housed in a concrete hut to further shield from unwanted backgrounds.

The target was a 34.4~cm liquid hydrogen (\LH) target with thin windows made of aluminum alloy. The \LHs was kept near 20~K with a heat exchanger supplied with liquid helium. A lead collimator selected scattered electrons in a range of angles from 5.8$^{\circ}$ to  11.6$^{\circ}$ in the low $Q^2$ region centered on 0.025~(GeV/c)$^2$ where hadronic effects are relatively insignificant. A series of three lead collimators were used to remove inelastic events upstream of the spectrometer. A torroidal spectrometer (\qtor) was used to focus elastic events onto the detector bars while keeping most of the inelastic events well out of the detector acceptance.

Drift chambers upstream of the spectrometer and directly in front of the main detector bars were used to reconstruct individual event trajectories for the determination of the average experimental $Q^2$. These were operated in extremely low current conditions (50~pA) so that individual events could be resolved. 

Beam position was monitored by a series of beam position monitors (BPM's) located every few meters along the beamline. A fast feedback system (FFB) reading from BPM's in the arc leading to Hall C  was used to stabilize beam energy as well as position and angle at the target.  

Luminosity monitors were placed at two locations downstream from the target near the beam pipe to monitor low angle scattering events such as might be produced by beam halo interacting with small apertures in the beam pipe and M\o ller scattering in the target. Detectors placed in various locations inside the shielded detector region outside the main detector acceptance were used to monitor unwanted backgrounds. 

Greater detail about the subsystems is presented in the sections ahead with emphasis placed upon both the procedures followed and the instrumentation utilized to minimize the uncertainty for \Q.

\section{\label{sctn:electron_source}Polarized Electron Source and Injector}
The program of increasingly sensitive parity-violation experiments at Jefferson Lab over the past decade, including G0, the HAPPEX experiments, PREX, PVDIS and finally Qweak, has created a focus on delivery of the highest quality electron beam with particular attention on the electron source. The electron beam is created by optically pumping a strained superlattice GaAs photo-cathode. The helicity of the laser is transferred to the electrons and the theoretical limit of 50\% polarization attainable with a simple GaAs photo-cathode is exceeded by introducing alternating layers of GaAs with lattice mismatched InGaAs to generate a strained superlattice\cite{Pierce1975}\cite{Maruyama}. Polarizations exceeding 88\% were routinely achieved during \Q.

\begin{figure}[!hbtbhbtb]
\centerline
{\includegraphics[width=\textwidth,angle=0]{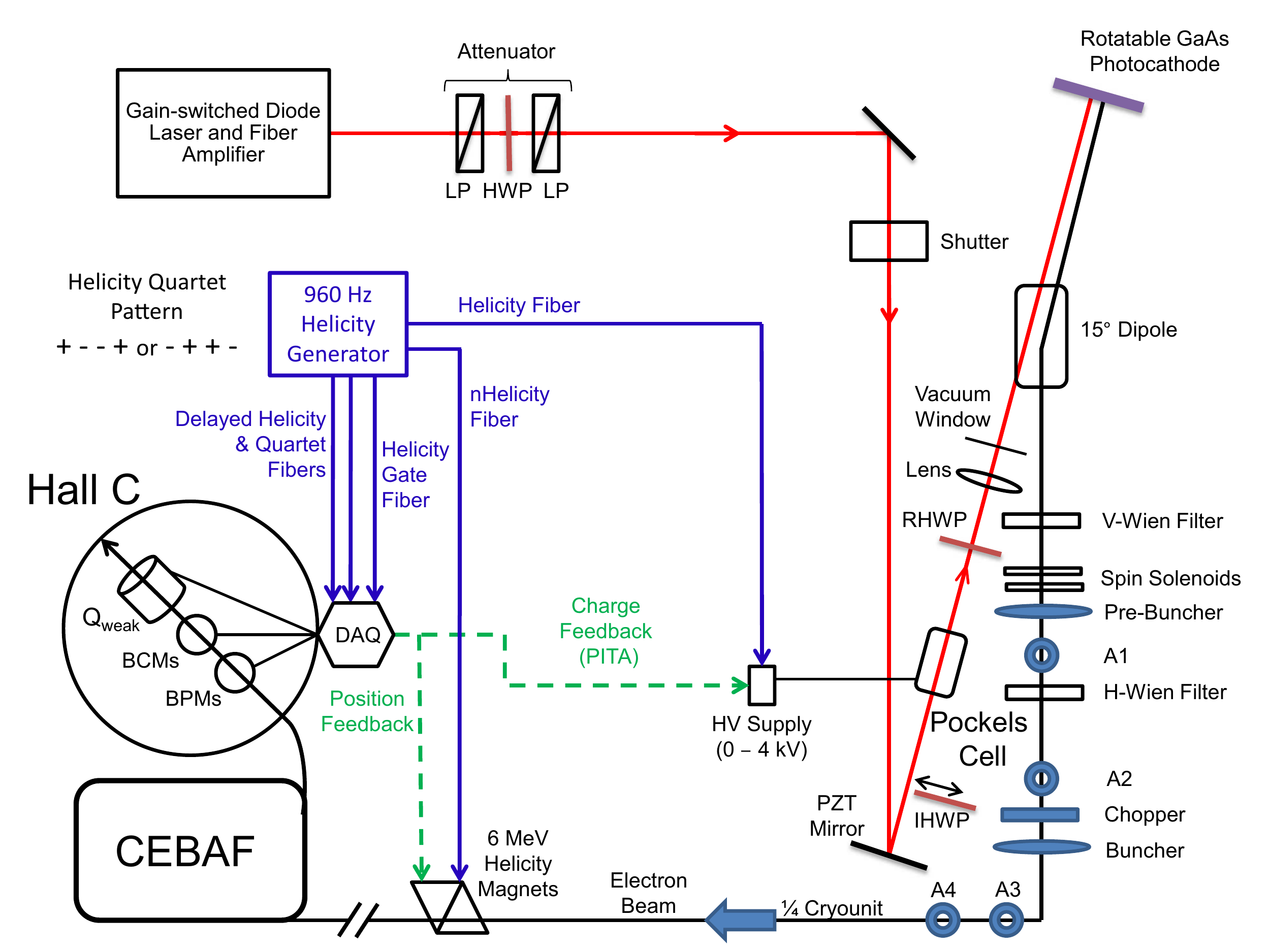}}
\caption{\label{fig:sourcetable} A schematic of the polarized source and injector components used for the \Qs experiment. An IR laser is circularly polarized using a linear polarizer (not shown) followed by a Pockels cell. An optical helicity signal at 960~Hz is used to switch the high voltage on the Pockels cell thus changing the laser helicity. Most of the electrons released from the GaAs photo-cathode by photo-emission carry the same helicity as the laser. A rotatable half-wave plate (RHWP) is used to rotate residual linear polarization to cancel polarization-analyzing gradients on the photo-cathode and polarization gradients across the laser spot.}
\end{figure}

\begin{figure}[ht]
\begin{center} 
\includegraphics[width=0.7\textwidth]{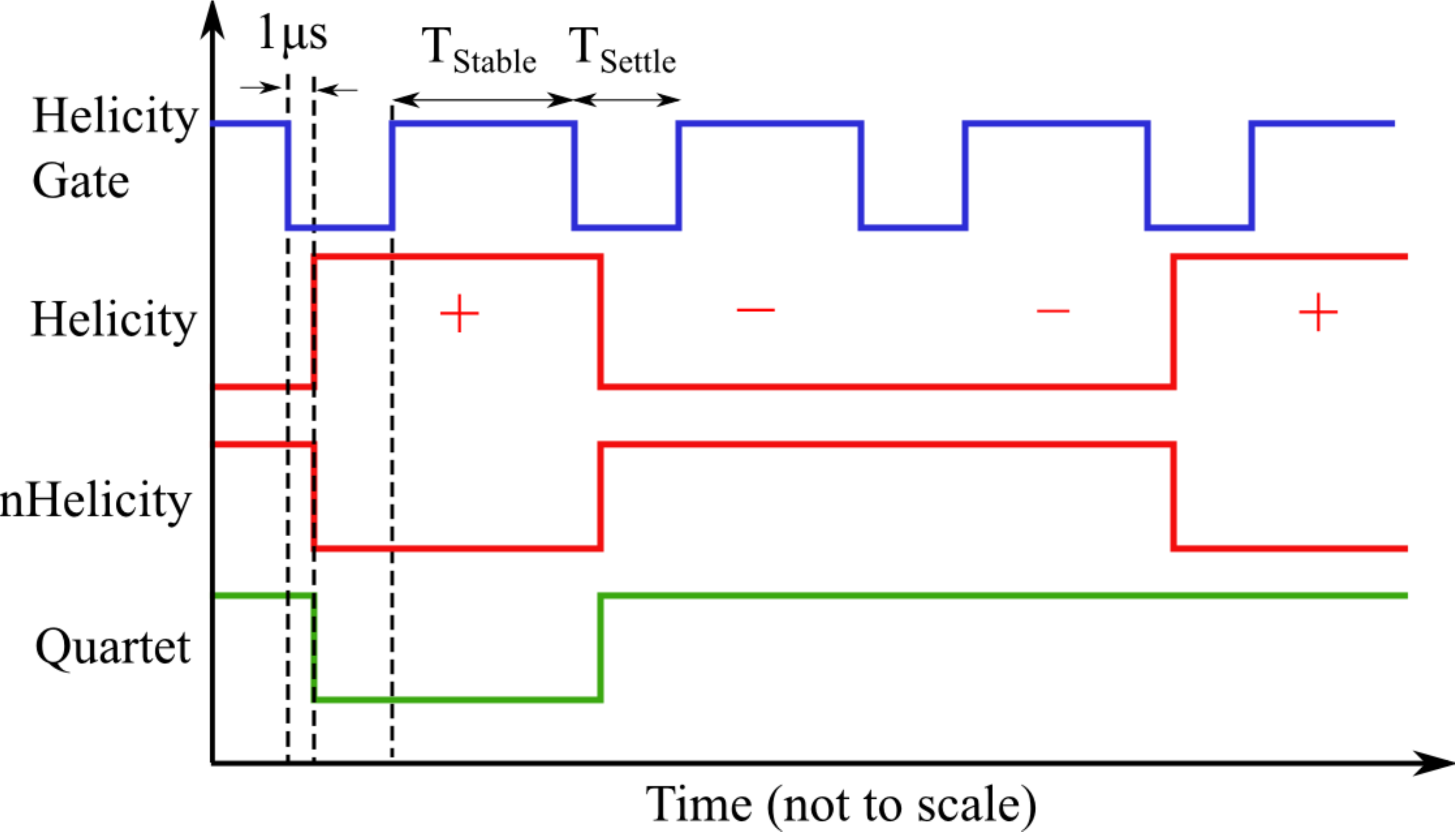}
\caption{ \label{fig:helicity_timing} 
Diagram illustrating timing of the helicity signals for the polarized source. The helicity signal (red) controlled the sign of the high voltage across the Pockels cell and always followed a $+ - - +$ or $- + + -$ pattern with the sign of the first element in each pattern chosen using a pseudo-random number generator. The nHelicity signal, the complement of Helicity, ensured that the timing board always drew the same current no matter which helicity state was being signaled. The Helcity Gate signal (blue) was used for timing data acquisition and was lowered $1~\mu$s before each helicity window and stayed low for time $T_{settle}=70~\mu$s to allow for the finite transition time to a new helicity state. Data acquisition was triggered whenever the Helicity Gate signal was high. Lastly, the Quartet signal (green) was used to mark the beginning of each new quartet pattern. }
\end{center}
\end{figure}

The laser is circularly polarized using a linear polarizer followed by a Pockels cell which utilizes an optically active crystal whose birefringence is proportional to the applied voltage across it. A fast switch is used to reverse the polarity of the high voltage across the Pockels cell and thus to flip the laser helicity. Figure \ref{fig:sourcetable} shows a schematic of the polarized electron source. An insertable half-wave plate before the Pockels cell allows the helicity of the laser to be reversed with respect to the high voltage state across the Pockels cell. This provides a sensitive test of systematic false asymmetries associated with high voltage state so that they can be minimized and/or canceled. 

A specially designed electrical timing board was used to control the helicity flip frequency during \Q. The electronics of this board were carefully isolated from other source and experimental electronics to eliminate contamination of detector signals and to minimize effects such as beam motion induced in the source by electronics cross-talk at the helicity reversal frequency\footnote{This was an issue in the HAPPEX-He experiment\cite{Paschke2007}.}. Four signals were generated by the timing board to control the timing of helicity reversal and to trigger data acquisition (see Figure \ref{fig:helicity_timing} for an explanation of the timing signals). The helicity signal was sent to a high voltage optical switch using LED's attached to fiber-optic cable. Although most previous experiments at Jefferson Lab flipped the helicity at 30~Hz, a much faster flip frequency of 961.015~Hz was chosen for \Qs to reduce the effects of drifts in beam, target and detector properties, requiring a redesigned high voltage switch with a rise/fall time of $\sim60 \mu$s \cite{Adderley2012}. 

Careful attention was given to the alignment and polarization of the source laser to minimize helicity-correlated charge and position differences. Further details about the source of these helicity-correlated beam asymmetries (HCBA's) and methods to reduce their effects can be found in the following reference \cite{Paschke2007} and specifics related to the \Qs experiment in a future thesis \cite{Kargiantoulakis}.

Upon emission from the photo-cathode, the electrons are accelerated by an electrostatic field inside what is called the ``electron gun''. The gun bias voltage was increased for \Qs from 100~keV to 130~keV to create a more compact beam and thus reduce aperture losses in the injector region which are known to create intensity asymmetries \cite{Adderley2012}.

\begin{figure}[ht]
\begin{center} 
\includegraphics[width=0.7\textwidth]{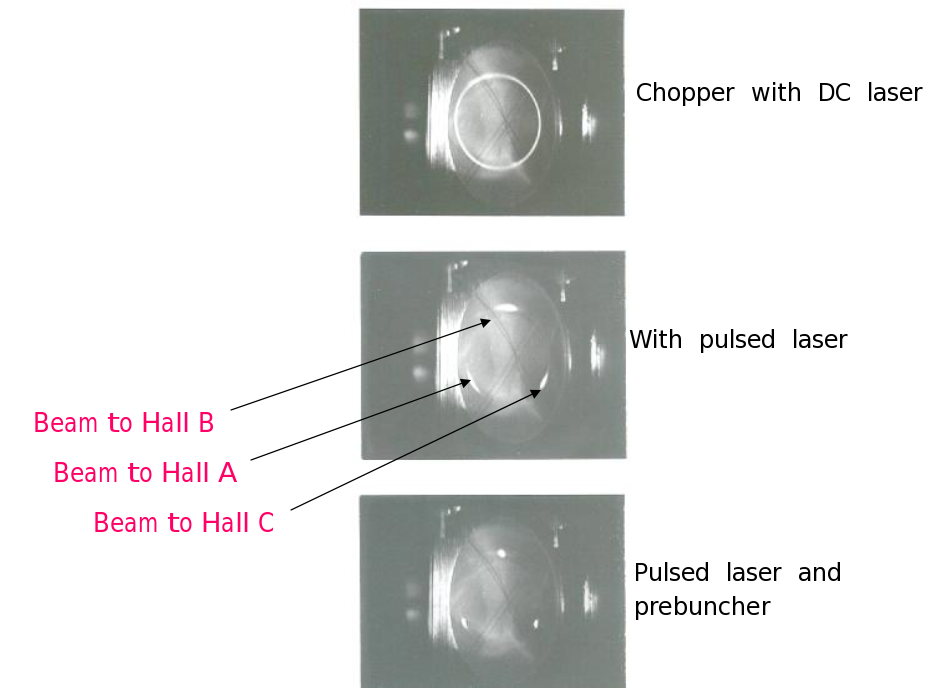}
\caption{ \label{fig:prebuncher} Chopper viewer with beams to three halls shown without pulsed laser, with pulsed laser and with pulsed laser and prebuncher. The small apertures in the chopper disk are not visible in the photograph but are smaller than the spot size in the bottom picture. 
}
\end{center}
\end{figure}

The accelerator RF cavities at Jefferson Lab are tuned to resonate at a frequency of 1497~MHz. Each experimental hall has its own source laser pulsed at 1/3 the accelerator frequency or 499~MHz so that all three halls can receive beam at the same time. An RF separator is frequency tuned to kick each of the bunches off the straight line trajectory through three small holes in a metal plate called the chopper. Figure \ref{fig:sourcetable} shows that the electron beam goes through a pre-buncher before the chopper. The pre-buncher is an accelerating cavity tuned such that the electron bunch spans a zero crossing in the RF standing wave making the two ends of the bunch experience opposite sign fields. This slows the frontmost electrons and accelerates the rearmost electrons creating a tighter bunch. Figure \ref{fig:prebuncher} shows a picture of the chopper with and without the prebuncher and how much cleaner the beam is when the prebuncher is used. The beam passes through another buncher after the chopper to further compact the electron pulses before entering cryogenic accelerating cavities further boosting the electrons to relativistic energies before entering the main accelerator. 
 
Apertures in the source region (labeled A1-A4 in Figure \ref{fig:sourcetable}) are used to clean up the beam. These apertures coupled with helicity-correlated beam position differences can be a significant source of charge asymmetry from differential chopping. 

\begin{figure}[ht]
\centering
\framebox{\includegraphics[width=4in]{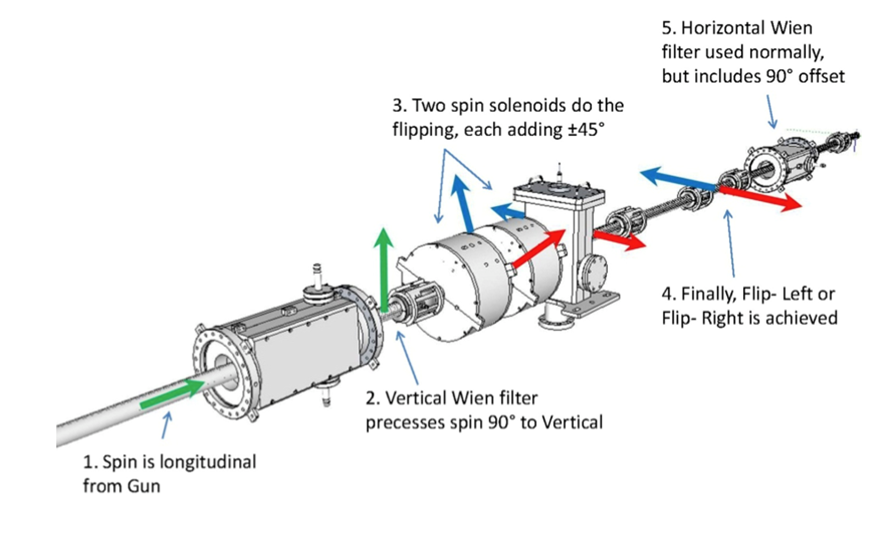}}
\caption{Double Wien filter schematic illustrating the reversal of electron beam helicity.}
\label{fig:double_wien}
\end{figure}

A double Wien spin-flipper with components labeled as ``H-Wien'', ``V-Wien'' and ``Spin Solenoids'' in Figure \ref{fig:sourcetable}, allows the helicity of the electron beam to be reversed relative to the laser helicity providing another level of diagnostics and cancelation for systematic helicity-correlated differences on the electron beam. A diagram of the double Wien spin-flipper is shown in Figure \ref{fig:double_wien}. The vertical Wien flips the spin from longitudinal to transverse pointing vertically upward. The spin solenoids then serve the dual purpose of focusing the beam and precessing the electron spin clockwise or counterclockwise to make them horizontal and transverse to the direction of propagation. Finally, the horizontal Wien filter rotates the spins in the horizontal plane to any desired angle relative to the direction of propagation such that the spins are properly aligned (typically fully longitudinal) when they enter the experimental hall. Due to inefficiencies associated with setting up the double Wien spin-flipper, during the entire \Qs experiment the spin was reversed only 8 times using the double Wien. A more detailed review of the double Wien filter at Jefferson Lab can be found at \cite{Adderley2011}.

\section{Beam Transport and Monitoring}
\label{sctn:btandm}
As mentioned earlier in this chapter, the accelerator at Jefferson Lab consists of two linacs (north and south) connected by recirculating arcs. Each pass gives an energy of approximately 1.1 GeV to the electrons, allowing certain beam energies to be selected for the experimental halls by choosing the number of passes around the accelerator. The collinear beams for the three halls are extracted at the end of the south linac by an RF extractor that selectively ``kicks'' the beam pulses for the three experimental halls into the direction of their respective beam extraction pipes. A septum magnet further separates the beam trajectories. The beam for Hall C then travels through an arc (to the left looking downstream) before entering the experimental hall. 
\begin{figure}[ht]
\centering
\framebox{\includegraphics[width=0.99\textwidth]{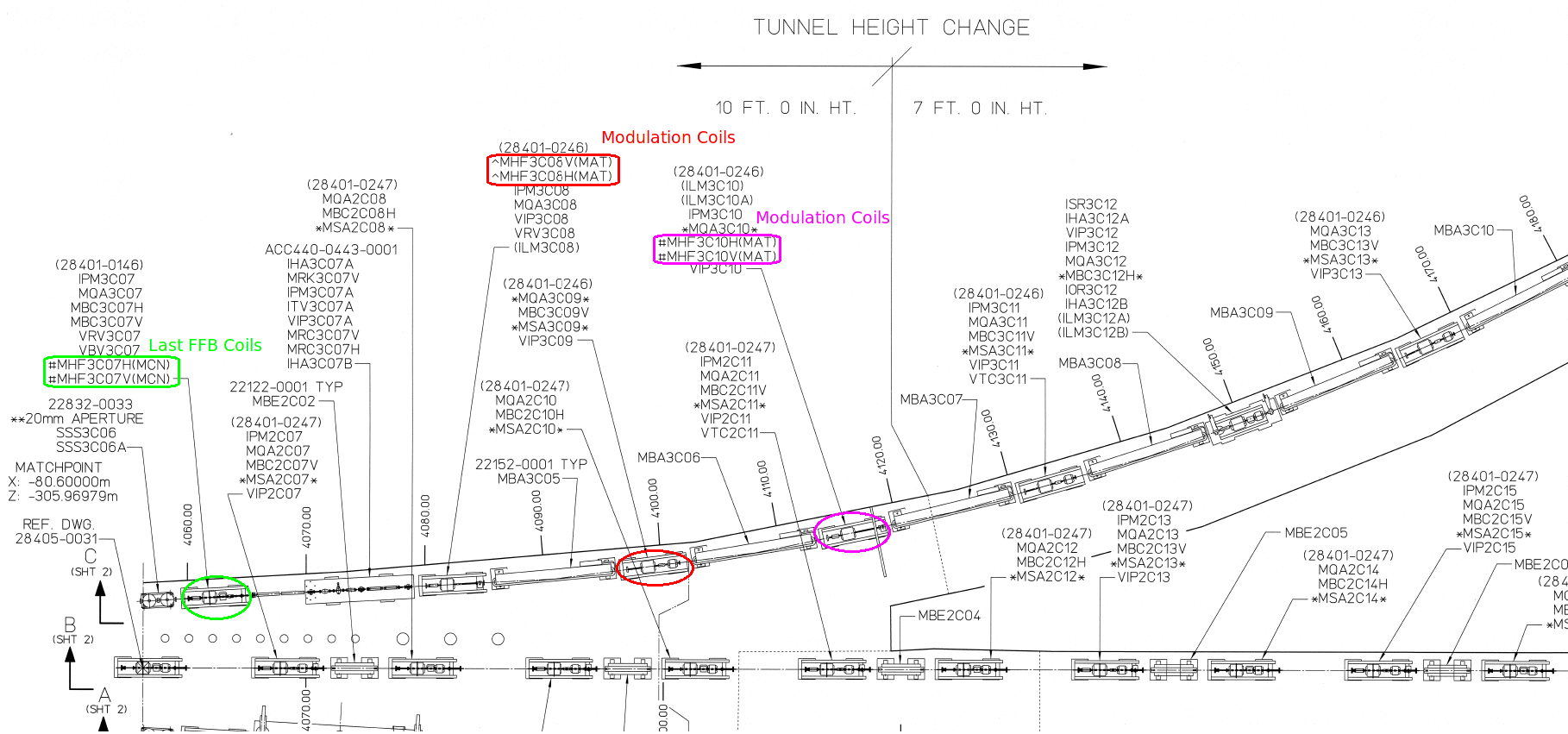}}
\caption{Partial engineering diagram of beamline arc heading into experimental Hall C. The most downstream set of fast feedback air-coil corrector magnets and the modulation coils are specifically labeled.}
\label{fig:HallC_arc}
\end{figure}

\begin{figure}[ht]
\centering
\framebox{\includegraphics[width=0.99\textwidth]{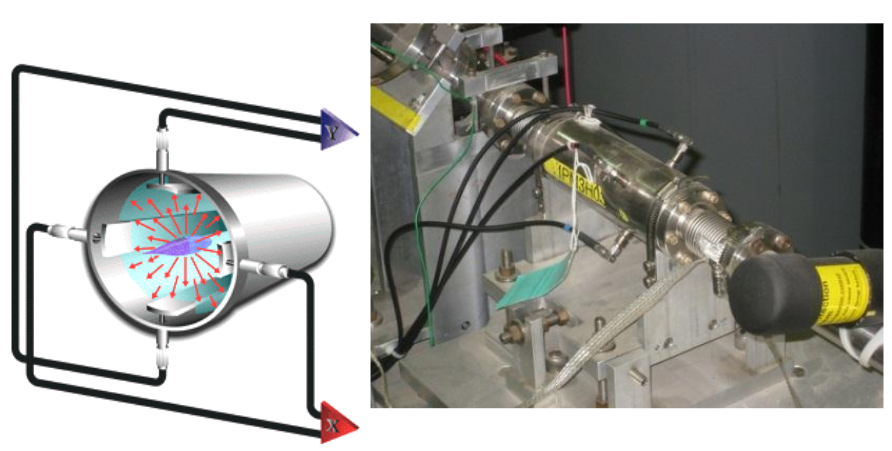}}
\caption{Left graphic illustrates stripline beam position monitor with four antennae. Position is calculated using the difference of the wire signals in a given direction. Picture on the right shows an actual BPM installed on the beamline. The antenna connections can be seen with cables going to readout electronics.}
\label{fig:bpm_attennae}
\end{figure}

The position of the electron beam is measured every few meters by beam position monitors (BPM's), which are metal canisters outfitted with four small RF antennae \cite{Barry1991}\cite{Powers1997}. A passing electron bunch creates signals in each antenna proportional to its distance from the antenna allowing the beam position to be measured relative the center of the canister. The signals are amplified, conditioned and averaged before the positions are calculated. Figure \ref{fig:bpm_attennae} shows an illustration of the BPM canister and antennae along with a picture of an installed BPM. BPM's at Jefferson Lab are typically installed such that the antennae are rotated $45^{\circ}$ relative to the horizontal and vertical planes to keep them out of direct line of synchrotron radiation on horizontal and vertical bends. The signals are then rotated in software to give true X and Y positions. The position in the rotated (primed) coordinate system can be calculated as \cite{Geoffrey1993}
\begin{equation}
X'=k\frac{(X^+-X_{off}^+)-\alpha_X(X^--X_{off}^-)}{(X^+-X_{off}^+)+\alpha_X(X^--X_{off}^-)},
\label{eq:x_prime}
\end{equation}
and
\begin{equation}
Y'=k\frac{(Y^+-Y_{off}^+)-\alpha_Y(Y^--Y_{off}^-)}{(Y^+-Y_{off}^+)+\alpha_Y(Y^--Y_{off}^-)},
\label{eq:y_prime}
\end{equation}
where $X^+$, $X^-$, $Y^+$, $Y^-$ are the signals from 4 antennae, $X_{off}^+$, $X_{off}^-$, $Y_{off}^+$, $Y_{off}^-$ are the antenna signals with no beam and $\alpha_{X(Y)}$ is a factor to account for gain differences between the + and -- antennae in the X' (Y') directions\footnote{The $\alpha$ scale factors are given explicitly as $\alpha_{X(Y)}=\frac{X(Y)_+-X(Y)_{off}^+}{X(Y)_-+X(Y)_{off}^-}$.}. The scale factor k is the sensitivity of the BPM at the accelerator frequency of 1497~MHz and depends upon the size of the canister.

  The readout system uses switched electrode electronics (SEE) so that the same amplifier-detector chain is used to read out the + and -- signals sequentially, greatly reducing sensitivity to differential gain and offset changes between readout channels. The electronics switch between the two wires every 4.2~$\mu s$ and integrate the position signal for a 2.9~$\mu s$ window.

The quoted resolution of the SEE BPM's has been $<100~\mu m$ from $1-1000~\mu A$ \cite{Powers1997}. A study of BPM resolutions in Hall C during the \Qs experiment, independently measured the resolution of the SEE BPM's as well as the dependence of this resolution on current. Under the assumption that all similar BPM's have the same intrinsic resolution, she used pairs of BPM's in a drift region of the Hall C beamline to predict the position of a third BPM and determined that the resolution of a single BPM at $180~\mu A$ is $\sim 0.9~\mu$m for each MPS which is about 1~ms ( see \cite{Waidyawansa} appendix A).

Superharps located at key positions along the beamline are used to measure the beam profile and absolute position. These consist of a motor-actuated triplet of thin wires ($22~\mu m$ diameter) at fixed angles relative to each other (two wires perpendicular to each other and a third wire in the same plane at 45$^{\circ}$ to the other two wires) which are passed through low current beam. The electrical signals generated in the wires as they pass through the beam are used to reconstruct the beam profile \cite{Yan1995}. Superharps are used to verify that the beam size and profile agree to within tolerance with the beam optics model. For example, having a beam spot size much smaller than the laser profile is critical in the region of the Compton polarimeter where the laser interacts with the electron beam. A superharp in the Compton region was used to verify that the beam spot size was close to the specified $1\sigma$ radius of $40~\mu$m. Superharps were also used to determine the shape of the beam and to check for the presence of asymmetric or unusual beam halo.

Superharps also serve a critical function in measuring the electron beam energy. The absolute beam position and angle is measured at three locations, upstream, at the center and downstream of the Hall C arc using pairs of superharps. The fields of dipole bending magnets in the arc have been mapped to good precision using a combination of Hall and NMR probes and all higher order magnets (sextapoles and quadrupoles) have their field integral set to zero. In this configuration, the beam dispersion at the center of the Hall C arc is large (12~cm/\%). The beam momentum can be calculated as 
\[
p=\frac{e}{\theta}\int B~dl,
\]
where $\theta$ is the net bend angle of the beam and $\int B~dl$ is the total magnetic field integral. The beam energy can then be obtained from $p$ with simple relativistic kinematics.

Differential changes in beam energy are monitored using BPM's located in the Hall~C arc. Higher order optical beamline elements (sextapoles and quadrupoles) create a dispersion in the arc which depends on beam tune. A typical tune for \Qs was around 4~cm/\% at the point of highest dispersion in the arc. The BPM located at this position of highest dispersion is called BPM3c12 and the horizontal beam position at this location, given by BPM3c12X, was used by \Qs as a relative energy monitor. 

A set of 4 air-coil dipole magnets (2 horizontal and 2 vertical) in the Hall C arc are used as part of a feedback system to remove beam position and angle motion at the target. This feedback system called ``Fast Feedback'' (FFB) utilizes BPM's in the Hall C arc and tunnel upstream of the experimental hall to determine frequency components of the beam's motion and drives the FFB dipole coils to null those frequencies. The most downstream set of FFB coils is labeled in Figure \ref{fig:HallC_arc}. The system is specifically tuned to null frequencies that are harmonics of the 60~Hz line noise \cite{Lebedev}. Because the accelerator mixes position and angle with beam energy, the FFB system also includes a BPM in the highly dispersive region of the Hall C arc as a relative energy measurement. Energy fluctuations are corrected by feeding back on a Vernier on one of the accelerating cavities in the south linac. This system was active throughout the \Qs experiment.

A second set of 4 air coil dipole magnets (2 horizontal and 2 vertical) in the Hall C arc were used to intentionally drive the beam in position and angle so that the sensitivity of the main detector to these parameters could be measured. The intentional driving of the beam for calibration of the main detector sensitivities is referred to as ``beam modulation''. The beam modulation  coils are labeled in Figure \ref{fig:HallC_arc}. In addition to these air coil magnets, the beam modulation system also utilized a Vernier on one of the south linac accelerating cavities for intentional modulation of energy. The use of the beam modulation system to calibrate the detector responses to beam energy, position and angle and to correct the main detector to remove these sensitivities is dealt with in detail in Chapters \ref{Chapter4} and \ref{Ch:BMod_correction}.

Beam intensity or current was measured during the \Qs experiment using six beam current monitors (BCM's) upstream of the target, although not all six were available for the entire experiment. The BCM's are cylindrical, stainless steel microwave cavities tuned to resonate at the accelerator frequency (1497~MHz). The $TM_{010}$ mode of the BCM's is proportional to beam current but insensitive to electron bunch length and position \cite{Denard2001}. The BCM's were enclosed in a temperature stabilized box to preserve their resonance tune. The BCM's were calibrated using a parametric current transducer (PCT), often referred to as the ``Unser''. Although the Unser provides a highly accurate current reading, it is subject to slow drifts which compromise its usefulness as a continuous monitor.

\begin{figure}[ht]
\centering
\framebox{\includegraphics[width=3in]{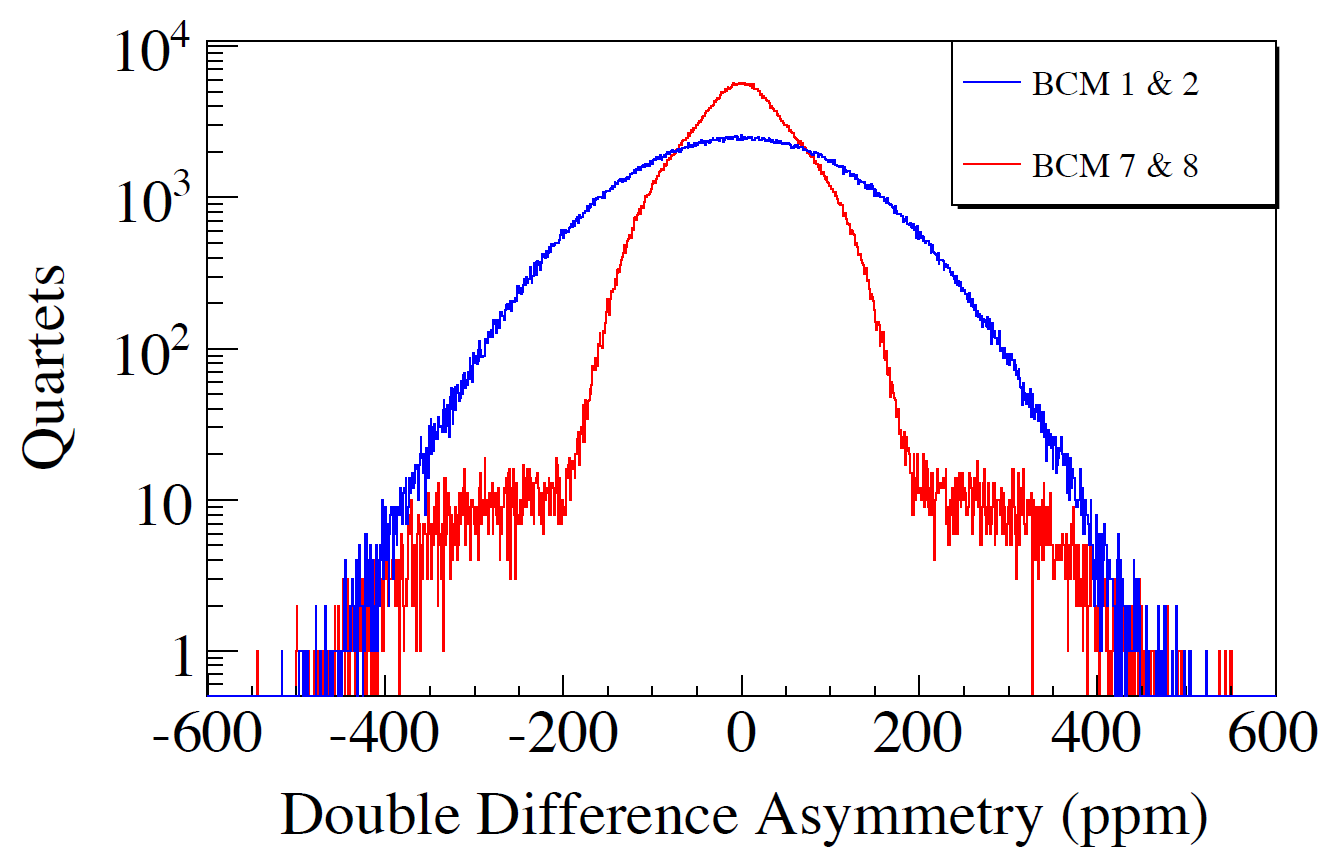}}
\caption{Charge asymmetry double difference (DD) plots (typical 1 hour run) for two pairs of BCM's used in the \Qs experiment. Double difference plots show half the difference in signal between two BCM's measuring the same beam current and as such they are a measurement of the intrinsic BCM noise. BCM's 1 and 2 (blue) have an RMS DD width of 115~ppm while BCM's 7 and 8 (red) have an RMS DD width of 57~ppm.}
\label{fig:BCM_dd_width}
\end{figure}

In Run 1 of the \Qs experiment, a pair of BCM's (BCM1 and BCM2) were used to normalize the main detector signals to charge. In Run 2 a new set of BCM's were used with updated digital electronics which increased the resolution of the beam current measurements. One method for determining BCM resolution involves taking the difference in the asymmetries measured in two BCM's located close to each other on the beamline. This difference between the two BCM's is termed the ``double difference'' since it is a difference between two measurements of asymmetry between states of opposite helicity. Since both BCM's are expected to be measuring exactly the same beam and have approximately the same intrinsic resolution, the width of the double difference distribution is $\sqrt{2}$ times the BCM resolution. Figure \ref{fig:BCM_dd_width} shows the double difference using BCM's 1 and 2 which were used in Run 1 and the double difference using BCM's 7 and 8 outfitted with new electronics.

\section{Polarimetry}
Electron beam polarization enters as a multiplicative factor in determining the parity-violating asymmetry from the measured raw asymmetry (Equation \ref{eq:raw_asymmetry}), requiring a heavy emphasis on polarimetry for \Q. Electron beam polarization was determined for \Qs using the existing M\o ller polarimeter as well as a new Compton polarimeter installed specifically to meet the demands of \Q. Although the M\o ller polarimeter can measure beam polarization to $<1\%$ within a few hours, it is limited to low currents $\sim 1~\mu$A and it is invasive, requiring hours each week of dedicated beam time. The Compton polarimeter can provide continuous, non-invasive polarization measurements at the current of the experiment. The operation of the two polarimeters as well as the results of each are discussed below.

\subsection{M\o ller Polarimeter}
Quantum electrodynamics (QED) accurately predicts a spin dependence in the cross section  of polarized M\o ller (electron-electron) scattering. The spin-dependent cross section for longitudinally polarized, ultra-relativistic electrons can be written at tree level as \cite{Hauger2001}
\begin{equation}
\frac{d\sigma}{d\Omega^{cm}}=\frac{d\sigma_0}{d\Omega^{cm}}\left(1+P_t^{||}P_b^{||}A_{zz}(\theta )\right),
\end{equation}
where the unpolarized cross section at high energy is given by \[\frac{d\sigma_0}{d\Omega^{cm}}=\left(\alpha(4-\sin^2\theta)/2m_e\gamma\sin^2\theta\right)^2,\]  and $P_{t(b)}^{||}$ is the longitudinal polarization of the target (beam). $A_{zz}$, the maximum scattering asymmetry as a function of angle is given by the following expression:\[A_{zz}(\theta )=-\sin^2\theta(8-\sin^2\theta )/(4-\sin^2\theta )^2.\] If the target polarization is known\footnote{Since only the two valence electrons of the 26 in iron are polarizable, the maximum effective target polarization $P_t$ is $\sim$8\%. A potentially significant uncertainty arising from Fermi motion of inner atomic electrons, was first described by Levchuk \cite{Levchuk1994}. The Hall C M\o ller was designed with a large acceptance to make this effect small enough to correct for it with a small uncertainty.}, the electron beam polarization can be obtained from the measured scattering asymmetry between beam polarization states of right and left helicities which is given as 
\begin{equation}
A_{M\o ller}=\frac{\frac{d\sigma}{d\Omega^{cm}}^{\uparrow\uparrow}-\frac{d\sigma}{d\Omega^{cm}}^{\downarrow\uparrow}}{\frac{d\sigma}{d\Omega^{cm}}^{\uparrow\uparrow}+\frac{d\sigma}{d\Omega^{cm}}^{\downarrow\uparrow}}=A_{zz}P_b^{||}P_t^{||},
\label{eq:moller_asym}
\end{equation}
where $\alpha$ is the fine structure constant, $\frac{d\sigma}{d\Omega^{cm}}^{\downarrow\uparrow}$ is the differential scattering cross section for target and beam spins aligned and $\frac{d\sigma}{d\Omega^{cm}}^{\uparrow\uparrow}$ is for the beam and target spins anti-aligned. In practice, $A_{M\o ller}$ is measured as the difference in the numbers of scattered electrons between the two helicity states. The kinematics of the M\o ller polarimeter in Hall C are set to select the maximum analyzing power with the center of mass scattering angle of $\theta_{cm}=90^{\circ}$ giving $A_{zz}\approx-\frac{7}{9}$. 

\begin{figure}[ht]
\centering
\framebox{\includegraphics[width=4in]{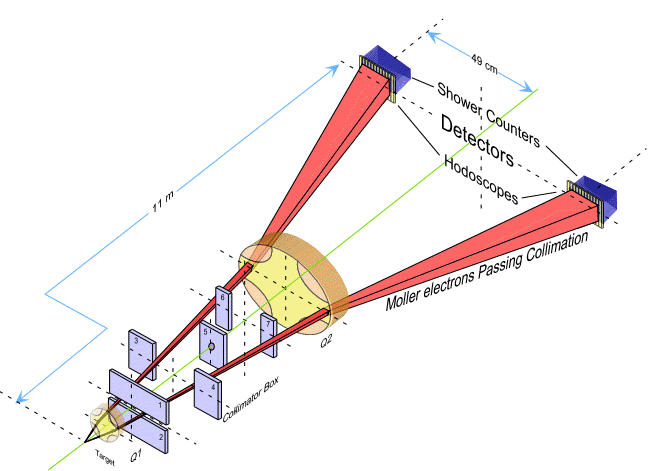}}
\caption{Diagram showing key components of the Hall C M\o ller polarimeter. }
\label{fig:Moller_diagram}
\end{figure}

The target for the Hall C M\o ller is a thin iron foil placed directly in the electron beam and polarized to the point of saturation in the direction of the electron momentum using a superconducting 3.5~T magnet.  Both the scattered electron and the dislocated atomic electron are detected in coincidence by symmetric detectors on both sides of the electron beam. A tight timing cut on the coincidence highly suppresses backgrounds primarily from Mott scattering. Two quadrupole magnets momentum analyze the scattered electrons so that the desired events are selected and the majority of background events blocked by a system of collimators. A diagram of the M\o ller polarimeter is shown in Figure \ref{fig:Moller_diagram}. The difference in the number of coincidences between right and left beam helicity states divided by the total number for both states gives the asymmetry $A_{M\o ller}$.

During \Qs the beam polarization was measured using the M\o ller 2-3 times per week. Beam current during these measurements was around $1~\mu$ A to keep uncertainty from target depolarization low. The list of systematic uncertainties associated with the M\o ller measurements is given in Table \ref{tab:moller_systematics}. During Run 1 an additional systematic uncertainty of 0.89\% was included to account for an intermittent short in one of the M\o ller quadrupoles which altered the analyzing power $A_{zz}$. 

\begin{table}[!h]
  \centering
    \begin{tabular}{|l|c|c|} \hline
                           & Uncer- & $dA/A$ \\ 
      Source & tainty & (\%) \\ \hline
      Beam position $X$     & 0.5 mm      & 0.17     \\
      Beam position $Y$     & 0.5 mm      & 0.28     \\
      Beam direction $X^\prime$    & 0.5 mrad    & 0.1      \\
      Beam direction $Y^\prime$    & 0.5 mrad    & 0.1      \\
      Q1 current                 & 2\%         & 0.07     \\
      Q3 current                 & 3\%         & 0.05     \\
      Q3 position                & 1 mm        & 0.01     \\
      Multiple scattering        & 10\%        & 0.01     \\
      Levchuk effect             & 10\%        & 0.33     \\
      Collimator position        & 0.5 mm      & 0.03     \\
      Target temperature         & 100\%       & 0.14     \\
      B-field direction          & 2$^\circ$   & 0.14     \\
      B-field strength           & 5\%         & 0.03     \\
      Spin depolarization        & --~         & 0.25     \\
      Electronic dead time       & 100\%       & 0.05     \\
      Solenoid focusing          & 100\%       & 0.21     \\
      Solenoid position($X,Y$)     & 0.5 mm       & 0.23     \\
      High current extrap. & --~         & 0.5      \\
      Monte Carlo statistics     & --~         & 0.14     \\ \hline
      ~                          & Total       & 0.83     \\ \hline
    \end{tabular}
  \caption{{Table of uncertainties for the Hall C M\o ller polarimeter for Run 2 of the experiment taken from the \Qs instrumentation paper \cite{QweakNIM}. During Run 1 an intermittent short in one of the polarimeter quadrupoles added an uncertainty not included here.}}
  \label{tab:moller_systematics}
\end{table}

\subsection{\label{sctn:compton}Compton Polarimeter} 
\begin{figure}[ht]
\centering
\includegraphics[width=0.99\textwidth]{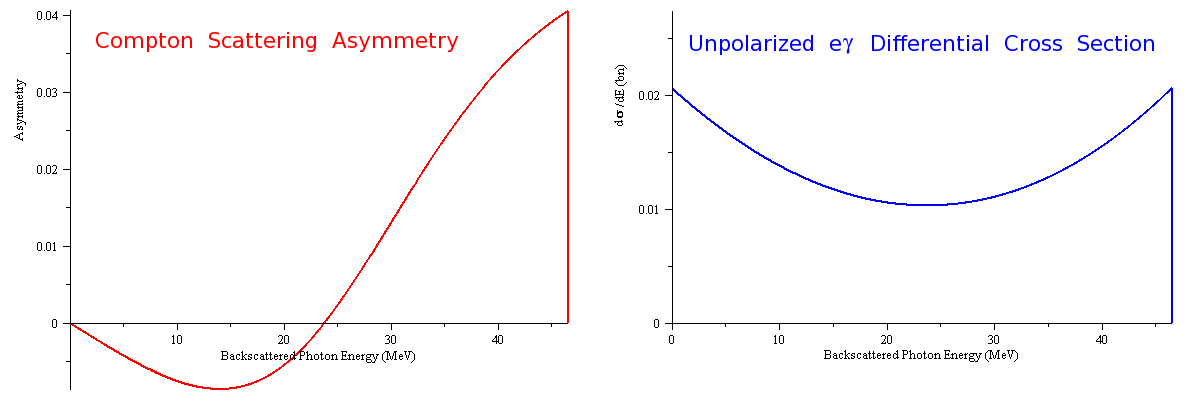}
\caption{Compton scattering asymmetry and differential cross section shown at the kinematics of the \Qs experiment using a green (532~nm) laser at 0$^{\circ}$ crossing angle. (Right)Average (unpolarized) differential cross section for electron-photon scattering. (Left) Scattering asymmetry for longitudinally polarized electrons on circularly polarized photons versus back-scattered photon energy. }
\label{fig:compton_cx}
\end{figure}

Compton polarimetry relies on electron-photon (e$\gamma$) scattering to determine electron beam polarization. A circularly polarized laser beam of photons intersects the electron beam at a small crossing angle. The e$\gamma$ cross section is small enough that its affect on the electron beam can be neglected; however, the cross section depends slightly upon the relative helicities of the electron and photon. The tree level spin-dependent cross section of longitudinally polarized electrons on circularly polarized photons with collinear (opposite direction) momenta\footnote{A zero angle crossing of the laser and electron beam is assumed. The Hall C Compton crossing angle is $1.33^{\circ}$ or $23.2~mrad$. The correction to account for the non-zero crossing angle is negligible for the kinematics of the Compton polarimeter. \cite{Denner1999}} is given by QED as \cite{Prescott1973}\cite{Tolhoek1956}
\begin{equation}
\frac{d\sigma}{d\rho}=\frac{d\sigma_0}{d\rho} \mp P_eP_{\gamma}\cos\theta\frac{d\sigma_1}{d\rho},
\label{eq:compton_cx}
\end{equation}
where $\frac{d\sigma_0}{d\rho}$ is the unpolarized cross section, $\frac{d\sigma_1}{d\rho}$ is the longitudinal spin-dependent piece of the cross section, $P_{e(\gamma)}$ is the electron (photon) polarization and $\theta$ is the angle between the electron spin and its momentum. The -- sign of the second term corresponds to the photon whose helicity is parallel to its momentum (opposite the electron momentum)\cite{Prescott1973}. The two components of the cross section can be shown to be
\begin{equation}
\frac{d\sigma_0}{d\rho}=2\pi r_0^2a\left[\frac{\rho^2(1-a)^2}{1-\rho(1-a)}+1+\left(\frac{1-\rho(1+a)}{1-\rho(1-a)}\right)^2\right],
\label{eq:unpol_compton_cx}
\end{equation}
and 
\begin{equation}
\frac{d\sigma_1}{d\rho}=2\pi r_0^2a\left[\left(1-\rho (1+a)\right)\left(1-\frac{1}{1-\rho(1-a)}\right)^2\right],
\label{eq:spin_dep_compton_cx}
\end{equation}
where $r_0$ is the classical electron radius, $\rho=k'/k_{max}$ is the ratio of the scattered photon energy to the maximum scattered photon energy in the lab frame and $a=1+4kE/m_e^2$ is a kinematic factor where $k$ is the incident photon energy and $E$ is the incident electron energy measured in the lab frame. The asymmetry then comes from the longitudinal spin dependent term of Equation \ref{eq:spin_dep_compton_cx} and takes the form
\begin{equation}
A_{l}=\frac{\frac{d\sigma}{d\Omega^{cm}}^{\uparrow\uparrow}-\frac{d\sigma}{d\Omega^{cm}}^{\downarrow\uparrow}}{\frac{d\sigma}{d\Omega^{cm}}^{\uparrow\uparrow}+\frac{d\sigma}{d\Omega^{cm}}^{\downarrow\uparrow}}=\frac{d\sigma_1}{d\rho}/\frac{d\sigma_0}{d\rho},
\label{eq:compton_asym}
\end{equation}
where the subscript $l$ emphasizes that this is only for longitudinal electron polarization.
Thus, if the photon polarization is known, the electron beam polarization can be determined from the measured asymmetry using the following relationship:
\begin{equation}
A_{measured}=P_eP_{\gamma}A_{Compton}\Longrightarrow P_e=\frac{1}{P_{\gamma}}\left(\frac{A_{measured}}{A_{l}}\right).
\label{eq:compton_pol}
\end{equation}
Plots of the unpolarized differential cross section and of the asymmetry $A_l$ as a function of back-scattered photon energy are shown in Figure \ref{fig:compton_cx}. The asymmetry changes sign in the center of the spectrum and reaches a maximum at the kinematic endpoint for maximum back-scattered photon energy.

\begin{figure}[ht]
\centering
\includegraphics[width=0.99\textwidth]{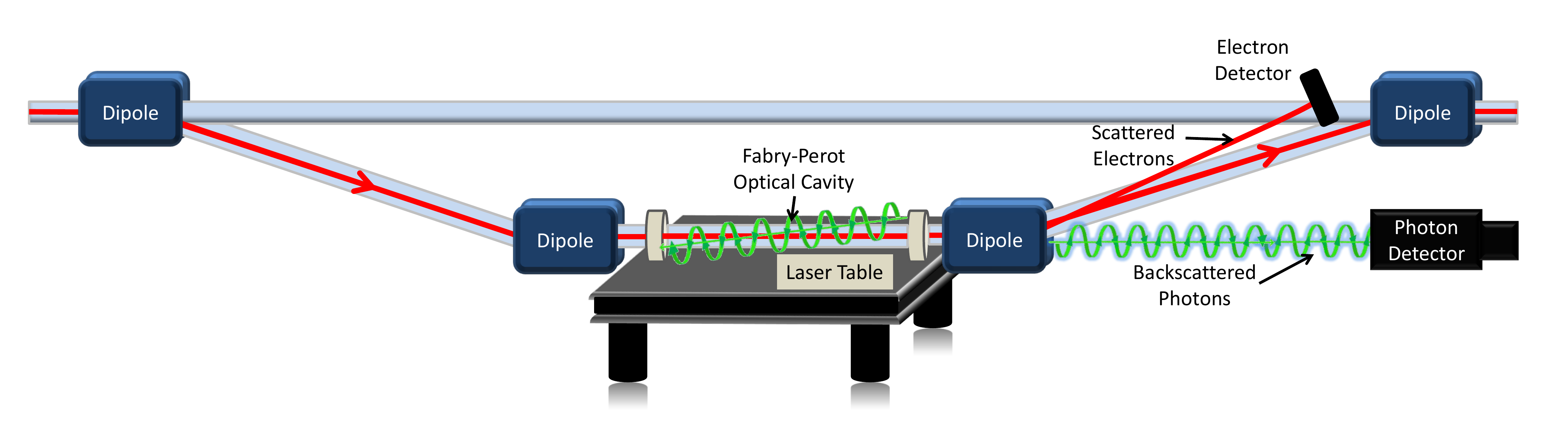}
\caption{Illustration of Compton polarimeter showing magnetic chicane, laser table, and electron and photon detectors. }
\label{fig:compton_layout}
\end{figure}

The Hall C Compton polarimeter was designed to measure the scattering asymmetry for both the back-scattered photons and the Compton scattered electrons. An illustration of the system is shown in Figure \ref{fig:compton_layout}. The electron beam is bent through a chicane by a set of 4 identical dipole magnets. At the center of the chicane the electron beam is 0.57~m lower than its usual trajectory and at this point it intersects a laser-fed Fabry-Perot optical cavity where the Compton interaction occurs. The scattered photons in the full $4\pi$ solid angle in the electron rest frame are contained in a small-angle cone in the lab frame centered on the electron beam trajectory and are captured by the photon detector, a scintillating crystal placed 4.3~m downstream of the Compton interaction point. The electron detector, located 5~mm above the electron beam directly upstream of the fourth dipole, measured the asymmetry versus energy spectrum of the scattered electrons momentum-analyzed by the third dipole of the chicane.
Design and construction of the photon target for the Compton polarimeter was one of the primary concerns of the author early in this experiment. A Coherent Verdi-V10 laser (details on the laser can be found in \cite{Verdi10}) delivered 10~W of green light (532~nm) that was mode-matched to an 84~cm long Fabry-Perot optical cavity. During Run 2 the optical cavity had high-reflectivity mirrors (R=99.5\%) yielding a cavity gain close to 200 with more than 1400-1800~W of stored power depending upon the quality of the mode-matching. Lower reflectivity mirrors used in Run 1 led to more than a factor of 2 reduction in stored power and a poorer signal to background ratio. 

Knowledge of laser polarization inside the optical cavity, housed inside the electron beam vacuum pipe, is essential to the integrity of the electron polarization measurement (see eq. \ref{eq:compton_pol}). A novel method used to determine intracavity laser polarization introduced during Run 2 made this key uncertainty almost negligible. Further details on the setup and analysis for the laser/photon target will be provided in Chapter \ref{Ch:Compton_Laser}.

\begin{figure}[ht]
\centering
\includegraphics[width=0.6\textwidth]{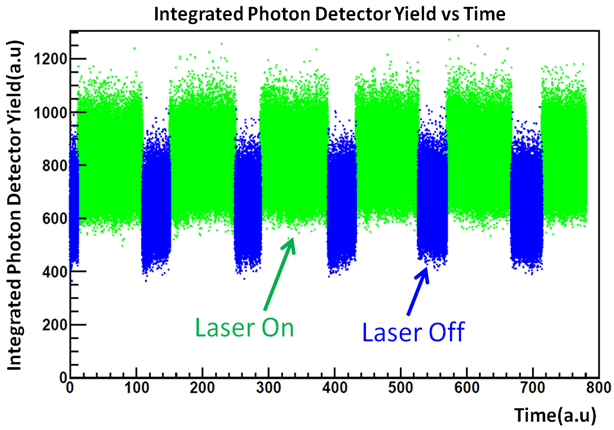}
\caption{Photon detector yield versus time clearly showing the transitions between laser on and off periods.}
\label{fig:laser_onoff}
\end{figure}

The photon detector consists of a square matrix of four 20~cm long lead tungstate (PbWO$_4$) scintillating crystals each with a 3~$\times$~3~cm$^2$ cross section attached to a single 7.6~cm Hamamatsu R4885 photo-multiplier tube (PMT) located on the downstream face. The photon detector was maintained at $14^{\circ}$C giving a light yield increase of 20\% relative to the usual $21^{\circ}-24^{\circ}$C temperature of Hall C. The signal from the photon detector was digitized using a flash analog-to-digital converter (FADC) with 200~MHz sampling. The signal was digitally integrated over each MPS and the accumulated sum recorded. In addition to the accumulator sum, a single self-triggered 250-sample waveform of a photon pulse in the detector was digitized and saved for each MPS allowing construction of the detected pulse spectrum for calibration purposes. The laser was continuously cycled on and off for background subtraction in a continuous pattern of 60 seconds on and 30 seconds off. ``Laser off'' means the optical cavity was unlocked and the laser beam blocked. Some periods during the experiment did not have the light blocked and a small residual beam of light leaked into the cavity, necessitating a small correction. Figure \ref{fig:laser_onoff} shows a graph of the photon detector yield (accumulated sums) versus time clearly demonstrating the increase of light yield during laser on periods.
Statistical uncertainty for the photon detector measurements remained relatively high throughout \Qs so that its usefulness in tracking polarization between M\o ller polarization measurements was not realized. The analyzing power and non-linearity of the photon detector data were determined by a combination of GEANT4 simulation and direct measurement (for details see \cite{Cornejo}). The combined statistical and systematic errors associated with the photon detector analysis were too large to provide useful polarization values for \Q.

The electron detector is composed of a multiple plane diamond microstrip detector. Each detector plane is a thin ($500~\mu$m thick) 21~cm~$\times$~21~cm diamond wafer. The front surface of each wafer has 96 metal microstrips each $180~\mu$m wide with a $20~\mu$m gap between them, while a single electrode covers the entire back of the wafer. When an electron travels through the diamond it creates a trail of electron-hole pairs. A high voltage applied across the electrodes makes the electron-hole pairs migrate towards opposite plates causing an electrical pulse in the nearest microstrip(s). Pulses are conditioned and summed individually for each strip over each MPS window. Three of four parallel detector planes were used during \Qs with an event trigger requiring a hit in at least two of the three planes. A small set subset of the data was taken requiring a coincidence in three planes which was useful for diagnostic purposes and for calculating dead time.

Two parameters required before beam polarization can be extracted from the asymmetry spectrum are the microstrip-to-energy conversion factor and the strip offset. Only when these are known can the asymmetry versus microstrip spectrum be properly mapped its corresponding energy spectrum. A calculation of the strip/MeV conversion factor was made using a magnetic map of the third dipole in the chicane combined with the measured electron path. A fit procedure for locating the ``Compton edge''\footnote{Compton scattering asymmetry spectra are typically shown versus back-scattered photon energy or electron energy lost to the photon. The ``Compton edge'' is the largest possible energy loss corresponding to the direct $180^{\circ}$ back-scattering of the photon which also corresponds to the electrons which are least energetic. The Compton edge is obvious in Figure \ref{fig:compton_cx} since no Compton scattering occurs at higher energies.} gave the necessary offset allowing the Compton asymmetry spectrum to be fit with only one additional parameter, the electron polarization.
\begin{figure}[ht]
\centering
\includegraphics[width=0.6\textwidth]{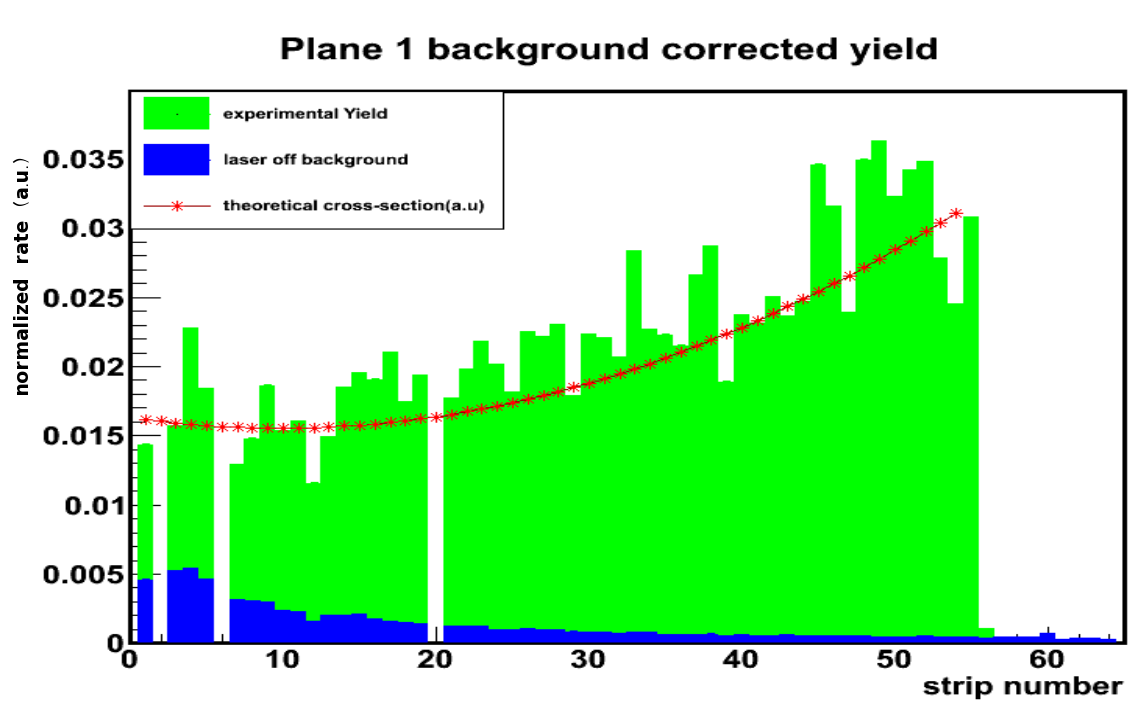}
\caption{Compton electron detector rate spectrum versus strip number (distance from electron beam). Blue is laser off background and green is background-subtracted laser on yield. Shape of theoretical Compton spectrum is shown by red curve.}
\label{fig:edet_spectrum}
\end{figure} 

\begin{figure}[ht]
\centering
\includegraphics[width=0.6\textwidth]{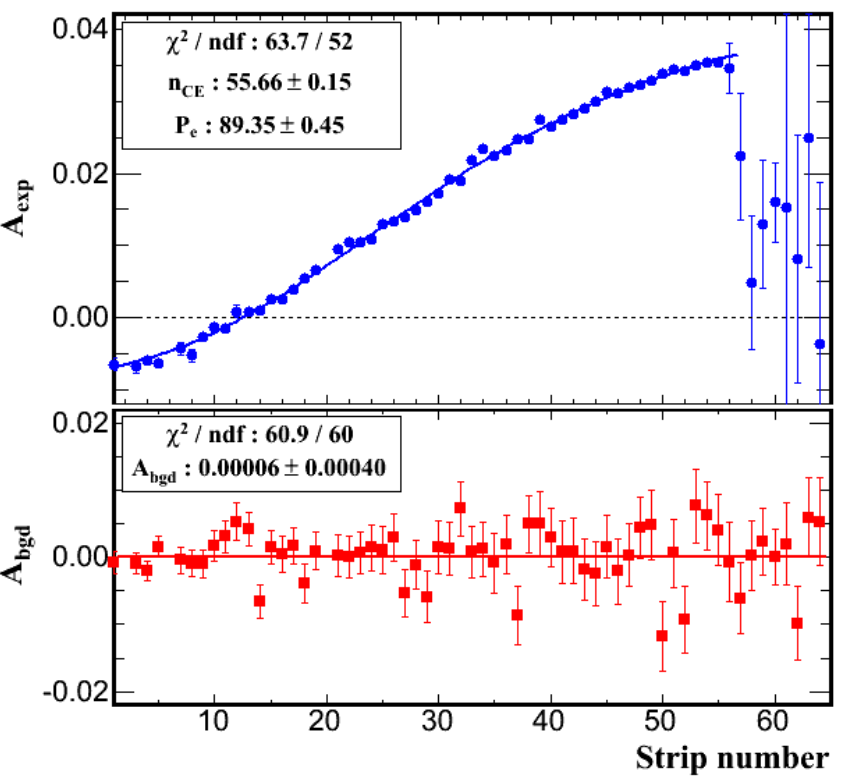}
\caption{Electron detector rate asymmetry versus strip number shown with fit to Compton spectrum with polarization of 89.4\%. Compton edge appears to be between strips 56 and 57. Background asymmetry is consistent with zero.}
\label{fig:edet_asym}
\end{figure}

Figure \ref{fig:edet_spectrum} shows the electron detector rate spectrum as a function of strip number or distance from electron beam. For the running conditions of \Q, the Compton edge was located $\sim$17~mm from the electron beam at the location of the electron detector. The electron detector was placed as close as 5~mm from the electron beam with the Compton edge around strip number 55.  Figure \ref{fig:edet_asym} shows a typical 1~hour asymmetry spectrum versus strip number. Variation in efficiencies from strip to strip create a jagged rate spectrum, but since the asymmetry is formed on a strip-by-strip basis, it is not affected  by efficiency to first order. Fitting the shape of the asymmetry spectrum maximizes the use of all the data and provides the highest precision measurement per unit time. During typical running conditions for Run 2 of \Q, the electron detector was able to reach a statistical precision of $\pm0.6\%$ in an hour.

\begin{figure}[ht]
\centering
\includegraphics[width=0.6\textwidth]{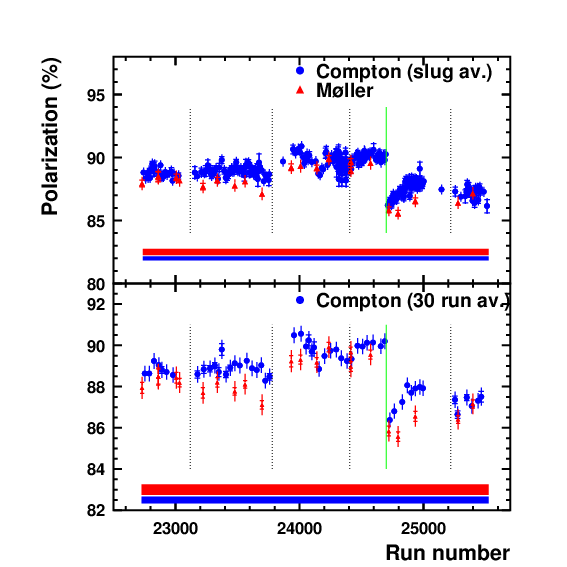}
\caption{Electron beam polarization versus ``time'' results compared for M\o ller and Compton electron detector for Run 2 of \Q. Inner error bars are statistical and outer are systematic. Dashed vertical lines mark changes in position of electron source laser on the photo-cathode. Solid green vertical line marks heat and reactivation of source photo-cathode to increase quantum efficiency.   }
\label{fig:Run2_edet_moller_pol}
\end{figure}

The quality of electron detector data was much higher in Run 2 than in Run 1 due to improvements made during the scheduled down time between the two run periods. During this down time the photon target power was doubled and the laser polarization increased. The electron detector pulse-conditioning electronics boards were replaced with improved, less noisy ones and a fourth electron detector plane was added. Analysis is ongoing to determine the electron detector polarizations and uncertainty for Run 1, while the Run 2 results are considered mature. A plot comparing the M\o ller and electron detector results for Run 2 is shown in Figure \ref{fig:Run2_edet_moller_pol}. A detailed account of the electron detector analysis can be found in \cite{Narayan}. Table \ref{tab:compton_systematics} lists the most significant contributions to the total systematic error for the Run 2 electron detector analysis.

\begin{table}[ttbh]
\scriptsize{
  \centering
    \begin{tabular}{|l|c|c|} \hline
                            & Uncer- & $\Delta P/P$ \\ 
      Source 				& tainty & (\%) \\ \hline
      Laser polarization   		& 0.14 \%      & 0.14     \\
      Plane-to-plane  			& secondaries      & 0.00     \\
      3$^{rd}$ Dipole field		& 0.0011 T    & 0.13      \\
      Beam energy    			& 1 MeV   & 0.08     \\
      Detector $Z$ position      & 1 mm         & 0.03     \\
      Trigger multiplicity       & 1-3 plane         & 0.19     \\
      Trigger clustering         & 1-8 strips  & 0.01     \\
      Detector tilt ($X$)        & 1$^\circ$   & 0.03     \\
      Detector tilt ($Y$)        & 1$^\circ$   & 0.02     \\
      Detector tilt ($Z$)        & 1$^\circ$   & 0.04     \\
      Strip eff. variation        & 0.0 - 100\%   & 0.1     \\
      Detector Noise	         & $\leq$20\% of rate  & 0.1     \\
      Fringe Field		         & 100\%       & 0.05     \\
      Radiative corrections      & 20\%		   & 0.05     \\
      DAQ ineff. correction & 40\%  & 0.3     \\
      DAQ ineff. pt-to-pt	 & 		& 0.3     \\ 
      helicity correl. beam pos. & 5~nm & $<0.05$\\ 
      helicity correl. beam pos. & 5~nm & $<0.05$\\ 
      vert. pos. variation & 0.5~mrad & 0.2\\
      chicane spin precession    & 20~mrad & $<0.03$ \\ \hline
      Total                          &        & 0.58     \\ \hline
    \end{tabular}
  \caption{The Hall C Compton polarimeter systematic uncertainties determined for Run 2 of the experiment.}
  \label{tab:compton_systematics}
}
\end{table}

\section{\label{sctn:target}Target}
\begin{figure}[ht]
\centering
\includegraphics[width=0.6\textwidth]{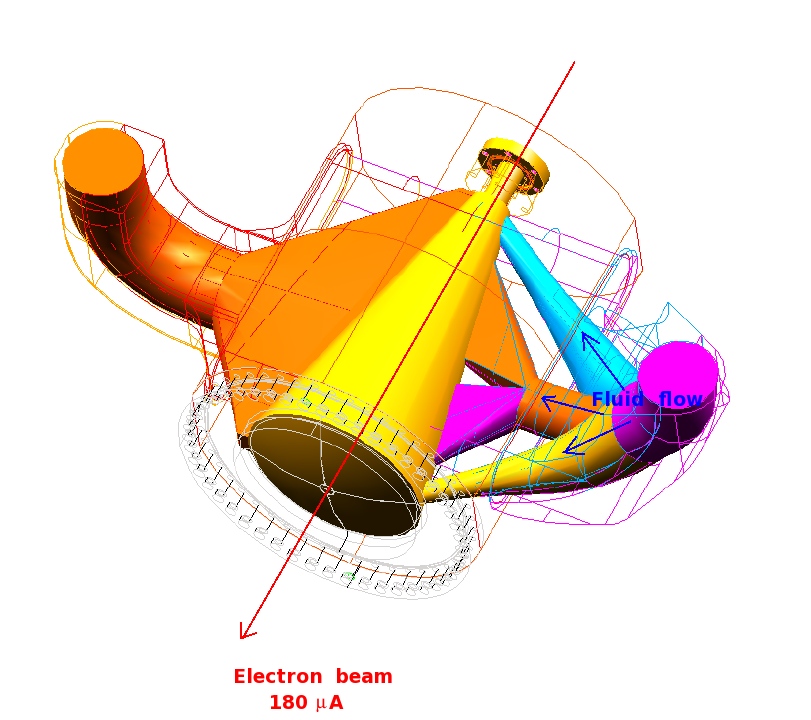}
\caption{Rendering of the target cell with the exterior shell shown only in outline to expose the inner structure designed to optimize fluid flow for elimination of hot spots and minimal target boiling. The conical shape with a larger downstream window allowed all elastic events within the detector acceptance to pass without interference from other target cell structures.}
\label{fig:target_cell}
\end{figure}

The main production target for \Qs was a 34.4~cm long conical cell with thin aluminum windows, filled with \LHs and designed to handle the heat load introduced by a $180~\mu$A electron beam. A heat exchanger maintained the liquid hydrogen at 20~K and a variable-speed recirculating pump continually cycled the hydrogen through the target cell. The target cell was designed using Computational Fluid Dynamics, modeling fluid flow and heat load to mitigate the effects of target boiling, especially near target windows (see Figure \ref{fig:target_cell}). The target windows were made of a strong aluminum alloy Al 75075-T6 with the 22.2~mm diameter entrance window being 97~$\mu$m thick and the 305~mm diameter exit window having a thickness of 640~$\mu$m. Under nominal running conditions during \Qs the target heat load from ionization was 2.1~kW. The target heat exchanger was designed to accommodate 2.8~kW to allow for other sources of heating (such as the pump) and still leave a comfortable reserve of 250~W. 

\begin{figure}[hhbt]
\hspace*{0.75cm}
\centerline{\includegraphics[width=0.75\textwidth,angle=0]{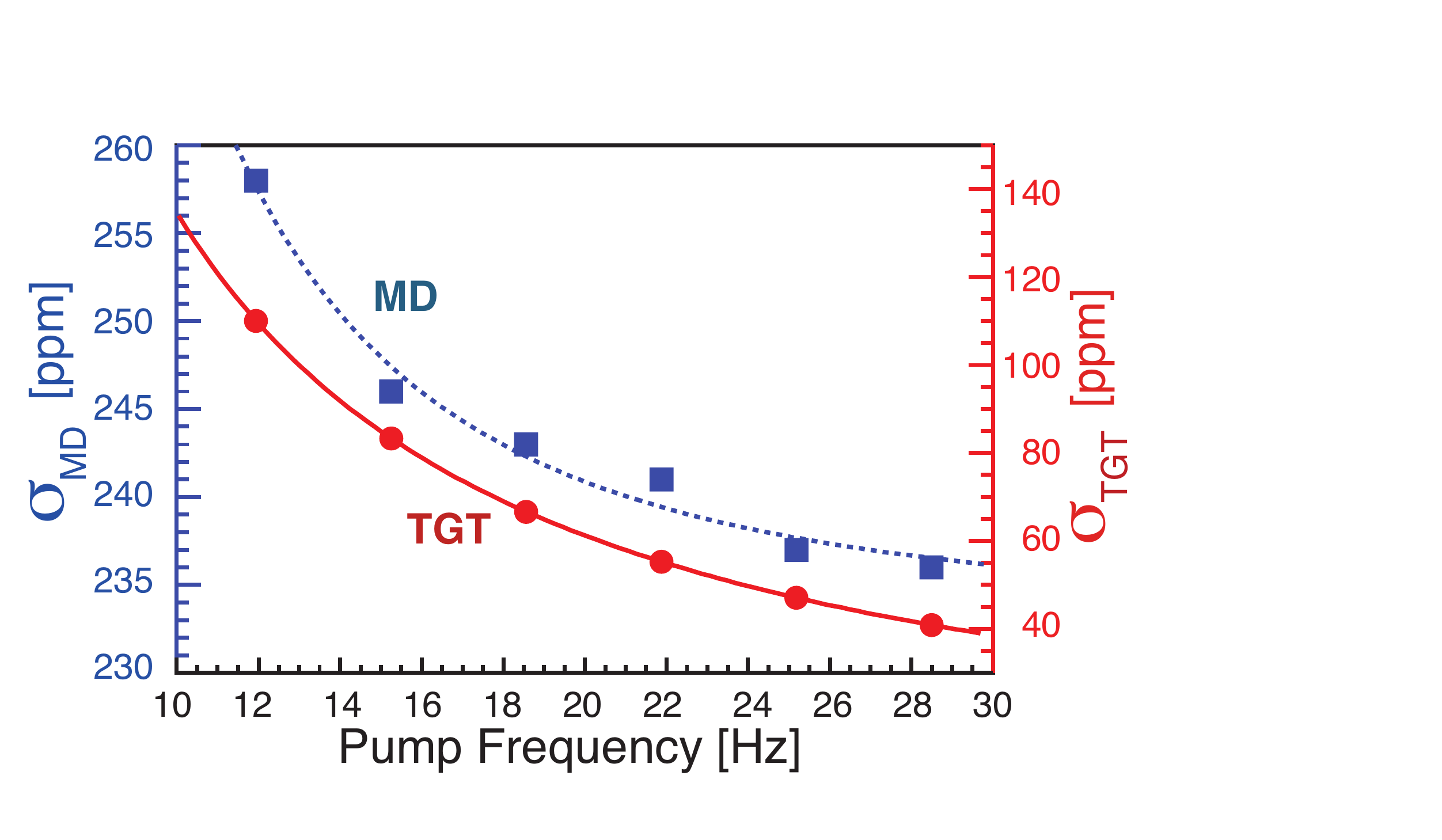}}
\caption{\label{fig:boiling} Measured quartet-level main detector asymmetry width (solid blue squares) versus target recirculation pump rotational speed with 169~$\mu$A current and a 4$\times$4~mm$^2$ raster. Contribution to main detector width from target noise (red solid circles and right hand vertical scale), calculated under the assumption that the change in main detector asymmetry width comes solely from varying target noise (see \cite{Smith} for details on the study).}
\end{figure}

Target boiling is a source of excess width -- often called ``common-mode noise'' because it is seen by all the detectors -- in the main detector signal. One of the primary metrics used to evaluate the performance of the target was the additional width to the main detector asymmetry distribution attributed to target boiling. During production running the electron beam was rastered in a square pattern typically 4~mm~$\times$~4~mm to reduce target boiling and prevent burning the aluminum windows. A series of studies were done varying the target recirculation pump speed, the raster size and beam current to test the contribution to the main detector width from target boiling. Figure \ref{fig:boiling} shows the published results of one such test done by varying pump speed \cite{QweakNIM}. At the nominal \Qs production running conditions of 180~$\mu$A beam of 1.16~GeV electrons rastered in a square 4~mm~$\times$~4~mm pattern with a 28.5~Hz pump speed, the contribution to the main detector asymmetry width from target boiling was $53\pm5$~ppm, near the design limit of $<$50~ppm. 

\begin{figure}[h]
\begin{minipage}{0.42\textwidth}
  \centering
  \includegraphics[width=0.9\textwidth]{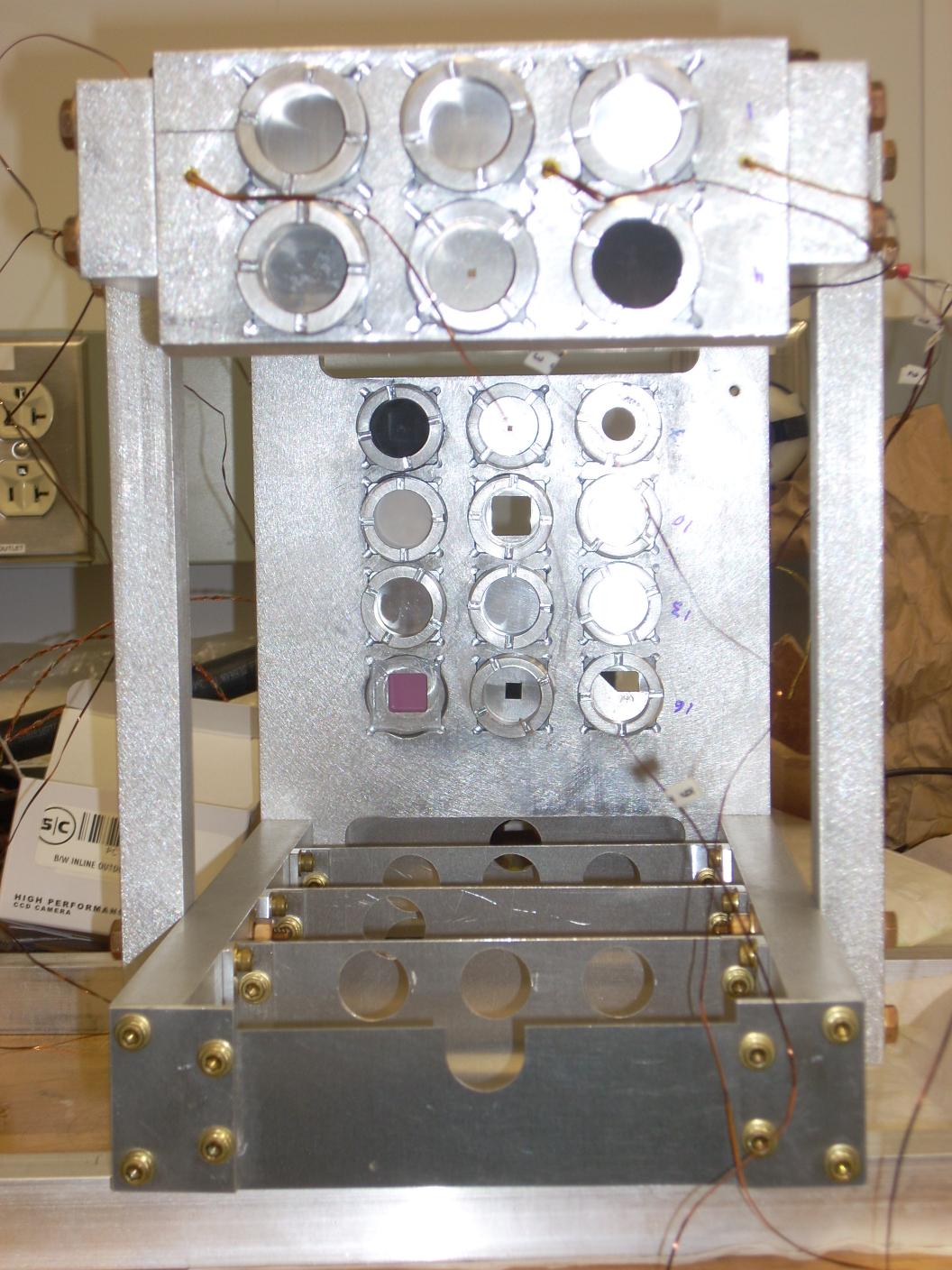}
\end{minipage}
\begin{minipage}{0.48\textwidth}
\centering
\includegraphics[width=1\textwidth]{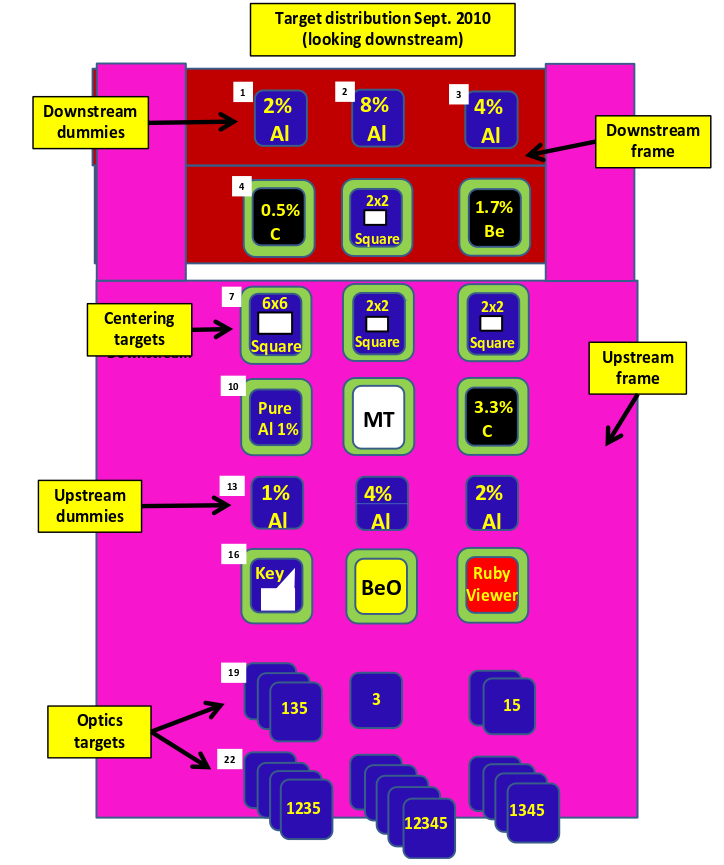}
\end{minipage}
\caption{\label{fig:dummy}Dummy target ladder for \Qs prior to Run 1. Picture on the left shows target ladder looking upstream. Schematic on right shows solid target configuration as it was before Run 1 looking downstream. Upstream and downstream targets of various materials were included. The most important solid targets were the aluminum targets at both upstream and downstream window locations used to measure the contribution of the aluminum target windows to the measured rates.}
\end{figure}

The target also included an array of solid targets on a ``ladder'' with locations along the beamline at the same position as the upstream and downstream aluminum windows. Aluminum solid targets of the same alloy as the target cell windows were used to measure the parity-violating asymmetry of aluminum  so that the window contribution could be removed from the main detector asymmetry measurement. Targets with various sizes of square holes were used to determine target alignment. Solid targets of carbon, beryllium and beryllium oxide were also included for possible systematic and calibration studies. A picture of the dummy target ladder in its September 2010 configuration is shown in Figure \ref{fig:dummy}. A few alterations to the targets were made between Runs 1 and 2 including the addition of more thin aluminum targets.

\section{Collimation and Shielding}
\begin{figure}[ht]
\centering
\includegraphics[width=0.8\textwidth]{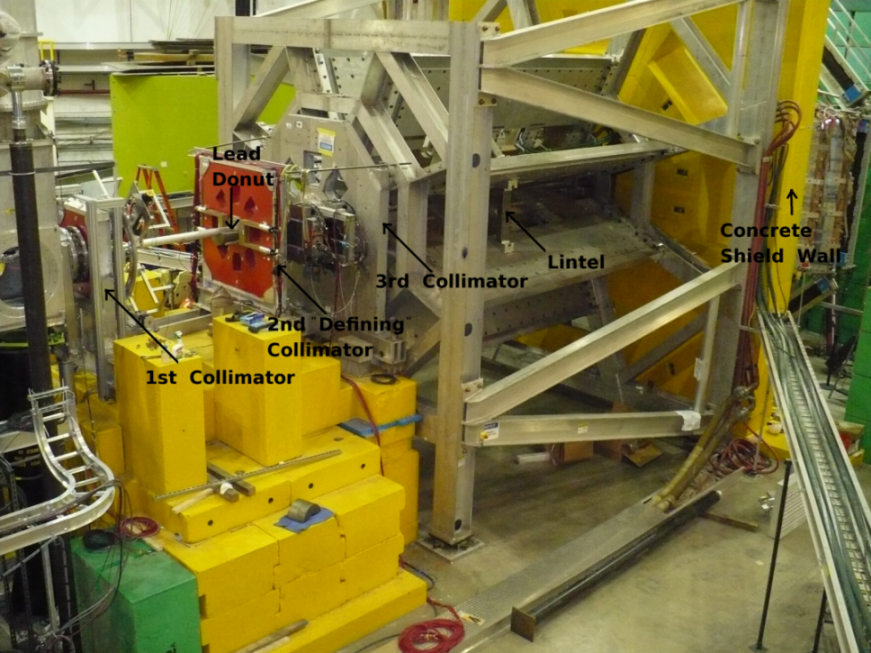}
\caption{Picture showing the collimation system used in the \Qs experiment during installation in Hall C at Jefferson Lab. Also labeled are the lintels for blocking line-of-sight photons from the target and the lead donut around the beam pipe on the primary collimator which reduced backgrounds from beamline scattering.}
\label{fig:collimators}
\end{figure}

Three lead collimators located between the target and spectrometer (see Figure \ref{fig:QweakApparatus}) formed the collimator system for the \Qs experiment. The first collimator located close to the target was a cleanup collimator used to reduce radiation for downstream equipment, whereas the second collimator defined the polar and azimuthal acceptance of the experiment. A third collimator located in the upstream fringe field of the spectrometer served to further reduce backgrounds before the spectrometer. Figure \ref{fig:collimators} shows the three main collimators during installation in Hall C. 

\begin{figure}[ht]
\centering
\includegraphics[width=0.6\textwidth]{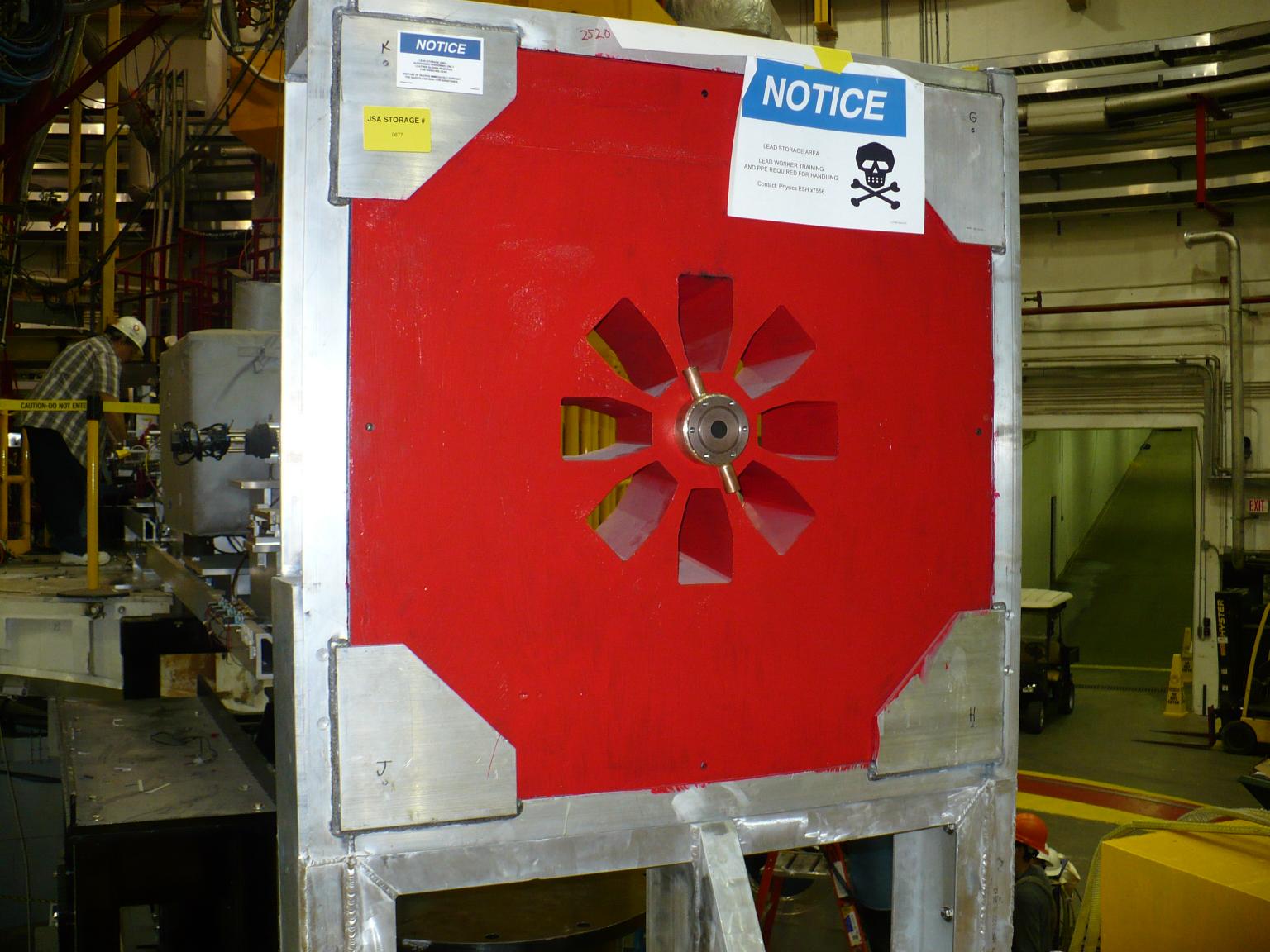}
\caption{First collimator shown with tungsten collimator installed in its central aperture. Perspective is looking upstream toward the target.}
\label{fig:tungsten_plug}
\end{figure}

A water-cooled, cylindrical tungsten collimator installed in the central aperture of the first collimator defined the maximum scattering angle that could pass through the beam enclosure and was designed to stop scattering with line-of-sight to the main detector. The tungsten collimator was machined to have a central conical shaped aperture with upstream and downstream diameters of 14.9~mm and 21.5~mm respectively. This collimator, located 47~cm downstream of the target exit window, was calculated to absorb a deposited power of 1.6~kW during typical production conditions. The tungsten collimator is associated with a significant source of helicity-correlated background seen uniformly by all the main detector bars and will figure prominently in discussions about beam corrections in following chapters. A picture of the tungsten collimator installed in the first collimator is shown in Figure \ref{fig:tungsten_plug}.

Lead ``lintels'' were installed inside the eight octants of the \qtor spectrometer to block line of sight photons to the main detector produced in the downstream edge of the defining collimator as well as in the tungsten collimator and the beam pipe just downstream of it. A lead donut was also installed around the beam pipe at the defining collimator (see Figure \ref{fig:collimators}) after it was determined to greatly reduce backgrounds in the main detector.

An 80~cm thick barite-loaded, stainless steel reinforced, high-density concrete wall was built just downstream of the detector to shield the main detector. The wall design was optimized using the GEANT3 simulation package to remove backgrounds associated with interactions of secondary photons and inelastic electrons in the main detector support structures. The apertures in this wall were designed to be well outside the elastic envelope defined by the second collimator and the \qtor optics.

During the scheduled accelerator down time between Runs 1 and 2, 5~cm thick insertable tungsten shutters were installed on the first collimator allowing octants 1 and 5 (octants located at 9 o'clock and 3 o'clock) to be completely blocked for dedicated background studies.  Further details of the shielding and collimation system and of the simulation efforts that went into their design can be found in \cite{Myers} as well as in the \Qs instrumentation publication \cite{QweakNIM}.

\section{\qtor Spectrometer}
The torroidal spectrometer (\qtor) employed in the \Qs experiment, in contrast to a pure momentum analyzing spectrometer, was designed to focus all elastic events inside the envelope defined by the collimator system into a mustache-shaped distribution on the detector bars. \qtor was composed of eight 3.7~m long, racetrack-shaped, double-pancake coils arranged symmetrically about a central axis to form eight octants. \qtor can be seen during installation in Figure \ref{fig:collimators}. The nature of \Qs required a highly degree of symmetry in the eight octants of the installed spectrometer. A precise mapping program that utilized shifts from predicted zero-crossings of the magnetic field in the fringe fields outside the ends of the spectrometer was used to determine coil misalignments. Details of this zero-crossing technique for determining coil misalignments are given in \cite{Wang}. The most important metric for quantifying design symmetry is the octant to octant variation of the total magnetic field seen by scattered electrons, often referred to as $\int B dl$. This variation was found to be within specified design tolerance of $\pm0.3\%$ (see page 28 \cite{QweakNIM}).

The \qtor spectrometer was composed of eight, water-cooled, resistive magnetic coils connected in series to a power supply. During nominal operating conditions \qtor ran at 8900~A and 123~VDC, generating a $\int B dl$ of 0.9~T-m at the mean scattering angle of $7.9^{\circ}$. Both the \qtor support structure and the coils were designed to be iron free to remove the possibility of helicity-correlated asymmetric scattering from magnetized iron reaching the detectors.

\section{Main Detector System}
\begin{figure}[ht]
\centering
\includegraphics[width=0.8\textwidth]{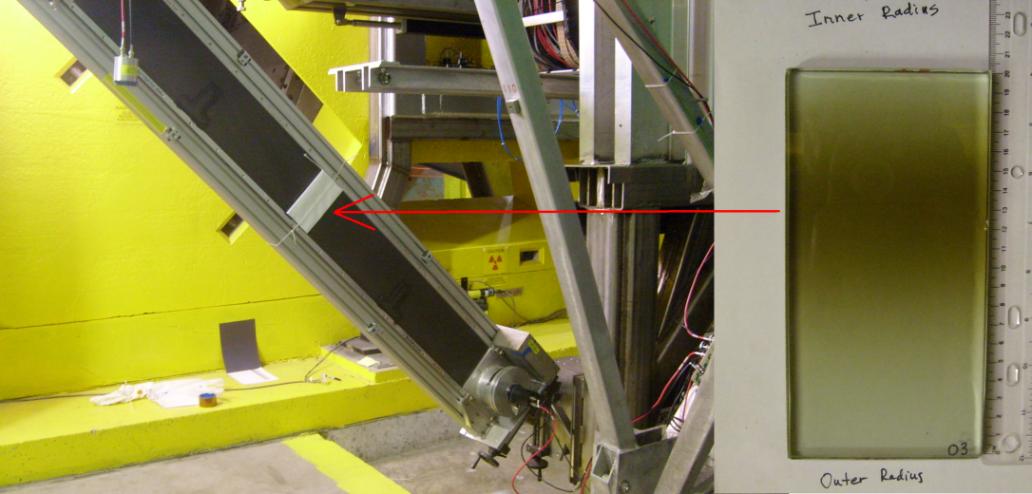}
\caption{Regular glass witness plates fastened to the downstream side of the main detector bar located and removed after only a few weeks of beam demonstrate clear darkening at the highest event region in the ``mustache'' distribution. Fused silica detector bars showed little radiation damage.}
\label{fig:witness_plate}
\end{figure}

\begin{figure}[ht]
\centering
\includegraphics[width=0.8\textwidth]{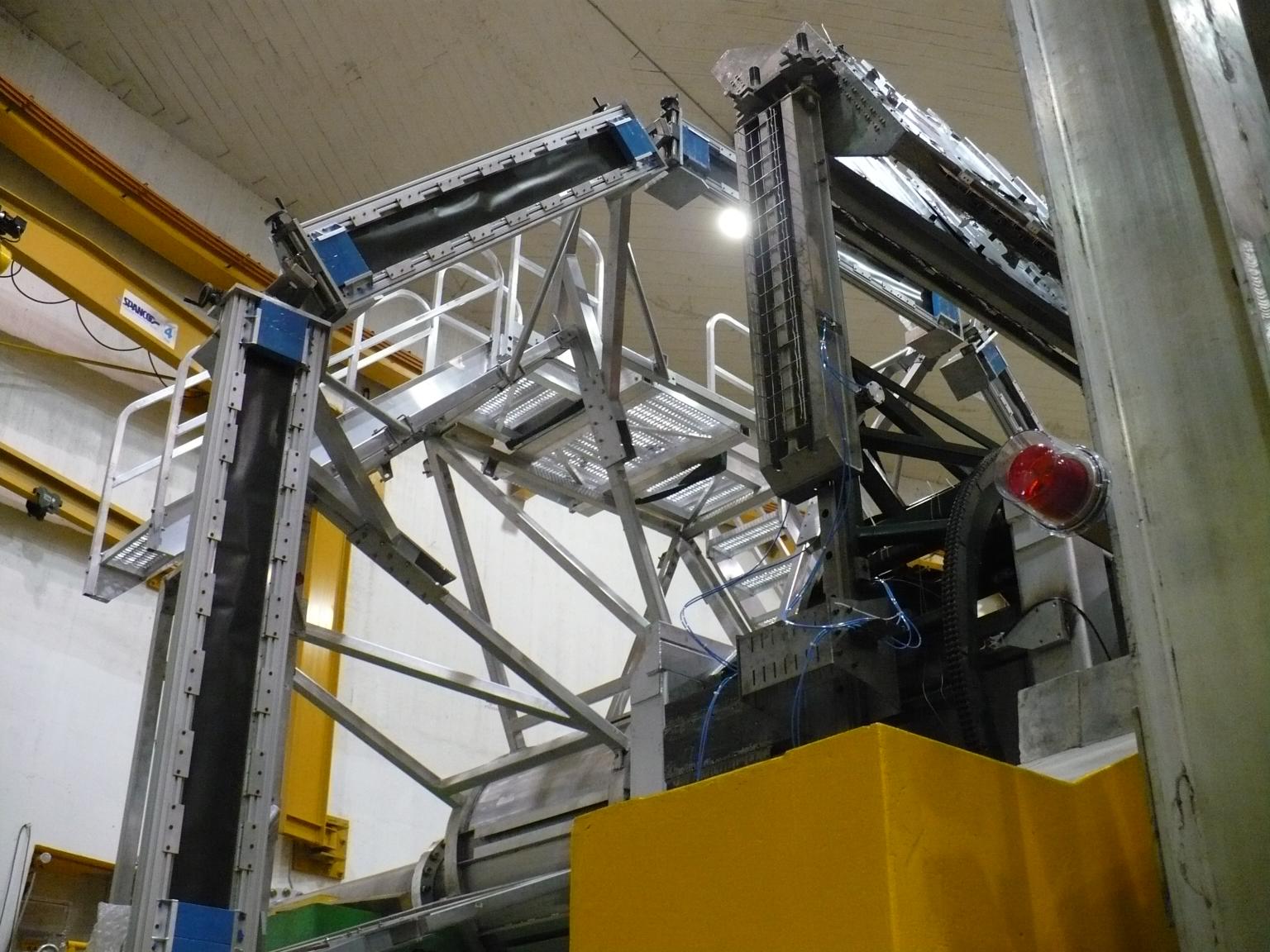}
\caption{Main detector bars seen during installation on the ``Ferris wheel'' support structure. Perspective is beam left looking downstream. Octants 1-3 are seen (see Figure \ref{fig:octant_coords} for explanation of coordinate terminology). Lead pre-radiators are not yet installed.}
\label{fig:md_ferris}
\end{figure}

The azimuthally symmetric main detector array for \Qs was designed to handle high rates with a high degree of linearity and low sensitivity to neutral backgrounds. The symmetric arrangement of the detectors also made their average response highly insensitive to shifts in beam position and angle. The main detector system comprised eight non-scintillating, quartz \v{C}erenkov bars, measuring 200~cm~$\times$~18~cm~$\times$~1.25~cm, arranged octagonally around the beam line at a radius of 335~cm. The quartz bars were sealed in light-tight enclosures and installed in a stiff aluminum support structure (see Figure \ref{fig:md_ferris}). The artificial fused-silica (Spectrosil 2000) bars were highly radiation hard and showed little yellowing over the course of the experiment even after sustaining a few thousand hours of rates as high as 900~MHz per bar. Figure \ref{fig:witness_plate} shows darkening of regular glass attached behind the main detector bars after only a few weeks of running. Each bar was composed of two 100~cm~$\times$~18~cm fused silica plates glued together in the center. An 18~cm~$\times$~18~cm~$\times$~1.25~cm waveguide glued to each end of the bar transported the light to 13~cm diameter PMT's which were glued to the downstream faces of the lightguides. 

A 2~cm thick lead pre-radiator was installed in front of each main detector bar to suppress low energy backgrounds. The pre-radiators increased light yield by a factor of 7 giving a signal to background improvement of 20 relative to test runs completed before the pre-radiators were installed. However, variation of shower size in the lead increased the main detector asymmetry width by 10\%. 

A bi-modal electronics readout chain allowed the main detector to operate in both high current (180~$\mu$A) and low current (50~pA) configurations. The high-current ``production mode'' for the experiment, designed for optimal accumulation of statistics, was an integrating configuration where each detector signal was digitally integrated and the average stored for each MPS. Low gain ($\sim10^3$) PMT bases with gain set close to 200, were used during high current running and the anode current was converted to voltage with a low noise, custom-made pre-amplifier. Low current or ``event mode'' was used with very low currents so that each event could be individually read out and the associated electron track from the target through the spectrometer could be reconstructed. Low current mode was used to check the alignment of the detectors and to verify simulated rates and event distributions on the bars. High gain ($\sim10^6$) PMT bases were installed for event mode running and individual pulses were digitized and stored. 

Digitization for the main detector signals (and many other diagnostic signals) was accomplished by custom built\footnote{Designed and built by TRIUMF in Vancouver, Canada.} 18-bit ADC's sampling at 500~kHz per channel.  

\begin{figure}[ht]
\centering
\includegraphics[width=0.6\textwidth]{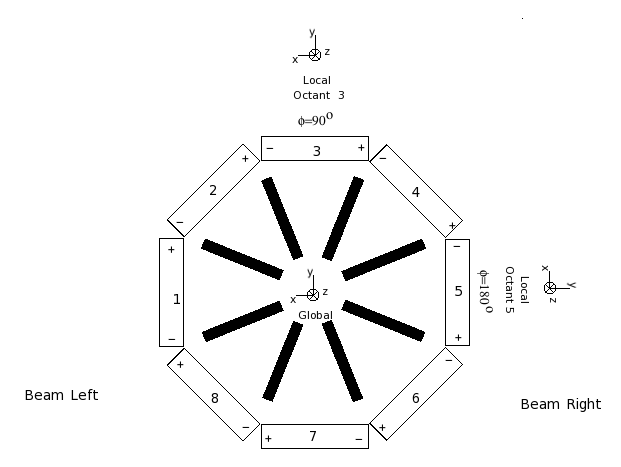}
\caption{Diagram of \Qs detector system looking downstream showing octant labeling. Numbered rectangles represent detector bars while solid spokes represent spectrometer coils.}
\label{fig:octant_coords}
\end{figure}

\section{Auxiliary Detectors}
A number of auxiliary detectors were added to the \Qs lineup to provide a variety of diagnostics. The main auxiliary detectors were the background detectors located inside the detector hut, a remotely controlled, movable focal plane scanner near main detector bar 7, upstream luminosity monitors (lumis) on the defining collimator and the downstream lumis 17~m downstream of the target. The design and use of each is discussed in the paragraphs ahead.
\begin{figure}[ht]
\centering
\includegraphics[width=0.7\textwidth]{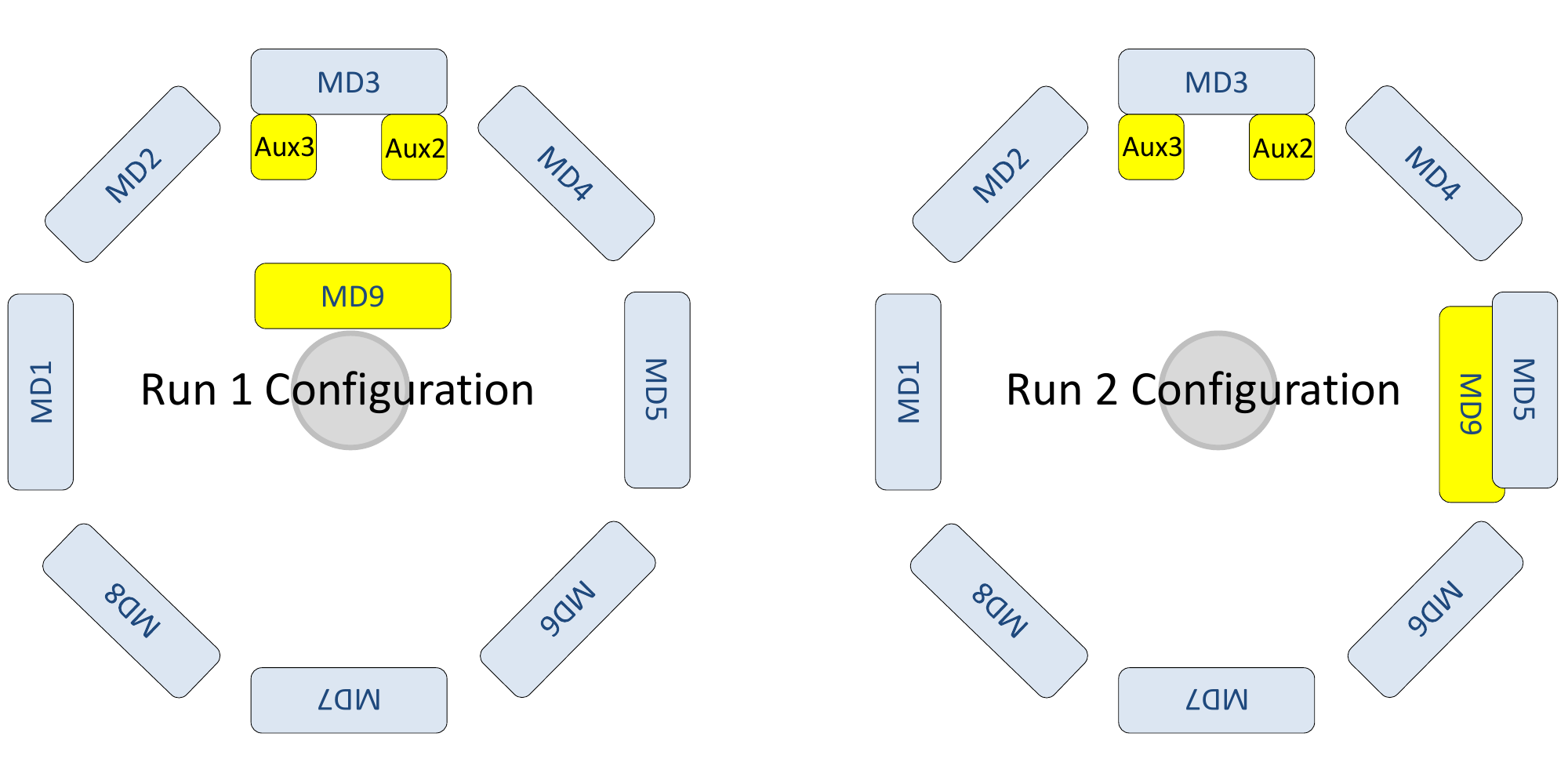}
\caption{Diagram showing the different positions of background detectors (yellow boxes) during Run 1 and Run 2. For a short period at the beginning of Run 1 the auxiliary detectors Aux2 and Aux3 were squeezed into the small spaces between detectors 5 and 6 and between detectors 1 and 8 respectively.}
\label{fig:background_det}
\end{figure}

\subsection{Background Detectors}
A set of detectors were installed inside the detector hut, outside the elastic envelope  to monitor backgrounds and to provide diagnostics for ruling out leakage of the helicity signal into the detector readout chain. The background detectors comprised a full \v Cerenkov detector bar assembly identical to the eight bars in the main detector array and three PMT's in dark boxes outfitted with LED light sources for testing, placed in various locations inside the detector shielding hut.

The background detectors were moved around during the early part of the \Qs experiment. Figure \ref{fig:background_det} summarizes the main configurations changes between Run~1 and Run~2. For the majority of the experiment (all of Run 2) the \v Cerenkov detector assembly called ``main detector 9'' or colloquially ``MD9'' since it was identical to the eight other main detector bars, was installed in the super-elastic region (smaller scattering angle than main detector acceptance) and slightly downstream of main detector bar 5 (MD5). The electronics and readout chain for MD9 were identical that of the main detector bars as well. Although MD9 was placed further into the super-elastic region than MD5 it also partially overlapped MD5 and derived part of its shower from events that passed through MD5, making its signal much more highly correlated with MD5 than with other main detector bars. 

The three ``dark box + PMT'' assemblies are referred to here as  Aux1-3. Aux1 remained near the floor in octant 7 and was illuminated with an LED to provide an anode current mimicking that of the main detectors during nominal production running. Aux1 is referred to colloquially as ``PMTLED''. PMTLED was fairly well shielded and was expected to have little response from scattering events. Its primary purpose was to verify that the main detector electronics chain was not picking up the helicity reversal signal. Aux2 was placed a meter downstream and on the super-elastic side of main detector 3 on beam right looking downstream, except for a short period at the beginning of Run 1 where it was installed in the gap between main detector bars 5 and 6. Aux2 was simply a PMT identical to the ones used on the main detector bars placed in a dark box and read out through the same signal chain as the main detectors. Aux2 was referred to colloquially as ``PMTONLY''. Aux3 was in the same position as Aux 2 except on beam left during most of the experiment with the exception of the same short period at the beginning of Run 1 when it was positioned in the space between main detector bars 1 and 8. It was composed of a PMT plus lightguide combination identical to those used in the main detector and was thus referred to as ``PMTLTG'' where ``LTG'' stands for ``lightguide''. Both PMTLTG and PMTONLY were used as background monitors for determining the background asymmetry contribution in the main detector. 

\begin{figure}[ht]
\centering
\includegraphics[width=0.6\textwidth]{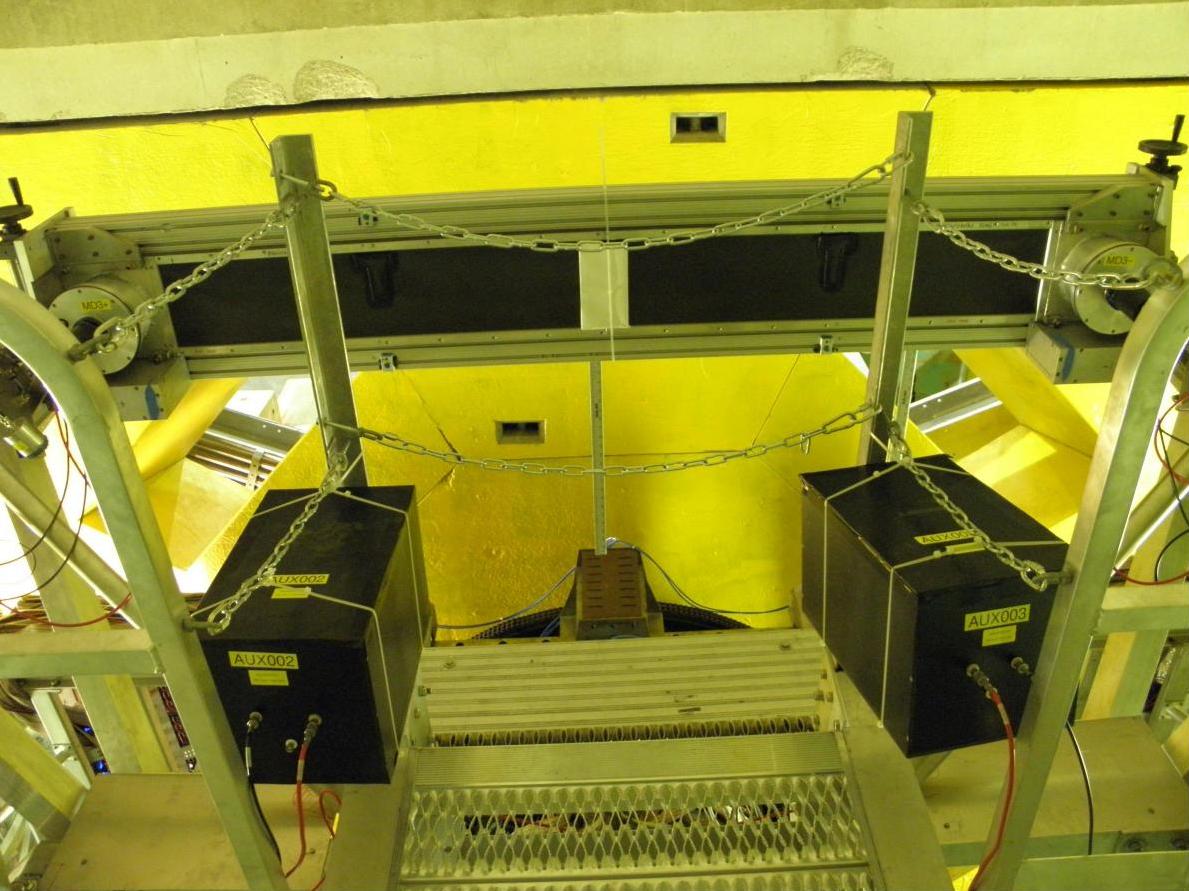}
\caption{Picture showing Aux2 (PMTONLY) and Aux3 (PMTLTG) installed behind and below main detector bar 3.}
\label{fig:ltg_and_onl}
\end{figure}

\subsection{Focal Plane Scanner}
\begin{figure}[h]
\begin{minipage}{0.45\textwidth}
\centering
\includegraphics[width=0.98\textwidth]{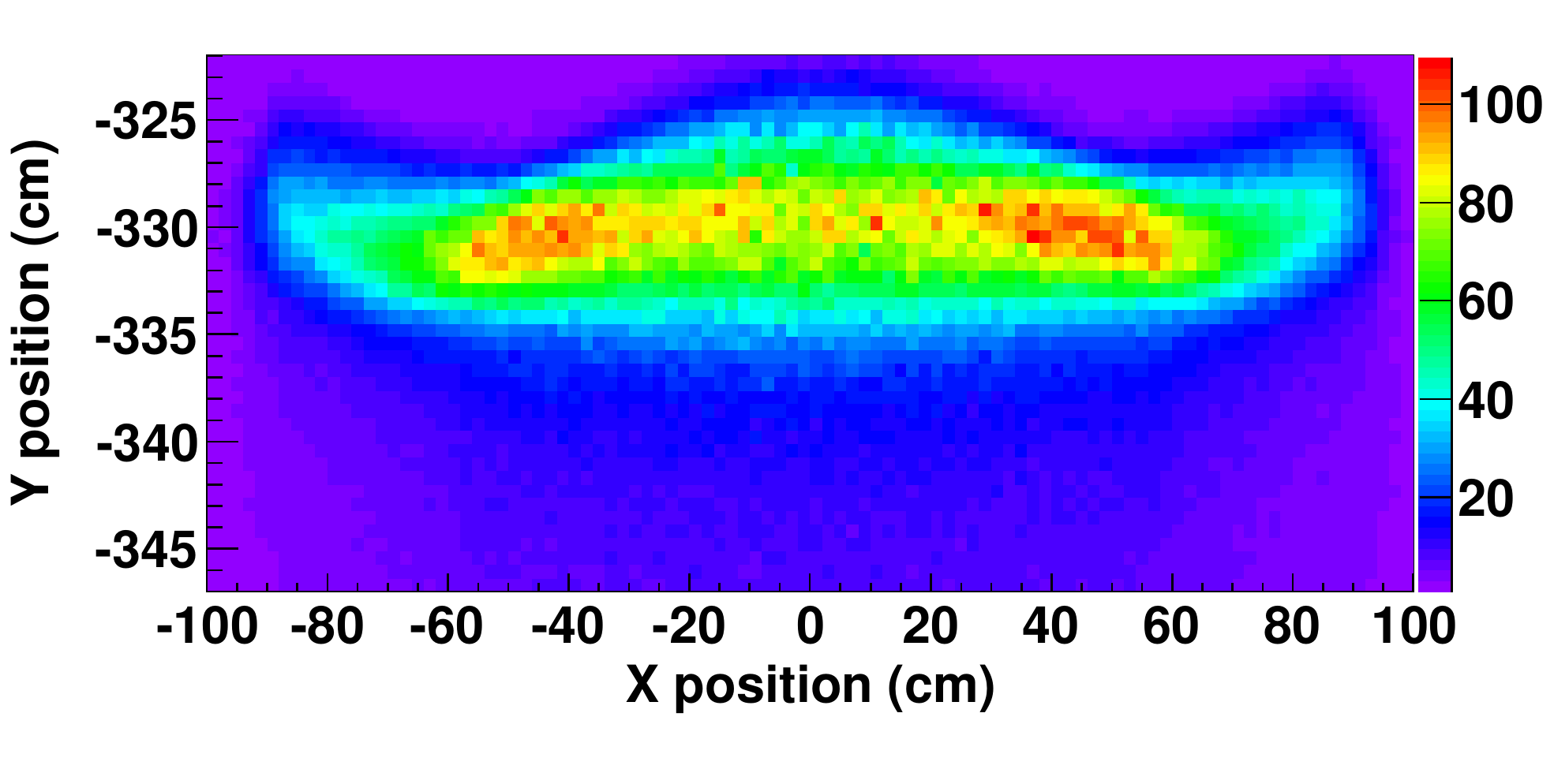}
\end{minipage}
\begin{minipage}{0.55\textwidth}
\centering
\includegraphics[width=0.98\textwidth]{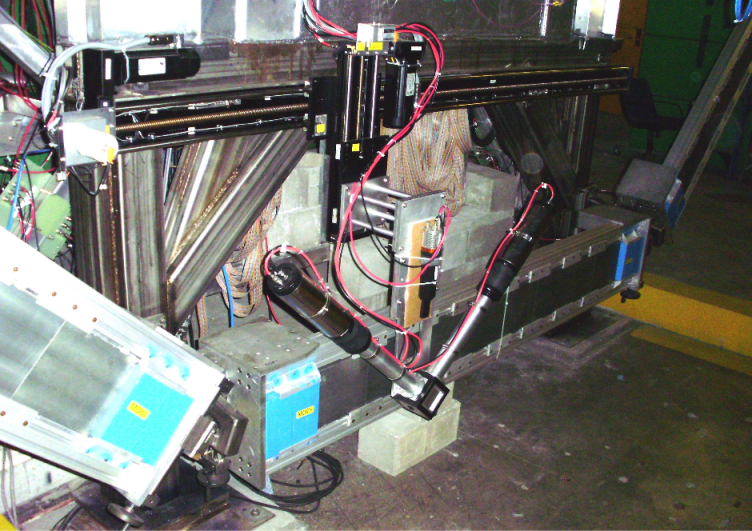}
\end{minipage}
\caption{(Left)Map of event distribution on detector bar 7 provided by the focal plan scanner showing ``mustache'' shape of elastic events focused by \qtor. (Right)Picture of the focal plane scanner shown installed in front of main detector bar~7. Two PMT and lightguide combinations were attached to overlapping fused silica crystals and the whole device could be scanned across the face of the detector bar to map out the distribution. A coincidence in both PMT's was required for an event to be recorded.}
\label{fig:scanner}
\end{figure}

A focal plane scanner was used to map the distribution of events on MD7 and served as a diagnostic tool for verifying expected rate distributions from simulation. This scanner utilized two \v Cerenkov detectors each with a 1~cm$^3$ artificial fused silica crystal and installed to overlap so a single electron would create a pulse in both detectors. The quartz crystal were attached to waveguides and PMT's read out in coincidence. With a maximum main detector flux estimated to be 1~MHz/cm$^2$, the scanner was designed to be run in pulse counting mode even during high current running. The light guides were arranged in a V pattern to minimize accidental coincidences. Motion controllers with position read-back allowed the scanner to be rastered in a pre-set pattern over the face of MD7. The scanner could be installed to move across either the front or rear faces of the detector bar in order to map out the rate distributions. Figure \ref{fig:scanner} shows a picture of the scanner installed upstream of main detector bar 7 and a typical rate distribution from the hydrogen target.

\subsection{Upstream Luminosity Monitors}
\begin{figure}[ht]
\centering
\includegraphics[width=0.5\textwidth]{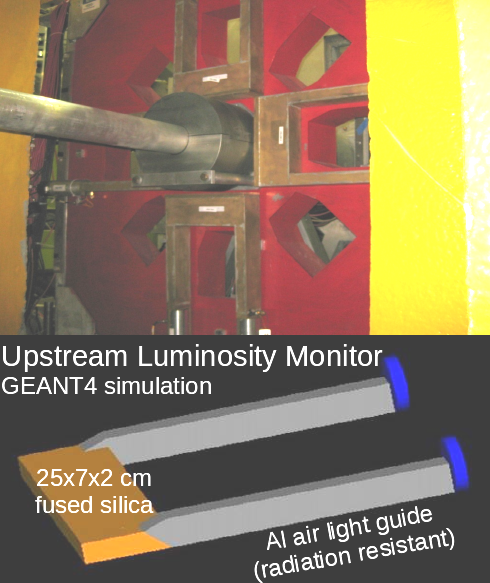}
\caption{Picture of four upstream lumis installed on the upstream face of the defining collimator. The lower section of the figure shows a GEANT4 generated drawing of the upstream lumi construction.}
\label{fig:uslumis}
\end{figure}

The upstream luminosity monitors (upstream lumis), installed 1.76~m downstream of the target on the upstream face of the defining collimator (see Figure \ref{fig:uslumis}), were originally designed to measure target density fluctuations (boiling) and to provide immediate beam diagnostics. In the end, other methods utilizing scans of raster size and pump speed (see Section \ref{sctn:target}) were considered to be effective at determining the effects of target boiling. The upstream lumis, positioned close to the beam line, were designed to measure low angle ($\sim 5^{\circ}$) scatterers (primarily Mott and M\o ller) in the target and were required to withstand much higher rates than the main detectors. By extrapolation from low current running they were determined to receive 115~GHz at nominal production current of 180~$\mu$A. 

The upstream lumis were composed of a 25~cm~$\times$ 7~cm~$\times$ 2~cm strip of Spectrosil 2000 fused silica connected to 35~cm, air-filled, reflective aluminum light guides on each end. The light guides were continuously flushed with nitrogen to prevent degradation of the aluminum reflective surfaces from moisture or contamination. A 5.1~cm Hamamatsu PMT, attached to each light guide, was coupled to a bi-modal electronics readout chain designed to run in either current or event mode similar to the main detector. In current mode, unity-gain PMT bases were attached to voltage pre-amplifiers providing a DC signal that was then digitally integrated by the same TRIUMF ADC modules used for the main detector. In event mode, medium-gain PMT bases were coupled to fast pre-amplifiers read out by scalers for individual pulse counting. In low-current/event-mode, the upstream lumi count rates were scaled to estimate beam currents well below the useful limit of the BCM's. 

Correlations between the upstream lumis and the main detector and background detectors provided evidence that a key background seen in the main detector originated in the tungsten collimator. The upstream lumis which turned out to receive a large component of their signal from the tungsten collimator were a critical tool in the diagnosis and removal of this unwanted background component\footnote{A detailed analysis of the removal of this background is expected in a future thesis \cite{Kargiantoulakis}.}. 

\subsection{Downstream Luminosity Monitors}
\begin{figure}[ht]
\centering
\includegraphics[width=0.5\textwidth]{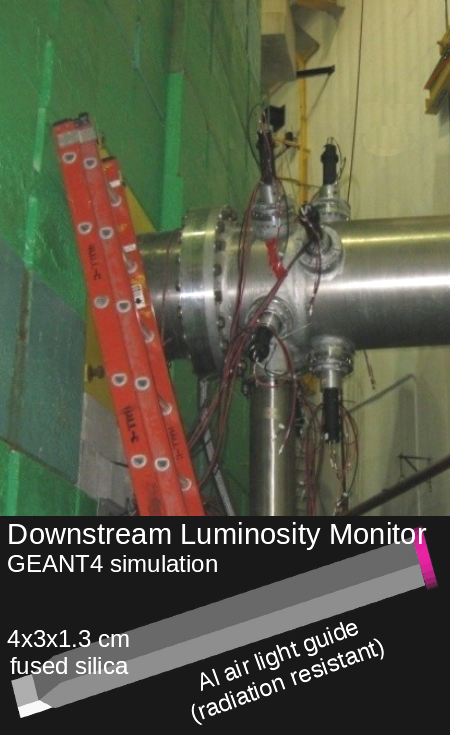}
\caption{Picture of the eight downstream lumis installed downstream of the main detector array. The lower section of the figure shows a GEANT4 rendering of the downstream lumi construction.}
\label{fig:dslumis}
\end{figure}

The eight downstream luminosity monitors (downstream lumis) were installed 17~m downstream of the target (see Figure \ref{fig:dslumis}) in an azimuthally symmetric pattern around the beamline. They were originally designed to be a measure of the main detector null asymmetry and to provide immediate beam diagnostics. The downstream lumis, which penetrated the beam enclosure, were designed to measure low angle ($\sim 0.5^{\circ}$) scatterers (about equally sensitive to Mott and M\o ller events) in the target. Their proximity to the electron beam combined with the tight acceptance of the tungsten collimator prevented the downstream lumis from sampling scattering events in the region near the upstream target window. Their sensitivity to beam position made the downstream lumis useful monitors of beam position after the target. They received rates of 150~GHz at nominal production current of 180~$\mu$A.

The eight downstream lumis were each composed of a 4~cm~$\times$ 3~cm~$\times$ 1.3~cm strip of Spectrosil 2000 fused silica connected to a single 35~cm, reflective, nitrogen-flushed, aluminum light guide. Like the upstream lumis, PMT's attached to each light guide were coupled to a bi-modal electronics readout chain designed to run in either high current or event mode. Further details of the design and operation of both the downstream and upstream luminosity monitors can be found in \cite{Leacock2012}. 

\section{Tracking System}
The tracking system for \Qs was a set of detectors designed to provide track reconstruction for individual scattered electrons from their scattering vertex in the target to the main detector bars. The tracking system was used to characterized the main detector and map out its response, compare with/verify simulation and to reconstruct the average $Q^2$ and scattering angle of the experiment. It was divided into three regions. Region I between collimators 1 and 2 was intended to house a gas electron multiplier (GEM) detector for use in accurate vertex reconstruction but it was never operational during the experiment. Region II, located between the defining and 3rd collimators, housed a pair of horizontal drift chambers (HDC's) for vertex reconstruction and determination of scattering angle. Region III, located downstream of the spectrometer in the shielded detector hut near the focal plane, housed the vertical drift chambers (VDC's) that were used to accurately determine particle position and angle on the detector bars. Also in Region III were a set of trigger scintillators used during event mode to trigger events that would hit the detector bars.

For elastic ep scattering, simple relativistic kinematics neglecting the electron mass and radiative effects, relates the incoming electron energy $E$ to the outgoing scattered electron energy $E^{\prime}$ in terms of the lab frame scattering angle $\theta$ and the proton mass $m_p$ as
\begin{equation}
E^{\prime} = \frac{E}{1+2\frac{E}{m_p}\sin^2\frac{\theta}{2}}.
\label{eq:Eprime}
\end{equation}
The four momentum transferred from the electron to the proton is given by
\begin{equation}
  Q^2=\frac{4E^2\sin^2\frac{\theta}{2}}{1+2\frac{E}{m_p}\sin^2\frac{\theta}{2}}.
\label{eq:Qsquared}
\end{equation}

From this equation it appears that the only information required from the tracking system to completely determine the $Q^2$, is the scattering angle $\theta$ since the beam energy is known accurately from dedicated energy measurements described in Section \ref{sctn:btandm}. However, $\theta$ must be accurately acceptance-averaged and corrections must be applied for radiative effects in the target both before and after scattering requiring that the final $\langle Q^2 \rangle$ for the experiment be derived from simulation. The tracking system measurements are used to verify that the simulation is both correct and well understood. 

Both the HDC's and the VDC's are multi-wire, drift chambers designed to reconstruct a 3-dimensional electron trajectory. A plane of taut, evenly spaced sensing wires are kept near zero potential in a uniform electric field produced by a cathode held at a large negative potential. The entire chamber is flushed with a gas mixture chosen, among other characteristics, for its ionizability. An electron passing through the chamber creates a track of positive ions and electrons which then migrate along the electric field lines with the electrons accelerating towards the sensing wires. The strong electric fields near the wires accelerate the low energy ionization electrons to the point where they ionize other gaseous atoms causing a shower of as many as $10^6$ electrons. A pulse of measurable size is produced by the shower on the sensing wire. Readout is triggered by a set of plastic scintillators sensitive only to charged particles located near the detector focal plane. Time-to-digital converters (TDC's) were used to measure the time delay between the trigger and the pulse arrival on a given wire. Known drift times for the chambers provide the needed temporal-to-spatial conversions.   

The terms ``horizontal'' and ``vertical'' are derived historically from their typical orientations in previous experiments and are not descriptive of their orientations during the \Qs experiment. There are a few distinguishing features between HDC's and VDC's. HDC's are designed such that the ions drift parallel to the wire planes whereas the ions drift nearly perpendicular to the wires planes in a VDC. VDC's are typically designed to have the incident electron angle near $45^{\circ}$ whereas HDC's are installed such that the incident electron track is nearly normal to the wire plane. Also, drift times in HDC's are typically shorter allowing higher incidence rates. 

\begin{figure}[ht]
\centering
\includegraphics{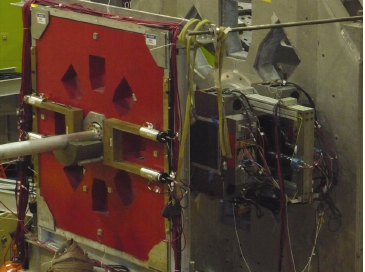}
\caption{Picture of HDC's during installation in Region II. In this orientation the HDC's would measure tracks in octants 1 and 5.}
\label{fig:HDCs}
\end{figure}

The HDC's used in \Qs were designed to have a large angular acceptance and to receive high rates since they were located in Region II before the spectrometer swept away the M\o ller events. During nominal running conditions in \Q, M\o ller rates in Region II were expected to exceed the ep scattering rates by a factor of 500. The HDC's consisted of two sets of drift chamber pairs each with an active area of 28~cm $\times$ 38~cm, and were located symmetrically on each side of the beam. The entire assembly of 4 drift chambers was attached to a mechanism designed to rotate by $\pm90^{\circ}$ to measure all octants. On either side of the beamline were two identical drift chamber assemblies with the second located 42~cm downstream from the first to provide angular resolution. Each drift chamber had 6 wire planes XUVX$^{\prime}$U$^{\prime}$V$^{\prime}$, with the wires in U and V at $\pm53.1^{\circ}$  relative to X. Measured position resolution for each chamber was in the 150-200~$\mu$m range giving an angular resolution of 0.6~mrad. A picture of the HDC's installed in Region II is shown in Figure \ref{fig:HDCs}.

\begin{figure}[ht]
\centering
\includegraphics[width=0.5\textwidth]{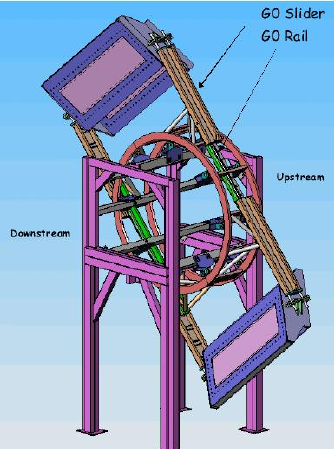}
\caption{Drawing of VDC's on rotator mechanism.}
\label{fig:VDC_rotator}
\end{figure}

The VDC's also consisted of two pairs or ``packages'' of drift chambers arranged symmetrically on opposite sides of the beamline fixed to a mechanism which allowed $\pm90^{\circ}$ rotation. The arms of the rotator also allowed for two radial positions, a retracted position close to the beamline for high current mode when the drift chambers were not being used and an extended position near the front of the main detector bars. A drawing of the rotator mechanism can be seen in Figure \ref{fig:VDC_rotator}. Each package has two drift chambers located 53~cm apart in the beam direction.

\begin{figure}[ht]
\centering
\includegraphics[width=0.8\textwidth]{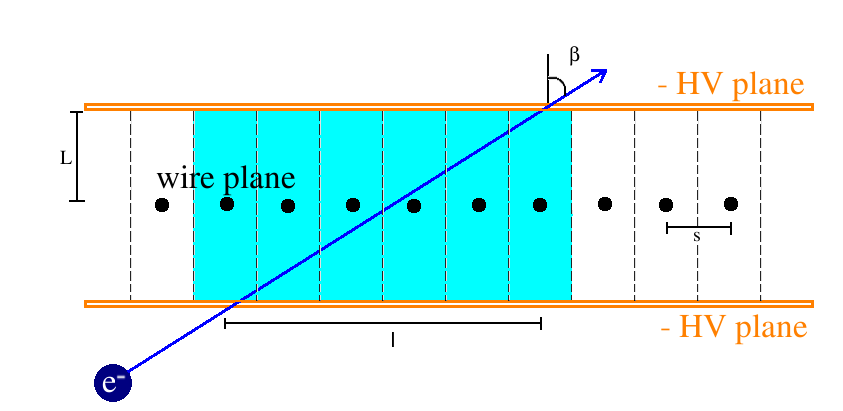}
\caption{Illustration of electron track intersecting a single wire plane. L, s, l and $\beta$ are parameters used in the tracking analysis. }
\label{fig:VDC_track}
\end{figure}

Each of the four VDC drift chambers housed two wire planes with the wires strung at $\pm26.6^{\circ}$ relative to the long axis of the chambers. The active area of each chamber was 53.3~cm~$\times$~204.5~cm. A typical electron track at $45^{\circ}$ to the chamber triggered 6 wires in a given plane (see Figure \ref{fig:VDC_track} for illustration) with a maximum of 8 wires at the largest possible angle accepted. The position resolution of a single wire was in the 265-295~$\mu$m range. Using a separation of 53~cm between the two chambers in a given package and an average of 12 hits per chamber gives a naive angular resolution of about 0.15~mrad.

The 2232 wires in the VDC's were read out using a custom-made multiplexing electronics readout which allowed sequential readout of multiple wires giving a factor nine fewer total TDC channels needed. Further details about the VDC's and track reconstruction can be found in \cite{Leckey2012}.

Track reconstruction was performed with an offline analysis using track pattern recognition algorithms. Straight line trajectories upstream and downstream of the spectrometer obtained from the HDC's and VDC's respectively, were crudely matched to each other using the known spectrometer field and an initial guess for the scattered electron energy $E^{\prime}$.  The guess for  $E^{\prime}$ was iteratively improved until the trajectories converged.
\begin{figure}[ht]
\centering
\includegraphics{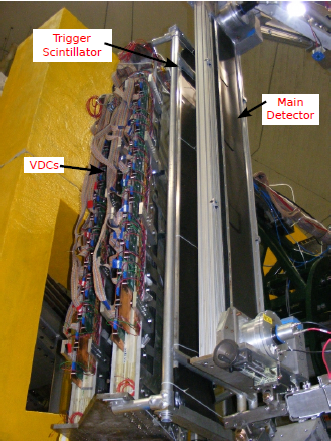}
\caption{Picture of VDC's installed in Region III looking upstream towards spectrometer.}
\label{fig:vdcs}
\end{figure}


\chapter{Beam Corrections} 
\captionsetup{justification=justified,singlelinecheck=false}

\label{Chapter4}

\lhead{Chapter 4. \emph{Beam Corrections}}

The parity-violating $ep$ scattering asymmetry is expected to be approximately 220~ppb at the kinematics of the \Qs experiment. The proposed goal of \Qs is to measure this asymmetry to within a few percent. This level of accuracy requires rigorous methods for distinguishing between the true parity-violating asymmetry and other sources of helicity-correlated asymmetric signals in the detectors. These false asymmetries must be measured and either removed or an error assigned to them. One of the dominant sources of false asymmetries arises from helicity-correlated beam properties. The focus of this chapter is to identify the false asymmetries arising from helicity-correlated properties of the electron beam and to explain how they are measured and removed in \Q.    

\section{Helicity-Correlated False Asymmetries}
Helicity-correlated (HC) beam properties are those properties of the electron beam whose measurement is correlated with the helicity state of the electrons. Any physical property of the electron beam other than helicity that changes with the electron helicity is a potential source of false asymmetry. As already mentioned in Section \ref{sctn:electron_source}, parity-violating experiments such as \Qs make use of a number of slow helicity reversals at different timescales in addition to the fast reversal to diagnose and remove false asymmetries. The fast reversal is accomplished using a Pockels cell to quickly reverse the source laser helicity and for \Qs the fast reversal was set at 960~Hz. A slow helicity reversal using an insertable half-wave plate (IHWP) before the Pockels cell in the source laser was used to reverse the electron helicity relative to the Pockels cell high voltage state and was changed on an 8~hour timescale. Finally, the double Wien filter was used to reverse the electron beam helicity relative to its state coming off the photo-cathode and was changed about once per month. 

Two categories of false asymmetries typically distinguished are those that cancel with IHWP reversal and those that do not. The false asymmetries that cancel with IHWP are those that have the same sign regardless of whether the IHWP is in or out of the beam, thus adding to one state and subtracting from the other and cancelling in the average. The false asymmetry that does change sign with IHWP is perhaps more subtle since it mimics the parity-violating physics asymmetry and does not cancel. False asymmetries arising from intensity asymmetries on the electron beam (beam current is systematically different between helicity states) can be readily measured and minimized. Residual intensity asymmetries are cancelled to first order by normalizing detector signals to current. False asymmetries from helicity-correlated trajectory and energy shifts on the beam depend upon individual detector sensitivities to the beam properties. Measurement of these sensitivities and correction of false asymmetries related to beam trajectory and energy constitute the subject material of the next two chapters. Further discussion of false asymmetries on the electron beam, especially those arising in the electron source, can be found in \cite{Paschke2007}.

Sensitive parity-violating experiments like \Qs employ a three-fold strategy for dealing with false asymmetries: 1) minimization, 2) cancellation and 3) correction for residual false asymmetries. The first two have to do with the experimental setup and procedure whereas the last is purely a set of analysis techniques. The main focus of this chapter is on 3, the analysis techniques used for removal of residual false asymmetries after techniques 1 and 2 have been implemented during the experiment.  
\subsection{Minimizing False Asymmetries}
Ideally, a parity-violating  experiment is set up so that false asymmetries are small relative to the physics asymmetry being measured so that additional means such as cancellation and first order corrections are effective. The following list, although not intended to be exhaustive, shows how a number of false asymmetries were minimized in both the design and operation of \Q.  
\begin{itemize}
\item {\it Alignment and azimuthal cancellation:} The \Qs experiment was designed to be azimuthally symmetric about the electron beam. The \qtor spectrometer focused elastic $ep$ scattering events onto eight quartz \v Cerenkov detector bars. Any single detector is highly sensitive to beam trajectory changes, but the average of the eight detector array cancels this sensitivity to first order due to its intrinsic azimuthal symmetry. Imagine, for example, the beam position moving to the left horizontally. The detectors on the left will experience a rate change that will be cancelled by the detectors on the right. Care was taken during the \Qs commissioning period to find the so called ``neutral axis'' of the \qtor spectrometer, that is, the beam trajectory was chosen such that the average detector sensitivity to position is minimized. Higher order effects do not necessarily cancel and any broken azimuthal symmetry such as octant to octant variations in the spectrometer field or small detector misalignments produce imperfect cancellation. 

\item {\it Source setup:} A schematic of the electron source can be seen in Figure \ref{fig:sourcetable}. High voltages across the Pockels cell used for flipping the helicity of the laser, create stresses on the crystal which can create helicity-correlated changes in the beam spot intensity, position and shape. Simply switching the direction of the high voltage across the Pockels cell does not produce exactly opposite helicity states. Furthermore, perfect circular polarization after the Pockels cell does not necessarily translate into perfect circular polarization at the cathode. Birefringent elements such as the vacuum window between the Pockels cell and the photo-cathode (see Figure \ref{fig:sourcetable}) create residual linear polarization at the photo-cathode. The strain on the GaAs crystal creates an analyzing power along the cathode strain axis for linearly polarized photons. A two-pronged approach minimizes effects from residual circulation polarization on the photo-cathode. First, a rotatable half-wave plate upstream of the vacuum window allows the axis of the residual polarization on the laser to be rotated relative to the photo-cathode analyzing axis. A prudent choice of RHWP angle will zero charge asymmetry on the beam from this effect. Second, a slightly different voltage can be set for each high voltage state of the Pockels cell. This relative small voltage offset termed a ``PITA'' (\textbf{P}olarization \textbf{I}nduced \textbf{T}ransport \textbf{A}symmetry) voltage is carefully selected to minimize charge asymmetries off the cathode. 

A series of Pockels cell position and angle scans were completed to find the optimal placement that minimizes HC deflections. Position differences on the laser are measured using a photodiode array and zeroed at the few micron level. With an ideal optical setup in the accelerator and injector, an effect called ``adiabatic damping'' reduces the beam emittance (a measure of the width of position and angle distributions in the electron beam) by a factor of $\sqrt{p_0/p}$ between the injector and the hall \cite{Paschke2007}. For the \Qs experiment with a 1.16~GeV beam and a 100~keV injector ($p_0 = 335$~keV) the position differences could be reduced in the hall by as much as a factor of 54 relative to the sourc
\item{\it Charge feedback:} During \Qs the charge asymmetry in the experimental hall was continuously monitored by a charge feedback system which adjusted the PITA voltage in real time to zero the charge asymmetry. The plot in Figure \ref{fig:charge_feedback} shows the charge asymmetry over time as the feedback system reduced the charge asymmetry.
\begin{figure}[ht]
\centering
\includegraphics[width=0.7\textwidth]{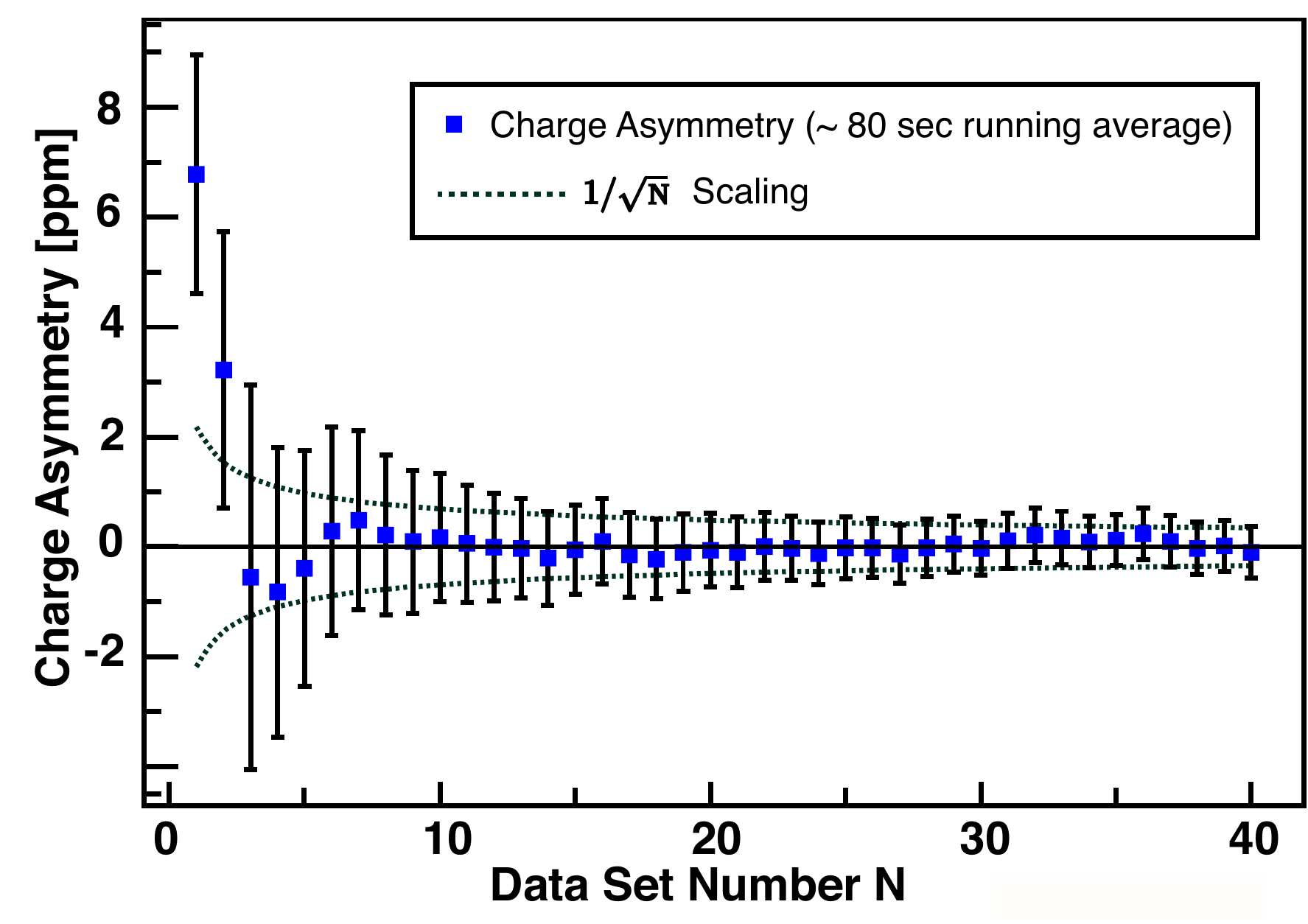}
\caption{\label{fig:charge_feedback}Plot of charge asymmetry versus time over typical 1 hour run during \Qs demonstrating the action of the charge feedback system.}
\end{figure}

\item {\it Elimination of electrical pickup:} Electronics carrying the helicity signal must be isolated from the data acquisition electronics to avoid contaminating the asymmetry measurements. The source electronics which create and utilize the helicity timing signals are carefully isolated from the electronics in the experimental halls at Jefferson Lab. The helicity of the electron beam is flipped in either a $+--+$ or $-++-$ pattern, chosen to cancel linear drifts, with the sign of the first event in a quartet being determined by a pseudo-random sequence. Given the potential for contaminating the asymmetry measurement with electronics pickup from the helicity signal, \Qs utilized delayed reporting of the helicity, that is, the sign of the helicity signal sent to the data acquisition electronics was delayed by eight quartet patterns to ensure that the reported helicity was entirely uncorrelated with the actual beam helicity.

To check that no helicity-correlated signals were getting to the DAQ electronics a battery was connected to one of the channels being read out. Since the battery signal was read through the same \Qs electronics chain as the physics measurement, any asymmetry on that channel would be the signal of a false asymmetry from electronics pickup. Figure \ref{fig:battery_asym} shows the asymmetry of the battery for a typical run.
\begin{figure}[ht]
\centering
\framebox{\includegraphics[width=3in]{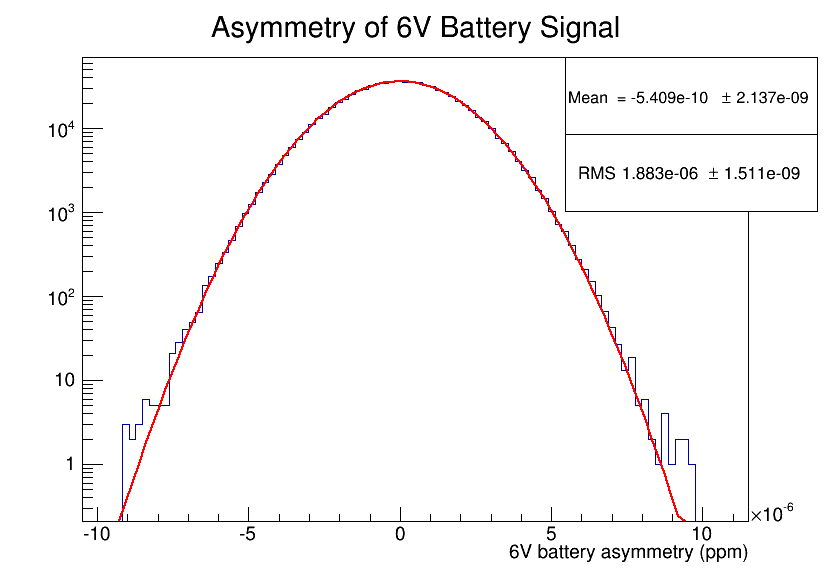}}
\caption{Asymmetry of a 6V battery for a typical run as read out by the \Qs DAQ electronics.}
\label{fig:battery_asym}

\end{figure}

\item {\it Helicity Magnets:} For part of Run 2 the HC electron beam position differences were also suppressed by utilizing a set of helicity magnets in the injector region (see Figure \ref{fig:sourcetable}). These magnets applied a carefully calibrated helicity-correlated ``kick'' to the electron beam at the fast reversal frequency.  

\end{itemize}  
 
\subsection{Cancelling False Asymmetries}   
\Qs was designed to provide cancellation of small residual false asymmetries by the slow reversals already discussed. The pseudo-random reversal sequence cancelled helicity-correlated effects near the quartet frequency (240~Hz). The IHWP slow reversal was used to cancel effects correlated with the high voltage state of the Pockels cell. Finally, the double Wien filter was used to cancel a family of false asymmetries that flip sign with IHWP.

\subsection{Correcting for False Asymmetries}   
In principle, any property of the beam that can change rapidly with time has the potential to create a false asymmetry. For example, beam position, angle, energy, halo, polarization, intrinsic spot size, shape and intensity can all, hypothetically, create helicity-correlated false asymmetries. In practice, the four main beam properties that can be measured with sufficient speed and accuracy to make useful helicity-correlated beam correction are beam energy, position, angle and intensity. As previously mentioned, issues with beam intensity asymmetries are minimized by feedback and normalization. Two correction techniques have been applied to the \Qs data set to remove false asymmetries from HC beam trajectory and energy differences. Multivariate linear regression measures and removes detector sensitivity to beam parameters using ``beam jitter'' or variations in the parameters that naturally exist on the electron beam. ``Dithering'' or beam modulation measures the detector sensitivities with a different technique that makes use of driven beam motion, that is, intentional driven fluctuations in the beam parameters\footnote{The terms ``modulation'', ``beam modulation'' and ``dithering'' are used interchangeably throughout the text and all refer to driven beam motion. The terms ``beam jitter'', ``jitter'' and ``natural beam motion'' are also used interchangeably and refer to naturally occurring motion on the beam as opposed to driven motion.}. Due to the importance of these topics a full section will be devoted to each.

The final asymmetry reported by the \Qs experiment will be given as the straight average of the asymmetries formed from the 16 main detector PMT's. This straight average, referred to as ``PMT Average Asymmetry'' will be used in the analysis ahead where convenient. However, no PMT average ``yields'' or ``differences'' exist since both of these require intrinsic PMT-weighting (see Appendix \ref{AppendixE} for detailed definitions of these terms). Whereas linear regression operates directly on asymmetries, the beam modulation analysis measures sensitivities on detector yields. For this reason, the beam modulation analysis uses a weighted average of the main detector PMT's called MDallbars. MDallbars yields are used to calculate asymmetries which, in general, are very close to the PMT Average asymmetries. For the beam modulation analysis, the final corrections are calculated for the MDallbars asymmetries for consistency. It is important to note that final corrections calculated for MDallbars asymmetries are very close to those found for PMT Average asymmetries differing by only 0.2~ppb for Run 1 and 0.1~ppb for Run 2.  Appendix \ref{AppendixE} provides definitions of ``yield'', ``asymmetry'', ``MDallbars'' and ``PMT Average''.


\section{\label{sctn:lin_reg}Linear Regression to Remove Helicity-Correlated False Asymmetries}
Linear regression utilizes the natural fluctuations in beam parameters (energy and trajectory) to determine and remove the sensitivity to these fluctuations from the data. As the name implies, any changes in the natural beam parameters are assumed to be linearly correlated to changes in the detectors. 

Linear regression in the context of \Qs is used to remove false asymmetries arising from changes in beam properties from the measured detector asymmetries. Correlations are found between detector asymmetries and helicity-correlated monitor differences, defined as the half the monitor difference between helicity states. The terminology used here is as follows: any given uncorrected detector asymmetry is $A_d$, the actual parity-violating asymmetry is $A_{PV}$, the helicity-correlated monitor differences are $\Delta X_i$ and the correction slopes are $B_i$.

Consider a model where a given detector asymmetry, $A_d$, includes the ``true'' detector parity-violating asymmetry, $A_{PV}$, that we want to measure plus other spurious signals that are linearly related to $N$ beam properties. We can express such a relationship as 
\begin{equation}
A_d=A_{PV} + \sum_{n=1}^N B_n(\Delta X_n).
\label{eq:lin_reg}
\end{equation}
Under this model, we can find the ``true'' detector asymmetry, $A_{PV}$, if we can determine the detector sensitivities, $B_n$, and the beam differences, $\Delta X_n$, as seen here:
\begin{equation}
 A_{PV}= A_d - \sum_n^k B_n(\Delta X_n).
\label{eq:lin_reg_correction}
\end{equation}
In principle, if we have $k$ orthogonal beam parameters which we can measure precisely, this is a trivial correction to make with the $B_i$ simply equal to the partial derivatives $\frac{\partial A_d}{\partial (\Delta X_i)}$. In practice, however, we are dealing with N correlated beam monitors that span the space of the $k$ orthogonal beam parameters but not equally well in all the  $k$ dimensions. In fact, some of the  $k$ parameters may not be well measured at all. For example, in the \Qs experiment, the resolution for natural beam position motion on target was much better than the resolution of natural beam angle shifts.

\begin{figure}[ht]
\centering

\framebox{\includegraphics[width=5.5in]{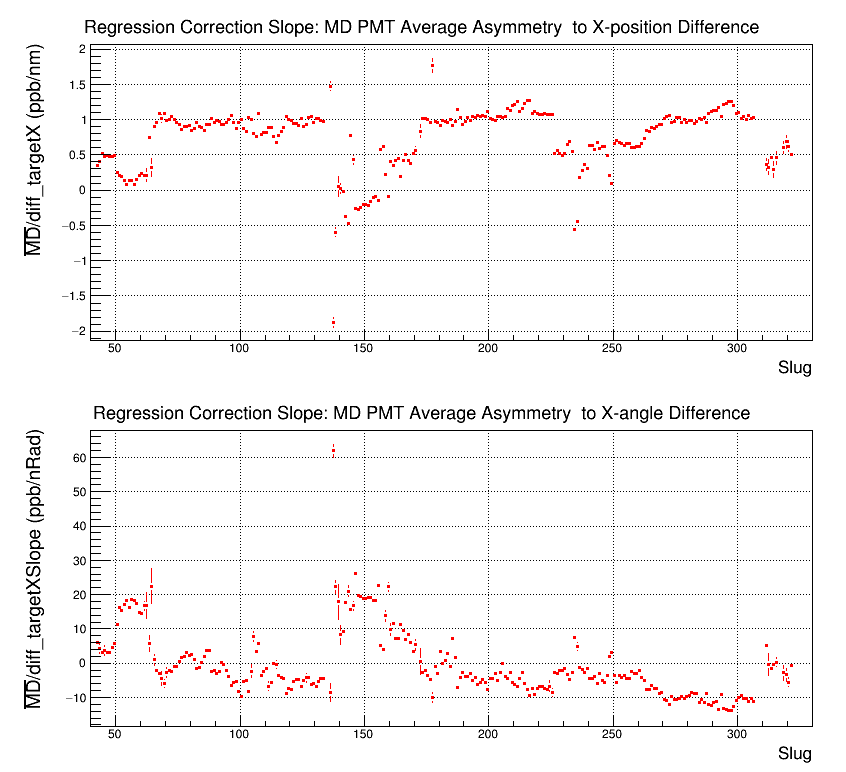}}
\caption{Set 11 regression correction slopes for horizontal position and angle on target averaged over slugs ($\sim 8~hrs$).}
\label{fig:Set11_X_slopes}
\end{figure}

\begin{figure}[ht]
\centering
\framebox{\includegraphics[width=5.5in]{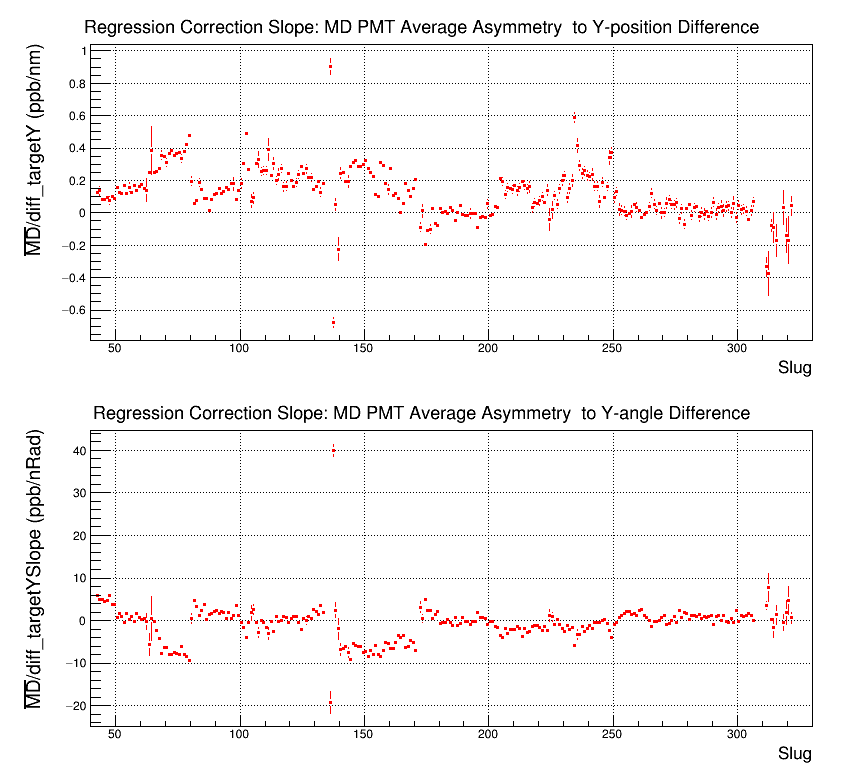}}
\caption{Set 11 regression correction slopes for vertical position and angle on target averaged over slugs ($\sim 8~hrs$).}
\label{fig:Set11_Y_slopes}
\end{figure}

\begin{figure}[h]
\centering
\framebox{\includegraphics[width=5.5in]{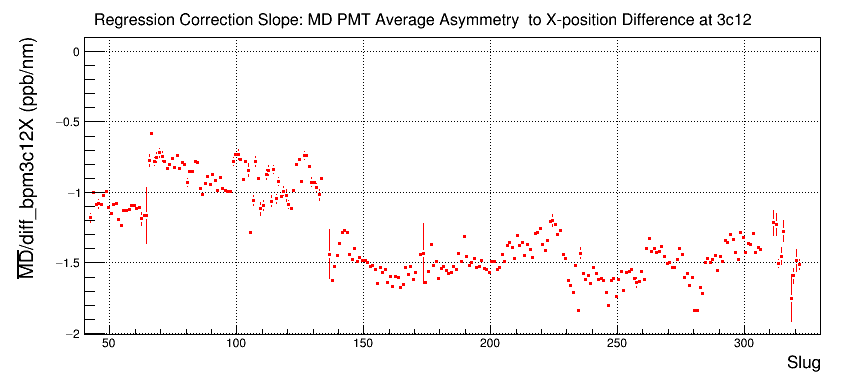}}
\caption{Set 11 regression correction slopes for horizontal position at bpm3c12 averaged over slugs ($\sim 8~hrs$). This is the most energy sensitive monitor on the beamline.}
\label{fig:Set11_E_slopes}
\end{figure}

One method for determining the slopes $B_n$ typically used in linear regression analyses is obtained by minimizing the $\chi^2$ statistic,
\begin{equation}
\chi^2=\sum_{i}\left(\frac{(A_{d}^0)_i- \sum_n^k B_n(\Delta X_n^0)_i}{\sigma_i}\right)^2,
\label{eq:lin_reg_chi_sqare}
\end{equation} 
with respect to the slopes, $B_n$, where the sum over $i$ goes over the measured quartet asymmetries and $\sigma_i^2$ is the variance of the i'th asymmetry measurement. Here the superscript 0's indicate that all detectors and monitors have been zero-centered\footnote{ $A_d^0=A_d-\overline{A_d}$ and $\Delta X_n^0=\Delta X_n-\overline{\Delta X_n}$} since the minimization is meant only to find the best slopes to model {\bf changes} in asymmetry with {\bf changes} in monitor differences. This minimization yields
\[
\frac{\partial \chi^2}{\partial B_m}=0\Longrightarrow2\sum_{i}\left(\frac{(A_{d}^0)_i-\sum_n^k B_n(\Delta X_n^0)_i}{\sigma_i}\right)\left(\Delta X_m^0\right)=0
\]
Assuming that the variance is constant for each of the  measurements, this reduces to 
\begin{equation}
\sum_{i=1}^N(A_{d}^0)_i\left(\Delta X_m^0\right)_i=\sum_{i=1}^N\sum_{n=1}^kB_n\left(\Delta X_n^0\right)_i\left(\Delta X_m^0\right)_i.
\label{eg:lin_matrix_equation}
\end{equation}
This can be expressed in matrix form as 
\[
{\mathbf D} = \mathbf{XB},
\]
where the entries are given by
\[
D_{m}=\mathrm{Cov}[A_{d},\Delta X_m]
\]
\[
X_{m,n}=\mathrm{Cov}[\Delta X_m,\Delta X_n],
\]
and $\mathbf{B}$ is a vector of detector to monitor correction slopes. The slopes are thus given by
\begin{equation}
\mathbf{B}=\mathbf{X^{-1}D}
\label{eq:lin_reg_slopes}
\end{equation}
For the \Q~experiment, detector to beam monitor slopes were calculated from approximately 5 minutes of data taken at 240 quartets per second. A number of regression schemes were chosen using different monitor sets against which to regress the main detector. The monitor set that best measures (or spans the vector space in the language of linear algebra) the five independent beam parameters will be expected to give the most accurate results. One particular monitor set, called set 11 in the jargon of \Q, uses four target variables and beam position monitor 3c12X defined as follows:\\
\begin{itemize}
\item {targetX(Y): electron beam horizontal(vertical) position measured by extrapolating positions from beam position monitors (BPM's) in the drift region before the target downstream to the target position.}
\item {targetXSlope(YSlope): electron beam horizontal angle(vertical angle) relative to the ideal beam axis and measured by finding differences between BPM's in the drift region before the target.}
\item{BPM3c12X: the X or horizontal measurement of the BPM located in the region of highest dispersion of the electron beam in the arc leading into Hall C. BPM3c12X is highly sensitive to energy shifts and is often referred to as our ``energy monitor''.}
\end{itemize}
This particular set of BPM's is interesting because it is the same set used in the beam modulation analysis, a parallel analysis using driven beam motion to determine the correction slopes. Figures \ref{fig:Set11_X_slopes}, \ref{fig:Set11_Y_slopes}  and \ref{fig:Set11_E_slopes}  show the Set 11 regression slopes used to correct the \Qs data set. Figure  \ref{fig:Set11_E_slopes} is of particular pedagogical interest because of its tight correlation with the sensitivity of the main detector to energy. The sensitivity of the main detector to a ``true'' energy variable would be expected to be relatively stable over the course of the experiment with shifts occurring, for instance, when the current in the \qtor spectrometer or the selected beam energy were changed significantly. In the absence of these changes, energy sensitivity should be stable at the few percent level. Instead what we see is large scale shifts in main detector sensitivity to BPM3c12X, showing that this BPM is sensitive to other beam parameters as well as energy. 
The correction slope plots in figures \ref{fig:Set11_X_slopes}--\ref{fig:Set11_E_slopes} can be compared with the monitor difference plots in Appendix \ref{AppendixB} to get an idea of the full correction being applied.

\FloatBarrier
\subsection{Bias in Multivariate Linear Regression}
In the context of the \Qs experiment linear regression minimizes noise in the detectors (dependent variables) by subtracting correlations to the beam monitors (independent variables). By definition all correlations of the detectors to monitors are removed. However, linear regression is susceptible to errors and biases. Perhaps the most glaring deficiency of linear regression arises when there is noise in the independent variables. This issue is familiar to statisticians and is called the problem of ``Errors in Variables'' (EIV). While noise in the dependent variable (detector) adds uncertainty, it does not bias the measured slope. On the other hand, noise in the independent variables (beam monitors) biases the regression correlation coefficients. In many cases where linear regression is utilized, the dependent variable is sensitive to a set of independent variables which cannot be determined accurately. Instead a measurable set of variables correlated to the set of true dependent variables is substituted. Often the substituted variables are not perfectly correlated with the true variables resulting in uncertainties and noise in the independent variables. In the case of electron scattering physics, the detectors are sensitive to beam position, angle and energy. We substitute combinations of BPM signals highly correlated to these true beam parameters. However, the BPM's introduce both correlated and uncorrelated noise which will bias the slopes.

To illustrate this point, consider a detector asymmetry $A_d$ with a sensitivity to horizontal helicity-correlated position differences in the electron beam. Regressing against position to remove the correlation using a noisy monitor $BPM$ to measure the true position differences $\Delta X$ gives 
\[
A_d=A_{PV}+\alpha\Delta X,
\]
\[
\Delta BPM = \Delta X + \delta,
\]
\[
\chi^2=\sum_{i}\left[(A_{d}^0)_i-a(\Delta BPM)_i\right]^2/\sigma_i^2
\]  
where $\delta$ is noise on the measurement of the helicity-correlated position difference $\Delta X$. Here, $\sigma_i$ is the precision on $A_d$ and $\sigma_{BPM}$ is a combination of actual beam motion (jitter) and the noise contribution from $\sigma_{\delta}$. Minimizing the $\chi^2$ statistic with respect to the desired regression slope $a$ assuming constant variance gives 
\[
a=\frac{\alpha(\sigma_{BPM}^2-\sigma_{\delta}^2)}{\sigma_{BPM}^2}=\alpha\left(1-\frac{\sigma_{\delta}^2}{\sigma_{BPM}^2}\right).
\]
This also assumes that the $\delta$ is uncorrelated with $\Delta X$ (which may or may not be true). Notice that coefficient, $a$, is diluted relative to the desired correlation coefficient, $\alpha$, to which the detector is sensitive. Statisticians refer to this effect as ``regression dilution'' or ``regression attenuation'' since the bias for a single regressor with noise is always toward zero. On the other hand, if $\delta$ and $\Delta X$ are correlated, the coefficient can be biased in either direction, although dilution is generally expected.\footnote{Economist Jerry Hausman calls this the ``Iron law of econometrics''--the magnitude of the estimate is usually smaller than expected.''\cite{Hausman}}

It is useful to point out an effect similar to this in the \Qs dataset. In order to correct for residual charge asymmetries on the electron beam, all detector measurements were normalized to the measured current. Of course, any noise in the BCM('s) used to normalize the detector will be introduced into the detector data. This added noise gives the appearance of detector sensitivity to charge when regressing against the same BCM(s) used in the normalization. Naively one might be tempted to regress against charge to remove this sensitivity. Linear regression schemes which include charge, do indeed find a residual detector to charge sensitivity and assign an additional correction. The effect of this can be easily seen in Table \ref{tab:regression_corrections_table}. The four columns show corrections assigned to the data using different monitor sets as regressors. Set 3 is the only set in the table that includes a sixth regression variable, charge. In particular, notice the variation in the corrections for the first 5 Wien states. Wien states from 6 to 10 have much less variation between the charge included and charge excluded regression schemes. One reason for this difference is bias introduced by the BCM's used in the regression scheme. For the first 5 Wien states, set 3 used BCM's 1 and 2, whereas the Wien states 6-10 used BCM's 7 and 8. BCM's 7 and 8 had new digital electronics which greatly reduced the width of their intrinsic noise by about a factor of two from the analog electronics used to read out BCM's 1 and 2 (see Figure \ref{fig:BCM_dd_width}).

\begin{table}[h]
\caption{Corrections for regression with various monitor sets shown averaged over Wien states. All corrections are weighted by the reciprocal of the square of the main detector error. {\it Note: this table is for purposes of comparison of various regression schemes and does not contain the actual correction and asymmetry values used for \Q. It contains only the data for which there are good regression slopes for all schemes shown.} The columns represent regression schemes using different monitor sets. Standard uses the target variables (targetX(Y), targetX(Y)Slope, and energy. Set 11 uses the target variables and the most energy-sensitive monitor BPM3c12X. Set 3 is the same as Set 11 except that charge is added as a sixth regressor. For definitions of these variables see Appendix \ref{AppendixE}.}
\begin{center}
\begin{tabular}[ht]{|l|c|c|c|c|}\hline
Wien & Raw Asymmetry & Standard &~~Set11~~&~~Set3~~~\\
~ & (ppb) & (ppb) & (ppb) & (ppb) \\\hline
  1  & -337.2 & +6.5  & +6.5 & +5.5 \\\hline
  2  & -192.2 & +38.0 & +38.0 & +50.6 \\\hline
  3  & -257.2 & -25.6 & -25.6 & -25.6 \\\hline
  4  & -270.7 & -6.1  & -6.1 & -3.5 \\\hline
  5  & -191.6 & -9.5  & -9.5 & -11.3 \\\hline
  6  & -207.0 & -25.6 & -25.6 & -25.5 \\\hline
  7  & -123.6 & +33.5 & +33.5 & +33.7 \\\hline
  8a & -184.5 & -12.9 & -12.9 & -12.8 \\\hline
  8b & -157.8 & -2.4  & -2.4 & -2.4 \\\hline
  9a & -137.9 & -0.5  & +0.3 & -0.5 \\\hline
  9b & -158.1 & +1.6  & +1.6 & +1.8 \\\hline
  10 & -222.7 & -6.7  & -6.7 & -4.7 \\\hline
\end{tabular}
\end{center}
\label{tab:regression_corrections_table}
\end{table}

\begin{figure}
\centering
\includegraphics[width=5.8in]{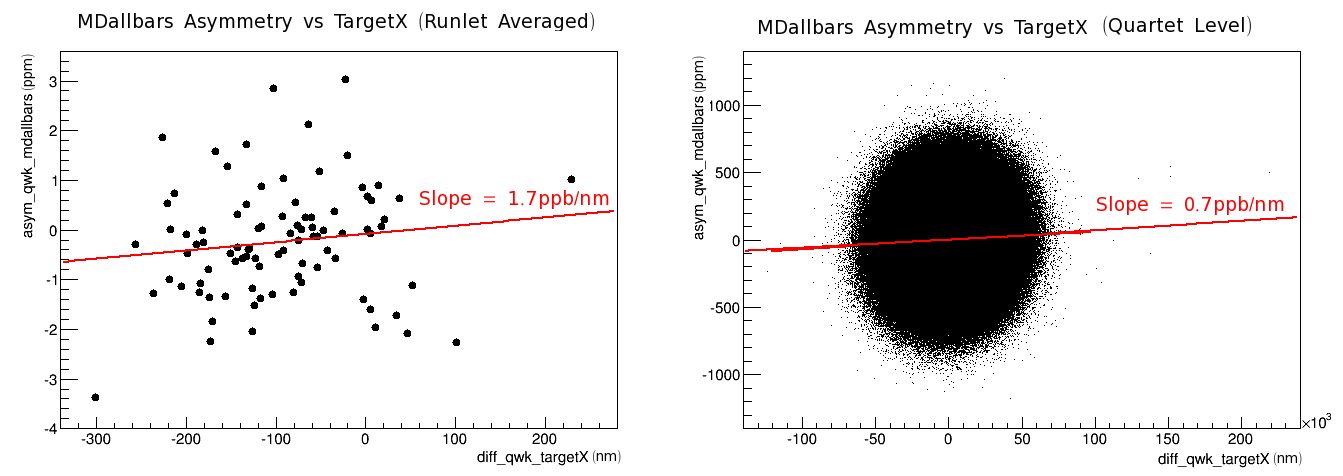}
\caption{\label{fig:regression_bias}Correlation between MDallbars asymmetry and targetX differences over a few hours during Run 2. Each point in the left plot is an average over a runlet (4-5 minutes) whereas each point in the right plot is an asymmetry measured at the quartet level (4~ms). The effect of regression dilution from noise in the position monitor is evident in the smaller measured slope for the quartet level plot.} 
\end{figure}

One evidence for bias arising from noisy monitors used in regression is a dependence of the correlations on the timescale of averaging utilized. The default method for determining regression correlations is to measure them at the quartet level, that is, at the smallest timescale for measured asymmetries in the \Qs experiment. If one chooses instead to average data over a longer period and then find the correlations between the averages, often one arrives at a different solution. One source of the difference is noise in the beam monitors that averages to zero over time. This effect can be seen in Figure \ref{fig:regression_bias} where the correlation of the MDallbars to targetX differences for the same data are determined with different averaging timescales. Although one might expect that longer averages might yield correlations closer to the desired ones, complications arise at longer timescale averaged from long term changes in beam conditions. Also, the longer average correlations have increased errors in the slopes due to decrease in the range of the independent variables and the number of data points. 

Another weakness in linear regression found to be an issue during \Qs concerns the use of natural beam motion. Although the beam is assumed to move in position, angle and energy, its motion may be small enough that it is below or close to the resolution of the beam monitors. When the natural motion of the beam is close to the monitor resolution limit in one or more of the parameters, those parameters will not be assigned a correct slope. In this case a ``quiet'' beam may be more difficult to correct than a noisy one. It is important to note that small differences in the beam at or near the monitor resolution limit does not mean that the detector system is not sensitive to those parameters, nor that the correction to the main detector from those parameters will be insignificant. An in depth study of the monitor sensitivities to beam motion during \Qs showed that, in fact, the differences in beam angle at the target were close to the monitor resolution level creating doubts as to the reliability of linear regression for assigning corrections for beam parameters. This study suggests that regression corrections are most reliable when the beam motion is more than 8 times the monitor resolution. The details of this study will be included in a future thesis \cite{Peng}.

Despite the weaknesses and biases inherent in the use of linear regression, it remains a useful tool. However, an independent method of determining corrections is highly desirable.

\section{Beam Modulation to Remove Helicity-Correlated False Asymmetries}
Beam modulation or beam dithering provides an alternate method to determine correction slopes for removing detector sensitivity to electron beam parameters. By deliberately modulating the five known beam parameters with some known drive signal and simultaneously measuring the detector and beam monitor responses to the modulation, it is possible to back out the requisite detector to monitor responses. Although it would be convenient to be able to modulate the beam purely in each of the five orthogonal beam parameters, in practice this is not easily accomplished. Instead five independent but not necessarily orthogonal modulations of the beam properties are used. It is helpful to think of a 5-dimensional parameter space composed of horizontal position (X) and angle (X'), vertical position (Y) and angle (Y') and energy (E). For a linear system, the five modulation types can be thought of as five independent straight line trajectories or vectors in this space. These modulation vectors are often referred to as ``coils'' since the position and angle trajectories are driven by air coil magnets. There are also five monitors whose sensitivities can be described by five independent vectors in this space. Finally, any component of the detector signal that can be described in this 5-D space must be characterized and removed. Given a set of coils $C_c$ driving the beam, the coils as well as monitor and detector responses can be expressed as
\begin{align}
Coil_c=C_c=\delta_C+\sum_{i=1}^{N}\gamma_{ic}X_i\\
Monitor_{m}=M_{mc}=\delta_M+\sum_{i=1}^{5}\alpha_{i}^{(m)}X_i\\
Detector_{d}=D_{dc}=\delta_D+\sum_{i=1}^{5}\beta_{i}^{(d)}X_i,
\label{eq:5parameter_space}
\end{align}
where the $X_i's$ are the five ideal orthogonal parameters (X,Y,X',Y',E). The coils $C_c$ are abstract quantities that refer to the effects of the modulation of magnetic coils and the RF cavity on the electron beam.\footnote{This picture of the coils, detectors and monitors as vectors in an abstract 5D space is an idea that was useful to the author but is a potential source of confusion. The idea being conveyed is that detector and monitor sensitivities ($\alpha_i,~\beta_i$) to these 5 dimensions can be thought of as vectors. The coils have the ability to move the electron beam along 5 independent straight-line trajectories in this 5D space but these trajectories are neither precisely orthogonal nor coincident with the 5 axes of this space. The coil vectors then, are inputs or forces, whereas the monitor and detector inputs are motions or responses. For example, in the \Qs setup, Coil 0 primarily moves the beam along some combination of horizontal position and angle change and might be represented as vector (a,0,b,0,0). The modulation analysis simply determines the dot products of the monitor and detector response vectors with the coil vectors by measuring the responses to the coils $\frac{\partial D_{dc}}{\partial C_c}$ and $\frac{\partial M_{mc}}{\partial C_c}$. For example, the detector response to coil $C_c$ is given by \[\frac{\partial D_{dc}}{\partial C_c}={\bf D_{d}\cdot C_c}=(\beta_{1},\beta_{2},\beta_{3},\beta_{4},\beta_{5})\cdot(\gamma_{0c},\gamma_{1c},\gamma_{2c},\gamma_{3c},\gamma_{4c})=\sum_{i=1}^5\beta_{ic}\gamma_{ic},\] where the detector response $\frac{\partial D_{dc}}{\partial C_c}$ is measured and the $\beta$'s and $\gamma's$ are not determined. It will be shown that as long the monitor set and coil set are chosen such that both span the 5D space, these detector and monitor responses to the five coils are sufficient to fully determine corrections to the five beam parameters. Knowledge of the composition of the coils, monitors and detectors in the ideal orthogonal basis is not required. } $M_{m}$ and $D_{d}$, on the other hand, are monitor and detector signals which include responses to beam parameters. The monitor and detector sensitivities to beam motion in the five dimensions are the $\alpha_i$'s and $\beta_i$'s respectively. The $\delta_C$ is a component of the modulation not well described by the five parameters and is expected to be 0 on average with noise that is small compared to the driven motion. A non-zero $\delta_C$ would mean that the coils are modulating the beam in a mode not described by the five expected parameters X, X', Y, Y' and E ({\it eg.} beam spot size). The $\delta_M$ term is a component of the monitor signal not described by the 5 parameters and may be attributed to effects such as electronics noise or resolution error. Finally, the $\delta_D$ term is the part of the detector signal which is insensitive to motion in the five beam parameters. The parity-violating asymmetry is part of the $\delta_D$ term. Here the detectors and monitors are actual measured yields not helicity-correlated differences as in the linear regression analysis. $M$ is the monitor reading and $D$ is the detector signal both averaged over an MPS window.  Rearranging the detector equation in \ref{eq:5parameter_space} gives
\[
\delta_D = D_d-\sum_{i=1}^{5}\beta_iX_i.
\]  
Taking the differential of the monitors and detectors with respect to the coils removes the independent $\delta$ terms giving
\begin{eqnarray}
\frac{\partial M_m}{\partial C_c}=\sum_{i=1}^{5}\frac{\partial M_m}{\partial X_i}\frac{\partial X_i}{\partial C_c}\\
\frac{\partial D_d}{\partial C_c}=\sum_{i=1}^{5}\frac{\partial D_d}{\partial X_i}\frac{\partial X_i}{\partial C_c}
\label{eq:5parameter_differential}
\end{eqnarray}
Now inserting an identity operator allows us to express Equation \ref{eq:5parameter_differential} as
\begin{equation}
\frac{\partial D_d}{\partial C_c}=\sum_{i=1}^{5}\frac{\partial D_d}{\partial X_i}\left(\frac{1}{5}\sum_{m=1}^5\frac{\partial X_i}{\partial M_m}\frac{\partial M_m}{\partial X_i}\right)\frac{\partial X_i}{\partial C_c}=\sum_{m=1}^{5}\frac{\partial D_d}{\partial M_m}\frac{\partial M_m}{\partial C_c},
\label{eq:det_to_coil}
\end{equation}
which we recognize as a simple change of basis from the five ideal, orthogonal parameters to our set of five monitors. This can be expressed as a matrix equation
\[
\mathbf{A=R\cdot S},
\]
where $\mathbf A$ is the vector of detector to coil sensitivities, $\mathbf R$ is the matrix of monitor to coil sensitivities and  $\mathbf S$ is the vector of detector-to-monitor correction slopes which can be solved for ${\bf S=R^-1A}$. 

Equation \ref{eq:det_to_coil} represents an exact solution assuming no error in the measured responses which is, of course, never the case. However, if the beam is driven in more than five ways, that is, if more than 5 coils are utilized, this extra information can be used to reduce the error in the solution, provided all coils yield constistent information about the beam/monitor/detector responses. One could imagine, for example, driving the beam with pure horizontal motion and then pure vertical motion and then along a straight line at 45$^{\circ}$. The extra driving mode (coil) should reduce noise in the solution but not give additional information. If N coils or driving modes are used, one ends up with N equations in 5 unknowns (5 correction slopes $\frac{\partial D_d}{\partial M_k}$). The best solution to this set of equations can be found by minimizing the $\chi ^2$ statistic independently for each detector $D_d$
\begin{equation}
\chi^2=\sum_{i=1}^N\left(\frac{\partial D_d}{\partial C_i} - \sum_{m=1}^{5}\frac{\partial D_d}{\partial M_m}\frac{\partial M_m}{\partial C_i} \right)^2/\sigma_i^2,
\label{eq:chisquare}
\end{equation}

where the sum is over N coils. Note that since the beam is still assumed to have five degrees of freedom, five monitors are still used. Taking the derivative with respect to the detector slopes, $\frac{\partial D_d}{\partial M_k}$, and setting the equation equal to 0 gives
\[
\frac{\partial\chi^2}{\partial\left( \frac{\partial D_d}{\partial M_m}\right)}=-2\sum_{i=0}^N\frac{1}{\sigma_i^2}\left(\frac{\partial D_d}{\partial C_i}- \sum_{m=1}^{5}\frac{\partial D_d}{\partial M_m}\frac{\partial M_m}{\partial C_i} \right)\frac{\partial M_m}{\partial C_i} = 0.
\]
If we assume that the detector noise is constant over all coil measurements,that is, that all responses are equally well determined the solution is given by
\footnote{Under the approximation of equal variance this procedure is the same as finding the Moore-Penrose matrix inverse often called the ``generalized inverse'' or ``pseudoinverse'' (see for example http://mathworld.wolfram.com/Moore-PenroseMatrixInverse.html), which is the ``shortest length least squares solution to'' ${\bf b=Mx}$. Multiplying both sides by ${\bf M^T}$ gives  ${\bf M^Tb=M^TMx}$ which is the same as Equation \ref{eq:chisquaresolution_no_variance}, with a solution  ${\bf x=(M^TM)^{-1}Mb}$. {$\bf M^+=(M^TM)^{-1}M$} is the definition of the Moore-Penrose matrix inverse for pure real matrices.}
\begin{equation}
\sum_{i=1}^N\left(\frac{\partial D_d}{\partial C_i}\frac{\partial M_k}{\partial C_i}\right) = \sum_{m=1}^5\frac{\partial D_d}{\partial M_m}\sum_{i=1}^N\left(\frac{\partial M_m}{\partial C_i}\frac{\partial M_k}{\partial C_i} \right). 
\label{eq:chisquaresolution_no_variance}
\end{equation}
Explicitly writing this for the set of five monitors and a single detector index yields
\[
\mathbf{R^TA=R^TR\cdot S}\longrightarrow\mathbf{B=M\cdot S},
\]
where ${\bf B=R^TA}$ and ${\bf M=R^TR}$ is a square symmetric matrix. Here ${\bf A, S}$ and ${\bf R}$ defined the same as in Equation \ref{eq:det_to_coil}.
\begin{equation}
\left(\begin{array}{c}\sum_i\frac{\partial D_d}{\partial C_i}\frac{\partial M_1}{\partial C_i}\\\sum_i\frac{\partial D_d}{\partial C_i}\frac{\partial M_2}{\partial C_i}\\\sum_i\frac{\partial D_d}{\partial C_i}\frac{\partial M_3}{\partial C_i}\\\sum_i\frac{\partial D_d}{\partial C_i}\frac{\partial M_4}{\partial C_i}\\\sum_i\frac{\partial D_d}{\partial C_i}\frac{\partial M_5}{\partial C_i}\end{array}\right)=\left(\begin{array}{ccccc} M_{11} & M_{12} & M_{13} & M_{14} & M_{15}\\ M_{21} & M_{22} & M_{23} & M_{24} & M_{25}\\M_{31} & M_{32} & M_{33} & M_{34} & M_{35}\\M_{41} & M_{42} & M_{43} & M_{44} & M_{45}\\M_{51} & M_{52} & M_{53} & M_{54} & M_{55}\end{array}\right) \left(\begin{array}{c}\frac{\partial D_d}{\partial M_1}\\\frac{\partial D_d}{\partial M_2}\\\frac{\partial D_d}{\partial M_3}\\\frac{\partial D_d}{\partial M_4}\\\frac{\partial D_d}{\partial M_5}\\\end{array}\right),
\label{eq:chisquare_matrix}
\end{equation}
where 
\[
M_{jk}=\sum_{i=1}^N\left(\frac{\partial M_j}{\partial C_i}\frac{\partial M_k}{\partial C_i}\right).
\]
The correction slopes are then given by 
\begin{equation}
\mathbf{S=M^{-1}\cdot B}.
\label{eq:solution_to_matrix}
\end{equation}

At this point it is helpful to point out the fundamental differences between linear regression and beam modulation for determining correction slopes. First, because the beam modulation analysis re-expresses detector and monitor sensitivities in the 5-parameter coil basis (which is designed to be over position, angle and energy) any correlations that exist outside this five parameter space will not be removed. Second, because all correlations are measured relative to a well-determined variable, that is, the phase of the coil being modulated, the issue of ``noise in independent variables'' or ``noise in regressors'' which creates biases in linear regression, is almost non-existent for the beam modulation analysis. Noise in the monitors will not bias the calculated slopes in the modulation analysis unless the noise happens to be at the precise frequency of the drive signal. Also, in practice, driven motion is much larger than natural motion so that measurement of the beam motion is not limited by the monitor resolution. Third, when beam modulation fails to find the proper correction slopes, its failure can be evidenced in residual sensitivity to the modulation coils if the system is over-determined\footnote{In this context over-determined means the system is driven by more modulation coils than there are expected degrees of freedom. For \Qs there were a total of 9 modulation modes used for only 5 expected degrees of freedom on the electron beam.}. Success in nulling sensitivity to the coils is not guaranteed by design unless the number of coils used in the analysis is equal to the number of degrees of freedom. Linear regression, on the other hand, is guaranteed by design to null first order correlations to the monitors at the sampling frequency. This is not to say that the failure of the modulation technique to give the proper slopes must be evident in residual sensitivity. There are subtle ways for the modulation analysis to fail without leaving residual sensitivity to the coils. The failure modes of the modulation analysis will be discussed in a future section. 

\subsection{Fast Feedback and Beam Modulation}
The beam modulation system, introduced in Chapter 3, was designed with a pair of air core dipole magnets to drive vertical motion and another pair for horizontal motion.  Sinusoidal waveforms created by function generators were sent to the modulation coils to drive the motion. The amplitudes of the waveforms sent to the  modulation coils were chosen to create four independent trajectory-related motions at the target position. Modulation of energy, the fifth degree of freedom, was accomplished via a sinusoidal waveform sent to a Vernier on an accelerating RF cavity in the south linac. Further details on the studies and models that went into the design, position and choice of waveform amplitudes can be found in \cite{Nur}.

\begin{figure}[ht]
\centering
\framebox{\includegraphics[width=5.6in]{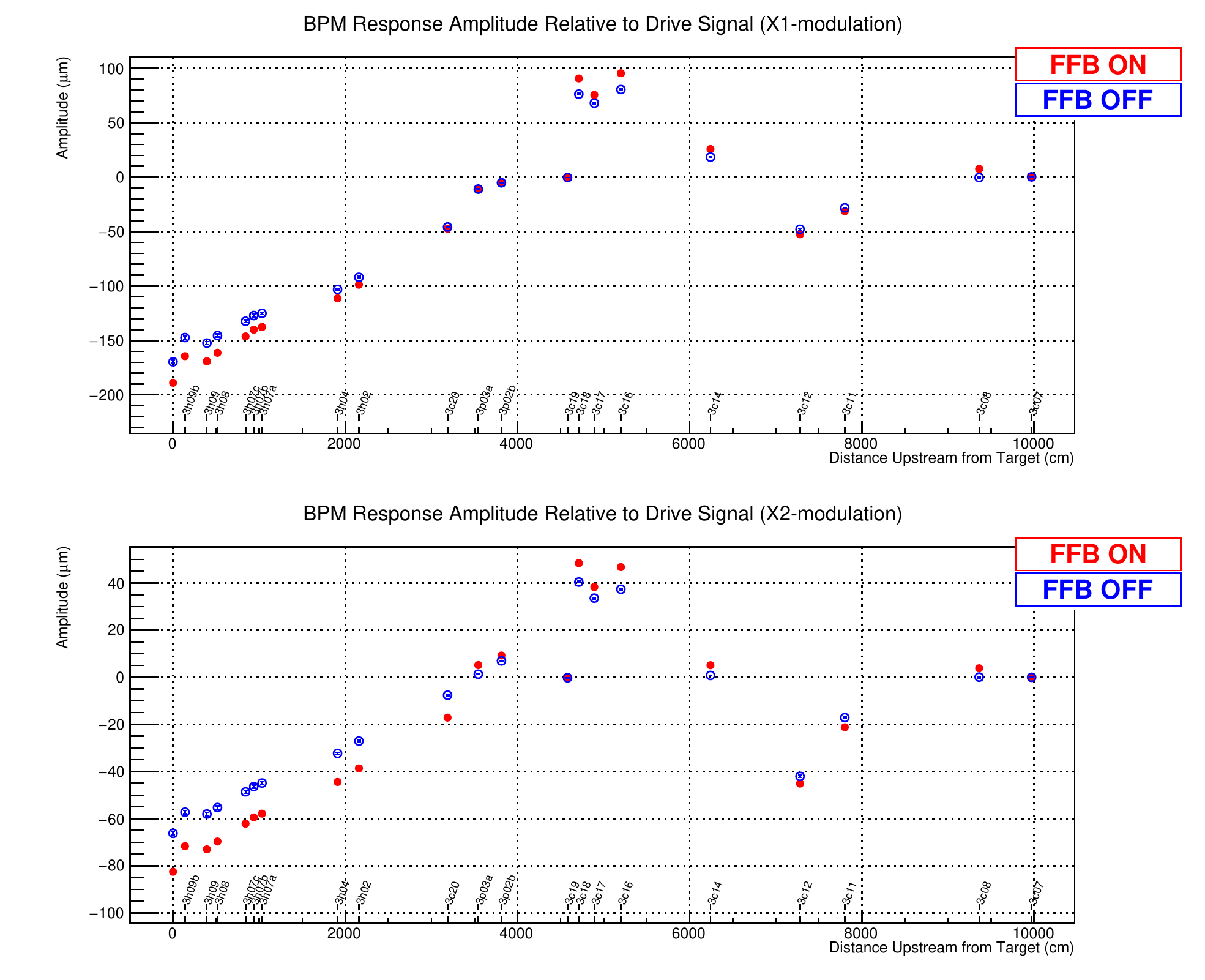}}
\caption{Amplitude of BPM response as a function of distance along beamline upstream from the target. Shown is X-response to two types of horizontal modulation. Notice the relatively small difference between FFB on and FFB off. This data corresponds to $A(z,\phi)$ in Equation \ref{eq:z_dep_response}. The data for FFB off came from 18445 and for FFB on from 17445 taken about 3 weeks apart.}
\label{fig:bpm_amplitude_vs_z_xmod}
\end{figure}

\begin{figure}[ht]
\centering
\framebox{\includegraphics[width=5.6in]{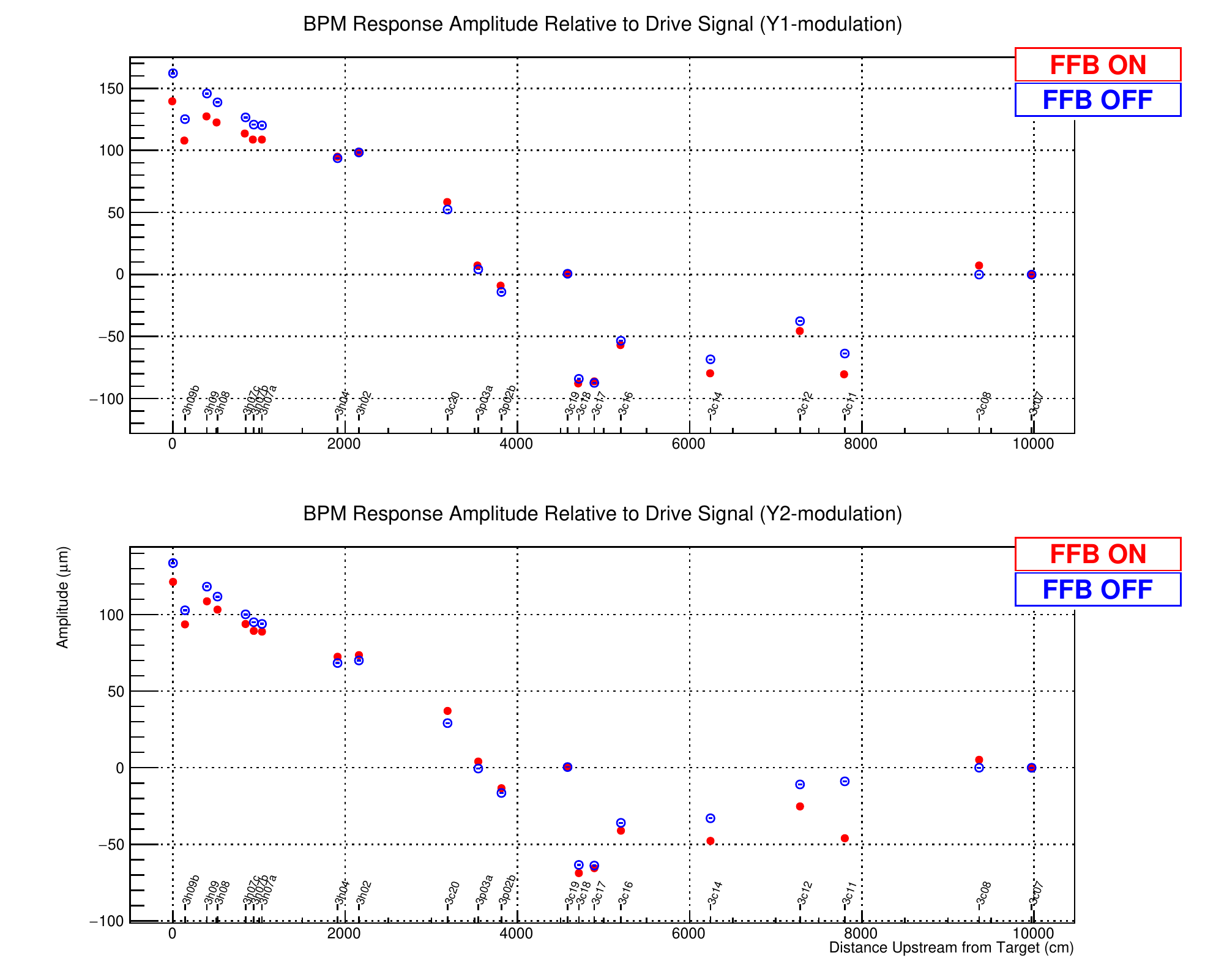}}
\caption{Amplitude of BPM response as a function of distance along beamline upstream from the target. Shown is Y-response to two types of vertical modulation. Notice the relatively small difference between FFB on and FFB off.  This data corresponds to $A(z,\phi)$ in Equation \ref{eq:z_dep_response}. The data for FFB off came from 18445 and for FFB on from 17445 taken about 3 weeks apart.}
\label{fig:bpm_amplitude_vs_z_ymod}
\end{figure}

The electron beam at Jefferson Lab has natural position and angle noise, termed ``beam jitter''. A Fast Feedback (FFB) system was designed to produce trajectory and energy shifts to cancel natural beam jitter at the target. The FFB system was designed particularly to remove low frequency motion of the beam ($<$80~Hz) as well as the first few harmonics of 60~Hz line noise \cite{Lebedev}. Like the beam modulation system, it utilizes air core dipole magnets and Verniers on RF accelerating cavities to manipulate the beam trajectory and energy. In principle, if the FFB system is operating properly during beam modulation it should null the effects of the modulation coils at the target by removing sensitivity to the modulation frequency. However, for the majority of the Qweak experiment, FFB was found to be ineffective at nulling driven beam position and angle motion at the target. As a result, the decision was made to leave FFB engaged during periods of position and angle modulation. The FFB system was paused during energy modulation because it was found to be very effective at nulling the beam energy shifts. Looking at Figures \ref{fig:bpm_amplitude_vs_z_xmod} and \ref{fig:bpm_amplitude_vs_z_ymod} we can see the response of the beam to the four different types of trajectory modulation both with FFB on and off as a function of distance along the beamline upstream from the target. Since the FFB on and off data shown were taken about 3 weeks apart some of the differences may be attributed to optics changes on the beamline; however, the plots clearly show that even with FFB on, the modulation system successfully drives large amplitude excursions at the target.

The effect of FFB on the beam modulation analysis was not understood at first. Since it is an active feedback system, its response to the beam motion may not necessarily be stable or exactly in phase with the modulation coils. The assumption of the simple modulation model given in Equation \ref{eq:det_to_coil} is that all detector and monitor responses at the modulation frequency and in phase with the modulation drive signal are created by actual beam trajectory and energy changes driven by the modulation coils.  An unknown FFB-driven response that is not stable over the modulation period could compromise the integrity of the analysis.
\begin{figure}[ht]
\begin{center}
\framebox{\includegraphics[width=5.6in]{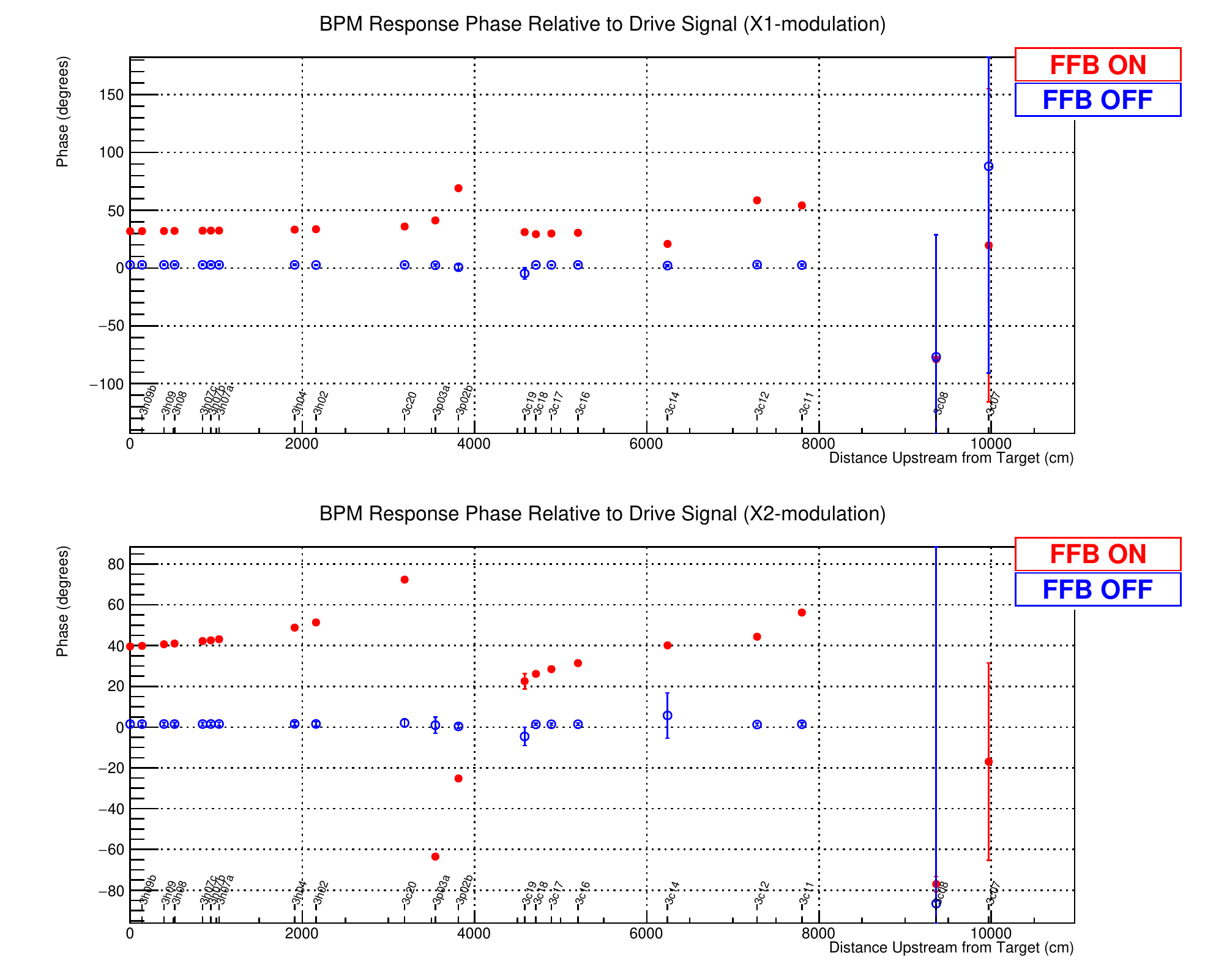}}
\end{center}
\caption{Phase of BPM response as a function of distance along beamline upstream from the target. Shown is $x$-position response to two types of horizontal modulation. This data corresponds to $\theta (z,\phi)$ in Equation \ref{eq:z_dep_response}.  Notice the phase dependence disappears with FFB off. The data for FFB off came from 18445 and for FFB on from 17445 taken about 3 weeks apart.}
\label{fig:bpm_phase_vs_z_xmod}
\end{figure}
\begin{figure}[ht]
\centering
\framebox{\includegraphics[width=5.6in]{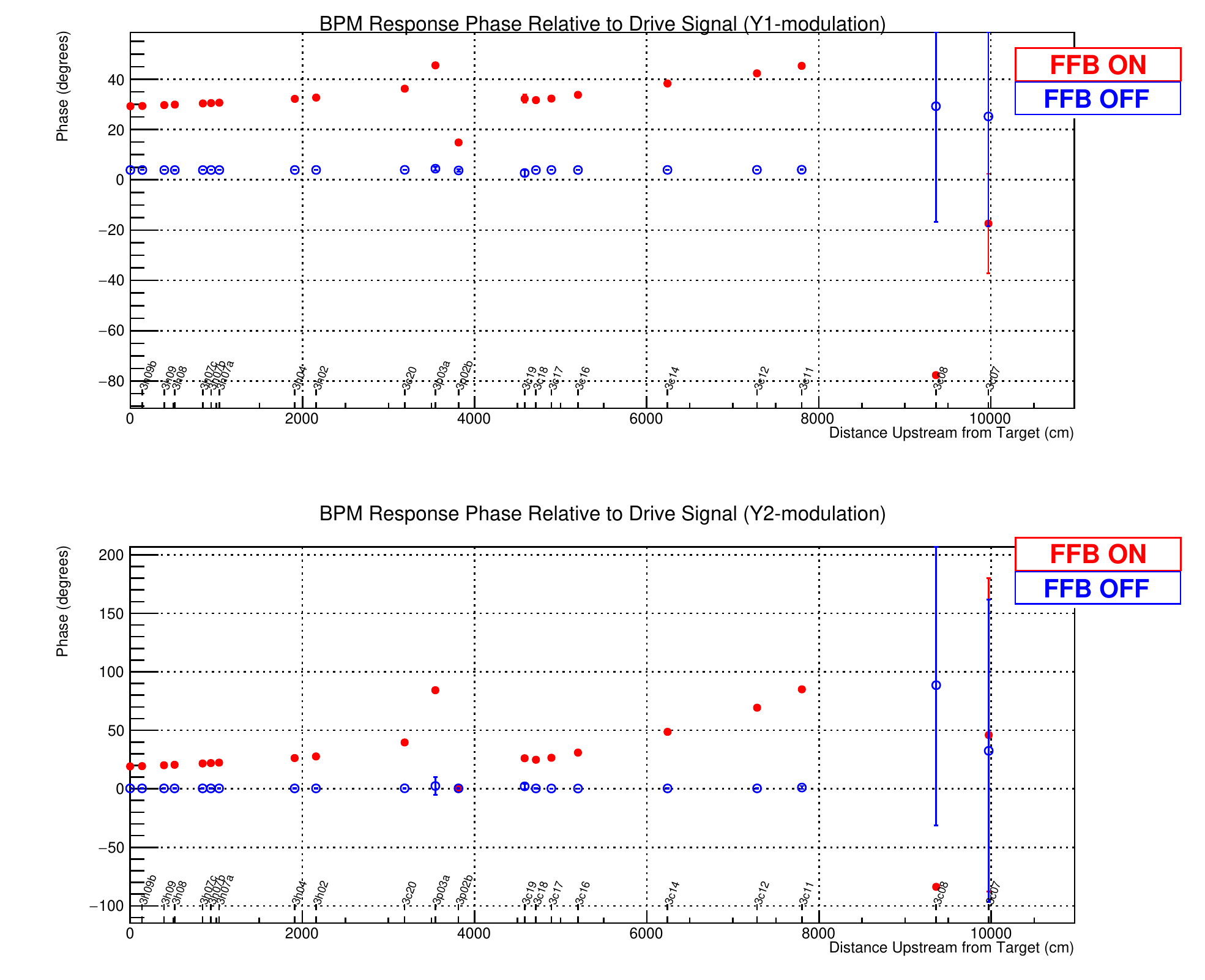}}
\caption{Phase of BPM response as a function of distance along beamline upstream from the target. Shown is $y$-position response to two types of vertical modulation.  Notice the phase dependence disappears with FFB off.  This data corresponds to $\theta (z,\phi)$ in Equation \ref{eq:z_dep_response}. The data for FFB off came from 18445 and for FFB on from 17445 taken about 3 weeks apart.}
\label{fig:bpm_phase_vs_z_ymod}
\end{figure}

The first indication that the system was not behaving as expected came from a study of BPM responses to beam modulation. The phase of the BPM response relative to the modulation driving signal has an obvious dependence on location along the beamline as seen in Figures \ref{fig:bpm_phase_vs_z_xmod} and \ref{fig:bpm_phase_vs_z_ymod}. This $z$-dependence of the phase can be understood in terms of the optics  of the electron beam. The beam is focused and defocused using quadrupole magnets along the beamline. The beam's size and shape is a function of $z$ (position along the beamline). The relative size and shape of the beam determines the distribution of angle and position trajectories it contains. For example, near a tight focus the electron beam will have a relatively greater spread in electron trajectory angles (measured relative to an ideal reference trajectory) and a relatively small spread in position. Therefore, BPM's positioned near beam focal points will tend to be more sensitive to beam angle shifts and BPM's positioned where the beam spot size changes little with $z$ will be relatively more sensitive to beam position shifts. The phase shifts observed in the BPM responses can be understood in terms of a $z$-dependent response of the electron beam to trajectory changes from the modulation coils, coupled  with a partially out-of-phase FFB response. We can think of a BPM as measuring the in-phase response to the modulation coils plus an out-of-phase response from FFB:
\begin{equation}
\Delta(z,t) = \alpha(z) sin(\omega t) + \beta(z) sin(\omega t + \phi), 
\label{eq:z_dep_response}
\end{equation}
where $\Delta$ is a measured transverse position shift from the ideal beam trajectory, $\alpha(z)$ is the amplitude of the response of the electron beam to the modulation coils at location $z$ along the beamline and $\beta(z)$ is the amplitude of the electron beam response to the FFB coils at z. Using simple trigonometry to recast this equation in terms of a z-dependent phase and amplitude gives
\begin{equation}
\Delta(z,t) = A(z,\phi) sin(\omega t + \theta(z,\phi)), 
\label{eq:z_dependent_phi}
\end{equation} 
where 
\[
A(z,\phi)=\sqrt{\alpha(z)^2+\beta(z)^2+2\alpha(z)\beta(z)\cos(\phi)}
\]
and 
\[
\theta(z,\phi) = tan^{-1}\left( \frac{\beta(z)\sin(\phi)}{\alpha(z)+\beta(z)\cos(\phi)}\right).
\]
In this form, the $z$-dependent phase of the beam response becomes obvious as does its origin in the FFB coil stimulus. Near the end of the \Qs experiment a set of data runs were taken with FFB paused during all modulation periods. The phase of the BPM responses relative to the modulation drive signal can be seen in Figures \ref{fig:bpm_phase_vs_z_xmod} and  \ref{fig:bpm_phase_vs_z_ymod}. The disappearance of the $z$-dependence of the phase when FFB is turned off confirms the hypothesis that it originates from the FFB response.

Having established that the FFB system is responding at the modulation frequency but not in phase with the modulation coils, the next question to address is the stability of this response and the importance of this stability. It is not obvious nor necessary that an active feedback response system will produce a somewhat stable response in time. A more detailed analysis of the stability of the FFB system and its affect on the modulation analysis can be found in Appendix \ref{AppendixB}. At this point it is sufficient to say that the FFB response is not in phase with the drive signal to the modulation coils and that its response is stable enough to be accurately determined over the course of a cycle ($\sim 4~s$). 

As previously mentioned, Equation \ref{eq:chisquaresolution_no_variance} allows for more coils to be used than expected degrees of freedom on the beam. We can make use of as many different linear combinations of the 5 expected degrees as we choose to measure. In this case, with FFB on we have access to the 5 modulation responses in phase with our driving coils and 4 orthogonal $\pi/2$ out-of-phase responses to the FFB coils.\footnote{To be clear, a single sinusoidal response is observed in the monitors and detectors, but it is not in phase with the modulation drive signal. The phase lag is interpreted as the net response to the modulation coils plus a slightly delayed response from FFB coils due to millisecond-scale time constants inherent in the FFB system. The result is a simultaneous driving of two independent modulation modes which can be temporally separated by finding sine and cosine amplitudes.} Of course, this is a simplification since the FFB coils need not be exactly out-of-phase, but we can expect that any out of phase component comes solely from FFB. We only have access to 4 true out-of-phase responses since FFB was paused during energy modulation. A full ``10-Coil'' analysis was performed using the solution of in Equation \ref{eq:chisquare_matrix} where we found 5 in phase and 5 out-of-phase responses which we called ``sine'' and ``cosine'' respectively. Using a full 10 coils instead of the 9 we really have, allows for a potential phase offset between the signal driving the coils and the ``ramp'' signal we fed to our data acquisition electronics to keep track of modulation phase.  

\begin{figure}[t]
\centering
\framebox{\includegraphics[width=5.5in]{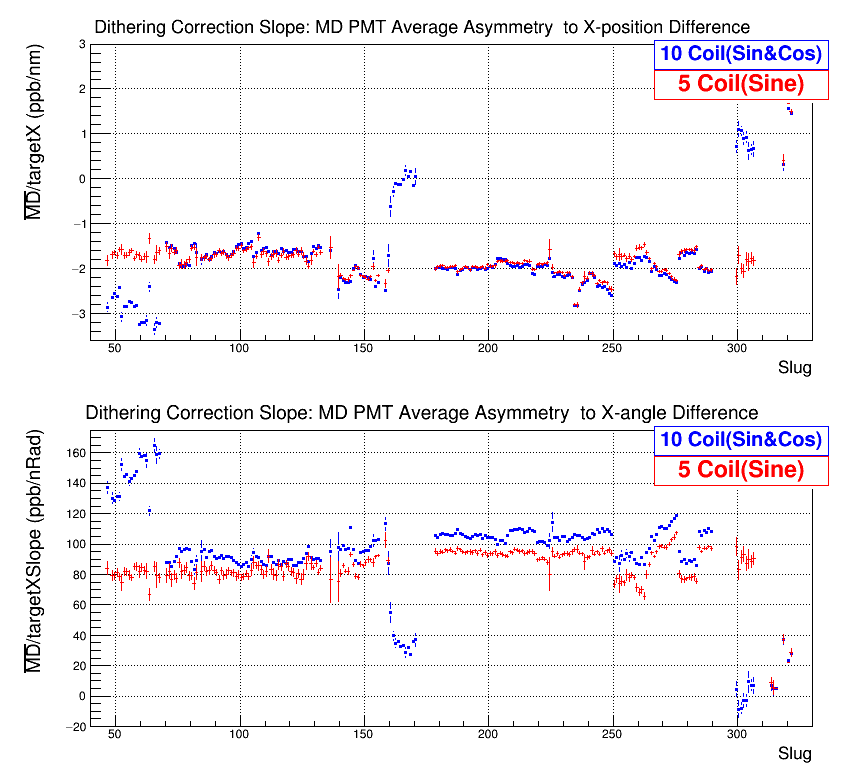}}
\caption{Dithering correction slopes for horizontal position and angle on target averaged over slugs ($\sim 8~hrs$) using all 10 coils.}
\label{fig:10Coil_bmod_X_slopes}
\end{figure}

\begin{figure}[ht]
\centering
\framebox{\includegraphics[width=5.5in]{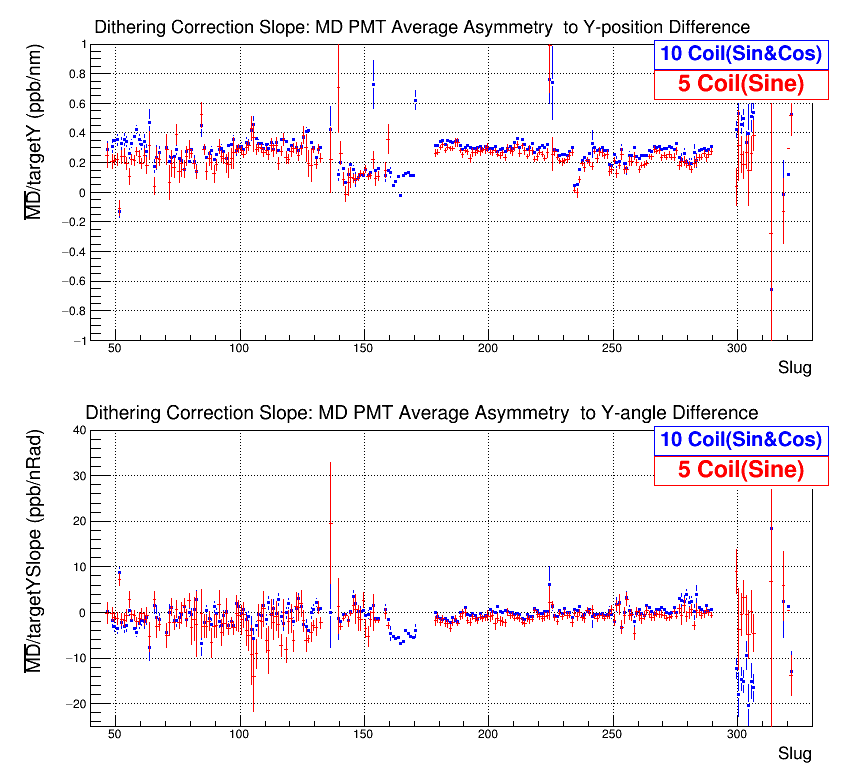}}
\caption{Dithering correction slopes for vertical position and angle on target averaged over slugs using all 10 coils.}
\label{fig:10Coil_bmod_Y_slopes}
\end{figure}

\begin{figure}[h]
\centering
\framebox{\includegraphics[width=5.5in]{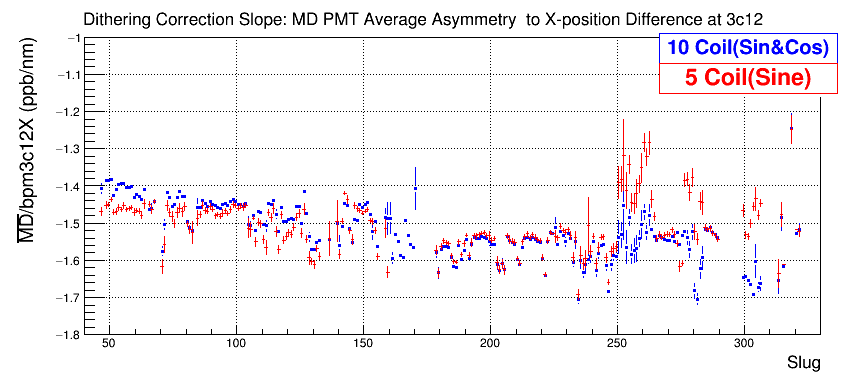}}
\caption{Dithering correction slopes for horizontal position at bpm3c12 averaged over slugs ($\sim 8~hrs$) using all 10 coils. This is the most energy sensitive monitor on the beamline.}
\label{fig:10Coil_bmod_E_slopes}
\end{figure}

Figures \ref{fig:10Coil_bmod_X_slopes} to \ref{fig:10Coil_bmod_E_slopes} compare the dithering correction slopes found using all 10 coils (Equation \ref{eq:chisquaresolution_no_variance}) and using only the 5 coils in phase with the modulation drive signal (eq. \ref{eq:chisquaresolution_no_variance}). Practically speaking, a 5-Coil ``sine only'' analysis means filling matrix and vector components in Equation \ref{eq:chisquaresolution_no_variance} using only the sine coefficients of the monitor and detector versus modulation phase fits. A 10-Coil analysis uses both the sine and cosine coefficients from the sinusoidal response. A quick comparison shows that these slopes differ greatly from the regression slopes using the same monitors. Perhaps this is not surprising given that the regression slopes minimize the collective noise of the detector by zeroing the detector to monitor correlation, while the slopes for dithering minimize the collective correlation to the coils (drive signals). This does not necessarily mean that dithering zeroes the correlation to coils.

Perhaps more troubling than the difference between regression and dithering slopes is the obvious systematic differences between the two dithering analyses. A few important observations must be made with respect to these differences:
\begin{itemize}
\item{As expected, the 10-Coil analysis with much more information and analyzed using least squares has much smaller error bars than the slopes found with the simple 5-Coil analysis.}
\item{There are large sections of data (see slugs 160-170 for example) where there is not enough information in the ``sine only'' analysis to create useful slopes. The in-phase monitor responses do not sufficiently span the space of beam distortions. However, with the extra information gleaned from the cosine terms, a clean solution for the slopes emerges.}
\item{Rather large inconsistencies in the solutions provided by the two analyses are evidence of an underlying problem in the data or in the analysis procedure. Although correlations and strength sharing between monitors play a role in the slopes ``chosen'' in a regression analysis, they are not expected to influence the dithering analysis very much if at all. The coils simply sample the phase space of possible beam trajectories to allow for measurement of accurate monitor responses. If the monitor responses are stable, the correction slopes found using any complete set of coils is expected to be consistent. A possible caveat to this statement arises if the coils modulate a beam property beyond the 5 in the model and to which the detectors are sensitive.}   
\item{Near the end of the Qweak dataset (after slug 306) the FFB system was paused for all types of modulation. After this, the analysis is reduced to 5 coils since there is little to no out-of-phase response. Because the coils do not sufficiently span the space of beam distortion modes during this period, dithering analysis fails to produce stable correction slopes.}
\end{itemize}

Different slopes do not necessarily mean different total corrections. In a 5-D space with non-orthogonal monitors, there can be many sets of slopes which yield the same total correction. Furthermore, cancellations over time may produce results that appear to be consistent. Table~\ref{tab:dithering_corrections_table} compares the total corrections for MDallbars prescribed by a dithering analyses using all 10 coils and using a 5-Coil ``sine only'' analysis. All corrections are averaged over a Wien states and weighted by the reciprocal of the variance of MDallbars.
\begin{table}[!h]
\caption{MDallbars dithering corrections compared for analyses with 10 coils and 5 coils (sine only) shown averaged over Wien states. All corrections are weighted by the reciprocal of the square of the main detector error. {\it Note: this table is for purposes of comparison of various regression schemes and does not contain the actual correction and asymmetry values used for \Q. It contains only the data for which there are good dithering slopes for all schemes shown.}}
\begin{center}
\begin{tabular}[h]{|l|c|c|c|}\hline
Wien & Raw Asymmetry & 10-Coil &~5-Coil~~\\
~ & (ppb) & (ppb) & (ppb)\\\hline
  1  & -328.5 & -24.5 & -10.9 \\\hline
  2  & -192.7 & +3.5 & +22.6 \\\hline
  3  & -256.6 & -37.8 & -30.2 \\\hline
  4  & -269.9 & -11.2 & -10.0 \\\hline
  5  & -190.2 & -17.5 & -16.6 \\\hline
  6  & -206.9 & -25.5 & -25.0 \\\hline
  7  & -153.6 & +56.3 & +51.7 \\\hline
  8a & -185.3 & -12.3 & -8.4 \\\hline
  8b & -156.8 & -4.2 & -5.0 \\\hline
  9a & -136.6 & +2.0 & +2.1 \\\hline
  9b & -157.7 & +1.3 & +1.2 \\\hline
  10 & -215.0 & -15.6 & -17.0 \\\hline
\end{tabular}
\end{center}
\label{tab:dithering_corrections_table}
\end{table}

Due to the length and complexity of the topic, the following chapter is devoted entirely to dealing with internal inconsistencies in the dithering dataset.



\chapter{Resolving Issues in the Beam Modulation Analysis}
\captionsetup{justification=justified,singlelinecheck=false}

\label{Ch:BMod_correction}

\lhead{Chapter 5. \emph{Resolving Issues}} 
A number of tests have been utilized to check the consistency and validity of the dithering analysis, focusing on three main questions. {\bf First}, is the analysis code working as expected and properly debugged? {\bf Second}, is the dithering solution self-consistent? It will be shown that a self-consistent analysis means: a) detector sensitivity to the modulation is removed and  b) the solution is independent of the selection of monitor or coil set used as long as both span the space of beam degrees of freedom. {\bf Third}, does the modulation solution reduce or remove correlations of the detectors to the beam monitors over long timescales?  

The first section of this chapter addresses each of these questions concerning the modulation analysis. Inconsistencies in the modulation analysis for Qweak are investigated and an attempt made to identify possible failure modes by looking at signatures of the failures. The last section of this chapter provides a method for dealing with the inconsistencies and assigning a systematic error. A  modulation correction and error is proposed for the subset of the full \Qs dataset for which valid modulation correction slopes were measured.

\section{Validating the Modulation Analysis}
\begin{figure}
\centering
\includegraphics[width=5.8in]{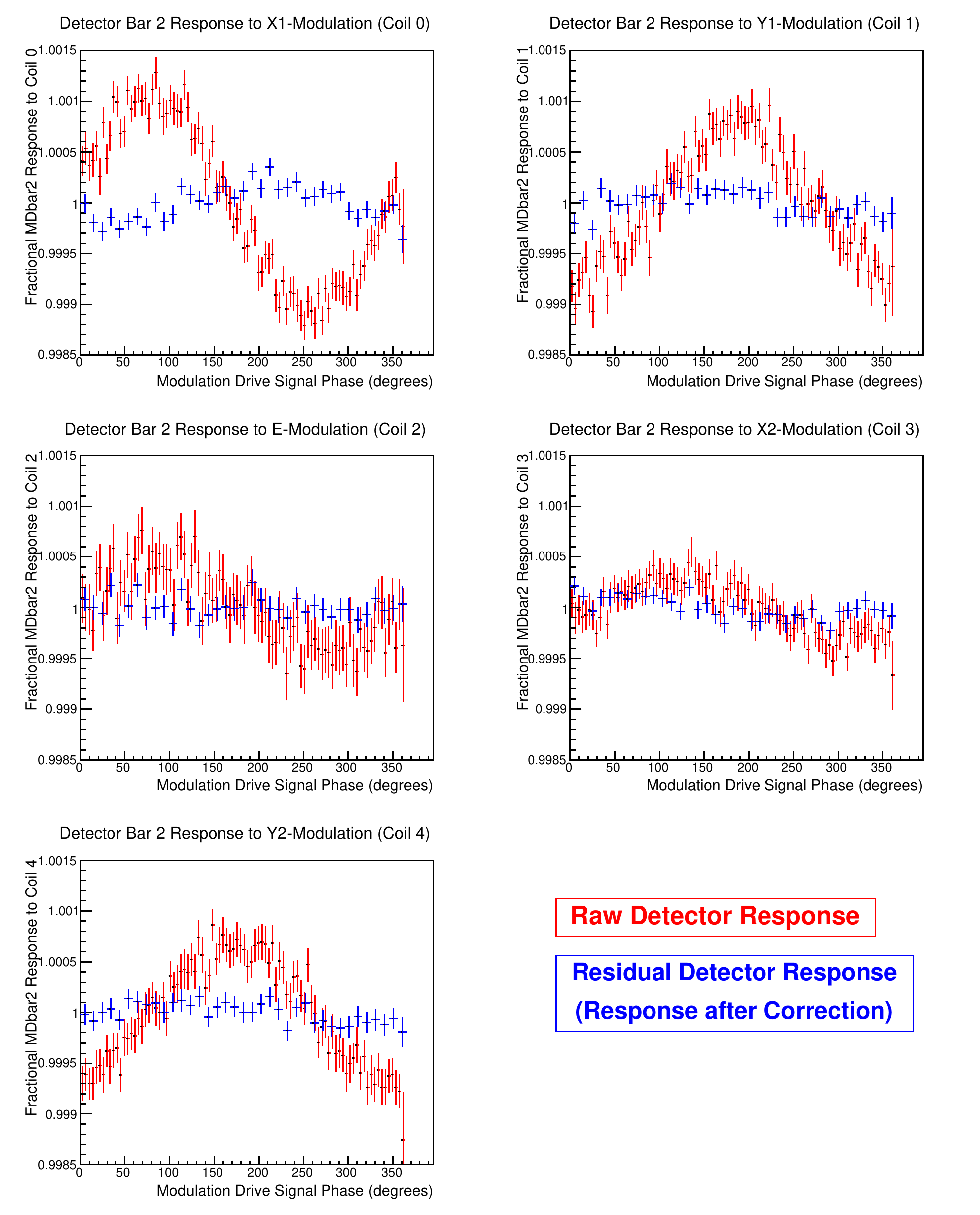}
\caption{\label{fig:det_vs_ramp}Normalized main detector bar 2 (chosen for its sensitivity to all modulation coils) responses to the five modulation coils shown versus drive signal phase. The responses before (red) and after (blue) correction are shown. A small residual response is evident in the plot of response to Coil 0 (upper left).}
\end{figure}

The modulation analysis involves measuring detector and monitor responses to coils, using these to determine correction slopes $\frac{\partial D}{\partial M_k}$ and finally applying these correction slopes to the detector data. If this procedure is successful, the corrected data should be insensitive to the driving signal. An unsuccessful or partially successful correction will see a residual correlation to the driving signal, which in this case is a sinusoidal waveform. The detector (monitor) responses can be found by fitting a sinusoidal function to the detector (monitor) versus modulation phase data. The residual detector response is found by doing the same fit after the correction has been applied to the detector data. Figure \ref{fig:det_vs_ramp} shows an example of main detector bar 2 responses to the coils before and after correction. A small residual sine response can be seen on the upper left plot of response to Coil 0. These response plots are intended to clarify the discussion ahead.

\subsection{Debugging the Analysis Procedure}

\begin{figure}[ht]

\centering
\framebox{\includegraphics[width=5.5in]{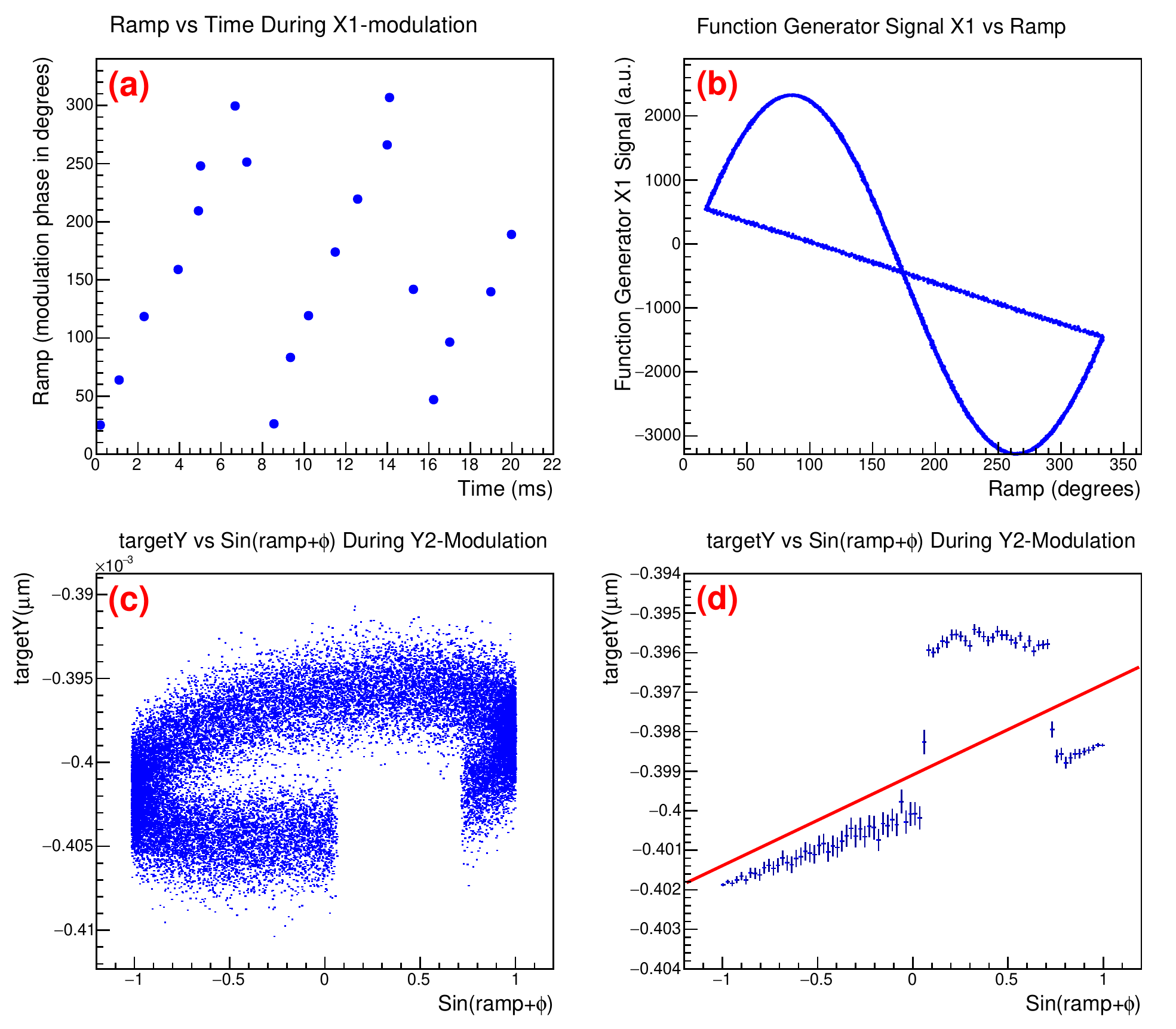}}
\caption{Various plots demonstrating ``ramp'' variable used to track modulation phase. (a). Ramp versus time showing the clear saw-tooth pattern. Signal integration ensures that phases near 0 and 360 are never reached. (b). Function generator signal driving the one of the horizontal modulation coils during X1 modulation plotted versus ramp variable. Signal integration improperly maps the return of the saw-tooth pattern making the line seen connecting 20 degrees and 340 degrees. (c). TargetY versus the sine of ramp plus and arbitrary phase offset. The combination of in and out-of-phase driving coils can produce elliptical responses in the monitors and detectors. When the improperly mapped phase in the saw-tooth return region is cut, a gap is left in the ellipse. (d). Profile plot (horizontal bin averages) of data from (c) with linear fit demonstrates how the coefficients of the sine and cosine responses can be biased by the missing region in ramp.}
\label{fig:ramp_plots}
\end{figure}

\begin{figure}[ht]

\centering
\framebox{\includegraphics[width=3in]{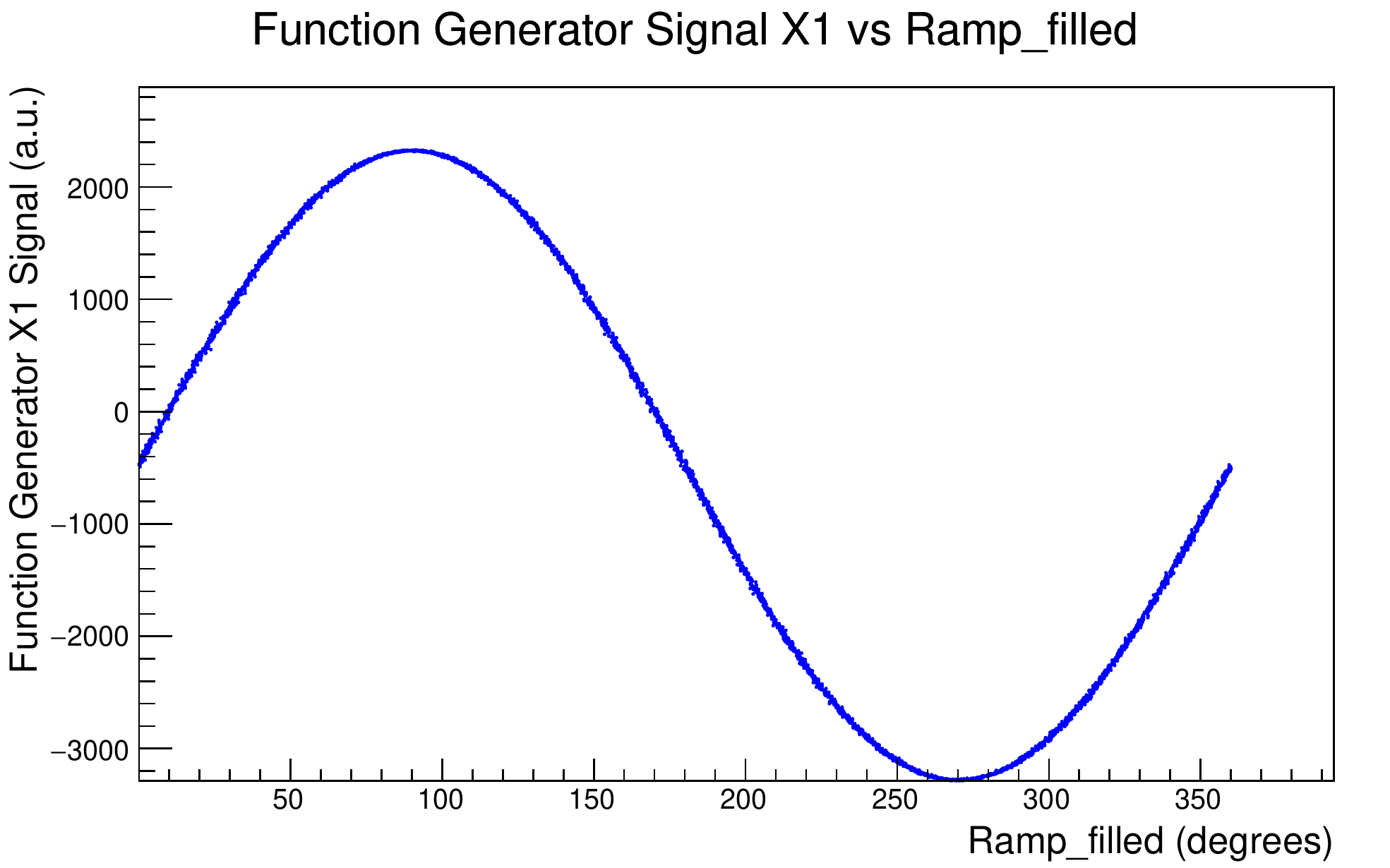}}
\caption {Function generator X1 driving signal plotted versus ``ramp\_filled'' where the incorrect phases in ramp have been properly mapped back to ``fill'' the gap region.}
\label{fig:ramp_filled}
\end{figure}

Early in the analysis, attempts to find a solution with 5 coils (sine-only) were only partially successful. One source of failure was found to be in the ``ramp'' variable used to keep track of the phase of the modulation drive signal. As previously mentioned, the ramp signal is just a saw-tooth signal sent to the DAQ in phase with and at the same frequency as the modulation signal. This saw-tooth signal was then pedestal subtracted and scaled such that its value was equal to the phase of the drive signal in degrees. A plot of ramp versus time can be seen in plot (a) of figure~\ref{fig:ramp_plots}. 
From the plot one can see that there are about 8 data points per cycle, consistent with a 125~Hz drive signal being sampled at 960~Hz. Because Qweak was an integrating experiment, the ramp signal was integrated over each MPS window (1~ms) of each sample, which ensured that points near 0 and 360 degrees could never be accessed. In fact, any ramp values which integrate over some portion of the fast return of the saw-tooth pattern will be incorrectly mapped to modulation phase. This can clearly be seen in plot (b) of  figure~\ref{fig:ramp_plots}. If one chooses to simply cut these incorrectly mapped events, a bias can be introduced into the sine and cosine response slopes found as seen in (c) and (d) of Figure \ref{fig:ramp_plots}. Early modulation results were subject to this bias until it was discovered and a simple solution implemented. A new variable ``ramp\_filled'' was created to replace ``ramp'' using a linear mapping from the ramp return region to the gap region of the ramp variable. Figure \ref{fig:ramp_filled} shows the same function generator signal seen in plot (b) of figure~\ref{fig:ramp_plots} but plotted against the new ``ramp\_filled'' variable.

An alternate solution to the problem of bias in the slopes from the gap region is to find both the sine and cosine coefficients simultaneously by fitting a function of
\[ 
\alpha+\alpha_{\sin}\sin(ramp)+\alpha_{\cos}\cos(ramp)
\]
to the detector and monitors versus ramp with the return region re-mapped or simply cut. An analytic solution to the coefficient values can be found making use of the known linearity of the monitor and detector response as follows:\\
\begin{equation*}\footnotesize
\left(\begin{array}{c}\sum_iQ_i\cos(\theta_i)\\\sum_iQ\sin(\theta_i)\\\sum_iQ\end{array}\right)=\left(\begin{array}{ccc}\sum_i\cos^2(\theta_i)&\sum_i\cos(\theta_i)\sin(\theta_i)&\sum_i\cos(\theta_i)\\\sum_i\sin(\theta_i)\cos(\theta_i)&\sum_i\sin^2(\theta_i)&\sum_i\sin(\theta_i)\\\sum_icos(\theta_i)&\sum_i\sin(\theta_i)&\sum_i1\end{array}\right)\left(\begin{array}{c}\alpha_{\cos}\\\alpha_{\sin}\\\alpha\end{array}\right),
\end{equation*}

with Q being the monitor or detector whose response is being measured and $\theta$ the modulation phase (ramp) and the sum going over the events in a given modulation cycle. In the final analysis, the exact analytic solution for the sine and cosine coefficients was utilized and the ``ramp\_filled'' variable was used to prevent unnecessary cutting of useful data. 

\begin{figure}[h]

\centering
\framebox{\includegraphics[width=5.5in]{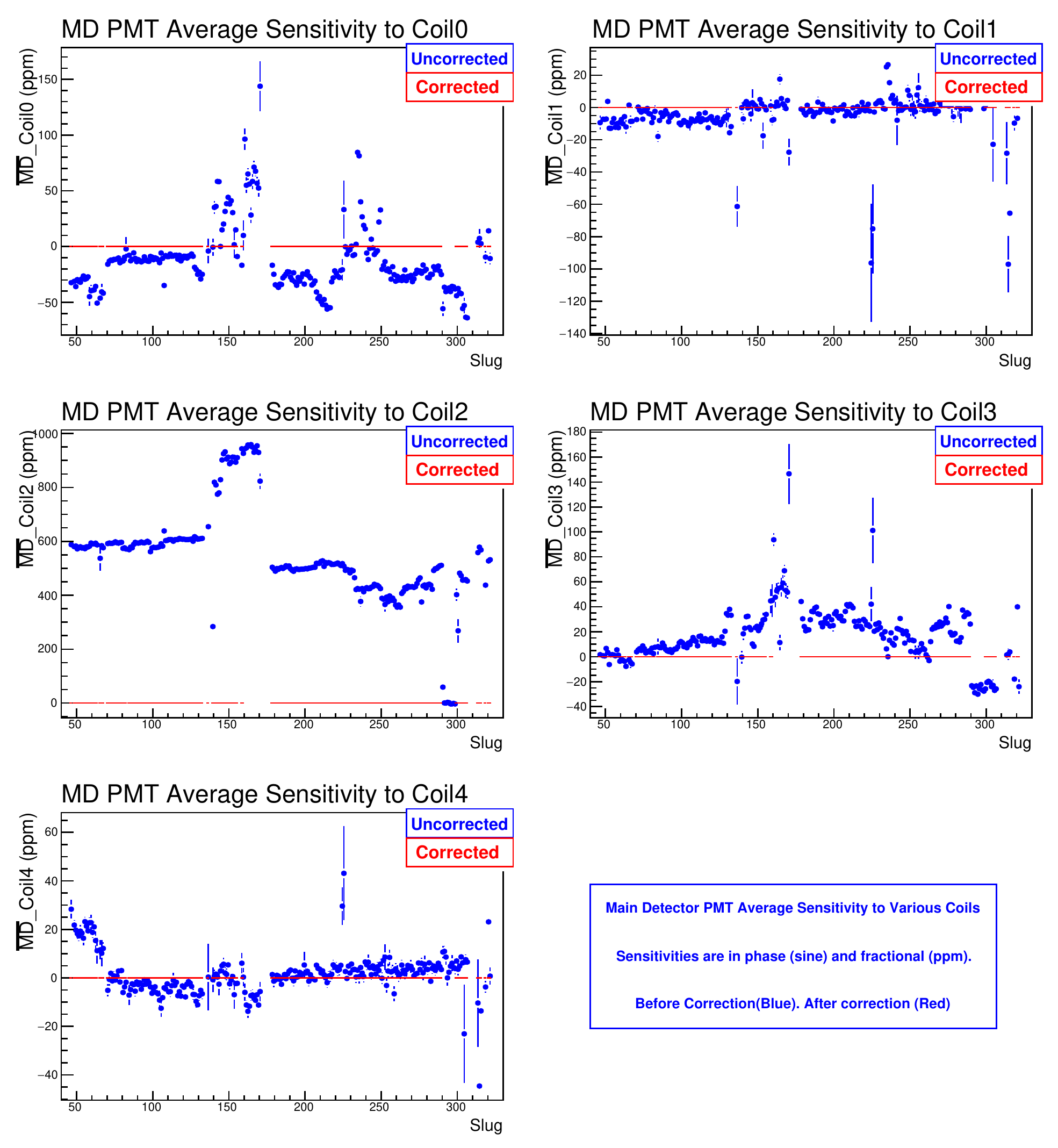}}
\caption{Main detector average in-phase (sine) response to the five different types of modulation. Blue is the raw responses and red is the residual response after the modulation corrections have been applied. Only the in-phase (sine) responses of detectors and monitors have been used in this analysis, so the residuals must be 0 by definition if the analysis code is working properly.}
\label{fig:sine_only_res}
\end{figure}

The function generator signals (see for example Figure \ref{fig:ramp_filled}) are expected to be proportional to the beam motion. Examples of detector responses to the drive signals were shown in Figure \ref{fig:det_vs_ramp}. A time history of the 5 raw sine response amplitudes for MDallbars is given in Figure \ref{fig:sine_only_res}. Also shown are the residual response amplitudes after correction using the 5-Coil sine-only analysis. The residual in-phase responses to modulation coils are indeed zero as required by linear algebra providing a first level cross-check of the analysis code.

\subsection{\label{sctn:inconsistency}Inconsistency Arising from Redundancy}
As previously mentioned, the subject of inconsistency in the modulation analysis naturally lends itself to two divisions:\\ 1). Inconsistency in the over-determined equation set manifesting itself as a residual sensitivity to modulation after correction.\\
2). Inconsistency in total correction to the dataset prescribed by parallel analyses using different coil selections.\\ Although these are related and it would appear that the first leads directly to the second, we will see that it is possible in some instances to have analyses whose differences in prescribed corrections cancel over time to produce a consistent net correction. The following two subsections individually deal with each of these two types of inconsistencies in the dataset. 
\subsubsection{Residual Sensitivity to Modulation}
\begin{sidewaysfigure}[t]

\centering
\framebox{\includegraphics[width=8.9in]{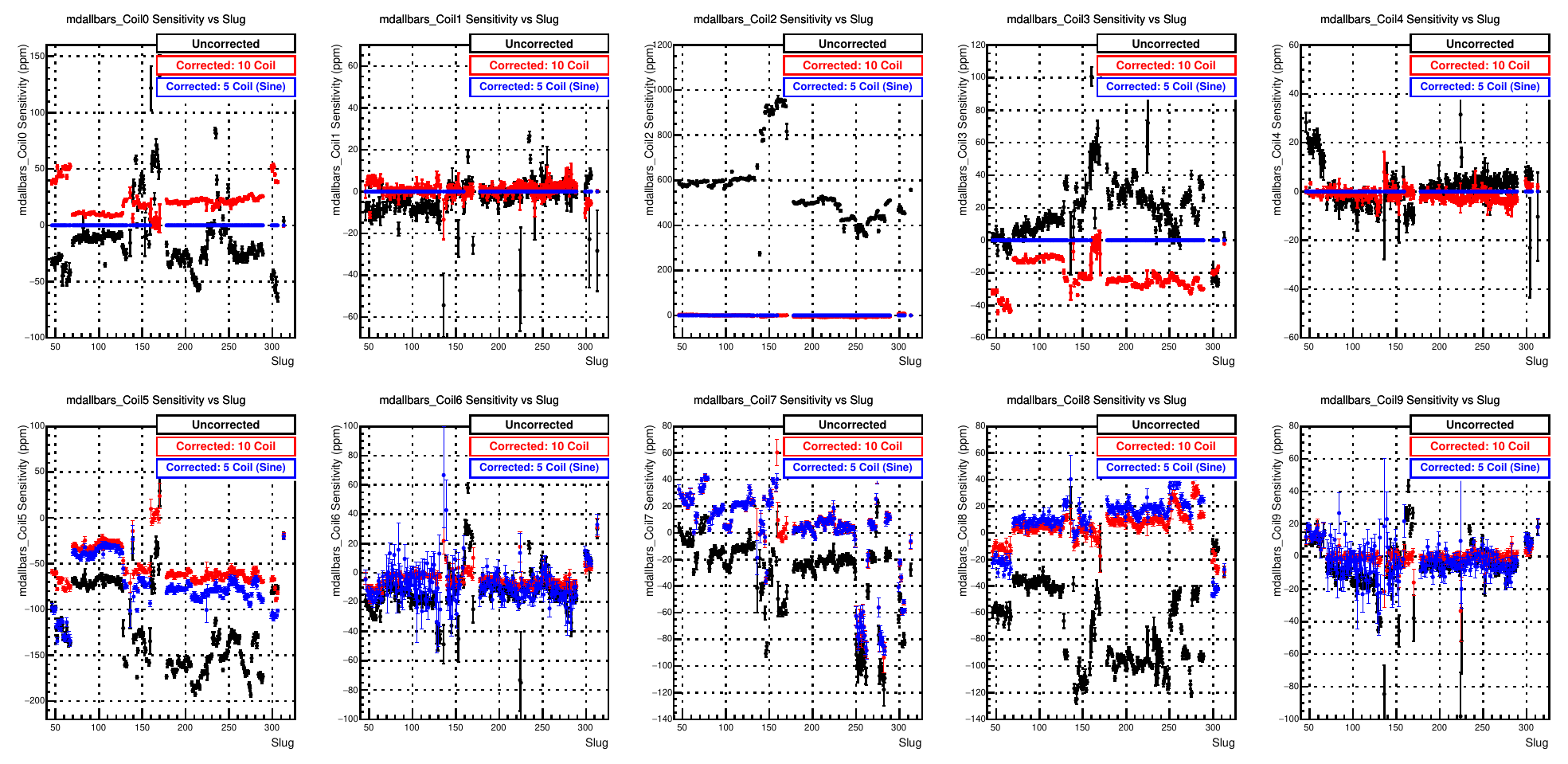}}
\caption{Comparison of residual MDallbars responses to the five different types of modulation. Black is the raw uncorrected responses and red is the residual response after correction using a 10-Coil analysis  and blue is the residual response after the 5-Coil (sine-only) modulation corrections have been applied. The 10-Coil analysis by definition gives the smallest collective least squares residual. Each plot represents a single term in the summation in equation~\ref{eq:total_residual}. It becomes apparent that the sine-only analysis creates a 0 in-phase residual response by increasing the residual out-of-phase response.}
\label{fig:10coil_and_sine_only_res}
\end{sidewaysfigure}

While introducing extra coils provides a way of reducing the uncertainty in the solution, it also provides a way of recognizing potential inconsistencies in the set of linear equations used.  Residual sensitivity of detectors to modulation after correction is the first place to look for this inconsistency. Figure \ref{fig:10coil_and_sine_only_res} compares the residual sensitivity of MDallbars to the modulation coils for a sine-only analysis and a 10 coil analysis. If one only had access to the 5 coils of the sine-only analysis, that is, if FFB were paused during modulation and only 5 modulation patterns were used, the inconsistency in the modulation system/analysis would be hidden. The residuals would not be a useful criterion for determining the effectiveness of the applied correction. With more than 5 coils, the total residual is minimized but not forced to be zero like the solution shown in figure~\ref{fig:sine_only_res}. Equation~\ref{eq:chisquaresolution_no_variance} shows that the ``total residual'' being minimized is given by
\begin{equation}
Total~Residual=\sqrt{\sum_{i=1}^{N}\left(\frac{\partial D_d}{\partial C_i} - \sum_{m=1}^{5}\frac{\partial D_d}{\partial M_m}\frac{\partial M_m}{\partial C_i} \right)^2}.
\label{eq:total_residual}
\end{equation}

\begin{figure}[t]

\centering
\framebox{\includegraphics[width=5in]{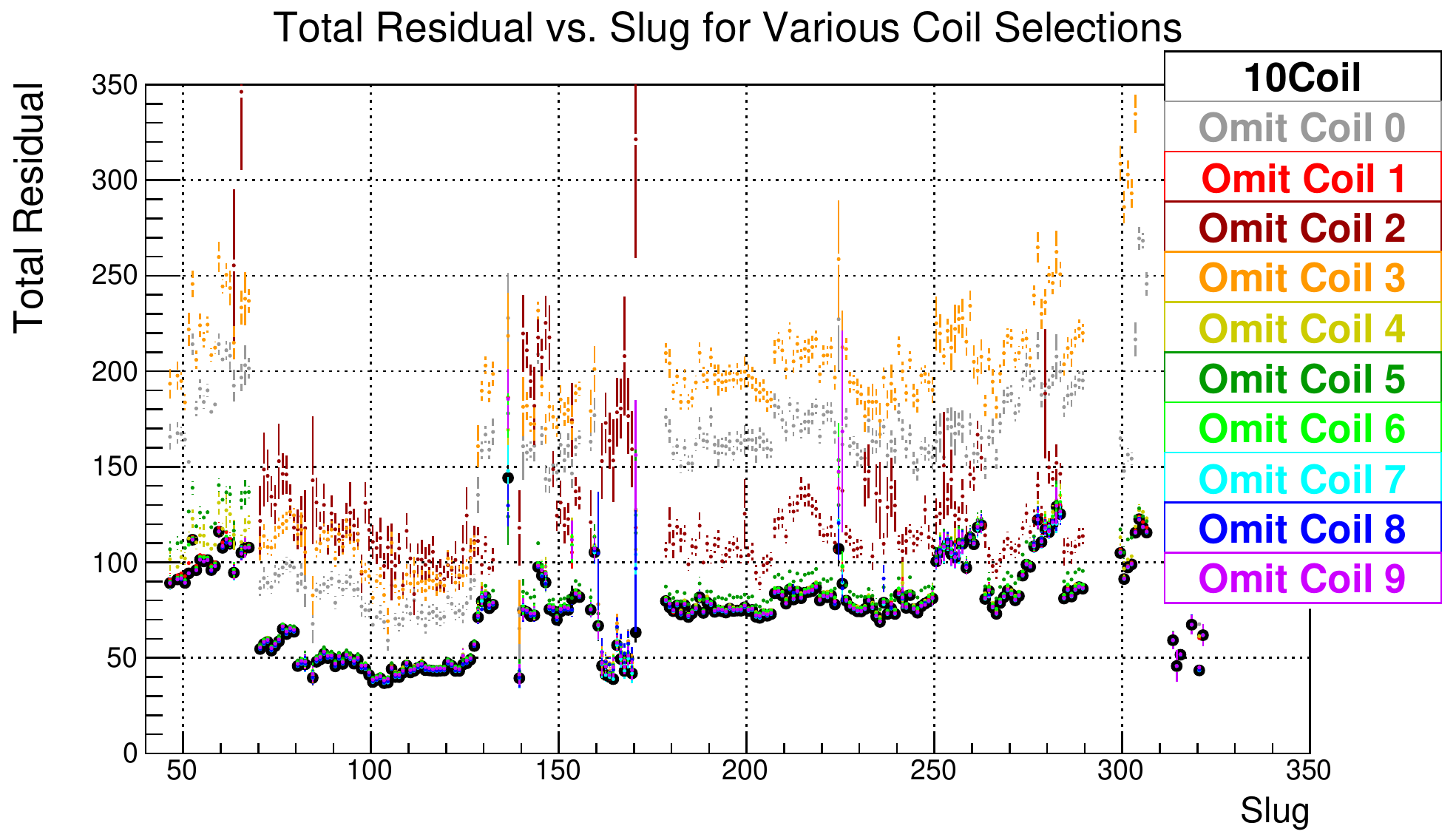}}
\caption{Comparison of total residual (Equation \ref{eq:total_residual}) versus slug for analyses with various coil selections.}
\label{fig:chisquare_vs_slug}
\end{figure}

For the \Qs modulation dataset, analyses with different complete coil sets yield systematically different solutions, pointing to a problem in either the data or the analysis procedure. With 10 coils (probably only 9 useful ones) available, there are dozens of sets of independent coils that adequately span the phase space, and while the author has not analyzed all of the possible sets, the 32 schemes he has completed present a not-so-consistent picture. Figure~\ref{fig:chisquare_vs_slug} compares the total residuals for a few different coil selections and clearly illustrates that the 10 coil analysis always produces the minimum total residual as defined by equation~\ref{eq:total_residual}.

If the residuals are a result of a problem in the analysis procedure, then the analysis should also fail to correct beam position monitors, that is, it should fail to remove monitor sensitivity to the modulation coils just as it fails in the detector correction. In a linear optics model for the electron beam a complete set of beam position monitors that are sufficiently sensitive to the phase space of beam position, slope and energy should be able to predict the response of any other monitor. Of course, given the modulation calibration procedure for determining monitor responses, a prediction will only be available for monitors downstream of the modulation coils. In principle, therefore, a complete\footnote{In this context ``complete'' means that you have a set of monitors that spans the 5 dimensional space of position, slope and energy''} set of 5 monitors should provide sufficient information to predict any other monitor and with properly measured correction slopes be able to completely remove sensitivity to modulation. 

\begin{sidewaysfigure}[t]

\centering
\framebox{\includegraphics[width=8.9in]{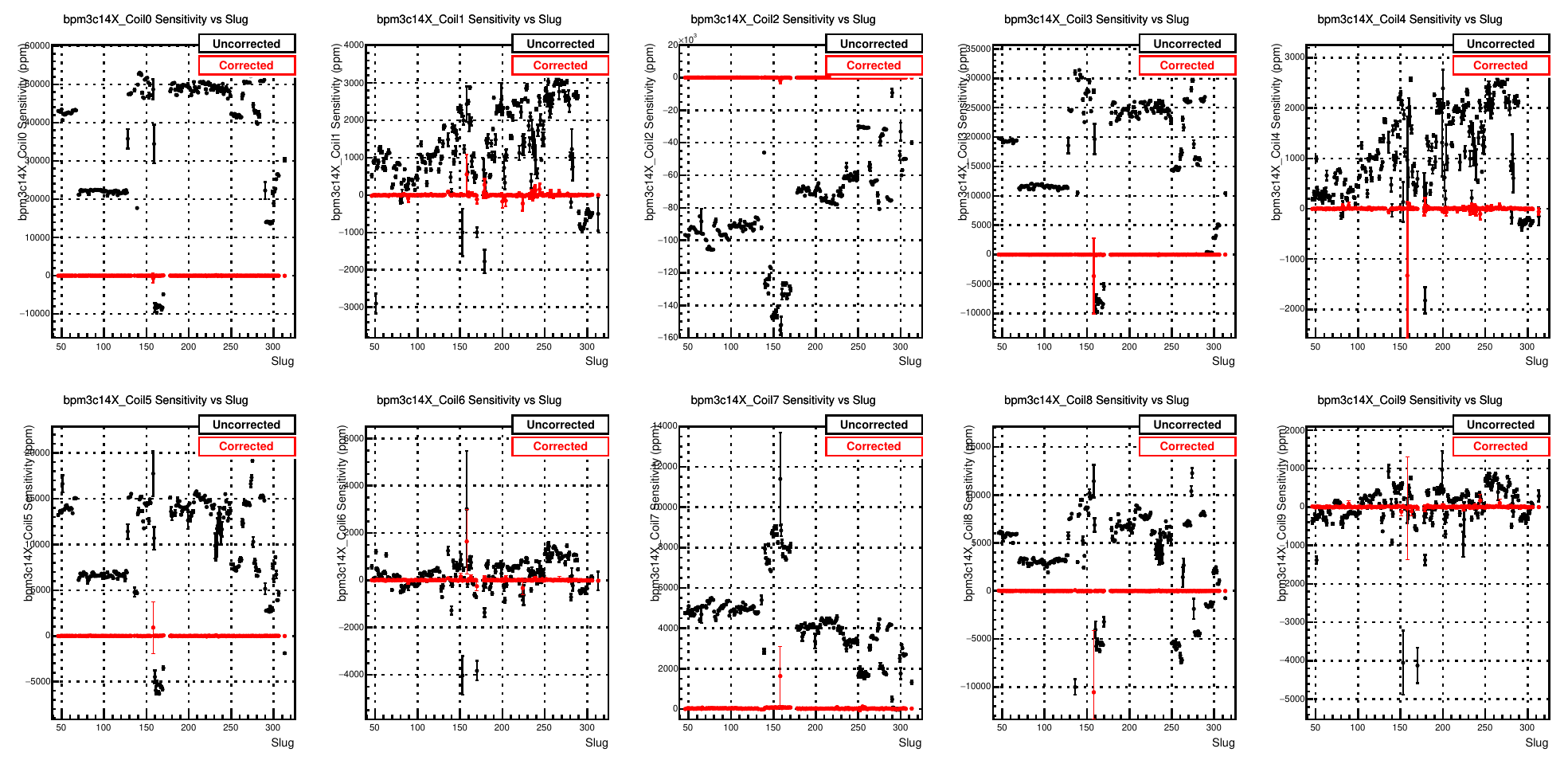}}
\caption{Comparison of residual bpm3c14X responses to the five different types of modulation. Black is the raw uncorrected responses and red is the residual response after correction using a 10-Coil analysis. The sensitivity to the modulation coils is almost completely removed by the correction using targetX, targetXSlope, targetY, targetYSlope, and bpm3c12X.}
\label{fig:residual_bpm3c14X}
\end{sidewaysfigure}
\begin{sidewaysfigure}[t]

\centering
\framebox{\includegraphics[width=8.9in]{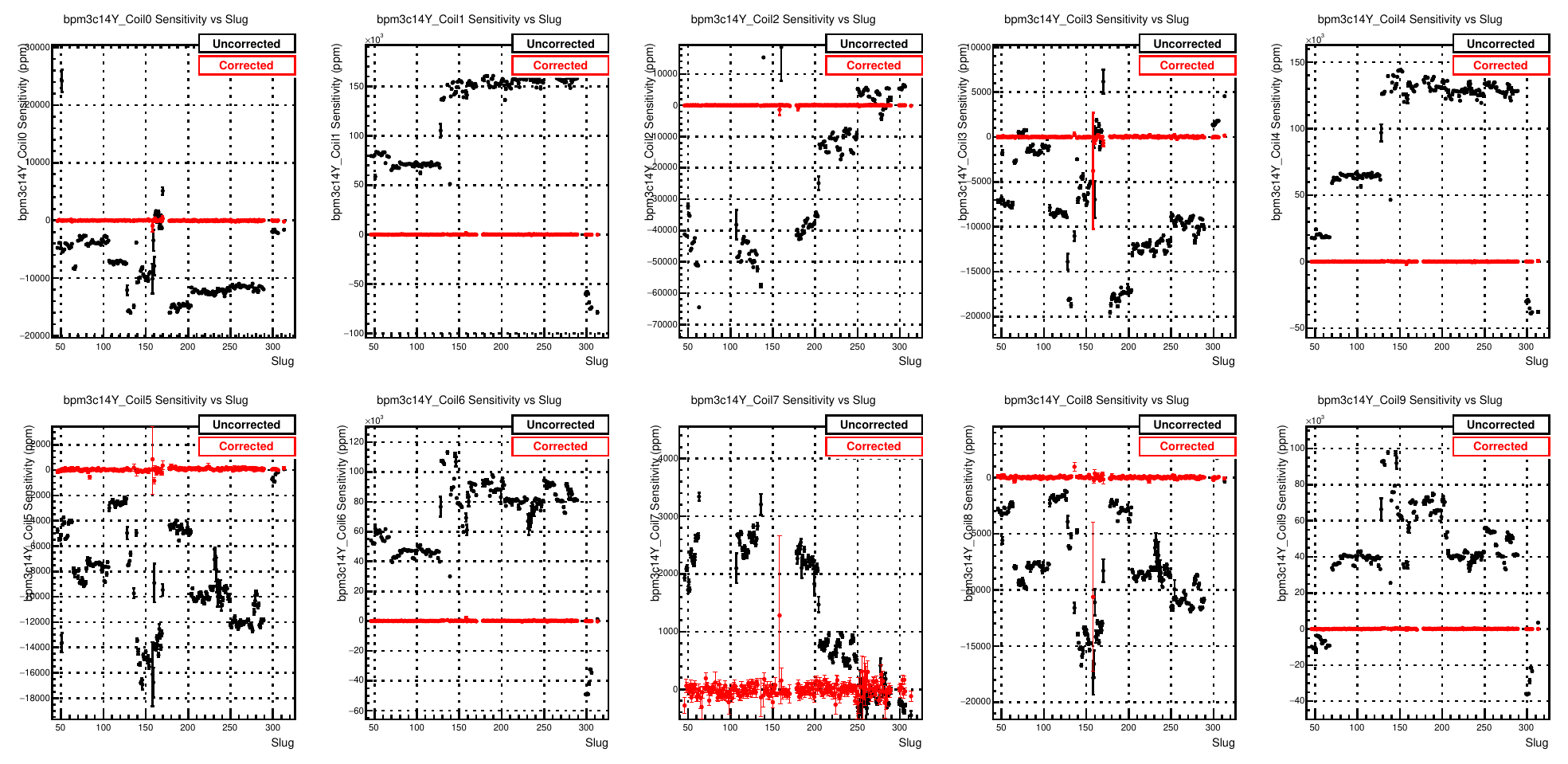}}
\caption{Comparison of residual bpm3c14Y responses to the five different types of modulation. Black is the raw uncorrected responses and red is the residual response after correction using a 10-Coil analysis. The sensitivity to the modulation coils is almost completely removed by the correction using targetX, targetXSlope, targetY, targetYSlope, and bpm3c12X.}
\label{fig:residual_bpm3c14Y}
\end{sidewaysfigure}

\begin{figure}[h]

\centering
\framebox{\includegraphics[width=3.5in]{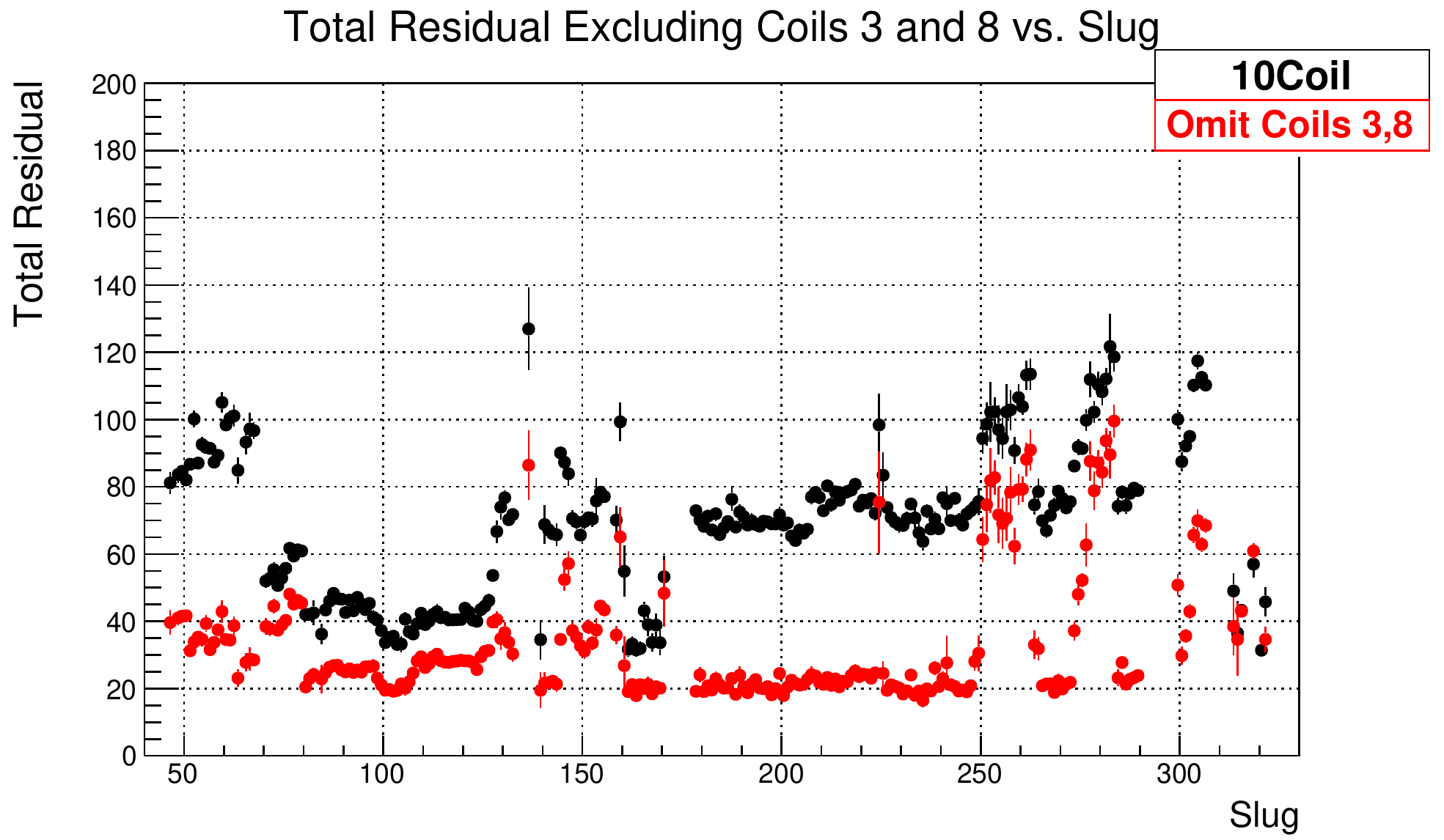}}
\caption{Comparison of total residual responses after correction using slopes from a 10-Coil analysis and those from an 8-Coil analysis where coils 3 and 8 are omitted. In this plot, the residual is calculated omitting the residuals from coils 3 and 8 as well. Coils 3 and 8 are associated with X2-modulation. If an issue were found with the X2-modulation coils, for example, this particular residual calculation would be of more interest than the full residual over 10 coils.}
\label{fig:residual_omit38}
\end{figure}

A set of monitors not included in the 5 used for the correction were treated as detectors in the analysis code and corrected to remove sensitivity to modulation. The analysis showed that the monitors chosen for the correction (4 target variables plus bpm3c12X) are effective at correcting other monitors. Thus, once the responses of the electron beam to position, angle and energy changes at a given location along the beamline have been calibrated to the responses of the five monitors to the same changes, then any given response at that location can be accurately predicted using only the five monitors. The modulation analysis is simply an accurate calibration procedure. Figures \ref{fig:residual_bpm3c14X} and \ref{fig:residual_bpm3c14Y} compare the uncorrected and corrected sensitivities of the beam position monitor at girder 3c14 on the Hall C beamline. The horizontal beam response is given by bpm3c14X and vertical response by bpm3c14Y. Notice how well the monitor response is zeroed after correction when compared with the MDallbars response in figure~\ref{fig:chisquare_vs_slug}. 
\FloatBarrier
A number of conclusions can be drawn from the study where monitors are corrected to remove modulation sensitivity:
\begin{itemize}
\item{This provides a second validation of the analysis procedure and code (the first being the residuals going to 0 with equal number of monitors and coils.)}
\item{The expected 5 degrees of freedom seem sufficient to describe the responses of the monitors.}
\item{Any model for the main detector residuals involving a failure of the monitors must specifically be able to explain their success in correcting other monitors. This observation narrows the space of allowed failure models.}
\item{It appears that the detectors are sensitive to some effect that the monitors do not measure, at least not well. This effect may be a real beam property like halo or spot size that the monitors do not resolve well. Alternately, it could be an artifact introduced to the detectors by the coils themselves such as noise or pedestal shifts coherent with the modulation signal.}
\end{itemize} 
Although equation~\ref{eq:total_residual} guarantees the 10-Coil analysis will provide the smallest total residual, it is not guaranteed to provide the optimal correction slopes. Furthermore, if one chooses to calculate the residual differently by only summing over a subset of the coils, the 10-Coil solution will not always be optimal in terms of residual size. Figure~\ref{fig:residual_omit38} demonstrates that when the residuals arising from coils 3 and 8 are omitted from the summation (see Table \ref{tab:coil_nomenclature} for definition of coils), the 10-Coil analysis yields a much larger residual than does the analysis where the same coils are omitted in the slope calculation. The residual shown is given as
\[
Residual=\sqrt{\sum\limits_{\substack{i=1\\i\neq 3,8}}^{10}\left(\frac{\partial D_d}{\partial C_i} - \sum_{m=1}^{5}\frac{\partial D_d}{\partial M_m}\frac{\partial M_m}{\partial C_i} \right)^2}.
\]

\begin{table}[!h]
\caption{Modulation coil nomenclature used in the analysis discussion.}
\begin{center}
\begin{tabular}[h]{|c||c|c|}\hline
Modulation&In-Phase Response&Out-of-Phase Response\\
Type&(Sine Amplitude)&(Cosine Amplitude)\\\hline\hline
X1-Type&Coil 0&Coil 5\\
X2-Type&Coil 3&Coil 8\\\hline\hline
Y1-Type&Coil 1&Coil 6\\
Y2-Type&Coil 4&Coil 9\\\hline\hline
E-Type&Coil 2&Coil 7\\\hline
\end{tabular}
\end{center}
\label{tab:coil_nomenclature}
\end{table}

One could imagine situations where cumulative evidence might force one away from the default 10-Coil choice.  Imagine, for example, a situation where it was found that whenever a particular modulation coil was active, an artifact appeared in the data that was inconsistent with what is known to be true about the system. One such example investigated was the introduction of noise coherent with the drive signal into the detector data. If a response to the modulation signal in any modulation period were found in a channel known to have no correlation with beam dynamics, this would signal a problem with electronics pickup possibly biasing the results. In that case, it would be wise to remove any coils associated with this artifact from the calculation of both the correction slopes and the residuals. 

\begin{figure}[ht]
\centering
\includegraphics[width=6in]{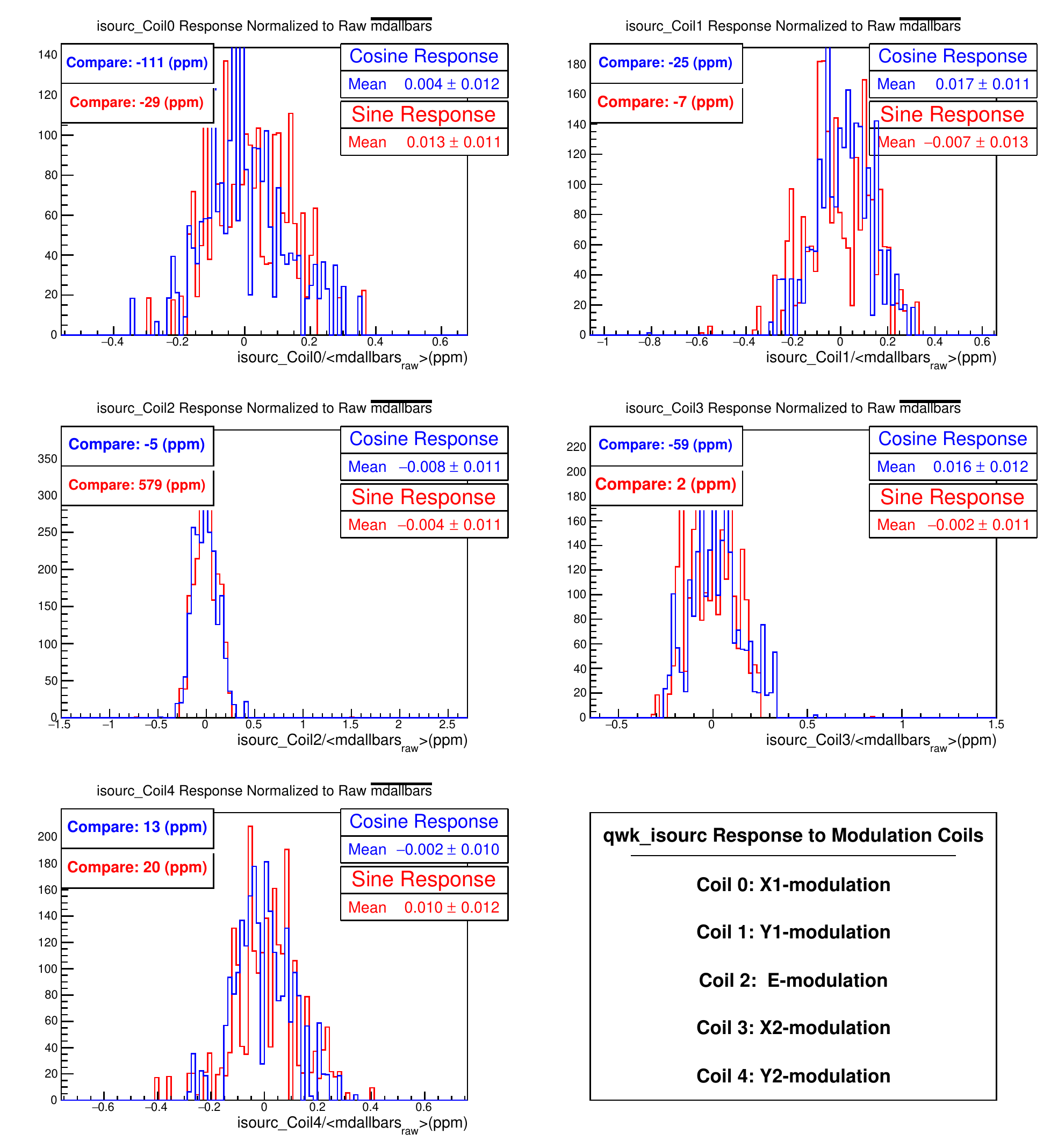}
\caption{\label{fig:isourc}Response of ``qwk\_isourc'', a current source located in detector hut in the experimental hall and fed through same pre-amplifier and electronics chain as the main detector. Responses shown are normalized to raw MDallbars mean (not normalized to current) for purposes of direct comparison with the MDallbars response given in the ``Compare'' text boxes. The average raw MDallbars value during this period was 5.3~V and the raw isourc value was 8.5~V. As usual, ``Sine'' and ``Cosine'' are the amplitudes of the responses in phase with the modulation coils and 90 degrees out of phase respectively.}
\end{figure}

\begin{figure}[ht]
\centering
\includegraphics[width=6in]{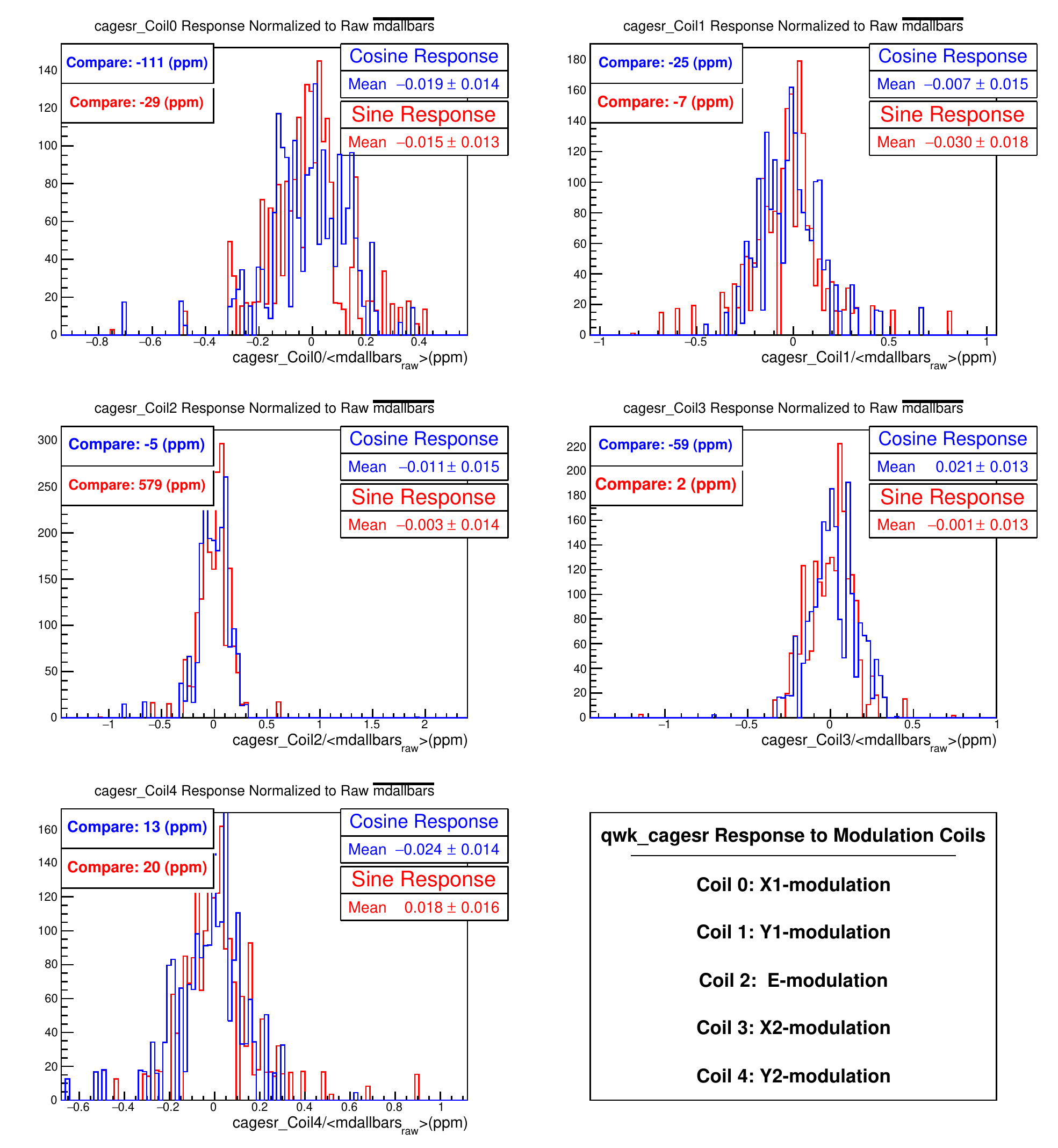}
\caption{\label{fig:cagesr}Response of ``qwk\_cagesr'', a battery located in the ``cage''in the Counting House that housed main experimental DAQ electronics. This battery was read out using one of the extra channels used to read out the main detector. The responses shown are normalized to raw MDallbars mean (not normalized to current) for purposes of direct comparison with the MDallbars response given in the ``Compare'' text boxes.  The average raw MDallbars value during this period was 5.3~V and the raw cagesr value was 5.8~V. As usual, ``Sine'' and ``Cosine'' are the amplitudes of the responses in phase with the modulation coils and 90 degrees out of phase respectively.}
\end{figure}

One detector channel in the \Qs analysis chain was fed by a signal from a battery and was setup to test for effects such as the one mentioned in the previous paragraph. Any response to modulation seen on this channel would certainly be a sign of electronics pickup. Although the data for this channel was not analyzed over the whole \Qs dataset, the response for a randomly chosen period was found to be nearly consistent with zero. Figures \ref{fig:isourc} and \ref{fig:cagesr} show the response of the null channels ``qwk\_isourc'' and ``qwk\_cagesr'' for a period near the beginning of Run 1 (between runs 10000 and 10200). Comparison of the responses of the null channels to the main detector response shows that they are typically 3 orders of magnitude apart, which is far too small an effect to account for the highly non-zero MDallbars residual response at the beginning of Run 1.
\FloatBarrier

\subsubsection{Differences in Prescribed Corrections}
With information from 10 coils and many beam monitors and with the analysis determining only 5 detector to monitor slopes, there are many combinations of coils and monitors from which the requisite information could be obtained (see Table \ref{tab:coil_nomenclature} for definition of terms). In this section the total correction prescribed by analyses using different coil and monitor sets is investigated. The following constraints must be kept in mind when selecting different sets:
\begin{itemize}
\item{Both monitors and coil sets must adequately span the five dimensions of beam distortion space. That is to say that any set of coils must adequately modulate the beam in all five dimensions and any set of monitors must be able to collectively resolve those five dimensions. }
\item{There are four X-type (horizontal direction) modulation coils and four Y-type (vertical direction) modulation coils. Any useful coil selection must have at a minimum two X-type coils and two Y-type coils. Leaving out any one X-type or Y-type coil gives a total of 8 possible sets. Leaving out any two X-type or two Y-type coils gives a further 12 sets for a total of 20 possible coil sets in addition to the full 10-Coil analysis. Not all of these are guaranteed to have adequate information to provide useful correction slopes.}
\item{There is one E-type (energy) modulation coil. A second but small out-of-phase sensitivity exists for energy as well, but is so small relative to the in-phase energy modulation coil that it is considered negligible and, in fact, its source is not well understood. Omitting the out-of-phase energy coil (Coil 7) makes almost no difference in the analysis but any analysis omitting the large in-phase energy modulation coil (Coil 2) should be considered critically due to lack of energy resolution. Three sets, omitting either in-phase or out-of-phase or omitting both can be considered keeping in mind that most of the energy information will be lost when the in-phase response is omitted. With FFB active during X and Y-type modulations, energy responses will still be measured in the other modulation coils although not nearly as well.}
\item{Although analyses removing 3, 4 and 5 coils are also possible, only a limited number of these will be investigated as evidence is provided for their utility.}
\item{When two coils are removed from either the X-type or Y-type directions, the utility of the redundancy cross-check is lost. In the 5-coil analysis this lack of redundancy was apparent in the precisely zero residuals. When the redundancy cross-check is lost, clues as to the success or failure of the analysis must be provided by other methods (see Section \ref{sctn:residual_correlations}).}
\item{Differences in monitor resolution of the beam parameters will mean that the width of the detector correction will depend upon the monitor set chosen. The monitor set of choice for the modulation analysis are the four target variables and the beam position monitor BPM3c12X because they are particularly well suited for resolving the five dimensions of beam modulation (see Section \ref{sctn:lin_reg} for definition of these monitors). Different monitor sets are expected to yield corrections consistent with each other within the expected statistical deviation from additional detector width associated with differences in monitor resolution.}
\end{itemize}

Tables \ref{tab:run1_dithering_corrections_table} and \ref{tab:run2_dithering_corrections_table} compare the corrections for Runs 1 and 2 for analyses utilizing different coil sets. Coil sets simultaneously omitting coils 0 and 3 and coils 5 and 8 were removed from the comparisons for Run 2 because omitting these coils from the analysis produced sections of data with insufficient information for extraction of useful correction slopes. The results shown are averaged over Wien states weighted by the main detector average asymmetry (MDallbars) errors\footnote{The error weights are the runlet-level (4-5 minutes of data) inverse variance of the MDallbars asymmetry divided by the number of quartets in the runlet: $w_{runlet}=\frac{1}{\sigma_{runlet}^2/N}$}. It becomes apparent, especially in Run 1, that there is a large disparity in the corrections prescribed by analyses with different coil sets. A careful look at the tables shows that omitting coil 3 creates the worst outliers. All coils sets considered invalid are shown as grayed-out in advance of evidence of their unreliability provided in Section \ref{sctn:residual_correlations}.

As previously explained in Section \ref{sctn:electron_source}, \Qs utilized two different slow reversals to cancel out certain types of false asymmetries: 1) half-wave plate insertion and removal on an 8 hour timescale to flip electron source laser helicity;  and  2) monthly double Wien filter reversal of the electron beam helicity relative to its helicity coming off the photo-cathode. With these slow reversal techniques in place, it is not inconceivable that inconsistencies apparent in the modulation could be created by an effect that largely cancels under these reversal, producing the proper average correction. The corrections for Run 2 in Table \ref{tab:run2_dithering_corrections_table} seem to indicate that this type of cancelation is happening and producing consistent average corrections at the $\pm1$~ppb level across all dithering schemes. However, for Run 1 the range of average corrections is a much larger $\pm 6$~ppb. This range helps provide a scale for the effect of inconsistencies in the modulation analysis on the physics asymmetry correction.  

\begin{table}[!h]

\caption{Run 1 dithering corrections for MDallbars asymmetry compared for different coil selections. Grayed-out sets often seen to be outliers are sets considered suspect for reasons investigated in Section \ref{sctn:residual_correlations}.}
\begin{center}
\begin{tabular}[h]{|l||c|c|c|c|c||c|}\hline
Dithering& Wien 1& Wien 2& Wien 3& Wien 4& Wien 5& Total\\
~Scheme&(ppb)&(ppb)&(ppb)&(ppb)&(ppb)&(ppb)\\\hline\hline
10-Coil& -24.1& +3.4& -44.6& -16.8& -17.5& -19.2\\\hline
{\color{Gray}Omit0,3}&{\color{Gray} -36.9}&{\color{Gray} -51.0}&{\color{Gray} -79.5}&{\color{Gray} -21.1}&{\color{Gray} -25.3}&{\color{Gray} -43.8}\\\hline
Omit0,5& -19.2& +21.6& -43.9& -16.9& -14.9& -13.6\\\hline
{\color{Gray}Omit0,8}&{\color{Gray} -16.1}&{\color{Gray} +21.8}&{\color{Gray} -68.8}&{\color{Gray} -24.6}&{\color{Gray} -16.0}&{\color{Gray} -20.2}\\\hline
{\color{Gray}Omit3,5}&{\color{Gray} -53.3}&{\color{Gray} -49.8}&{\color{Gray} -59.0}&{\color{Gray} -16.4}&{\color{Gray} -16.8}&{\color{Gray} -38.9}\\\hline
{\color{Gray}Omit3,8}&{\color{Gray} -88.7}&{\color{Gray} -102.7}&{\color{Gray} -77.6}&{\color{Gray} -22.0}&{\color{Gray} -19.3}&{\color{Gray} -62.3}\\\hline
{\color{Gray}Omit5,8}&{\color{Gray} -10.0}&{\color{Gray} +22.3}&{\color{Gray} -37.9}&{\color{Gray} -16.1}&{\color{Gray} -17.2}&{\color{Gray} -11.0}\\\hline
{\color{Gray}Omit1,4}&{\color{Gray} -28.3}&{\color{Gray} +2.6}&{\color{Gray} -61.2}&{\color{Gray} +11.7}&{\color{Gray} -99.2}&{\color{Gray} -34.1}\\\hline
Omit1,6& -19.5& +7.4& -44.5& -15.8& -17.1& -17.3\\\hline
Omit1,9& -23.9& +3.8& -44.6& -17.2& -17.2& -19.1\\\hline
Omit4,6& -46.0& +4.5& -44.5& -15.3& -17.8& -22.0\\\hline
Omit4,9& -17.3& +1.7& -44.7& -17.4& -18.0& -18.9\\\hline
Omit6,9& -23.6& -1.3& -41.9& -7.4& -12.1& -16.7\\\hline
Omit2,7& -16.6& -2.9& -53.1& -17.9& -19.3& -22.0\\\hline
Omit 0& -16.4& +19.3& -55.9& -20.5& -16.9& -17.4\\\hline
Omit 1& -24.2& +3.6& -44.8& -16.9& -17.3& -19.2\\\hline
Omit 2& -14.8& -6.2& -47.3& -15.2& -16.6& -20.1\\\hline
{\color{Gray}Omit 3}&{\color{Gray} -84.0}&{\color{Gray} -95.8}&{\color{Gray} -75.4}&{\color{Gray} -20.7}&{\color{Gray} -19.2}&{\color{Gray} -59.2}\\\hline
Omit 4& -23.7& +2.5& -44.7& -16.7& -18.2& -19.5\\\hline
Omit 5& -11.5& +20.4& -38.7& -16.1& -17.0& -11.8\\\hline
Omit 6& -24.0& +3.3& -44.5& -15.6& -17.2& -18.9\\\hline
Omit 7& -24.1& +3.4& -44.7& -17.1& -17.5& -19.3\\\hline
Omit 8& -23.5& +4.3& -44.3& -16.9& -17.7& -18.9\\\hline
Omit 9& -23.8& +3.1& -44.6& -17.2& -17.5& -19.3\\\hline
\end{tabular}
\end{center}
\label{tab:run1_dithering_corrections_table}
\end{table}

\begin{table}[!h]

\caption{Run 2 dithering corrections for MDallbars asymmetry compared for different coil selections. Grayed-out sets often seen to be outliers are sets considered suspect for reasons investigated in Section \ref{sctn:residual_correlations}.}
\begin{center}
\begin{tabular}[h]{|l||c|c|c|c|c|c||c|}\hline
Dithering& Wien 6& Wien 7& Wien 8a& Wien 8b& Wien 9a& Wien 9b& Total\\
~Scheme&(ppb)&(ppb)&(ppb)&(ppb)&(ppb)&(ppb)&(ppb)\\\hline\hline
10-Coil& -25.5& +35.6& -12.3& -4.3& +2.0& +1.3& -1.7\\\hline
Omit0,5& -24.7& +36.0& -13.9& -5.3& +1.7& +0.2& -2.4\\\hline
{\color{Gray}Omit0,8}&{\color{Gray} -27.2}&{\color{Gray} +36.4}&{\color{Gray} -33.8}&{\color{Gray} +0.9}&{\color{Gray} -0.1}&{\color{Gray} +0.2}&{\color{Gray} -5.0}\\\hline
{\color{Gray}Omit3,5}&{\color{Gray} -26.3}&{\color{Gray} +36.3}&{\color{Gray} -17.4}&{\color{Gray} -5.6}&{\color{Gray} +2.9}&{\color{Gray} +4.5}&{\color{Gray} -1.8}\\\hline
{\color{Gray}Omit3,8}&{\color{Gray} -29.3}&{\color{Gray} +38.5}&{\color{Gray} -23.6}&{\color{Gray} -1.2}&{\color{Gray} +3.0}&{\color{Gray} +13.9}&{\color{Gray} +0.2}\\\hline
{\color{Gray}Omit1,4}&{\color{Gray} -14.2}&{\color{Gray} +31.3}&{\color{Gray} -40.8}&{\color{Gray} -11.0}&{\color{Gray} -2.2}&{\color{Gray} -1.0}&{\color{Gray} -8.0}\\\hline
Omit1,6& -25.7& +37.6& -12.1& -4.4& +1.9& +1.3& -1.6\\\hline
Omit1,9& -25.5& +35.6& -12.2& -4.2& +2.0& +1.3& -1.7\\\hline
Omit4,6& -25.4& +37.8& -12.2& -4.2& +1.9& +1.5& -1.5\\\hline
Omit4,9& -25.1& +35.7& -12.5& -4.4& +1.9& +1.4& -1.8\\\hline
Omit6,9& -27.3& +36.1& -10.8& -3.7& +1.9& +1.5& -1.5\\\hline
Omit2,7& -28.8& +53.2& -13.0& -5.4& +2.3& +3.3& -0.5\\\hline
{\color{Gray}Omit 0}&{\color{Gray} -26.2}&{\color{Gray} +37.8}&{\color{Gray} -23.4}&{\color{Gray} -1.8}&{\color{Gray} +0.4}&{\color{Gray} +1.0}&{\color{Gray} -3.3}\\\hline
Omit 1& -25.3& +36.5& -12.3& -4.2& +2.0& +1.3& -1.6\\\hline
Omit 2& -27.0& +33.1& -12.9& -5.3& +2.2& +2.7& -2.0\\\hline
{\color{Gray}Omit 3}&{\color{Gray} -29.1}&{\color{Gray} +37.9}&{\color{Gray} -22.5}&{\color{Gray} -1.8}&{\color{Gray} +2.8}&{\color{Gray} +7.9}&{\color{Gray} -1.2}\\\hline
Omit 4& -25.1& +36.1& -12.6& -4.3& +1.9& +1.4& -1.7\\\hline
Omit 5& -24.6& +35.4& -9.3& -5.2& +2.2& +1.0& -1.4\\\hline
Omit 6& -25.4& +37.3& -12.1& -4.2& +1.9& +1.4& -1.6\\\hline
Omit 7& -25.5& +35.7& -12.3& -4.3& +2.0& +1.4& -1.7\\\hline
Omit 8& -25.5& +37.7& -12.0& -4.1& +2.0& +1.4& -1.5\\\hline
Omit 9& -25.2& +35.7& -12.2& -4.2& +2.0& +1.3& -1.7\\\hline
\end{tabular}
\end{center}
\label{tab:run2_dithering_corrections_table}
\end{table}

Another useful set of cross-checks used in parity-violation experiments are null asymmetries which are a measurement of the differences cancelled by slow reversals. Null asymmetries that are small or consistent with zero are a good indication that there are not significant sources of systematic false asymmetries being cancelled by the slow helicity reversal. If the source of the cancellations is understood, it is not necessary that the null be consistent with 0. After all, one of the purposes of slow helicity reversal is to cancel small false asymmetries. However, large cancellations of effects not well understood sometimes requires the addition of a systematic error. For this reason it is appropriate to look at beam modulation corrections prescribed to null asymmetries as well. For the purposes of \Qs there are three slow helicity-reversal mechanisms: the insertable half-wave plate (IHWP), the double Wien filter, and the g-2 reversal from traveling around the accelerator.

The two most important null asymmetries relevant to this study are defined as follows:
\begin{itemize}
\item{{\bf ``IHWP+Spin-reversal'' Null}~~~~~$\left[\frac{A_{IN}^{raw}+A_{OUT}^{raw}}{2}\right]$:\\ This null is the average asymmetry of both half-wave plate (HWP) states with no sign corrections applied to the asymmetries for any of the three slow reversals. The parity-violating physics asymmetry is expected to cancel leaving false asymmetries due to effects such as electrical pickup of the helicity reversal signal or mechanical effects such as lensing in the electron source Pockels cell. }
\item{{\bf ``IHWP-only'' Null}~~~~$\left[\frac{Sign(A_{IN}^{raw})-Sign(A_{OUT}^{raw})}{2}\right]$:\\ This null is the difference between average asymmetries of IHWP states, each individually sign-corrected for all slow reversals. By sign-correcting for all slow reversals and then taking the difference between the two IHWP states, the effect of the IHWP reversal is isolated. This null reveals false asymmetries that cancel with the IHWP reversal but survive the other slow reversal cancelations.}
\end{itemize}

Tables \ref{tab:run1_null_corrections_table} and \ref{tab:run2_null_corrections_table} show the range of corrections for Runs 1 and 2 respectively to the ``IHWP+Spin-reversal'' null asymmetry provided by analyses with various coil selections. Tables \ref{tab:run1_ihwp_only_null_corrections_table} and \ref{tab:run2_ihwp_only_null_corrections_table} show the analogous range of corrections to the ``IHWP-only'' null. From these tables one can easily observe: 1) that the inconsistency in the various modulation corrections is more apparent in the null asymmetry corrections than in the physics asymmetry corrections; and 2) variations in the null asymmetries are at least as large if not larger in Run 2 than they are in Run 1 even though the spread in the physics asymmetry corrections is much smaller.

Inconsistencies on the order of 30~ppb in the null asymmetry corrections using various coil selections help give a scale to the issue with the beam modulation analysis. If \Qs were not set up to benefit from slow reversal cancellation or if a single IHWP state were modulation-corrected, there would be discrepancies of tens of ppb between different schemes. This does not mean that there is an error in the modulation correction slopes that is IHWP-correlated. The modulation correction comes from the product of the correction slopes and monitor differences and this discrepancy in the null mainly points to the slow-reversal cancellation of monitor differences, that is, the helicity-correlated beam properties measured in the monitors. Utilizing the full cancellation of slow reversals allows the assignment of a much smaller systematic error for this discrepancy.

\begin{table}[!h]
\caption{Run 1 dithering corrections to ``IHWP-only'' null asymmetry for various coil selections. Grayed-out sets considered suspect  for reasons investigated in Section \ref{sctn:residual_correlations}.}
\begin{center}
\begin{tabular}[h]{|l||c|c|c|c|c||c|}\hline
Dithering& Wien 1& Wien 2& Wien 3& Wien 4& Wien 5& Total\\
~Scheme&(ppb)&(ppb)&(ppb)&(ppb)&(ppb)&(ppb)\\\hline\hline
10-Coil& -14.6& -13.1& +38.5& -115.8& +6.3& -19.0\\\hline
{\color{Gray}Omit0,3}&{\color{Gray} -14.3}&{\color{Gray} -76.2}&{\color{Gray} +76.8}&{\color{Gray} -312.0}&{\color{Gray} -203.6}&{\color{Gray} -107.1}\\\hline
Omit0,5& -6.3& -9.7& +37.2& -102.1& +8.8& -13.7\\\hline
{\color{Gray}Omit0,8}&{\color{Gray} -6.4}&{\color{Gray} -15.1}&{\color{Gray} +57.3}&{\color{Gray} -128.9}&{\color{Gray} -113.6}&{\color{Gray} -39.6}\\\hline
{\color{Gray}Omit3,5}&{\color{Gray} -27.1}&{\color{Gray} -58.8}&{\color{Gray} +53.6}&{\color{Gray} -206.5}&{\color{Gray} -81.6}&{\color{Gray} -65.7}\\\hline
{\color{Gray}Omit3,8}&{\color{Gray} -42.9}&{\color{Gray} -115.6}&{\color{Gray} +72.9}&{\color{Gray} -274.6}&{\color{Gray} -181.9}&{\color{Gray} -112.3}\\\hline
{\color{Gray}Omit5,8}&{\color{Gray} -9.4}&{\color{Gray} +10.4}&{\color{Gray} +32.0}&{\color{Gray} -91.4}&{\color{Gray} +47.0}&{\color{Gray} -0.5}\\\hline
{\color{Gray}Omit1,4}&{\color{Gray} -16.2}&{\color{Gray} -30.3}&{\color{Gray} +33.7}&{\color{Gray} -409.9}&{\color{Gray} -74.7}&{\color{Gray} -97.2}\\\hline
Omit1,6& -36.3& +2.9& +38.5& -113.6& +6.1& -17.6\\\hline
Omit1,9& -14.1& -13.1& +38.5& -114.8& +6.4& -18.7\\\hline
Omit4,6& -16.7& -13.7& +38.6& -116.3& +5.6& -20.5\\\hline
Omit4,9& -15.3& -13.4& +38.5& -118.7& +5.9& -19.6\\\hline
Omit6,9& -15.8& -17.5& +39.3& -85.1& +1.2& -15.2\\\hline
Omit2,7& -7.5& -18.4& +45.8& -124.5& +19.8& -16.6\\\hline
Omit 0& -6.6& -13.0& +47.2& -119.0& -56.7& -28.3\\\hline
Omit 1& -14.3& -13.3& +38.7& -114.8& +6.4& -18.8\\\hline
Omit 2& -6.1& -19.9& +40.9& -119.0& +6.7& -19.4\\\hline
{\color{Gray}Omit 3}&{\color{Gray} -40.0}&{\color{Gray} -108.7}&{\color{Gray} +70.7}&{\color{Gray} -264.0}&{\color{Gray} -169.2}&{\color{Gray} -105.8}\\\hline
Omit 4& -15.0& -13.4& +38.5& -118.4& +5.8& -19.7\\\hline
Omit 5& -10.1& +8.3& +32.7& -95.1& +41.3& -2.9\\\hline
Omit 6& -15.7& -13.1& +38.6& -114.0& +5.9& -18.9\\\hline
Omit 7& -14.6& -13.1& +38.6& -115.6& +6.4& -18.9\\\hline
Omit 8& -14.3& -12.2& +38.2& -114.2& +8.8& -18.0\\\hline
Omit 9& -14.6& -13.3& +38.5& -115.6& +6.2& -19.0\\\hline
\end{tabular}
\end{center}
\label{tab:run1_ihwp_only_null_corrections_table}
\end{table}

\begin{table}[!h]

\caption{Run 2 dithering corrections to ``IHWP-only'' null asymmetry for various coil selections. Grayed-out sets considered suspect  for reasons investigated in Section \ref{sctn:residual_correlations}.}
\begin{center}
\begin{tabular}[h]{|l||c|c|c|c|c|c||c|}\hline
Dithering& Wien 6& Wien 7& Wien 8a& Wien 8b& Wien 9a& Wien 9b& Total\\
~Scheme&(ppb)&(ppb)&(ppb)&(ppb)&(ppb)&(ppb)&(ppb)\\\hline\hline
10-Coil& +24.6& +129.1& -35.9& -22.6& -0.9& -8.4& +0.6\\\hline
Omit0,5& +24.1& +127.6& -39.9& -23.9& -3.9& -17.3& -3.3\\\hline
{\color{Gray}Omit0,8}&{\color{Gray} +19.1}&{\color{Gray} +142.8}&{\color{Gray} -145.4}&{\color{Gray} -93.4}&{\color{Gray} -38.6}&{\color{Gray} -35.3}&{\color{Gray} -44.0}\\\hline
{\color{Gray}Omit3,5}&{\color{Gray} +25.2}&{\color{Gray} +129.1}&{\color{Gray} -75.4}&{\color{Gray} -41.2}&{\color{Gray} -5.2}&{\color{Gray} -24.4}&{\color{Gray} -13.4}\\\hline
{\color{Gray}Omit3,8}&{\color{Gray} +28.7}&{\color{Gray} +137.1}&{\color{Gray} -132.3}&{\color{Gray} -77.2}&{\color{Gray} -14.8}&{\color{Gray} -58.3}&{\color{Gray} -37.8}\\\hline
{\color{Gray}Omit1,4}&{\color{Gray} -2.4}&{\color{Gray} +129.5}&{\color{Gray} -16.9}&{\color{Gray} -26.7}&{\color{Gray} -7.6}&{\color{Gray} -6.6}&{\color{Gray} -1.1}\\\hline
Omit1,6& +24.8& +130.4& -35.3& -21.2& -0.5& -8.7& +1.1\\\hline
Omit1,9& +24.6& +129.3& -35.8& -22.6& -1.0& -8.7& +0.5\\\hline
Omit4,6& +25.0& +133.1& -35.4& -22.0& -0.6& -8.5& +1.2\\\hline
Omit4,9& +24.8& +131.5& -35.7& -23.2& -1.3& -8.6& +0.6\\\hline
Omit6,9& +28.7& +131.3& -32.9& -19.5& -0.2& -8.0& +2.4\\\hline
Omit2,7& +29.2& +100.3& -34.0& -20.0& -0.4& -5.2& +0.5\\\hline
{\color{Gray}Omit 0}&{\color{Gray} +20.8}&{\color{Gray} +135.7}&{\color{Gray} -89.8}&{\color{Gray} -59.5}&{\color{Gray} -18.1}&{\color{Gray} -24.1}&{\color{Gray} -21.9}\\\hline
Omit 1& +24.8& +130.0& -36.0& -22.7& -0.9& -8.7& +0.6\\\hline
Omit 2& +27.8& +129.1& -34.1& -20.3& -0.8& -6.6& +2.1\\\hline
{\color{Gray}Omit 3}&{\color{Gray} +28.6}&{\color{Gray} +135.1}&{\color{Gray} -123.0}&{\color{Gray} -72.2}&{\color{Gray} -12.4}&{\color{Gray} -37.7}&{\color{Gray} -30.2}\\\hline
Omit 4& +24.8& +129.5& -35.9& -23.2& -1.1& -8.6& +0.4\\\hline
Omit 5& +24.5& +126.1& -15.9& -10.2& +3.0& -4.2& +7.5\\\hline
Omit 6& +25.0& +130.4& -35.5& -21.7& -0.6& -8.2& +1.1\\\hline
Omit 7& +24.6& +129.1& -35.9& -22.6& -0.9& -8.4& +0.6\\\hline
Omit 8& +24.6& +132.8& -33.7& -21.5& -0.1& -6.9& +1.9\\\hline
Omit 9& +24.8& +131.7& -35.8& -22.7& -0.9& -8.4& +0.8\\\hline
\end{tabular}
\end{center}
\label{tab:run2_ihwp_only_null_corrections_table}
\end{table}

\begin{table}[!h]
\caption{Run 1 dithering corrections to ``IHWP+Spin-reversal'' null asymmetry for various coil selections. Grayed-out sets considered suspect  for reasons investigated in Section \ref{sctn:residual_correlations}.}
\begin{center}
\begin{tabular}[h]{|l||c|c|c|c|c||c|}\hline
Dithering& Wien 1& Wien 2& Wien 3& Wien 4& Wien 5& Total\\
~Scheme&(ppb)&(ppb)&(ppb)&(ppb)&(ppb)&(ppb)\\\hline\hline
10-Coil& -14.6& +13.1& +38.5& +115.8& +6.3& +33.4\\\hline
{\color{Gray}Omit0,3}&{\color{Gray} -14.3}&{\color{Gray} +76.2}&{\color{Gray} +76.8}&{\color{Gray} +312.0}&{\color{Gray} -203.6}&{\color{Gray} +56.1}\\\hline
Omit0,5& -6.3& +9.7& +37.2& +102.1& +8.8& +30.9\\\hline
{\color{Gray}Omit0,8}&{\color{Gray} -6.4}&{\color{Gray} +15.1}&{\color{Gray} +57.3}&{\color{Gray} +128.9}&{\color{Gray} -113.6}&{\color{Gray} +18.2}\\\hline
{\color{Gray}Omit3,5}&{\color{Gray} -27.1}&{\color{Gray} +58.8}&{\color{Gray} +53.6}&{\color{Gray} +206.5}&{\color{Gray} -81.6}&{\color{Gray} +46.7}\\\hline
{\color{Gray}Omit3,8}&{\color{Gray} -42.9}&{\color{Gray} +115.6}&{\color{Gray} +72.9}&{\color{Gray} +274.6}&{\color{Gray} -181.9}&{\color{Gray} +55.7}\\\hline
{\color{Gray}Omit5,8}&{\color{Gray} -9.4}&{\color{Gray} -10.4}&{\color{Gray} +32.0}&{\color{Gray} +91.4}&{\color{Gray} +47.0}&{\color{Gray} +30.4}\\\hline
{\color{Gray}Omit1,4}&{\color{Gray} -16.2}&{\color{Gray} +30.3}&{\color{Gray} +33.7}&{\color{Gray} +409.9}&{\color{Gray} -74.7}&{\color{Gray} +83.6}\\\hline
Omit1,6& -36.3& -2.9& +38.5& +113.6& +6.1& +26.3\\\hline
Omit1,9& -14.1& +13.1& +38.5& +114.8& +6.4& +33.3\\\hline
Omit4,6& -16.7& +13.7& +38.6& +116.3& +5.6& +32.4\\\hline
Omit4,9& -15.3& +13.4& +38.5& +118.7& +5.9& +34.2\\\hline
Omit6,9& -15.8& +17.5& +39.3& +85.1& +1.2& +27.3\\\hline
Omit2,7& -7.5& +18.4& +45.8& +124.5& +19.8& +42.0\\\hline
Omit 0& -6.6& +13.0& +47.2& +119.0& -56.7& +24.7\\\hline
Omit 1& -14.3& +13.3& +38.7& +114.8& +6.4& +33.3\\\hline
Omit 2& -6.1& +19.9& +40.9& +119.0& +6.7& +37.9\\\hline
{\color{Gray}Omit 3}&{\color{Gray} -40.0}&{\color{Gray} +108.7}&{\color{Gray} +70.7}&{\color{Gray} +264.0}&{\color{Gray} -169.2}&{\color{Gray} +54.4}\\\hline
Omit 4& -15.0& +13.4& +38.5& +118.4& +5.8& +33.9\\\hline
Omit 5& -10.1& -8.3& +32.7& +95.1& +41.3& +30.6\\\hline
Omit 6& -15.7& +13.1& +38.6& +114.0& +5.9& +32.8\\\hline
Omit 7& -14.6& +13.1& +38.6& +115.6& +6.4& +33.4\\\hline
Omit 8& -14.3& +12.2& +38.2& +114.2& +8.8& +33.3\\\hline
Omit 9& -14.6& +13.3& +38.5& +115.6& +6.2& +33.4\\\hline
\end{tabular}
\end{center}
\label{tab:run1_null_corrections_table}
\end{table}

\begin{table}[!h]

\caption{Run 2 dithering corrections to ``IHWP+Spin-reversal'' null asymmetry for various coil selections. Grayed-out sets considered suspect for reasons investigated in Section \ref{sctn:residual_correlations}.}
\begin{center}
\begin{tabular}[h]{|l||c|c|c|c|c|c||c|}\hline
Dithering& Wien 6& Wien 7& Wien 8a& Wien 8b& Wien 9a& Wien 9b& Total\\
~Scheme&(ppb)&(ppb)&(ppb)&(ppb)&(ppb)&(ppb)&(ppb)\\\hline\hline
10-Coil& +24.6& -129.1& -35.9& -22.6& +0.9& +8.4& -13.9\\\hline
Omit0,5& +24.1& -127.6& -39.9& -23.9& +3.9& +17.3& -11.8\\\hline
{\color{Gray}Omit0,8}&{\color{Gray} +19.1}&{\color{Gray} -142.8}&{\color{Gray} -145.4}&{\color{Gray} -93.6}&{\color{Gray} +38.6}&{\color{Gray} +35.3}&{\color{Gray} -27.9}\\\hline
{\color{Gray}Omit3,5}&{\color{Gray} +25.2}&{\color{Gray} -129.1}&{\color{Gray} -75.4}&{\color{Gray} -41.3}&{\color{Gray} +5.2}&{\color{Gray} +24.4}&{\color{Gray} -18.1}\\\hline
{\color{Gray}Omit3,8}&{\color{Gray} +28.7}&{\color{Gray} -137.1}&{\color{Gray} -132.3}&{\color{Gray} -77.4}&{\color{Gray} +14.8}&{\color{Gray} +58.4}&{\color{Gray} -22.5}\\\hline
{\color{Gray}Omit1,4}&{\color{Gray} -2.4}&{\color{Gray} -129.5}&{\color{Gray} -16.9}&{\color{Gray} -26.7}&{\color{Gray} +7.6}&{\color{Gray} +6.6}&{\color{Gray} -13.1}\\\hline
Omit1,6& +24.8& -130.4& -35.3& -21.3& +0.5& +8.7& -13.7\\\hline
Omit1,9& +24.6& -129.3& -35.8& -22.6& +1.0& +8.7& -13.8\\\hline
Omit4,6& +25.0& -133.1& -35.4& -22.0& +0.6& +8.5& -14.0\\\hline
Omit4,9& +24.8& -131.5& -35.7& -23.2& +1.3& +8.6& -14.0\\\hline
Omit6,9& +28.7& -131.3& -32.9& -19.5& +0.2& +8.0& -13.0\\\hline
Omit2,7& +29.2& -100.3& -34.0& -20.0& +0.4& +5.2& -11.4\\\hline
{\color{Gray}Omit 0}&{\color{Gray} +20.8}&{\color{Gray} -135.7}&{\color{Gray} -89.8}&{\color{Gray} -59.6}&{\color{Gray} +18.1}&{\color{Gray} +24.1}&{\color{Gray} -21.1}\\\hline
Omit 1& +24.8& -130.0& -36.0& -22.7& +0.9& +8.8& -13.9\\\hline
Omit 2& +27.8& -129.1& -34.1& -20.3& +0.8& +6.6& -13.4\\\hline
{\color{Gray}Omit 3}&{\color{Gray} +28.6}&{\color{Gray} -135.1}&{\color{Gray} -123.0}&{\color{Gray} -72.4}&{\color{Gray} +12.4}&{\color{Gray} +37.8}&{\color{Gray} -25.7}\\\hline
Omit 4& +24.8& -129.5& -35.9& -23.2& +1.1& +8.6& -13.9\\\hline
Omit 5& +24.5& -126.1& -15.9& -10.1& -3.0& +4.2& -10.6\\\hline
Omit 6& +25.0& -130.4& -35.5& -21.8& +0.6& +8.2& -13.9\\\hline
Omit 7& +24.6& -129.1& -35.9& -22.6& +0.9& +8.4& -13.9\\\hline
Omit 8& +24.6& -132.8& -33.7& -21.5& +0.1& +6.9& -14.2\\\hline
Omit 9& +24.8& -131.7& -35.8& -22.7& +0.9& +8.4& -14.1\\\hline
\end{tabular}
\end{center}
\label{tab:run2_null_corrections_table}
\end{table}
\FloatBarrier
To study the effect of varying the set of {\bf monitors} used in the modulation analysis on the detector correction, a full analysis was completed with three different sets of monitors chosen. The first set, hereafter referred to as ``Set A'', was the original default set with the target variables plus BPM3c12X. A second set, hereafter referred to as ``Set B'', comprised the target variables plus an alternate energy-sensitive beam position monitor BPM3p02bY. This BPM is located in the vertically-dispersive region of the Hall C Compton polarimeter and is about 7 times less sensitive to energy shifts than BPM3c12X since its dispersion is only 57~cm compared to the 4.4~m of the Hall C arc at girder 3c12. This choice has a two-fold advantage. First, the dispersive direction of the BPM3p02bY is vertical whereas that of  BPM3c12X is horizontal which might allow potential issues associated with strength sharing between energy and position to be disentangled. Second, this BPM is located 35 meters downstream of BPM3c12 and might be sensitive to additional effects to which BPM3c12X is blind. An example of such an hypothesis which was investigated as a possible source of the modulation inconsistency is electronics pickup of the modulation signal by an air coil magnet on the beamline downstream of BPM3c12. This would create an effect on the beam coherent with the modulation driving signal which would be invisible to BPM3c12X and would likely compromise the validity of the energy correction. Finally, a third set ``Set C'' is composed of five single BPM's named according to the beamline girder on which they are located: BPM3c11X (energy),  BPM3c14X and BPM3h02X (X and X angle), and BPM3c14Y and BPM3h02Y (Y and Y angle). ``Set C'' represents a completely independent set of monitors none of which are included in the Sets A and B. Differences in the widths of the corrections can be expected due to variations in the resolution of the different monitor sets. Set A which uses the average of many BPM's to calculate position and angle on target along with the most sensitive energy BPM is expected to give the smallest correction distribution width since its collective resolution of the five beam parameters is smallest. 

\begin{figure}[ht]

\centering
\framebox{\includegraphics[width=5in]{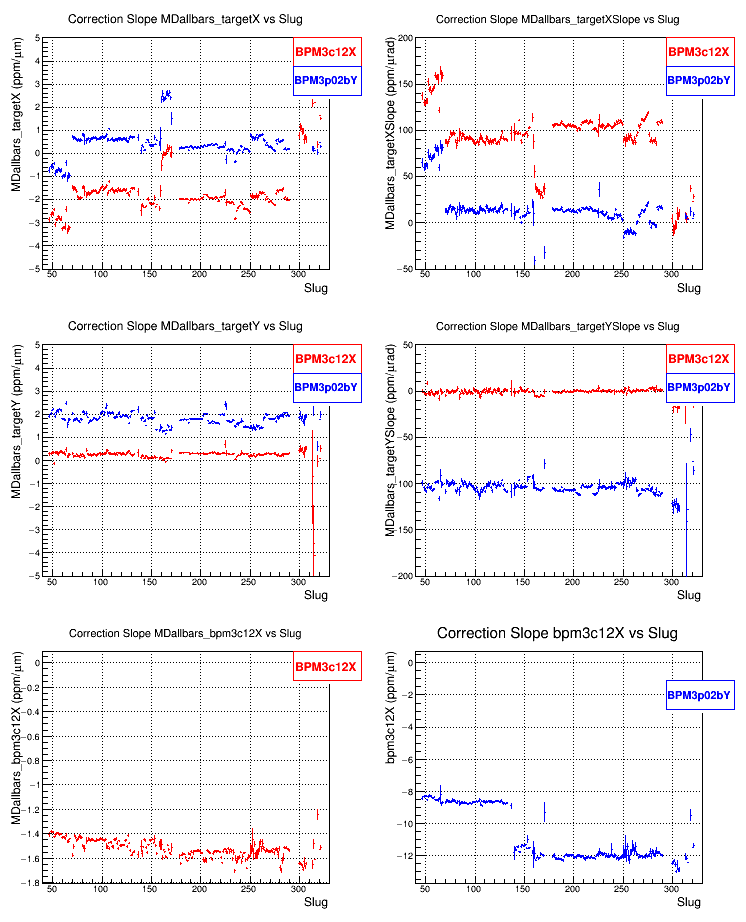}}
\caption{Comparison of correction slopes from two analyses, one using BPM3c12X as energy monitor and the other using BPM3p02bY.}
\label{fig:compton_bpm_slopes}
\end{figure}

The plots in Figure \ref{fig:compton_bpm_slopes} compare the correction slopes given by analyses using BPM3c12X (Set A) and BPM3p02bY (Set B) as energy monitor. It can easily be seen that monitors sensitive to the dispersive direction of the energy monitor used have much larger slopes due to mixing of energy and position/slope in that direction. During Run 2 (after slug 136), energy slopes using BPM3p02bY are 7-8 times larger than those from BPM3c12X which is roughly the scaling one would expect from the difference in dispersion. During Run 1, the slopes do not appear to scale the same way with dispersion. Since there are no common monitors between Set C and the other two sets, comparison of the slopes for Set C is irrelevant.

\begin{figure}[ht]

\centering
\framebox{\includegraphics[width=5in]{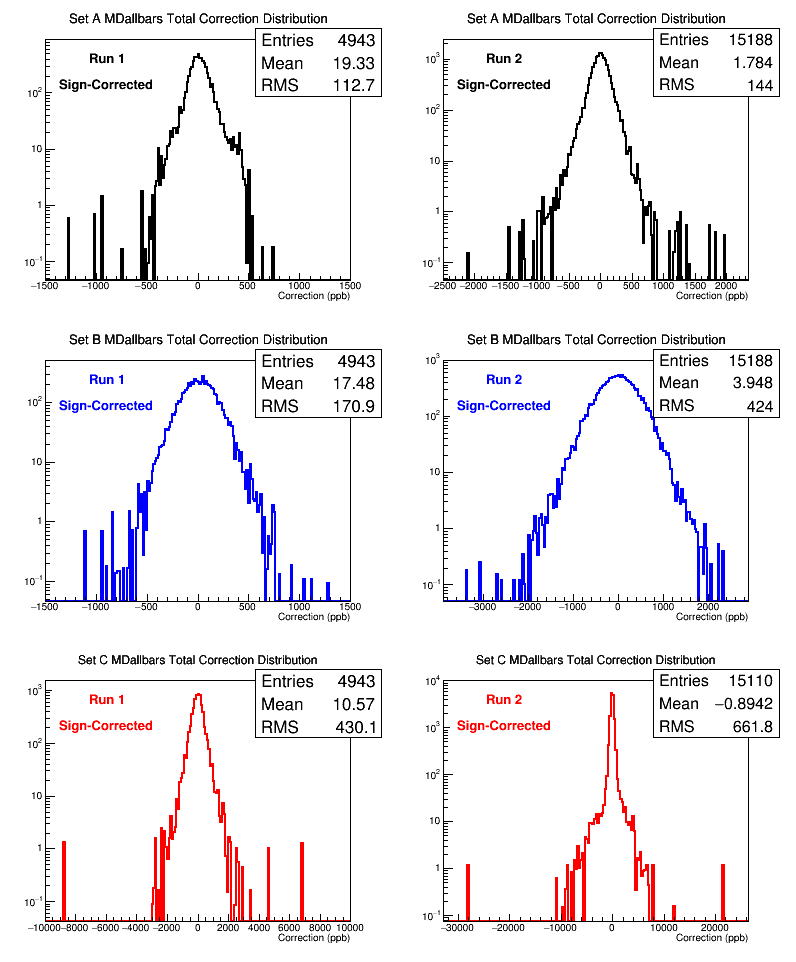}}
\caption{Comparison of total correction  distributions for monitor Sets A, B and C. Each entry in the distributions represents the correction for approximately 5 minutes of data. Distributions are sign-corrected to remove the sign of the slow helicity reversals and are all weighted by the uncorrected ``MDallbars'' inverse variance for purposes of comparison.}
\label{fig:correction_distributions}
\end{figure}

The key issue is whether or not the three sets give statistically consistent results for the magnitude of the corrections. The total correction distributions for the three monitor sets are shown in Figure \ref{fig:correction_distributions}. The width increase from Set A to the other monitor sets is obvious. These distributions are not expected to be normal, so direct statistical comparison of the distributions is not meaningful.
 However, one might expect that the difference between corrections of two Sets to arise solely from differences in monitor resolution or noise which {\it is} statistical. Of course, monitor resolution of the position and angle on target as well as energy can change with beam tune so each difference in the distribution must be normalized to the time-dependent standard deviation of the difference distribution. An estimate of this standard deviation for a given time can be obtained from the quadrature difference of the corrected main detector error given by the two sets for that time period.
 The distribution of these normalized differences is called a pull plot and each entry is given by
\[
\frac{Correction_{(Set~i)}-Correction_{(Set~j)}}{\sqrt{\left|\frac{\sigma_{(Set~i)}^2}{N}-\frac{\sigma_{(Set~j)}^2}{N}\right|}},
\]
where $\sigma_{(Set~i)}$ is the main detector average (MDallbars) asymmetry width after modulation correction with monitor Set~i and $N$ is the number of samples averaged. If the differences in the applied corrections between the two analyses are consistent with statistics, the pull-plot of the distribution of the difference between the two sets will be consistent with a normal distribution with mean 0 and standard deviation 1.\begin{figure}[ht]

\centering
\framebox{\includegraphics[width=4.5in]{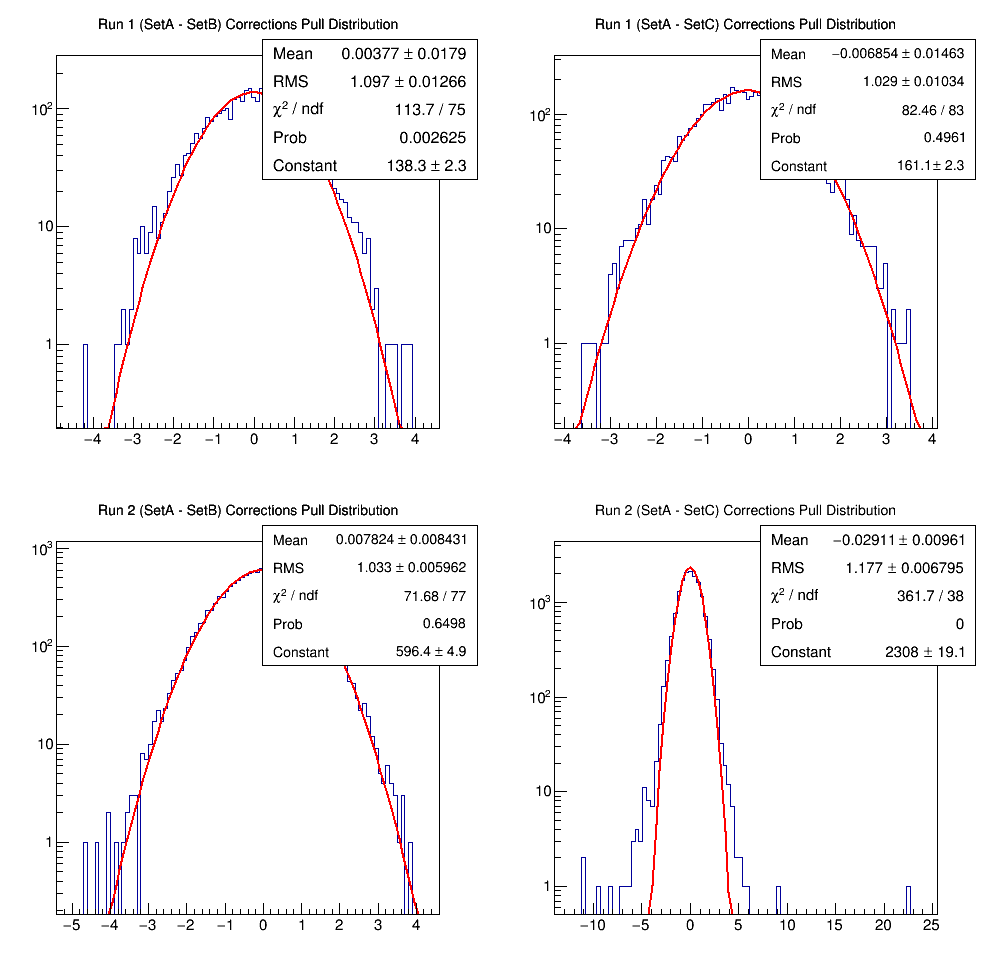}}
\caption{Pull plot distributions for difference in corrections provided by monitor Set A and those of Sets B and C. Each entry in the distributions is the average over a ``runlet'' and represents about 5 minutes of data and no sign-correction has been applied. The fit shown is for a Gaussian distribution with mean 0 and standard deviation 1 and with the scale factor as the only free parameter. From the probabilities it is clear that during Run 1 Sets A and C are statistically consistent whereas Sets A and B are not. In Run 2 the opposite appears to be true with Sets A and B statistically consistent and Sets A and C inconsistent.}
\label{fig:runlet_pull_plots}
\end{figure}
 Figure \ref{fig:runlet_pull_plots} shows the pull plot distributions for comparing corrections using monitor Set A with Sets B and C. The pull distributions have been fit with normal distribution curves. The probabilities clearly point to inconsistencies beyond statistics between Sets A and B in Run 1 and between Sets A and C in Run 2 at least at this averaging timescale ($\sim 5$~minutes). On the other hand, the results of Sets A and C are statistically consistent during Run 1 as are Sets A and B during Run 2. The pull plots further indicate that the non-normality of the pull distributions arises from an under-estimated width whereas their means are nearly consistent with 0. 

\begin{figure}[ht]

\centering
\framebox{\includegraphics[width=5in]{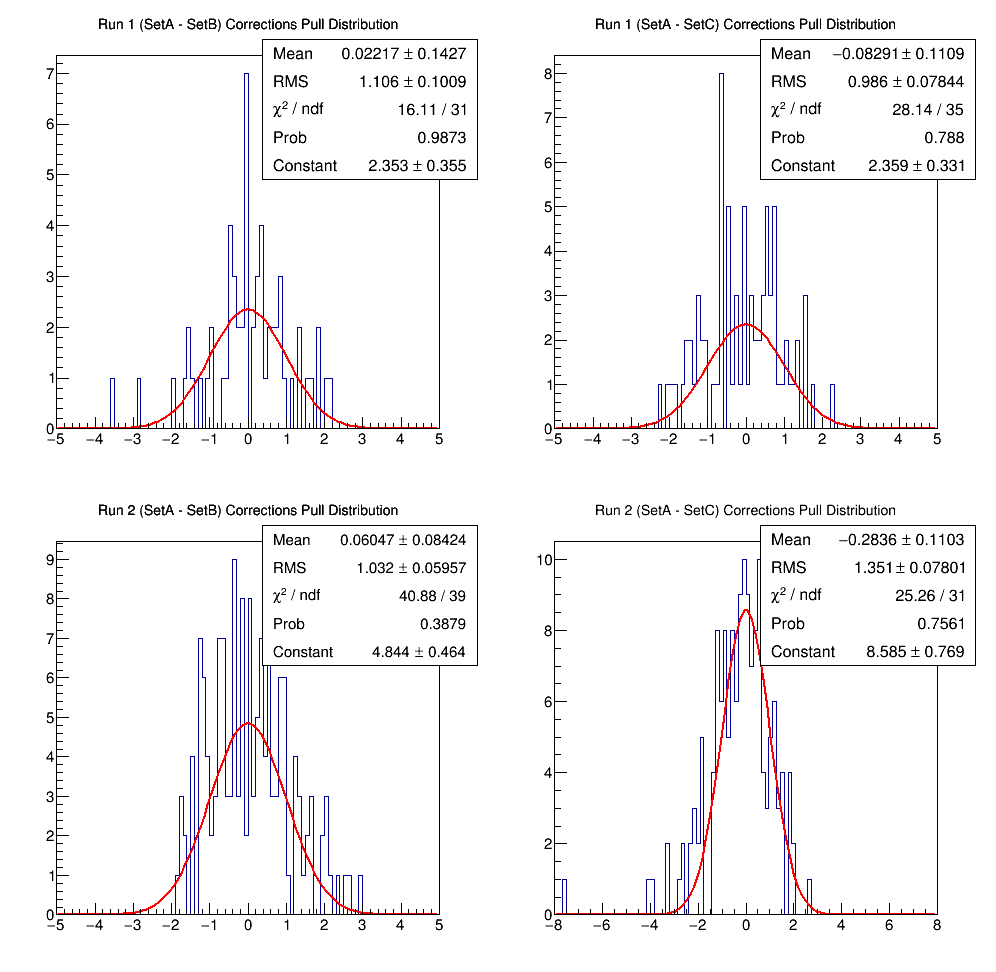}}
\caption{Pull plot distributions for difference in corrections provided by monitor Set A and those of Sets B and C.  Each entry in the distributions is the average over a ``slug'' and represents about 8 hours of data. No sign-correction has been applied. The fit shown is for a Gaussian distribution with mean 0 and standard deviation 1 and with the scale factor as the only free parameter. At this averaging timescale, all monitor sets appear to give statistically consistent corrections. }
\label{fig:slug_pull_plots}
\end{figure}

Over longer averaging periods the inconsistencies between monitor sets disappears. This can be clearly seen in Figure \ref{fig:slug_pull_plots} which shows a set of pull plots similar to those in Figure \ref{fig:runlet_pull_plots} but with each entry in the pull histograms representing the average over  $\sim 8$~hours. The pull distributions indicate that at this timescale all three monitor sets give consistent results within statistics.  
 
A final test of the consistency can be made by comparing the fully corrected main detector averages using the three monitor sets. Once again the expected statistical variation between analyses can be estimated using the quadrature difference of the main detector average asymmetry error between analyses with different monitor sets. Table \ref{tab:physics_diff_mon_sets} shows the variation along with the expected statistical variation consistent with the asymmetry error change. Differences in the final physics asymmetries using the three monitor sets appear to be statistically consistent. 

\begin{table}[!h]
\caption{\label{tab:physics_diff_mon_sets}Comparison of average corrected physics asymmetries as given using the three different monitor sets. All values are in parts per billion (ppb). ``Statistical Difference'' refers to the quadrature difference between the asymmetry errors in the first two columns and is an estimate of the expected statistical variation consistent with the change in the asymmetry distribution width between two analyses. Asymmetry values shown for purposes of comparison and are not expected to precisely match those of the final \Qs dataset.  }
\begin{center}
\begin{tabular}[h]{|l|c|c|c|c|}\hline
~&Set A&Set B & Difference & Statistical \\ 
~&Asymmetry&Asymmetry&Set A - Set B&Difference\\ \hline
Run 1 & -218.2$\pm$12.7 & -220.1$\pm$12.9 & 1.9 &  $\pm$2.3 \\ \hline
Run 2 & -156.8$\pm$8.2  & -154.4$\pm$8.8&-2.4   & $\pm$3.2 \\ \hline
~&~&~&~&~\\
~&Set A&Set C & Difference & Statistical \\ 
~&Asymmetry&Asymmetry&Set A - Set C&Difference\\ \hline
Run 1 & -218.2$\pm$12.7 & -227.4$\pm$14.1 & 7.3 & $\pm$6.1\\ \hline
Run 2 & -156.8$\pm$8.2  & -159.2$\pm$9.7  & 4.8 & $\pm$5.2\\ \hline
\end{tabular}
\end{center}
\end{table}

The conclusion of this study is that short-timescale non-statistical differences between modulation analyses using different monitor set disappear over longer timescale averaging and are not apparent in the final corrected physics asymmetries. The most accurate results come from the default monitor set, Set A, which uses the target variables and BPM3c12X. 

\subsection{Long Timescale Detector to Monitor Sensitivity}
\label{sctn:residual_correlations}
The modulation analysis is meant to remove false asymmetries associated with helicity-correlated beam properties from the main detector data. The final corrected main detector data is expected to be uncorrelated with the monitors sensitive to those beam properties. However, noise such as electronics pickup, common to both the monitors and detectors will create false correlations that disappear with sufficient averaging. Therefore, one place to look for the success or failure of the modulation correction is in correlations between the main detector and beam monitors averaged over long timescales. Slug-level averaging was chosen for this correlation study where a ``slug'' is the amount of data taken on a specific HWP slow reversal state and is about 8 hours of data. Correlations of the main detector average (MDallbars) asymmetry were found by plotting the slug averages of the main detector versus the slug averages of the monitors over all Run 1 and Run 2. All asymmetries and monitor differences have been sign-corrected to remove the unwanted effects of the slow reversals. It is important to point out that this method is subject to false correlations arising from long-timescale changes in beam conditions. Long timescale drifts in charge or background asymmetries could create false long-timescale correlations. Since changes in beam conditions such as these are expected to be small, the residual correlations are a useful source of information; however, it is important to keep in mind that this method of determining residual correlations is intended only as a tool for finding large and well-determined residuals, not as a fine-resolution tool for distinguishing between plausible datasets. Tables \ref{tab:run1_residual_correlations_table} and  \ref{tab:run2_residual_correlations_table} show the residual correlations of the MDallbars asymmetry to the five monitors of interest to the modulation analysis as well as to the average current or charge measurement. A few schemes show large residual correlations to the beam monitors and can quickly be eliminated as unreliable. 

Fortunately, tables \ref{tab:run1_residual_correlations_table} and  \ref{tab:run2_residual_correlations_table} paint a rather unequivocal picture. Schemes that produce significant residual correlations to monitors are generally quite obvious with $>4\sigma$ significance and with more than one monitor showing large correlations. The residual charge correlation is not considered in this cutoff since it is not part of the modulation correction.  Cutting only the obviously problematic schemes eliminates the following schemes: ``Omit 0,3'', ``Omit 0,8'', ``Omit 3,5''\footnote{Scheme ``Omit 3,5'' has hints of residual correlations in Run 2 but the evidence for its removal is not conclusive. Choosing to remove or retain it does not change any of the arguments ahead.}, ``Omit 3,8'', ``Omit 5,8''\footnote{Scheme ``Omit 5,8'' is removed from Run 1 due to high residual correlations to monitors and from Run 2 because it does not contain sufficient information to find useful slopes for slugs 160-170 in Run 2.} and ``Omit 3''.  ``Omit 0'' must be removed from from Run 2 only, due to high residual correlations. Scheme ``Omit 1,4'' must also be eliminated from consideration for the entire dataset due the high level of instability in the correction slopes for this analysis. These eliminated schemes are shown as grayed out in tables \ref{tab:run1_dithering_corrections_table} - \ref{tab:run2_null_corrections_table}. 

It is tempting to try to further distinguish between modulation schemes using residual correlations as a measure of ``correctness''. For example, one might be tempted to choose the ``Omit 0,5'' scheme for Run 1 and the ``Omit 5'' scheme for Run 2 since the results of these alternate schemes appear to be better consistent with 0 correlations than the full 10-Coil scheme. As previously mentioned, these long-timescale residual correlations are not reliable as a fine-resolution tool for distinguishing the quality of schemes.

A case may be made for choosing the Omit 0,5 scheme based upon residual correlations to beam modulation and to the final corrected main detector width. When coils 0 and 5 are omitted from the analysis, the residual correlations for the other modulation coils are nearly consistent with 0 (see Figure \ref{fig:md_omit05_residuals}). This means that evidence for inconsistency between modulation coils is absent in this scheme. Although it could be argued that the number of X-type modulation coils has now been reduced to 3 which forces a 0 residual on the system, this argument does not stand up to scrutiny. This idea suggests that the system can be modeled by a block-diagonal matrix structure with essentially a $3\times3$ block with $X$, $X^{\prime}$ and $E$ mixing only with each other and a $2\times2$ block with $Y$ and $Y^{\prime}$ mixing. Thus using only 3 modulation coils for the X-type block forces the system of equations to have 0 residual sensitivity to X-type modulation coils. This is not the case. There is mixing between X-type and Y-type sensitivities meaning that X-type monitors are slightly sensitive to Y-type modulations and vice versa. Furthermore, not all schemes that remove two X-type coils successfully zero the residual sensitivities to the remaining X-type coils. So the argument for using the Omit 0,5 scheme is as follows: ``Inconsistency between modulation coils evidenced by residual correlations to the coils shows that there is a problem with one or more coils. This inconsistency mainly affects the X-type modulation and goes away for some schemes by removing two X-type coils. Using long timescale residual correlations of the main detector to beam monitors to further inform the choice of schemes leaves the Omit 0,5 as the only scheme with both residual correlations to monitors and residual sensitivity to modulation coils (when only the coils used in the analysis are considered) consistent with zero." 

\begin{figure}[ht]

\centering
\framebox{\includegraphics[width=5in]{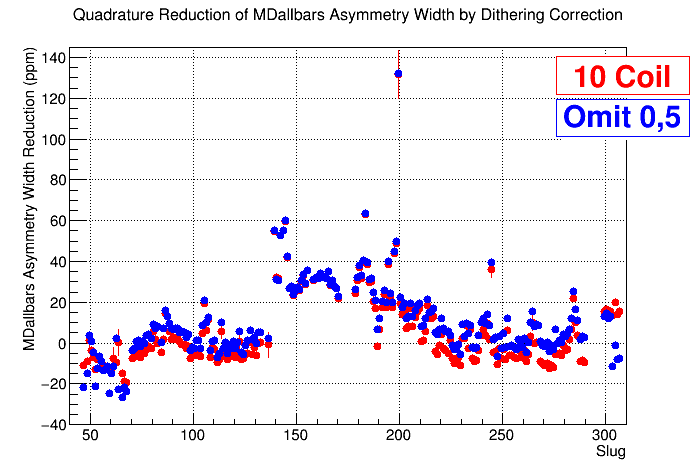}}
\caption{Shown is the width reduction of MDallbars asymmetry quartet distribution for ``10-Coil'' and ``Omit 0,5'' schemes. Width reduction is calculated as the quadrature difference of the width of the uncorrected and modulation corrected MDallbars asymmetry distributions. The sign of the quadrature difference is assigned such that negative numbers means the modulation correction increased the width.}
\label{fig:md_width_compared}
\end{figure}

Furthermore, the ``Omit 0,5'' scheme slightly reduces the corrected main detector asymmetry width beyond what the 10-Coil scheme does. Figure \ref{fig:md_width_compared} compares the width reduction of the MDallbars asymmetry distributions given by the two modulation schemes. The 10-Coil corrections increase the width of the MDallbars asymmetry for a large fraction of the \Qs dataset, whereas after slug 70 the Omit 0,5 corrections almost always reduce the width or leave it unchanged.
\begin{landscape}
\begin{table}[!h]

\caption{Run 1 residual correlations of MDallbars asymmetry to various beam monitors after dithering corrections. Blue colored correlations are $>3\sigma$ from 0 and Red are  $>4\sigma$.}
\begin{center}
\begin{tabular}[h]{|l|c|c|c|c|c|c|}\hline
Dithering&targetX&targetY&targetXSlope&targetYSlope&bpm3c12X&charge\\
~Scheme&(ppb/nm)&(ppb/nm)&(ppb/nrad)&(ppb/nrad)&(ppb/nm)&(ppb/ppb)\\\hline
10-Coil& +0.53$\pm$0.20& -0.65$\pm$0.30& +7.7$\pm$5.4& -10.9$\pm$6.4& +0.06$\pm$0.15& \color{red}+0.30$\pm$0.06\\\hline
Omit 0,3& \color{red}-1.07$\pm$0.20& \color{red}-2.55$\pm$0.30& \color{red}-48.7$\pm$5.4& \color{red}-60.2$\pm$6.4& \color{red}-1.30$\pm$0.15& \color{red}+0.27$\pm$0.06\\\hline
Omit 0,5& +0.31$\pm$0.20& -0.62$\pm$0.30& +5.8$\pm$5.4& -9.9$\pm$6.4& +0.15$\pm$0.15& \color{red}+0.29$\pm$0.06\\\hline
Omit 0,8& \color{blue}-0.61$\pm$0.20& \color{red}-1.25$\pm$0.30& \color{blue}-20.0$\pm$5.5& \color{red}-26.9$\pm$6.4& -0.43$\pm$0.15& \color{blue}+0.21$\pm$0.06\\\hline
Omit 3,5& -0.07$\pm$0.20& \color{red}-1.64$\pm$0.30& \color{blue}-17.4$\pm$5.4& \color{red}-34.9$\pm$6.4& \color{red}-0.68$\pm$0.15& \color{red}+0.30$\pm$0.06\\\hline
Omit 3,8& -0.55$\pm$0.20& \color{red}-2.27$\pm$0.30& \color{red}-39.1$\pm$5.4& \color{red}-52.2$\pm$6.4& \color{red}-1.41$\pm$0.15& \color{red}+0.27$\pm$0.06\\\hline
Omit 5,8& \color{red}+0.83$\pm$0.20& -0.32$\pm$0.30& \color{blue}+18.4$\pm$5.4& -2.7$\pm$6.4& +0.36$\pm$0.15& \color{red}+0.32$\pm$0.06\\\hline
Omit 1,4& -0.30$\pm$0.20& \color{red}-2.65$\pm$0.30& \color{red}-34.1$\pm$5.4& \color{red}-80.8$\pm$6.4& \color{red}-0.90$\pm$0.15& \color{red}+0.54$\pm$0.06\\\hline
Omit 1,6& \color{blue}+0.72$\pm$0.20& -0.54$\pm$0.30& +11.1$\pm$5.4& -9.1$\pm$6.4& +0.06$\pm$0.15& \color{red}+0.29$\pm$0.06\\\hline
Omit 1,9& +0.53$\pm$0.20& -0.64$\pm$0.30& +7.8$\pm$5.4& -10.7$\pm$6.4& +0.06$\pm$0.15& \color{red}+0.30$\pm$0.06\\\hline
Omit 4,6& +0.52$\pm$0.20& -0.59$\pm$0.30& +7.4$\pm$5.4& -9.8$\pm$6.4& +0.05$\pm$0.15& \color{red}+0.30$\pm$0.06\\\hline
Omit 4,9& +0.52$\pm$0.20& -0.69$\pm$0.30& +7.3$\pm$5.4& -11.8$\pm$6.4& +0.04$\pm$0.15& \color{red}+0.30$\pm$0.06\\\hline
Omit 6,9& \color{blue}+0.60$\pm$0.20& -0.29$\pm$0.30& +10.7$\pm$5.4& -2.9$\pm$6.4& +0.10$\pm$0.15& \color{red}+0.28$\pm$0.06\\\hline
Omit 2,7& \color{blue}+0.63$\pm$0.20& -0.67$\pm$0.30& +9.4$\pm$5.4& -11.5$\pm$6.4& +0.10$\pm$0.15& \color{red}+0.33$\pm$0.06\\\hline
Omit 0& -0.15$\pm$0.20& \color{blue}-0.95$\pm$0.30& -7.7$\pm$5.4& -18.9$\pm$6.4& -0.17$\pm$0.15& \color{red}+0.25$\pm$0.06\\\hline
Omit 1& +0.53$\pm$0.20& -0.64$\pm$0.30& +7.8$\pm$5.4& -10.7$\pm$6.4& +0.06$\pm$0.15& \color{red}+0.30$\pm$0.06\\\hline
Omit 2& +0.58$\pm$0.20& -0.66$\pm$0.30& +7.9$\pm$5.4& -11.1$\pm$6.4& +0.00$\pm$0.15& \color{red}+0.31$\pm$0.06\\\hline
Omit 3& -0.53$\pm$0.20& \color{red}-2.20$\pm$0.30& \color{red}-37.4$\pm$5.4& \color{red}-50.3$\pm$6.4& \color{red}-1.34$\pm$0.15& \color{red}+0.28$\pm$0.06\\\hline
Omit 4& +0.52$\pm$0.20& -0.67$\pm$0.30& +7.3$\pm$5.4& -11.5$\pm$6.4& +0.05$\pm$0.15& \color{red}+0.30$\pm$0.06\\\hline
Omit 5& \color{blue}+0.79$\pm$0.20& -0.38$\pm$0.30& \color{blue}+16.9$\pm$5.4& -4.0$\pm$6.4& +0.32$\pm$0.15& \color{red}+0.32$\pm$0.06\\\hline
Omit 6& +0.54$\pm$0.20& -0.62$\pm$0.30& +7.9$\pm$5.4& -10.3$\pm$6.4& +0.06$\pm$0.15& \color{red}+0.30$\pm$0.06\\\hline
Omit 7& +0.53$\pm$0.20& -0.65$\pm$0.30& +7.7$\pm$5.4& -10.8$\pm$6.4& +0.06$\pm$0.15& \color{red}+0.30$\pm$0.06\\\hline
Omit 8& +0.55$\pm$0.20& -0.62$\pm$0.30& +8.4$\pm$5.4& -10.2$\pm$6.4& +0.08$\pm$0.15& \color{red}+0.30$\pm$0.06\\\hline
Omit 9& +0.53$\pm$0.20& -0.65$\pm$0.30& +7.7$\pm$5.4& -10.9$\pm$6.4& +0.05$\pm$0.15& \color{red}+0.30$\pm$0.06\\\hline
\end{tabular}
\end{center}
\label{tab:run1_residual_correlations_table}
\end{table}

\begin{table}[!h]

\caption{Run 2 residual correlations of MDallbars asymmetry to various beam monitors after dithering corrections.}
\begin{center}
\begin{tabular}[h]{|l|c|c|c|c|c|c|}\hline
Dithering&targetX&targetY&tgtXSlope&tgtYSlope&bpm3c12X&charge\\
~Scheme&(ppb/nm)&(ppb/nm)&(ppb/nrad)&(ppb/nrad)&(ppb/nm)&(ppb/ppb)\\\hline
10-Coil& -0.14$\pm$0.14& -0.42$\pm$0.30& -4.6$\pm$5.5& -24$\pm$11& -0.00$\pm$0.29& +0.02$\pm$0.05\\\hline
Omit 0,3& \color{red}-1.00$\pm$0.17& \color{red}-2.72$\pm$0.45& \color{red}-37.7$\pm$6.1& \color{red}-93$\pm$15& \color{red}-1.83$\pm$0.34& -0.08$\pm$0.05\\\hline
Omit 0,5& -0.18$\pm$0.14& -0.47$\pm$0.30& -6.1$\pm$5.4& -26$\pm$11& -0.03$\pm$0.29& +0.02$\pm$0.05\\\hline
Omit 0,8& \color{red}-1.28$\pm$0.15& \color{red}-2.09$\pm$0.30& \color{red}-48.5$\pm$5.5& \color{red}-94$\pm$11& \color{red}-1.73$\pm$0.30& -0.10$\pm$0.05\\\hline
Omit 3,5& \color{blue}-0.45$\pm$0.14& -0.88$\pm$0.30& \color{blue}-16.9$\pm$5.5& \color{blue}-42$\pm$11& -0.50$\pm$0.30& -0.02$\pm$0.05\\\hline
Omit 3,8& \color{red}-1.00$\pm$0.15& \color{red}-1.70$\pm$0.30& \color{red}-38.3$\pm$5.5& \color{red}-75$\pm$11& \color{red}-1.37$\pm$0.30& -0.08$\pm$0.05\\\hline
Omit 5,8& +0.24$\pm$0.17& +0.42$\pm$0.44& +9.1$\pm$6.0& -2$\pm$14& +0.33$\pm$0.34& +0.08$\pm$0.05\\\hline
Omit 1,4& -0.18$\pm$0.15& -0.54$\pm$0.30& -5.2$\pm$5.5& \color{blue}-38$\pm$11& +0.05$\pm$0.30& +0.04$\pm$0.05\\\hline
Omit 1,6& -0.13$\pm$0.14& -0.42$\pm$0.30& -4.4$\pm$5.5& -23$\pm$11& +0.02$\pm$0.29& +0.02$\pm$0.05\\\hline
Omit 1,9& -0.14$\pm$0.14& -0.42$\pm$0.30& -4.6$\pm$5.5& -24$\pm$11& -0.00$\pm$0.29& +0.03$\pm$0.05\\\hline
Omit 4,6& -0.14$\pm$0.14& -0.44$\pm$0.30& -4.7$\pm$5.5& -24$\pm$11& +0.00$\pm$0.29& +0.02$\pm$0.05\\\hline
Omit 4,9& -0.15$\pm$0.14& -0.44$\pm$0.30& -4.8$\pm$5.5& -24$\pm$11& -0.01$\pm$0.29& +0.02$\pm$0.05\\\hline
Omit 6,9& -0.11$\pm$0.14& -0.36$\pm$0.30& -3.4$\pm$5.5& -21$\pm$11& +0.05$\pm$0.29& +0.02$\pm$0.05\\\hline
Omit 2,7& -0.05$\pm$0.14& -0.17$\pm$0.30& -1.6$\pm$5.5& -14$\pm$11& +0.04$\pm$0.30& +0.02$\pm$0.05\\\hline
Omit 0& \color{red}-0.71$\pm$0.14& \color{red}-1.26$\pm$0.30& \color{red}-26.7$\pm$5.5& \color{red}-59$\pm$11& -0.88$\pm$0.29& -0.04$\pm$0.05\\\hline
Omit 1& -0.14$\pm$0.14& -0.43$\pm$0.30& -4.7$\pm$5.5& -24$\pm$11& -0.01$\pm$0.29& +0.02$\pm$0.05\\\hline
Omit 2& -0.12$\pm$0.14& -0.38$\pm$0.30& -3.7$\pm$5.5& -22$\pm$11& +0.13$\pm$0.30& +0.02$\pm$0.05\\\hline
Omit 3& \color{red}-0.90$\pm$0.15& \color{red}-1.55$\pm$0.30& \color{red}-34.7$\pm$5.5& \color{red}-69$\pm$11& \color{red}-1.24$\pm$0.30& -0.07$\pm$0.05\\\hline
Omit 4& -0.14$\pm$0.14& -0.43$\pm$0.30& -4.7$\pm$5.5& -24$\pm$11& -0.00$\pm$0.29& +0.02$\pm$0.05\\\hline
Omit 5& +0.05$\pm$0.14& -0.13$\pm$0.30& +3.0$\pm$5.5& -12$\pm$11& +0.31$\pm$0.29& +0.05$\pm$0.05\\\hline
Omit 6& -0.14$\pm$0.14& -0.42$\pm$0.30& -4.5$\pm$5.5& -23$\pm$11& +0.01$\pm$0.29& +0.02$\pm$0.05\\\hline
Omit 7& -0.14$\pm$0.14& -0.42$\pm$0.30& -4.6$\pm$5.5& -24$\pm$11& -0.00$\pm$0.29& +0.02$\pm$0.05\\\hline
Omit 8& -0.13$\pm$0.14& -0.42$\pm$0.30& -4.2$\pm$5.5& -23$\pm$11& +0.03$\pm$0.29& +0.03$\pm$0.05\\\hline
Omit 9& -0.15$\pm$0.14& -0.43$\pm$0.30& -4.8$\pm$5.5& -24$\pm$11& -0.01$\pm$0.29& +0.02$\pm$0.05\\\hline
\end{tabular}
\end{center}
\label{tab:run2_residual_correlations_table}
\end{table}
\begin{figure}[ht]
\centering
\framebox{\includegraphics[width=8.9in]{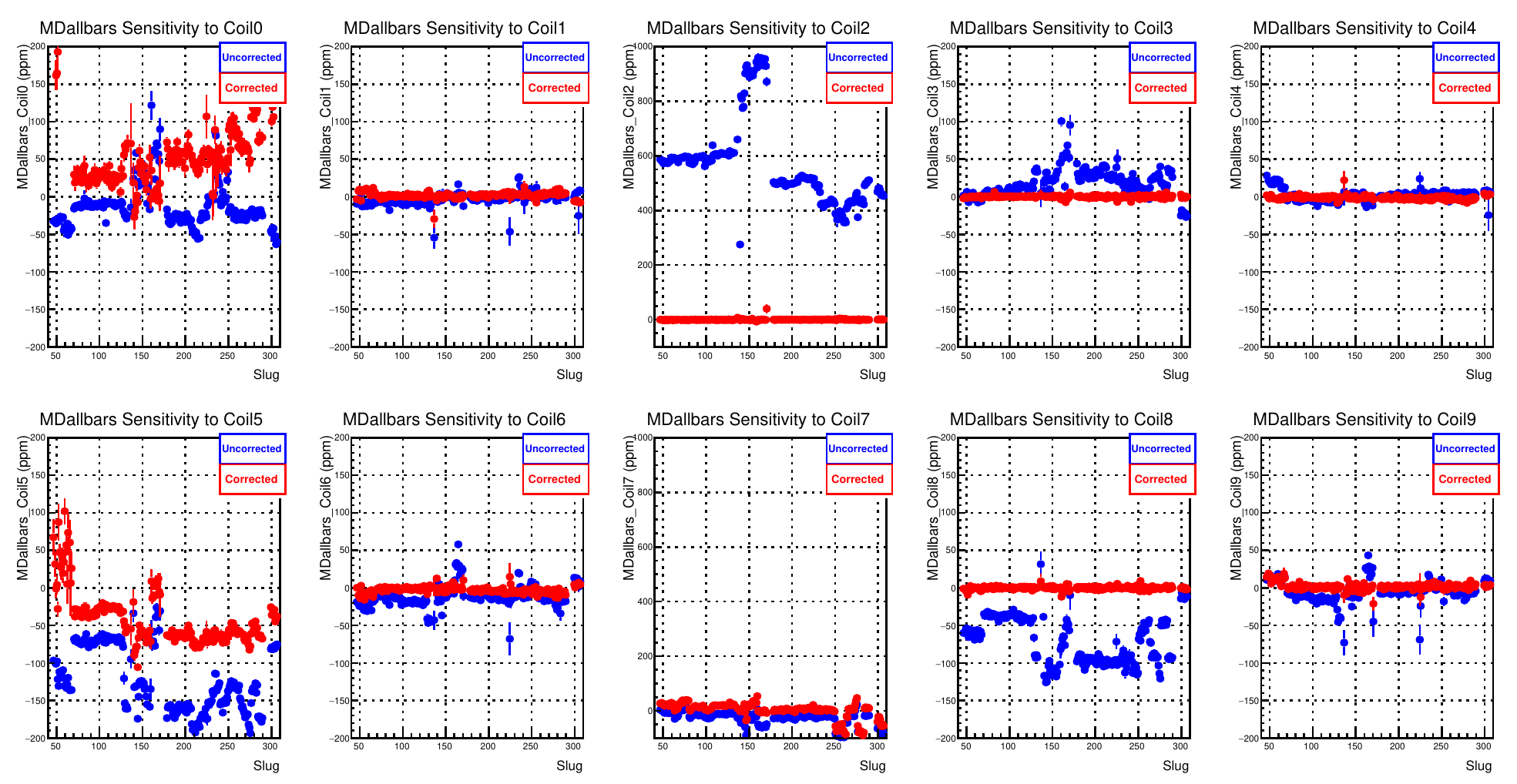}}
\caption{Residual sensitivity of MDallbars to modulation coils for modulation scheme omitting coils 0 and 5 (``Omit 0,5'' scheme) from the analysis. The residual sensitivities remain large to  coils 0 and 5 as expected, but go to nearly 0 for the other X-type coils 3 and 8. Sensitivities to Y-type coils are small before correction and remain small after correction}
\label{fig:md_omit05_residuals}
\end{figure}
\end{landscape}

Having established at least a plausible argument for using the ``Omit 0,5'' corrections perhaps it would be prudent to look at variations around the ``Omit 0,5'' results that occur from omitting various Y-like coils in addition to coils 0 and 5. 

Tables \ref{tab:run1_om05_residual_correlations_table} and \ref{tab:run2_om05_residual_correlations_table} show the residual correlations for various Y-type coil omissions paired with omitting coils 0 and 5. The only elimination that can be made at the $3\sigma$ level is the scheme that omits coils 0, 1, 4 and 5 (``Omit 0,5,1,4'') and that only during Run 1. The last row of these tables shows an example of a five coil analysis where Coil 7 (out-of-phase energy modulation) is also omitted in addition to two X-type and two Y-type coils, which produces zero-residuals in the five coils used by definition. No further distinction can be made between the remaining schemes using these residual correlations.

\begin{table}[!h]
\caption{Run 1 residual correlations of MDallbars asymmetry to various beam monitors after dithering corrections using various coil selections. Blue-colored cells have residual correlations with $3\sigma$ significance and  red-colored cells have residual correlations with $4\sigma$ significance.}
\begin{center}
\begin{tabular}[h]{|l|c|c|c|c|c|}\hline
Dithering&targetX&targetY&targetXSlope&targetYSlope&bpm3c12X\\
~Scheme&(ppb/nm)&(ppb/nm)&(ppb/nrad)&(ppb/nrad)&(ppb/nm)\\\hline
Omit0,5& 0.31$\pm$0.20& -0.62$\pm$0.30& 5.9$\pm$5.4& -9.9$\pm$6.4& 0.15$\pm$0.15\\\hline
Omit0,5,1,4&\color{blue} 0.60$\pm$0.20& 0.18$\pm$0.29& \color{red}21.5$\pm$5.4&\color{blue} 20.7$\pm$6.4& 0.37$\pm$0.15\\\hline
Omit0,5,1,6& 0.15$\pm$0.20& -0.66$\pm$0.30& 3.5$\pm$5.4& -10.3$\pm$6.4& 0.18$\pm$0.15\\\hline
Omit0,5,1,9& 0.30$\pm$0.20& -0.63$\pm$0.30& 5.9$\pm$5.4& -9.8$\pm$6.4& 0.15$\pm$0.15\\\hline
Omit0,5,4,6& 0.31$\pm$0.20& -0.59$\pm$0.30& 5.7$\pm$5.4& -9.4$\pm$6.4& 0.14$\pm$0.15\\\hline
Omit0,5,4,9& 0.29$\pm$0.20& -0.67$\pm$0.30& 5.3$\pm$5.4& -11.1$\pm$6.4& 0.14$\pm$0.15\\\hline
Omit0,5,6,9& 0.27$\pm$0.20& -0.54$\pm$0.30& 5.1$\pm$5.4& -8.7$\pm$6.4& 0.12$\pm$0.15\\\hline
Omit0,5,6,7,9& 0.26$\pm$0.20& -0.55$\pm$0.30& 5.0$\pm$5.4& -8.8$\pm$6.4& 0.12$\pm$0.15\\\hline
\end{tabular}
\end{center}
\label{tab:run1_om05_residual_correlations_table}
\end{table}
\begin{table}[!h]
\caption{Run 2 residual correlations of MDallbars asymmetry to various beam monitors after dithering corrections using various coil selections.}
\begin{center}
\begin{tabular}[h]{|l|c|c|c|c|c|}\hline
Dithering&targetX&targetY&targetXSlope&targetYSlope&bpm3c12X\\
~Scheme&(ppb/nm)&(ppb/nm)&(ppb/nrad)&(ppb/nrad)&(ppb/nm)\\\hline
Omit0,5& -0.18$\pm$0.14& -0.47$\pm$0.30& -6.1$\pm$5.5& -26$\pm$11& -0.03$\pm$0.29\\\hline
Omit0,5,1,4& -0.16$\pm$0.14& -0.42$\pm$0.30& -5.4$\pm$5.5& -23$\pm$11& -0.01$\pm$0.30\\\hline
Omit0,5,1,6& -0.19$\pm$0.14& -0.50$\pm$0.30& -6.4$\pm$5.5& -27$\pm$11& -0.04$\pm$0.29\\\hline
Omit0,5,1,9& -0.18$\pm$0.14& -0.48$\pm$0.30& -6.1$\pm$5.5& -26$\pm$11& -0.04$\pm$0.29\\\hline
Omit0,5,4,6& -0.19$\pm$0.14& -0.50$\pm$0.30& -6.5$\pm$5.5& -27$\pm$11& -0.04$\pm$0.29\\\hline
Omit0,5,4,9& -0.19$\pm$0.14& -0.48$\pm$0.30& -6.2$\pm$5.5& -26$\pm$11& -0.04$\pm$0.29\\\hline
Omit0,5,6,9& -0.19$\pm$0.14& -0.49$\pm$0.30& -6.5$\pm$5.5& -26$\pm$11& -0.05$\pm$0.29\\\hline
Omit0,5,6,7,9& -0.19$\pm$0.14& -0.48$\pm$0.30& -6.4$\pm$5.5& -26$\pm$11& -0.05$\pm$0.29\\\hline
\end{tabular}
\end{center}
\label{tab:run2_om05_residual_correlations_table}
\end{table}

Tables \ref{tab:run1_dithering_corrections_table} and \ref{tab:run2_dithering_corrections_table} show that omitting various combinations Y-type coils (1, 4, 6, and 9) while using all X-type coils creates a spread of total corrections around the 10-Coil values. For Run 1 the 10-Coil correction is -19.2~ppb. The values for omitting various Y coils are approximately centered on this value with a range of -16.7~ppb to  22.0~ppb. For Run 2 there is a similar spread centered on the 10-Coil correction of -1.7~ppb with a range from -1.5~ppb to -1.8~ppb from analyses omitting different combinations of Y-type coils. A similar analysis ``centered on'' the ``Omit 0,5'' scheme reveals a similar spread. Tables \ref{tab:run1_om05_dithering_corrections} and \ref{tab:run2_om05_dithering_corrections} show similar results for omitting Y-coils in addition to omitting coils 0 and 5. For Run 1 the total correction for ``Omit 0,5'' is -13.6~ppb with a range from -12.4~ppb to -16.1~ppb for schemes with additional Y-coil omissions. Similarly for Run 2 the ``Omit 0,5'' correction is -2.4~ppb with a range from -2.1~ppb to -2.6~ppb for schemes with further Y-coil omissions.
\begin{table}[!h]

\caption{\label{tab:run1_om05_dithering_corrections}Run 1 dithering corrections for MDallbars asymmetry compared for different coil selections. ``Omit0,5,1,4'' results grayed out due to large residual correlations seen in Table \ref{tab:run1_om05_residual_correlations_table}}
\begin{center}
\begin{tabular}[h]{|l||c|c|c|c|c||c|}\hline
Dithering& Wien 1& Wien 2& Wien 3& Wien 4& Wien 5& Total\\
~Scheme&(ppb)&(ppb)&(ppb)&(ppb)&(ppb)&(ppb)\\\hline\hline
Omit0,5& -19.2& +21.6& -43.9& -16.9& -14.9& -13.6\\\hline
{\color{Gray}Omit0,5,1,4}&{\color{Gray} -30.6}&{\color{Gray} +35.1}&{\color{Gray} -36.9}&{\color{Gray} +46.2}&{\color{Gray} -8.9}&{\color{Gray} +3.4}\\\hline
Omit0,5,1,6& -17.9& +26.1& -45.1& -16.3& -14.7& -12.4\\\hline
Omit0,5,1,9& -19.2& +20.2& -45.1& -17.5& -15.2& -14.3\\\hline
Omit0,5,4,6& -35.5& +22.0& -45.2& -15.8& -15.7& -16.1\\\hline
Omit0,5,4,9& -13.4& +18.8& -45.2& -17.6& -16.1& -14.0\\\hline
Omit0,5,6,9& -19.2& +20.6& -44.8& -14.5& -15.1& -13.6\\\hline
Omit0,5,6,7,9& -19.2& +20.3& -44.6& -14.4& -15.4& -13.6\\\hline
\end{tabular}
\end{center}
\end{table}
\begin{table}[!h]

\caption{\label{tab:run2_om05_dithering_corrections}Run 2 dithering corrections for MDallbars asymmetry compared for different coil selections.}
\begin{center}
\begin{tabular}[h]{|l||c|c|c|c|c|c||c|}\hline
Dithering& Wien 6& Wien 7& Wien 8a& Wien 8b& Wien 9a& Wien 9b& Total\\
~Scheme&(ppb)&(ppb)&(ppb)&(ppb)&(ppb)&(ppb)&(ppb)\\\hline\hline
Omit0,5& -24.7& +36.0& -13.9& -5.3& +1.7& +0.2& -2.4\\\hline
Omit0,5,1,4& -31.3& +31.6& -11.8& -1.1& +2.1& +0.9& -2.1\\\hline
Omit0,5,1,6& -24.5& +38.2& -13.9& -4.7& +0.8& +0.2& -2.3\\\hline
Omit0,5,1,9& -24.4& +35.9& -13.9& -4.7& +0.9& +0.1& -2.5\\\hline
Omit0,5,4,6& -24.4& +38.1& -14.1& -4.7& +0.7& +0.3& -2.4\\\hline
Omit0,5,4,9& -24.2& +36.0& -14.3& -5.0& +0.8& +0.2& -2.6\\\hline
Omit0,5,6,9& -24.9& +36.6& -13.8& -4.7& +0.8& +0.3& -2.5\\\hline
Omit0,5,6,7,9& -24.9& +36.7& -13.8& -4.6& +0.7& +0.3& -2.4\\\hline
\end{tabular}
\end{center}
\end{table}

Thus it appears that there are two clusters of modulation results-- a cluster formed by various Y-coil omissions centered on the 10-Coil result and a cluster formed by various Y-coil omissions centered on the ``Omit 0,5'' result. The ``10-Coil'' and ``Omit 0,5'' results are good representatives of these two clusters both of which produce MDallbars results with residual correlations to beam monitors that are consistent with zero. 

\subsection{Signature of the Inconsistency}
\begin{figure}[!ht]
\begin{center}
\includegraphics[width=5.9in]{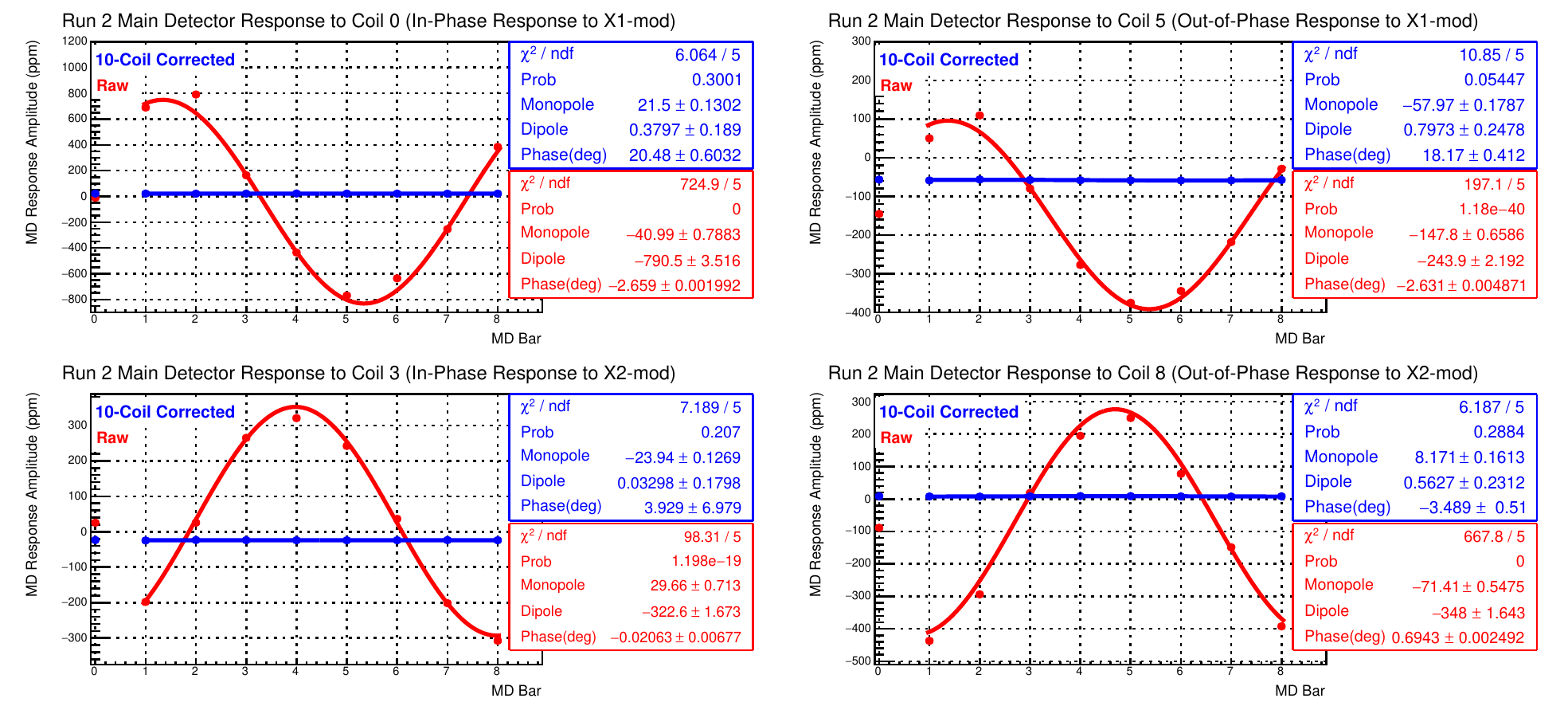}
\caption{\label{fig:Run2_10coil_Xdipoles}Responses of individual main detector bars averaged over Run 2 to X-type modulation coils before correction and after correction using a full 10-Coil analysis. The dipole response is given by the amplitude of the sinusoid while the monopole is the offset. The response of MDallbars is given by the point at Bar 0.}
\end{center}
\end{figure}

\begin{figure}[!ht]
\begin{center}
\includegraphics[width=5.9in]{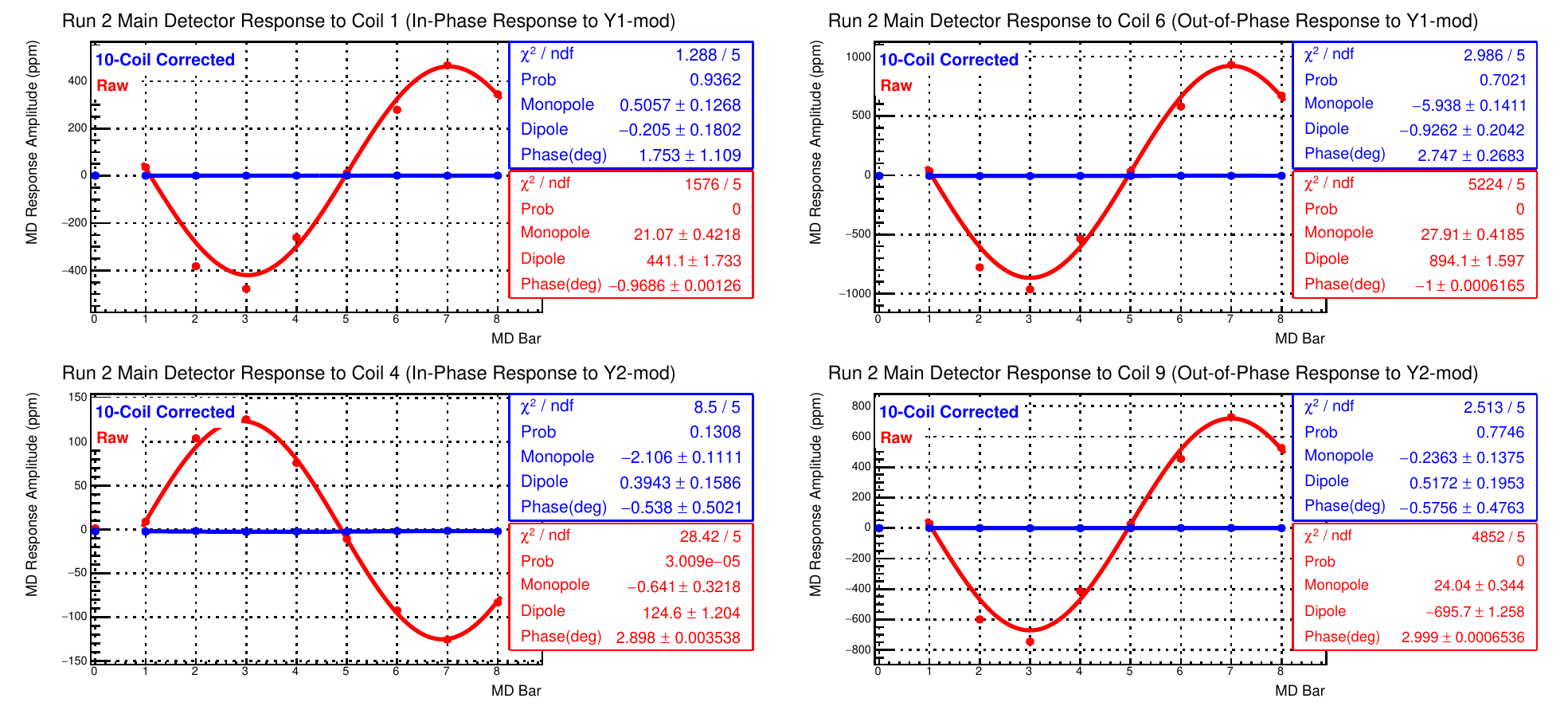}
\caption{\label{fig:Run2_10coil_Ydipoles}Responses of individual main detector bars averaged over Run 2 to Y-type modulation coils before correction and after correction using a full 10-Coil analysis. The dipole response is given by the amplitude of the sinusoid while the monopole is the offset. The response of MDallbars is given by the point at Bar 0.}
\end{center}
\end{figure}

Before attempting to determine a final modulation correction and assign a systematic error it is important to look at a key feature of the failure or inconsistency of the modulation analysis. Considering the azimuthal arrangement of the main detector bars (see Figure \ref{fig:octant_coords}) allows one to qualitatively predict responses of individual bars. For example, the response of bar 1 should be equal and opposite of that of bar 5 to horizontal position and angle modulation, whereas bars 3 and 7 should see small but equal responses to horizontal modulations. The response of the main detector to position or angle modulation as a function of the octant angle around the azimuth should have an approximately sinusoidal pattern. These sinusoidal patterns are referred to as ``dipoles'' and in the jargon of \Qs when the maximum response is in the vertical direction (maximum response on bars 3 and 7) it is referred to  as a ``vertical dipole''. Likewise when the maximum response is in the horizontal direction (maximum response on bars 1 and 5), it is called a ``horizontal dipole''. Thus, X-type modulations will create horizontal dipole responses and Y-type modulations are expected to produce vertical dipole responses. Any effect that creates equal responses in all detector bars is called a ``monopole response''. Shifts in energy and charge are candidates for monopole responses. Examples of dipole and monopole responses to the various types of X and Y modulations before and after correction  can be seen in figures \ref{fig:Run2_10coil_Xdipoles} and \ref{fig:Run2_10coil_Ydipoles}. The dipole response is given by the amplitude of the sinusoid, while the monopole is the offset. While a more detailed analysis of the main detector dipole and monopole response patterns is investigated in Appendix \ref{AppendixD}, the most significant observations are as follows:\\
1. The failure mode of the beam modulation analysis affects all the main detectors equally. A monopole residual sensitivity is evident whereas the dipoles are removed.\\
2. The successful removal of dipoles is a feature of even of coil schemes deemed untrustworthy due to large residual correlations to monitors they produce. Expected geometric sensitivities are removed but failure to remove the source of monopole sensitivity is evident for all schemes, although not equally.\\
3. The residual monopole responses are largest for the X-coils. Omitting coils 0 and 5 shrinks the monopole responses in the other two X-coils (3 and 8) to be consistent with zero while enlarging the monopole response in the omitted coils (coils 0 and 5). However, even for coils omitted from the analysis, the residual dipole response is consistent with zero.\\ 

\begin{figure}[!ht]
\begin{center}
\includegraphics[width=4in]{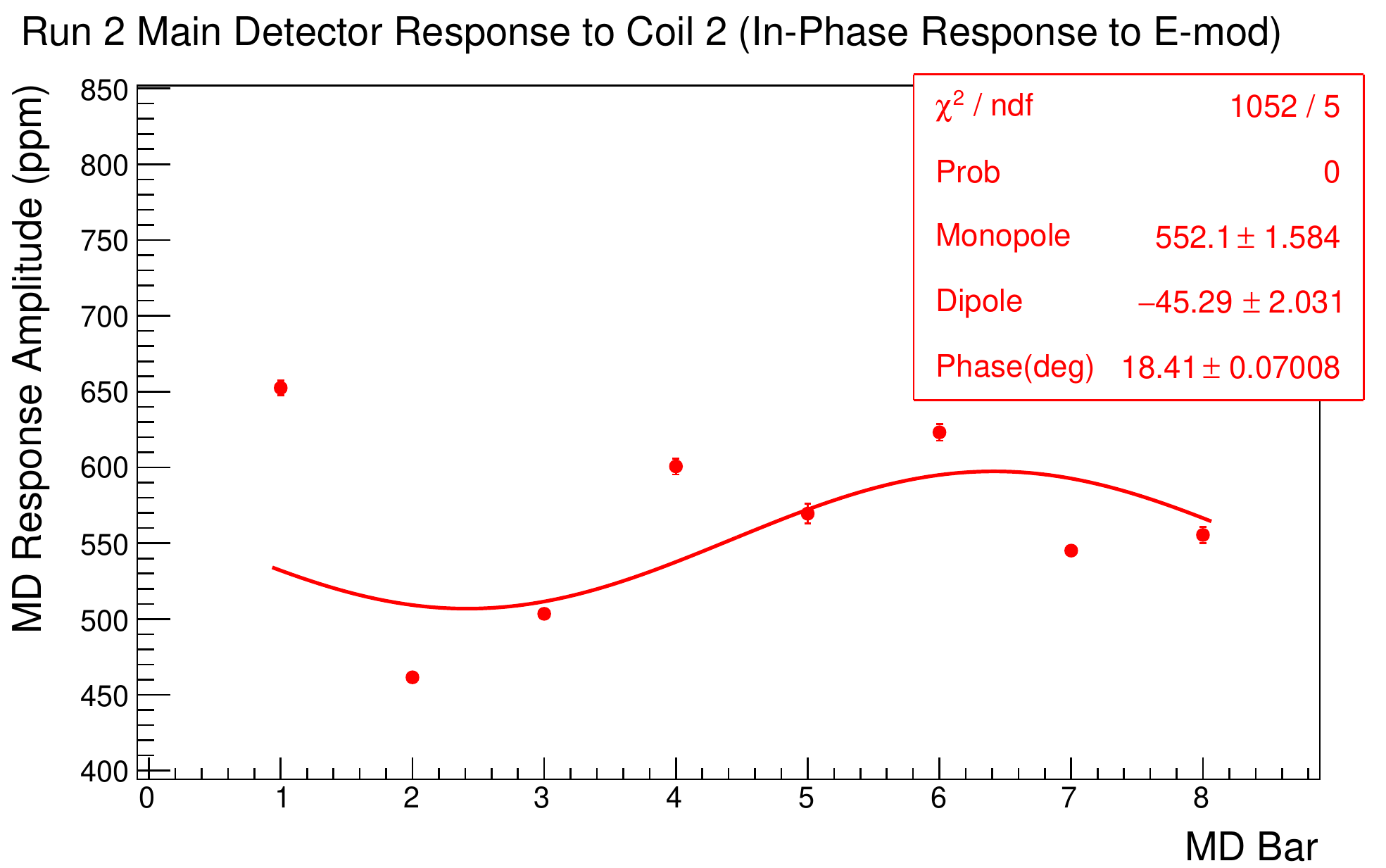}
\caption{\label{fig:Run2_10coil_Edipole}Responses of individual main detector bars averaged over Run 2 to energy modulation. Although dipole and monopole values are shown, the shape is not well characterized by either and instead gets its characteristic pattern from broken azimuthal symmetry in the electron beam/spectrometer/detector system.}
\end{center}
\end{figure}

It is tempting to consider the energy correction as a possible source of this monopole residual, since the energy response in a detector system with perfect azimuthal symmetry would be monopole. However, the evidence against this hypothesis is threefold. First, in the analysis where beam monitors were corrected using the same modulation analysis toolset, energy sensitive BPM's did not show residual sensitivities. Second, the \qtor spectrometer and detector system do not manifest perfect azimuthal symmetry. The combination of mean beam position and angle, differences in the magnetic field from octant to octant in the spectrometer, as well as small errors in the positioning of the detectors, conspire to create an energy response that is far from monopole in the main detector. The characteristic response pattern of the main detector to energy modulation is seen in Figure \ref{fig:Run2_10coil_Edipole}. Third, changing the BPM used in the modulation analysis from one in a horizontally dispersive region to one with less sensitivity and in a vertically dispersive region showed no evidence of statistically different corrections averaged over slug timescales or over the entire dataset.


A final potential failure mode that needs to be ruled out as a candidate for producing the monopole residual responses is charge or current. Although all detectors are normalized to the charge and are not expected to be sensitive to charge fluctuations to first order, there are subtle ways in which charge could still create monopole residuals. Imagine, for example, the BCM(s) used for charge normalization somehow picking up the modulation signal. This would not be visible on the battery or source channels previously monitored since they are not normalized to charge. To test the ``monopole residuals from charge'' hypothesis, a full analysis was done using main detector signals that were not normalized to charge. Figure \ref{fig:no_Q_norm_slopes}  compares the correction slopes found using normalized and unnormalized detectors and although small shifts are evident, these do not remove the residual sensitivities to modulation coils (see Figure \ref{fig:no_Q_norm_residuals}). Apparently charge is not the culprit either.
\begin{landscape}
\begin{figure}[!ht]
\begin{center}
\includegraphics[width=9in]{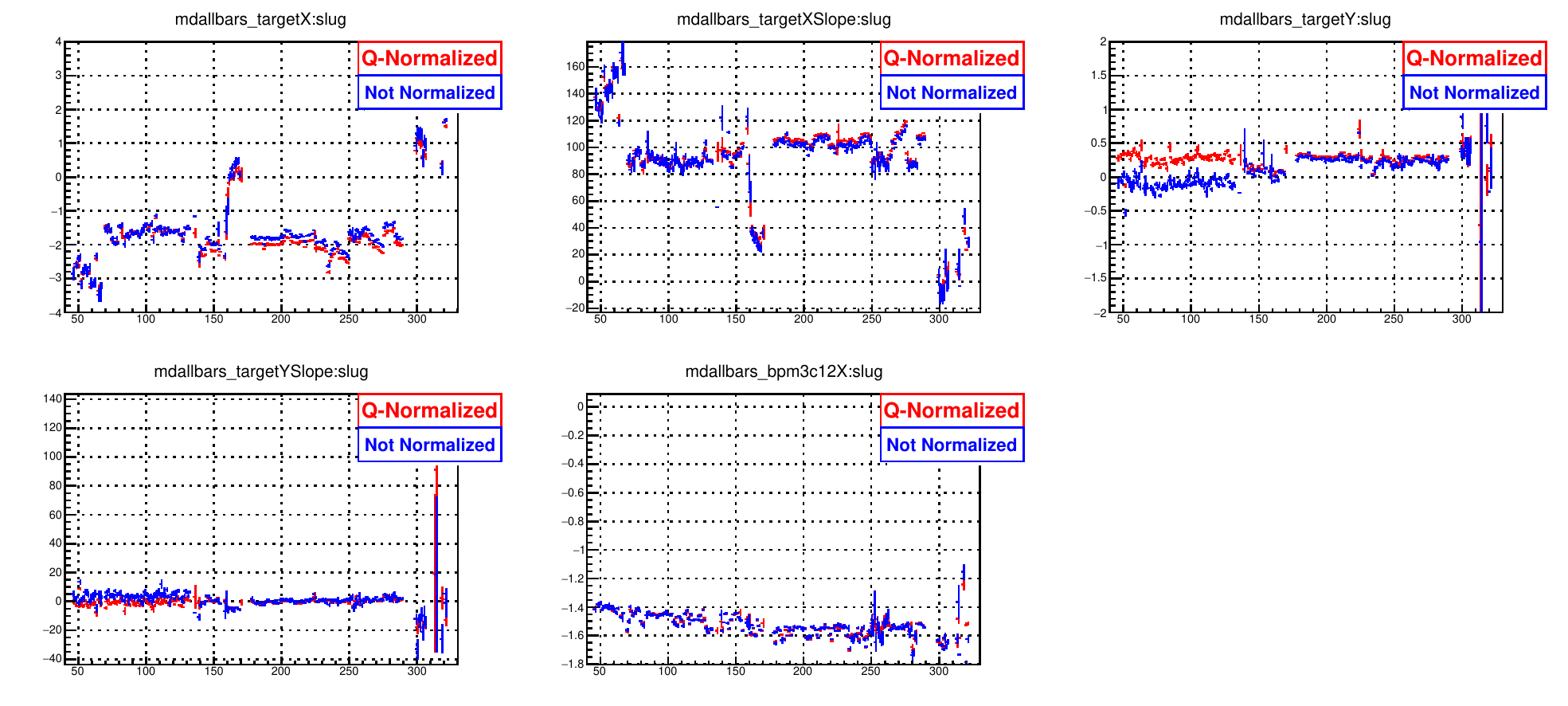}
\caption{\label{fig:no_Q_norm_slopes}Comparison of MDallbars correction slopes using a charge-normalized main detector and an unnormalized main detector signal. Both show equal residual sensitivities.}
\end{center}
\end{figure}

\begin{figure}[!ht]
\begin{center}
\includegraphics[width=8.9in]{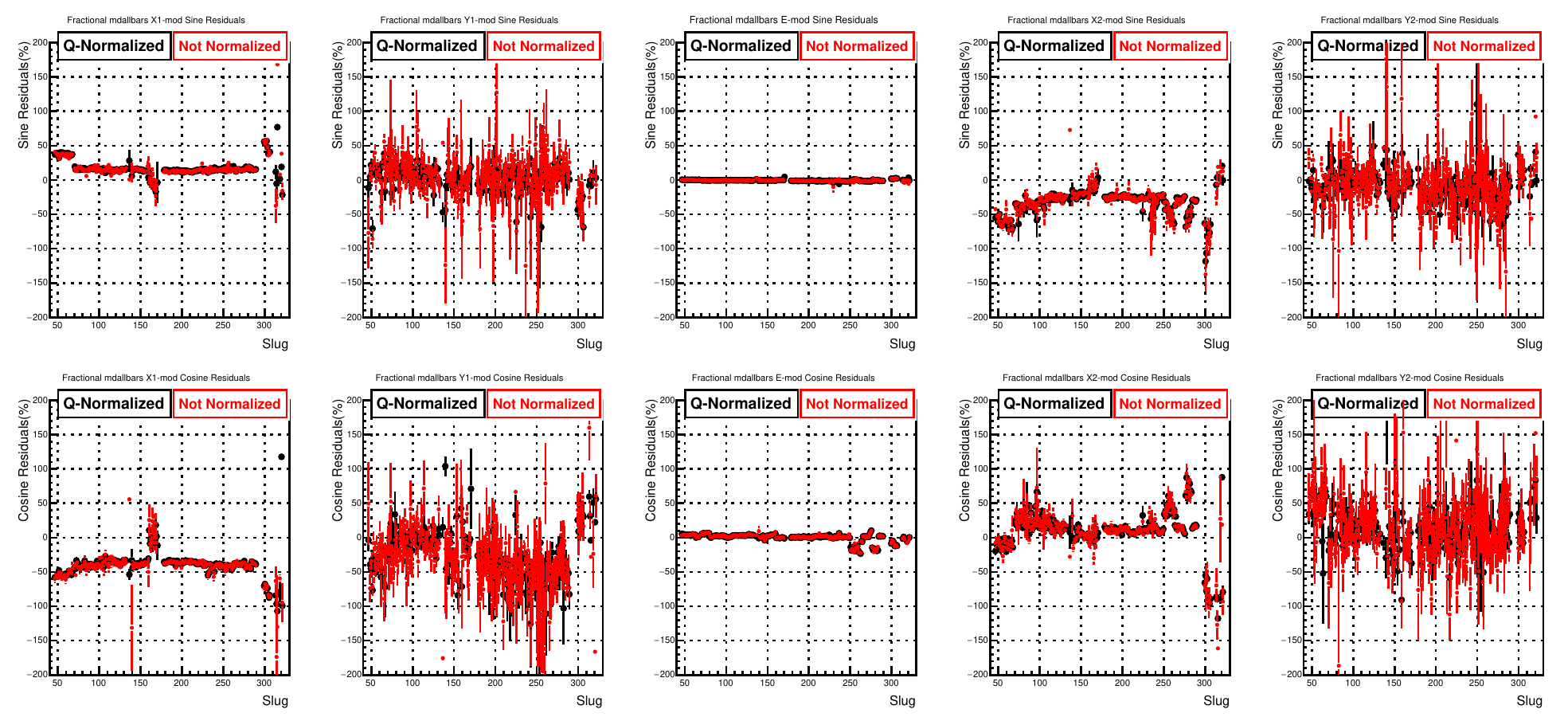}
\caption{\label{fig:no_Q_norm_residuals}Comparison of residual responses of MDallbars to modulation coils after correction using a charge-normalized main detector and an un-normalized main detector signal. Both show equal residual sensitivities.}
\end{center}
\end{figure}
\end{landscape}

\section{Determining a Beam Modulation Correction and Error}
Although the source of monopole residual sensitivity to coils after beam modulation correction remains undetermined, the signature of the inconsistency greatly limits the possible failure modes. Whatever is causing it must equally affect all main detector bars and be invisible to the beam monitors since they correct well with the same modulation analysis procedure. This signature points to an effect beyond the 5 degrees of freedom originally included in the simple linear beamline model. Something like an intrinsic beam spot size or halo modulation would seem like good candidates. These models are not completely unrealistic\footnote{Imagine a small quadrupole or higher order component to the dipole magnets driving the modulation or even electronics pickup of the modulation signal by a focusing quadrupole. These could easily change the beam spot size but it would be very difficult to detect.} but are difficult to prove. Since the modulation analysis is only correcting detectors using beam monitors, it is impossible to remove an effect to which the monitors are not sensitive. At this point it is important to find a way of deciding what correction to apply and what error to assign that reflects the uncertainty associated with inconsistencies in the analysis. The remainder of this section is devoted to developing a method for choosing a correction and assigning a systematic and statistical error to that correction.

In Section \ref{sctn:residual_correlations} almost two dozen beam modulation schemes using different coils selections were compared. It was determined that a few of these needed to be eliminated for creating residual correlations to beam monitors over long timescales or because they produced unreliable/unstable correction slopes. Much larger variations were observed in the subset of schemes where only X-type coils were omitted than in the subset omitting Y-like coils. Similarly, the residual correlations to modulation coils is much larger in the X-type coils. One scheme ``Omit 0,5'' stood out because it was the only scheme which simultaneously showed no evidence of residual sensitivity to X-type modulation coils used in the analysis and had no significant long-timescale residual correlations to beam monitors. However, the default ``10-Coil'' analysis also showed no evidence of significant long-timescale residual correlations to beam monitors. When various Y-coils were omitted from the ``10-Coil'' and ``Omit 0,5'' schemes there appeared to be two families of solutions more or less centered on the ``10-Coil'' and ``Omit 0,5'' results. Without further evidence favoring one or the other of these solutions the author proposes a straight average of the two. An estimate of the systematic error arising from the inconsistency in this analysis can be made using the spread of the data. The author proposes a conservative estimate provided by the difference from the average of the ``10-Coil'' and ``Omit 0,5'' corrections to the correction with the largest variation from this average. Looking back at tables \ref{tab:run1_dithering_corrections_table} and \ref{tab:run1_om05_dithering_corrections} for Run 1 shows the spread in total corrections to be  -11.8~ppb to -22.0~ppb. Similarly for Run 2, the spread from tables \ref{tab:run2_dithering_corrections_table} and \ref{tab:run2_om05_dithering_corrections} is -1.4~ppb to -2.6~ppb. Table \ref{tab:total_corrections} below summarizes the proposed corrections and systematic error.

\begin{table}[ht]
\caption{\label{tab:total_corrections}Proposed corrections using the average of the corrections prescribed by the ``10-Coil'' and ``Omit 0,5'' schemes. The systematic error comes from the maximum variation of other schemes from this average. Statistical error calculated from statistical spread of the correction slopes.}
\begin{center}
\Large
\begin{tabular}{|l||c|}\hline
~&~\\
\large Run 1&$\frac{-13.6-19.2}{2}=-16.4\pm 0.8(stat)\pm 5.6(syst)$~ppb\\
~&~\\
\large Run 2&$\frac{-1.7-2.4}{2}=-2.1\pm 0.3(stat)\pm 0.7(syst)$~ppb\\~&~\\\hline
\end{tabular}
\end{center}
\end{table}

In addition to the systematic error quoted above, an additional statistical error associated with the correction slopes must also be assigned. Correction slopes $\frac{\partial Detector}{\partial Monitor}$ were determined approximately every 10 minutes or as soon as sufficient information was collected to determine all the monitor and detector sensitivities to the modulation coils. One type of modulation was active for four seconds of each minute with an attempt made to cycle through all the modulation types successively. Each time a completely independent set of coil sensitivities was collected a new set of correction slopes was calculated. These short-timescale slopes were then averaged over slugs (HWP states lasting about 8~hours) and a statistical error calculated from the standard deviation divided by the square root of the number of slopes in the slug. 

The modulation correction is given by $C=\sum_j\sum_{i=1}^5w_j\left(\frac{\partial D}{\partial X_{ij}}\right)\Delta X_{ij}$ where $C$ is the correction $\frac{\partial D}{\partial X_i}$ is the correction slope of the detector to the $i$'th monitor in the $j$'th slug and $\Delta X_{ij}$ is the average of the $i$'th monitor differences over $j$'th slug. The $w_j's$ are weights applied to the corrections by slug and are the same as the main detector error weights used for calculating the average main detector asymmetry. These weights are the inverse variance of the corrected main detector asymmetry distributions divided by the number of quartet measurements in the slug ($\frac{1}{\sigma_{MD}^2/N}$). 

Using standard propagation of errors along with the main detector asymmetry error weights assigned to the main data set for each slug, allows one to calculate an estimate for the statistical error arising from the slopes. Although usual propagation of errors would involve two terms one with the error in slope times the monitor difference plus one with the error in the monitor times the slope only the former of these needs to be calculated since the latter is already included in error associated with the main detector statistical width. Propagation of errors for the first term gives
\begin{equation}
\label{eq:stat_error_cor}
\sigma_C^2=\frac{\sum_jw_j^2\sum_i\left(\delta_{ij}^{(s)} \Delta X_{ij}\right)^2}{\left(\sum_jw_j\right)^2},
\end{equation}
where $\delta_i^{(s)}$ are the statistical errors in the detector to monitor slopes at the slug level. Equation \ref{eq:stat_error_cor} assumes that the correction slopes are uncorrelated. This is clearly not the case for the monitor set used in this analysis. Slopes to targetX and targetXSlope as well as targetY and targetYSlope are highly correlated. For the purposes of estimating this error, linear combinations of these monitors were used which produced slopes that were much less correlated. Table \ref{tab:uncorrelated_monitors} shows the linear combinations used. Using slopes and statistical errors from these uncorrelated monitors gives a statistical error of 0.8~ppb for the Run 1 correction and 0.3~ppb for the Run 2 correction. These statistical errors are taken from the ``Omit 0,5'' scheme which has a larger statistical spread in its slopes. Details of a more sophisticated analysis with completely uncorrelated variables will be provided in a future thesis showing these statistical errors to be conservative \cite{Peng}.

\begin{table}[h]\
\caption{\label{tab:uncorrelated_monitors} Linear combinations of monitors used to produce semi-uncorrelated correction slopes. Units of $\mu$m for targetX(Y) and $\mu$rad for targetXSlope(YSlope) are assumed.}
\begin{center}
\begin{tabular}{l|c|c}
~&Monitor Name&Linear Combination\\\hline
Run 1 & MX1 & targetX$+35\times$targetXSlope\\
~     & MX2 & targetX$-35\times$targetXSlope\\
~     & MY1 & targetY$+29\times$targetYSlope\\
~     & MY2 & targetY$-29\times$targetYSlope\\\hline
Run 2 & MX1 & targetX$+37\times$targetXSlope\\
~     & MX2 & targetX$-37\times$targetXSlope\\
~     & MY1 & targetY$+29\times$targetYSlope\\
~     & MY2 & targetY$-29\times$targetYSlope\\
\end{tabular}
\end{center}
\end{table}
 

\chapter{Compton Photon Target}
\captionsetup{justification=justified,singlelinecheck=false}

\label{Ch:Compton_Laser}

\lhead{Chapter 6. \emph{Compton Photon Target}} 
The Compton polarimeter, built in Hall C for the \Qs experiment, measured the polarization of the electron beam by probing it with a circularly-polarized green laser and measuring the difference of e$\gamma$ scattering rates between the two helicity states of the electron beam. Figure \ref{fig:compton_layout} shows the basic components of the polarimeter and its operation is explained in Section \ref{sctn:compton}. The photon target, produced and controlled on the laser table and injected into the electron beam pipe at the interaction region, is a critical subsystem common to both photon and electron detector polarimetry measurements. The production, control and measurement of the properties of the photon target, one of the key contributions of the author to the \Qs experiment, is the subject of this chapter. Although Compton polarimetry data exists for part of Run 1, issues in the analysis for both photon and electron detectors produce larger uncertainties, and at the time of this writing, it appears that the electron beam polarization for Run 1 will be determined by the M\o ller polarimeter. A number of hardware, software and operational changes made before Run 2 make this a much higher quality dataset and although occasional mention of methods and issues during Run 1 will be made, all analyses and calculations shown in this chapter will be for the Run 2 data set.

The Compton scattering asymmetry between the two electron helicity states is calculated as 
\begin{equation}
A_{meas}=\frac{Y_R-Y_L}{Y_R+Y_L},
\label{eq:ameas}
\end{equation}
where $Y_{R(L)}$ are the background subtracted scattering rates for right and left electron spin states. This measured asymmetry is a product of three components given by
\begin{equation}
A_{meas}=P_{\gamma}P_eA_{Compton},
\label{eq:asymtopol}
\end{equation}
where $P_{\gamma (e)}$ are the photon (electron) polarizations and $A_{Compton}$ is the analyzing power precisely determined from the expected asymmetry from quantum electrodynamics convoluted with a carefully measured detector response. The electron beam polarization is calculated from this equation with all other terms measured.

Since the yields (integrated rates) in Equation \ref{eq:ameas} are background subtracted, backgrounds in the Compton detectors are a source of potentially large systematic errors. The larger the signal to background ratio, the smaller the contribution of this error. Equation \ref{eq:asymtopol} clearly illustrates the importance of a highly polarized light source and an accurate determination of its polarization. The remainder of this chapter is devoted to these two issues: the creation of a high power target for increasing signal to background\footnote{Higher target power also allows faster measurements of statistically accurate polarizations necessary for real time polarization tracking.} and the setup and accurate determination of its polarization.

\section{The Fabry-Perot Optical Cavity}
The photon target for the Compton polarimeter was produced using a Coherent Verdi-V10 (\cite{Verdi10}) laser delivering $>$10~W of green light (532~nm) mode-matched to an 84.2~cm long Fabry-Perot optical cavity. Mode-matching an optical cavity refers to the process of shaping and aligning the laser to match the allowed transverse resonating modes of the optical cavity. The Coherent Verdi-V10 produced a high quality Gaussian beam with $M^2<1.1$. Although the propagation of Gaussian beams is beyond the scope of this thesis, a few specifics will be provided to facilitate the discussion of optical mode-matching.

Gaussian beams have a transverse intensity profile that is approximately normal, meaning that most of their power propagates in the TEM$_{00}$ mode. Perfect Gaussian beams can be characterized completely by two parameters: $w(z)$, the $2\sigma$ width of the transverse intensity distribution, and $R(z)$, the radius of curvature of the laser-beam wavefront where $z$ is the distance along the beam propagation path. The two parameters expressed in terms of the laser wavelength $\lambda$ and waist $w_0$ (the minimum laser width $w$) are given as
\begin{equation}
R(z)=z\left[1+\left(\frac{\pi w_0^2}{\lambda z}\right)^2\right]
\label{eq:R}
\end{equation}
and
\begin{equation}
w(z)=w_0\sqrt{1+\left(\frac{\lambda z}{\pi w_0^2}\right)^2}.
\label{eq:w}
\end{equation}
From these equations it is obvious that $z$ is measured relative to where the beam wavefront is flat $R(0)=\infty$ and that at this position the beam is also at its minimum width $w_0$. From Equation \ref{eq:w} one can see that $w(z)$ asymptotically approaches a linear divergence in $z$ with an angle $\theta=\lambda/\pi w_0$ with respect to $z$ (small angle approximation).

Realistic, non-ideal, Gaussian beams require a third parameter, $M^2>1$, which measures the deviation of a laser from the ideal Gaussian TEM$_{00}$ propagation mode. $M^2=1$ describes an ideal Gaussian beam. The presence of a non-unity $M^2$ modifies the equations \ref{eq:R} and \ref{eq:w} giving 
\begin{equation}
w_M(z)=w_{0M}\sqrt{1+\left(\frac{\lambda z M^2}{\pi w_{0M}^2}\right)^2}
\label{eq:w_m}
\end{equation}
and
\begin{equation}
R_{M}(z)=z\left[1+\left(\frac{\pi w_{0M}^2}{\lambda z M^2}\right)^2\right],
\label{eq:R_m}
\end{equation}
where $w_{0M}$ is the measured minimum waist. The Coherent Verdi-V10 laser output was a 10~W highly Gaussian beam with quoted $M^2<1.1$, although measurements of the laser $M^2$ on the table often gave higher values in the 1.2 - 1.3 range. This decrease in laser-beam quality is not surprising given the observation of thermal lensing in optical elements on the laser table.

One final thing to note is, that for a beam that is not circular, two independent sets of equations are required: a $w_M(z)$ and an $R_M(z)$ for two orthogonal transverse dimensions ($x$ and $y$ for example). This means that the horizontal profile of the beam will propagate independently of the vertical profile and they may not even come to a waist at the same $z$ location. 

\begin{figure}
\begin{center}
\includegraphics[width=3.2in]{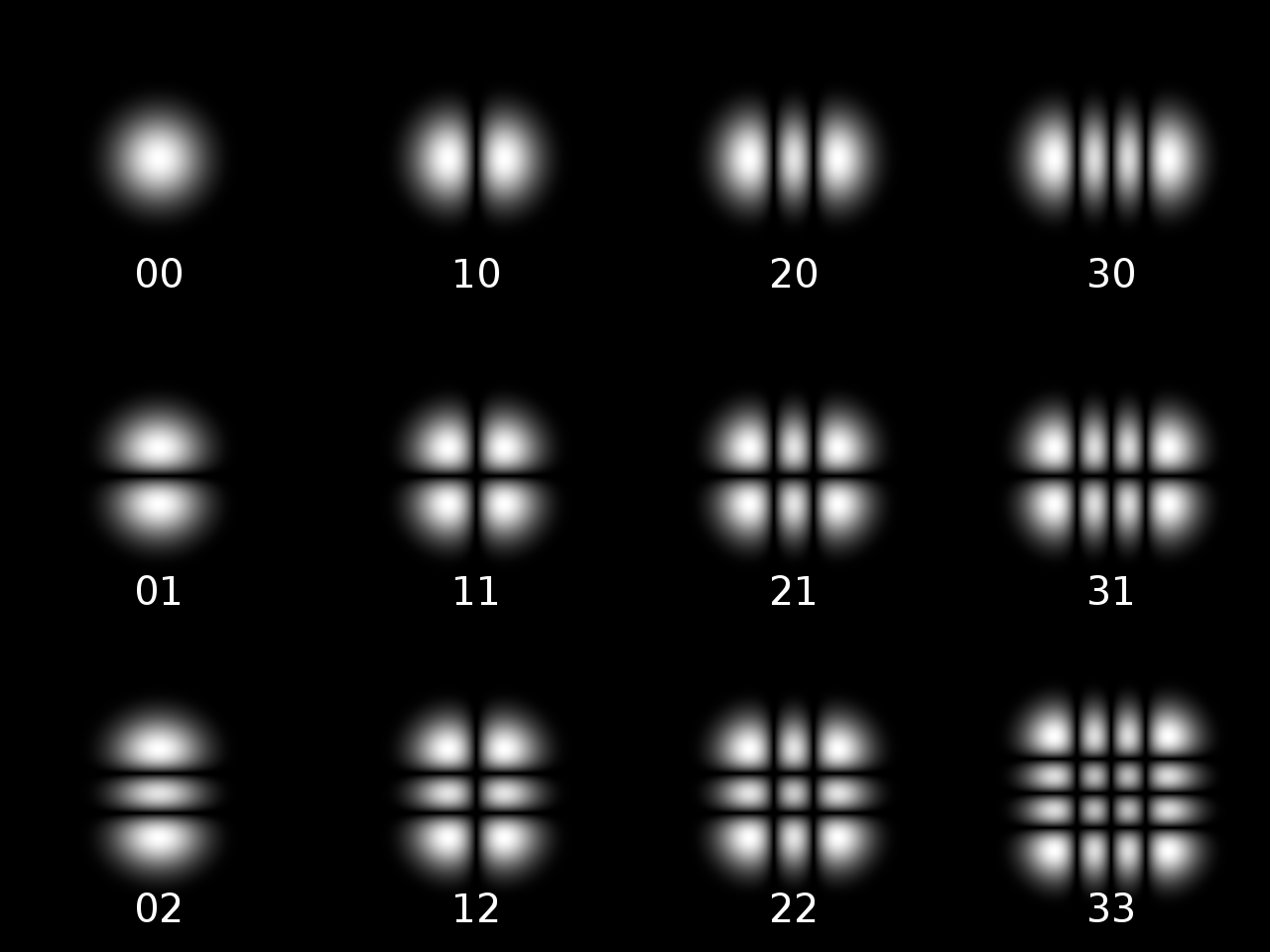}
\caption{\label{fig:hermite_modes}Hermite-Gaussian resonator modes for an optical cavity. Figure courtesy of {\it Wikimedia Commons}.}
\end{center}
\end{figure}

Gaussian beams are shaped using lenses, and linear transformation rules similar to those of geometric optics can be formulated to model the beam transport (see reference for details \cite{Kogelnik}). Mode-matching to an optical cavity is the process of shaping and aligning a laser such that its electric field distribution matches a particular resonating mode of the optical cavity. Although a laser is typically matched to the fundamental Gaussian mode of the optical cavity, misalignments and poor mode-matching solutions will allow other modes to propagate. Cavity resonating modes can be expanded in either cylindrical or Cartesian bases; however, the simplest basis depends upon the type of mode mismatch as explained below\cite{Anderson}. \\
1. Transverse Cartesian Basis:\\
This basis is made up of a Gaussian multiplied by Hermite polynomials and produces an intensity pattern given by 
\[I_{mn}(x,y)=I_0\left[H_m\left(\frac{\sqrt{2}x}{w}\right)e^{-x^2/w^2}\right]^2\left[H_n\left(\frac{\sqrt{2}n}{w}\right)e^{-y^2/w^2}\right]^2,
\] 
where $w$ is the mode transverse spot size and $H_{m(n)}$ are the Hermite polynomials for the $x~(y)$ dimensions. The indices $m$ and $n$ indicate the number of nulls in the intensity distribution in the $x$ and $y$ directions respectively. This basis is useful for describing resonator modes that exhibit broken cyclindrical symmetry. Examples of this type of broken symmetry are mirror angle misalignment, laser position or angle offsets from the cavity axis and the presence of linear polarization. The lowest order Hermite-Gaussian cavity modes can be seen in Figure \ref{fig:hermite_modes}.

\begin{figure}
\begin{center}
\includegraphics[width=3.2in]{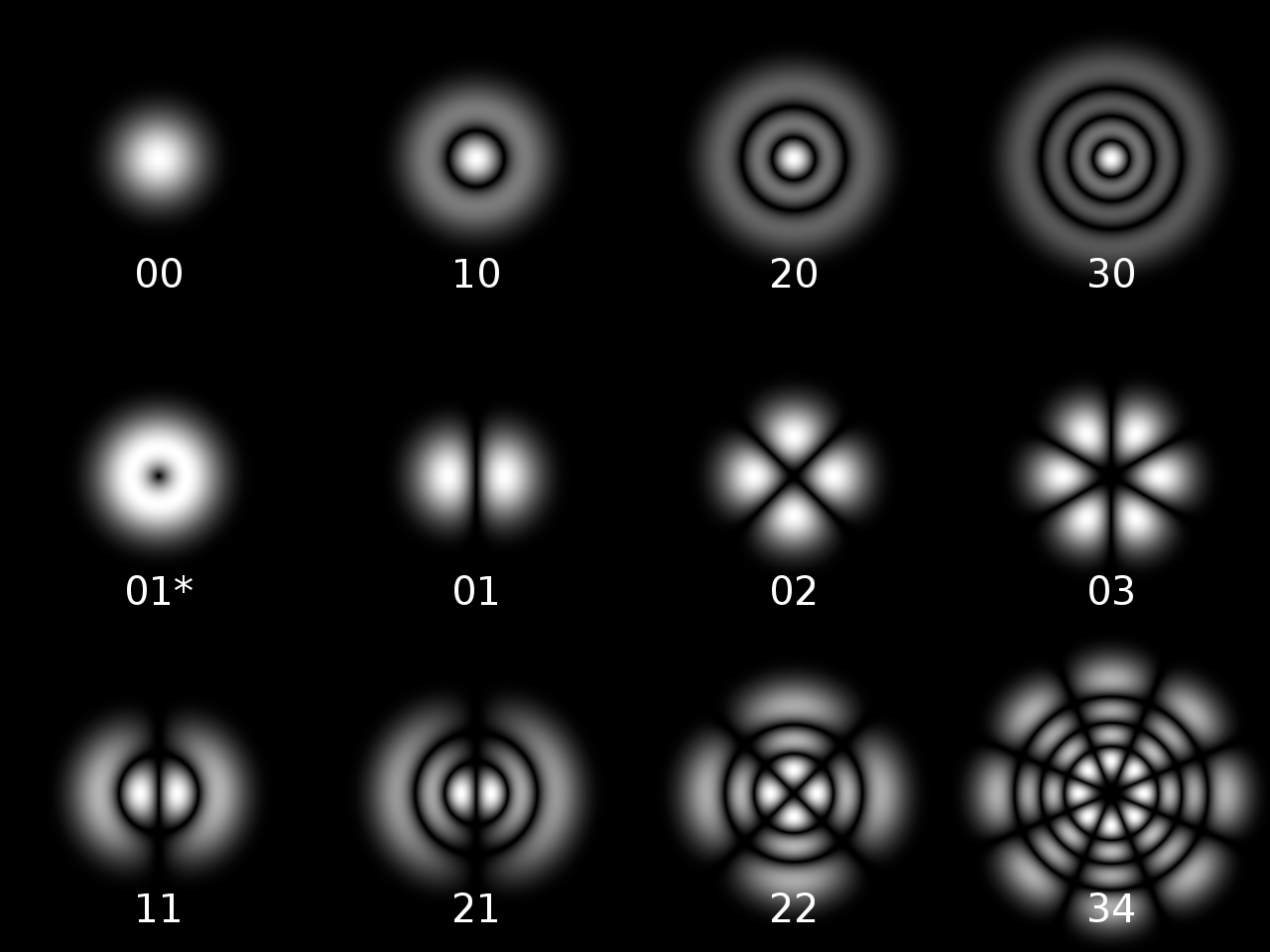}
\caption{\label{fig:laguerre_modes}Laguerre-Gaussian resonator modes for an optical cavity. Figure courtesy of {\it Wikimedia Commons}.}
\end{center}
\end{figure}

2. Transverse Cylindrical Basis:\\
This basis is composed of a Gaussian multiplied by Laguerre polynomials and produces a transverse intensity distribution given by
\[
I_{\rho l}(\rho ,\phi )=I_0\rho^l\left(L_p^l(\rho)\right)^2\cos^2(l\phi)e^{-\rho},
\]
where $\rho$ and $\phi$ are polar coordinates, $L_p^l$ is the order-$p$ Laguerre polynomial with index $l$, and $w$ is the transverse spot size of the resonator mode. This basis is useful for describing mismatches that preserve cylindrical symmetry such as a mismatch of the radius of curvature of the laser to the cavity mirrors or a laser waist that is not centered on the cavity resonator mode waist. The lowest order Laguerre-Gaussian cavity modes can be seen in Figure \ref{fig:laguerre_modes}.

\begin{figure}
\begin{center}
\includegraphics[width=4in]{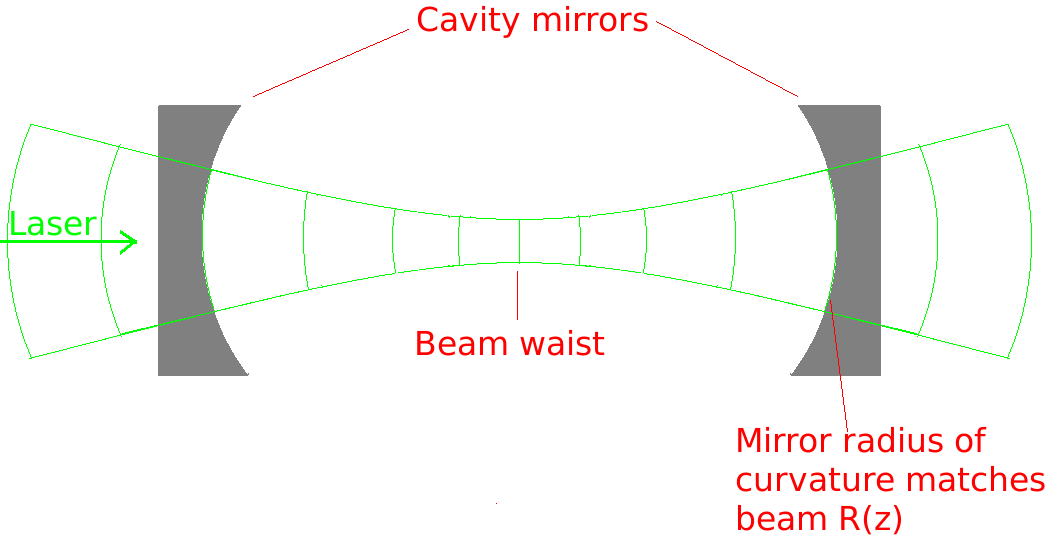}
\end{center}
\caption{\label{fig:mode_matching}Illustration of symmetric optical cavity mode-matched to TEM$_{00}$ mode of laser. The laser wavefront radius of curvature is matched to the surface of cavity mirrors.}
\end{figure}

Mode-matching the laser to the fundamental TEM$_{00}$ mode requires matching the radius of curvature $R_M(z)$ (equation \ref{eq:R_m}) to the radius of curvature of the mirrors in the optical cavity. This is illustrated in Figure \ref{fig:mode_matching}. The better mode-matched a laser is to an optical cavity, the more power will be coupled into the fundamental cavity oscillating mode and the less will be reflected backward. The Gaussian fundamental mode is matched to the cavity by finding lens positions that create the beam radius of curvature $R_M$ that matches the radius of curvature of the cavity mirrors.

\begin{figure}[h]
\begin{center}
\includegraphics[width=5in]{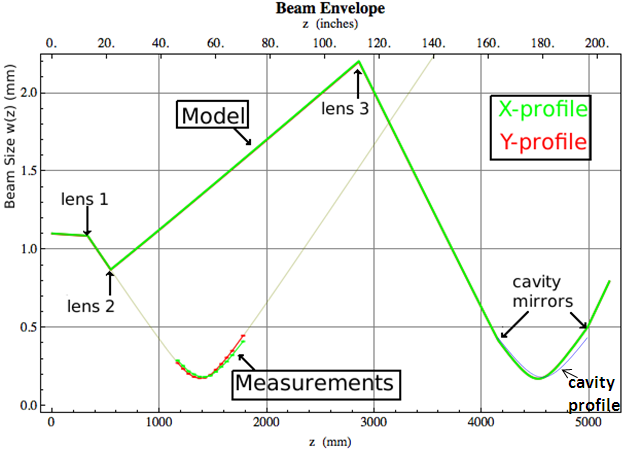}
\end{center}
\caption{\label{fig:mode_match}Laser size as a function of $z$ along its path. To mode-match the laser to the optical cavity careful measurements were taken around the waist with only lens 1 installed. Notice that X and Y profiles do not have a waist at the same z-location. Fitting these measurements yields $w_M(z)$ and $R_M(z)$ which are then fed to an optimization routine for determining the best locations for lenses 2 and 3. The cavity TEM$_{00}$-mode profile for mirrors with radius of curvature $R=50$~cm and laser $w_{0M}=184~\mu$m is also shown, indicating a small mismatch with the predicted laser profile. }
\end{figure}

Before an optimal mode-matching solution can be found, the laser's propagation equations, $w_M(z)$ and $R_M(z)$, must be determined. Measurements of the beam profile as a function of $z$ allows one to fit $w_M(z)$ to determine the parameters $w_{0M}$ and $M^2$, which are sufficient to fully define Equations \ref{eq:w_m} and \ref{eq:R_m}. Beam profile measurements on the Compton laser were performed using a camera for imaging the transverse dimensions of the beam, with software for calculating estimates of the beam radius\footnote{The beam was not actually assumed to be circular. The software determined beam widths for the two transverse dimensions of the beam.}. The camera was moved along the beam in discrete steps to measure the profile as a function of $z$. Measurements were taken on both sides of a waist and extended to the far-field region where the beam divergence is nearly linear in $z$. These measurements were then used to find the laser propagation equations. A linear optics model with the system described in terms of matrices\cite{Kogelnik}, was then used to determine the optimal positions for a given set of lenses to produce the desired beam shape matched to the optical cavity. Figure \ref{fig:mode_match} shows an example of a particular solution used on the Hall C Compton laser table. 

\begin{figure}
\begin{center}
  \includegraphics[width=4in]{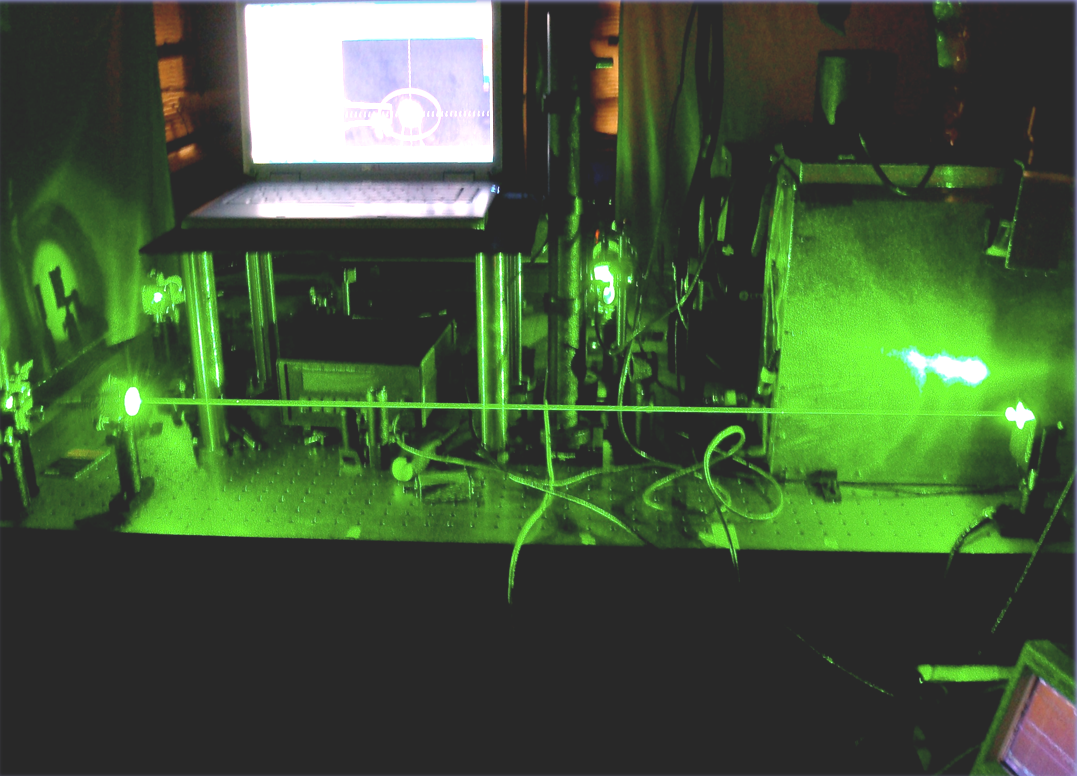}
\caption{Locked optical cavity shown during testing on an optics table at the University of Virginia. During the research phase shown, cavity lock was maintained by feeding back on the external cavity length. Cavity length was changed by attaching one of the cavity mirrors to a PZT stack. }
\label{fig:lockedcavity}
\end{center}
\end{figure}

In order to build up power in a Fabry-Perot optical cavity, the length of the cavity must be an integer number of wavelengths of the incident light. This condition, called the resonance condition for constructive interference, can be expressed as
\[
 2L=n\lambda,
\]
where $L$ is the cavity length, $\lambda$ is the wavelength of the light and $n=1,2,3...$ is an integer greater than zero. With a wavelength of 532~nm, small vibrations are sufficient to take a macroscopic optical cavity in and out of resonance, requiring a continuous feedback on either cavity length or on laser wavelength to maintain the resonance condition. The Verdi V-10 laser was retro-fitted with actuators to allow fine adjustment of the laser wavelength around the 532~nm mean. This was accomplished by mounting one of the internal resonator mirrors on two stacks of piezo-electric (PZT) material\cite{Coherent}. A short stack with low inertial mass was used for small cavity length changes at frequencies up to 20~kHz, typically associated with corrections for acoustic noise. A tall stack enabled larger mirror displacements to correct for slow drifts such as thermal changes. Both the slow and fast channels were adjusted with independent PID feedback loops with outputs in the 0-100~V range. Figure \ref{fig:lockedcavity} shows a picture of a Fabry-Perot optical cavity locked on the cavity resonant frequency during the research phase at the University of Virginia.

\begin{figure}
\begin{center}
\includegraphics[width=5in]{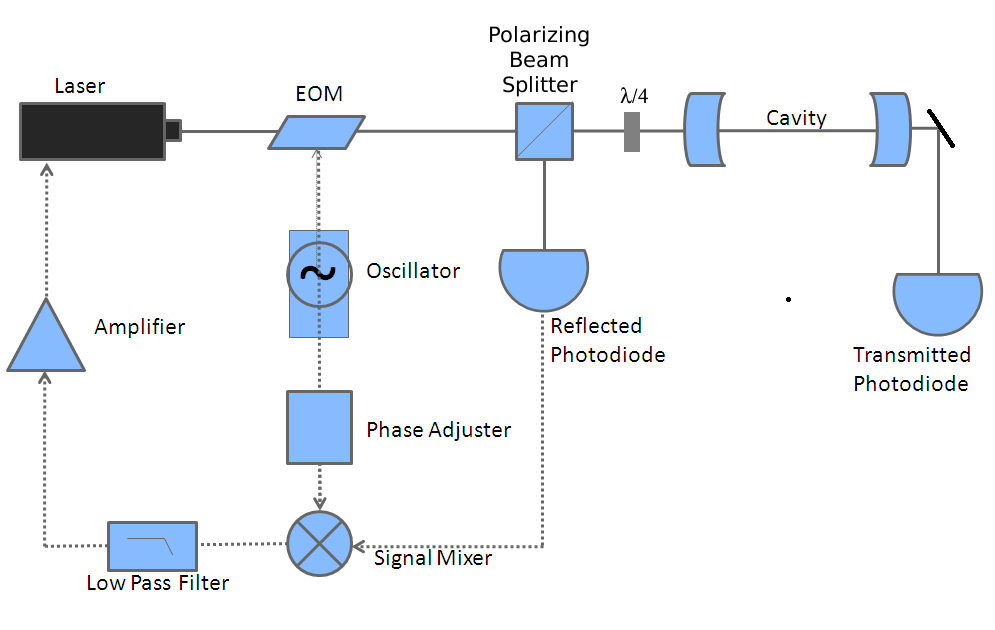}
\caption{Basic hardware setup for Pound-Drever-Hall cavity locking technique. Transmitted photodiode shown but not used in this setup.}
\label{fig:PDH_setup}
\end{center}
\end{figure}

The feedback signal for maintaining cavity resonance was created using the Pound-Drever-Hall (PDH) technique for frequency stabilization. The setup for this technique is illustrated in Figure \ref{fig:PDH_setup}. For the Hall C Compton, the technique was employed as follows. The laser was phase-modulated at 6.25~MHz using an electro-optical modulator (EOM). The EOM served as the capacitor in an oscillating resistor-capacitor (RC) circuit, tuned to resonate at 6.25~MHz\footnote{Phase-modulating the light produces a frequency distribution with most of the power near the main laser frequency and a small fraction in ``sidebands'' at the sum and difference of the laser and modulation frequencies. Near resonance these sidebands are reflected from the cavity since they are far from the resonance condition. The interference between these sidebands and the central laser frequency is used to produce the error signal for feeding back on laser wavelength to maintain cavity lock. Although a detailed analysis of the PDH locking technique is beyond the scope of this discussion, a conceptual overview can be found at \cite{Black}.}. The modulated light then passed through a polarizing beam splitter and quarter wave plate oriented to produce circularly polarized light. Circularly polarized light reflected from the entrance cavity mirror was deflected by the polarizing beam splitter into a photodiode termed the reflected photodiode (RPD). The signal from the RPD was then mixed with a phase-adjusted oscillator frequency for demodulation, producing sum and difference frequencies. The difference frequency (difference between the oscillator and RPD) produces a DC signal approximately proportional to the frequency offset from resonance for small offsets. This comparatively low frequency signal is filtered out using a low pass filter and then amplified and used as the error signal for a PID feedback loop connected to the laser frequency adjustment actuators. A general purpose, digital locking module, the Digilock 110 Feedback Controlyzer built by Toptica Photonics, was used to maintain cavity lock and greatly simplified the laser locking setup, providing a method for remote control of the electronics. Details on the Digilock module can be found in \cite{Digilock}. 

\begin{figure}
\begin{center}
\includegraphics[width=4in]{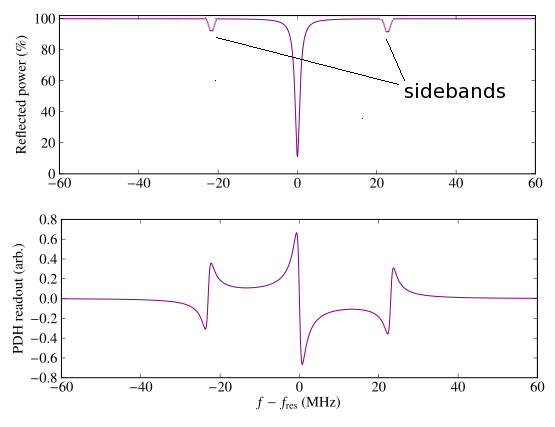}
\caption{Curves drawn from expected functional form of PDH error signal and the readout from the reflected photodiode as a function of frequency offset from resonance. {\it Plots courtesy of Wikimedia Commons.} Plot edited to include exaggerated sidebands for purposes of illustration.}
\label{fig:PDH_signals}
\end{center}
\end{figure}

\begin{figure}
\begin{center}
\includegraphics[width=4in]{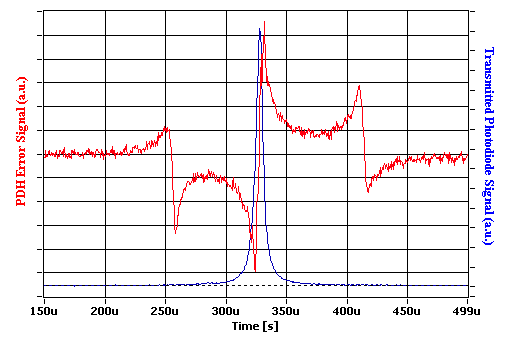}
\caption{Digital trace of PDH error signals  and transmitted photodiode signal from Compton polarimeter laser electronics while scanning through resonance. Since the scan is a linear scan of the laser output wavelength, the horizontal time axis can also be thought of as frequency as in Figure \ref{fig:PDH_signals}. Notice that the transmitted signal is nearly the inverse of the reflected signal.}
\label{fig:PDH_digilock}
\end{center}
\end{figure}

Figure \ref{fig:PDH_signals} gives an example of the PDH error signal as a function of frequency offset from resonance clearly showing the approximately linear region near resonance. Near resonance the power in the reflected photodiode decreases steeply while the transmitted photodiode signal rises. For a lossless cavity the sum of the signals in the reflected and transmitted photodiodes is constant. Figure \ref{fig:PDH_digilock} shows actual signals measured with the Compton laser setup.

\begin{figure}[!h]
\begin{center}
\includegraphics[width=3in]{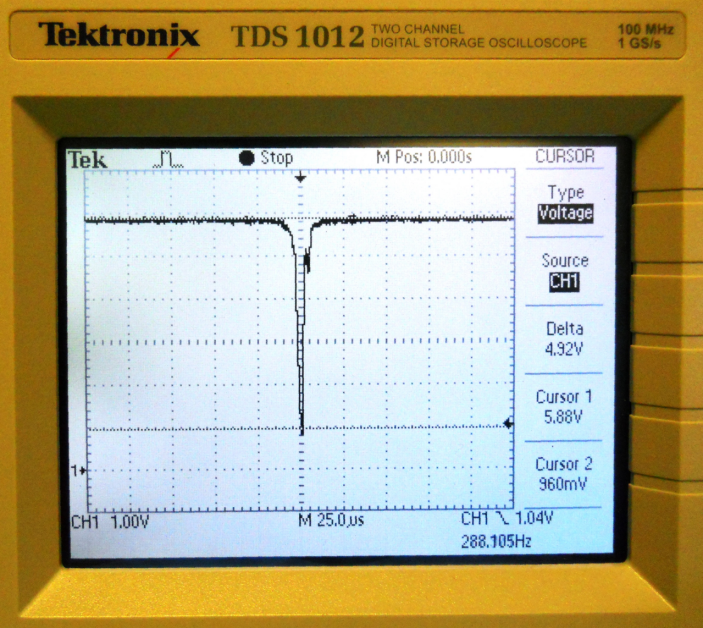}
\caption{Photograph of oscilloscope trace of reflected photodiode signal while scanning through resonance. Reading the cursor values printed on the screen to gauge the depth of the dip in the signal passing through resonance gives 4.92/5.88=84\% of the laser power coupled into the cavity.}
\label{fig:RPD_signal}
\end{center}
\end{figure}

The quality of cavity mirror alignment and laser mode-matching can be determined by the fraction of the light still detected in the reflected photodiode at resonance. A poorly aligned and/or mode-matched cavity may lock to cavity oscillating modes other than the TEM$_{00}$ mode where most of the laser power resides. In this case little power will be coupled into the cavity and most will be reflected. Even if the cavity is locked to the TEM$_{00}$ mode, the coupling of power into the cavity will be a function of the mode-match and alignment quality. During the \Qs experiment the mode-matching and cavity alignment were typically maintained to couple 60-80\% of the laser power into the cavity. The oscilloscope trace of the reflected photodiode signal seen in Figure \ref{fig:RPD_signal} demonstrates a configuration that yields 84\% power coupling.

The Fabry-Perot optical cavity for the Compton polarimeter was a symmetric design with both mirrors having a 50~cm radius of curvature, an overall cavity length of 84.2~cm and an electron beam crossing angle of $1.3^{\circ}$. This design produced a waist with a radius $w_{0M}$  of 180--190~$\mu$m at the center of the cavity\footnote{$w_{0M}$ is the radius that of the circular beam cross section that encloses $1/e^2$ of the laser power at its minimum size.}. High-reflectivity mirrors (R=99.5\%) yielded a cavity gain close to 200 with 1400--1800~W of stored power depending upon the quality of the mode-matching and laser alignment. Cavity finesse $\mathcal{F}$ is given by
\begin{equation}
\mathcal{F}=\frac{\pi}{2\sin^{-1}\left(\frac{1-R}{2\sqrt{R}}\right)}\approx \frac{\pi \sqrt{R}}{1-R}=627,
\end{equation}
for mirror reflectivity of 99.5\%. The cavity free spectral range, the frequency change between successive longitudinal resonant modes is given in terms of the cavity length $L$ as, $\nu_{fsr}=c/2L=178$~MHz. The cavity FHWM (full width at half maximum) linewidth, which is approximately the ratio of the free spectral range to the finesse, is $\Delta \nu_{_{FWHM}}=\frac{\nu_{fsr}}{\mathcal{F}}=284$~kHz.

 The electron beam optics were set up to have a much smaller electron beam spot size (1~$\sigma$ radius of 40~$\mu$m) at the point of interaction with the laser to avoid systematic errors associated with partial sampling of electron beam \footnote{It is possible to have polarization gradients across the cross section of the electron beam which could potentially bias polarization measurements. It would be useful to measure the sensitivity to the overlap of the laser and electron beams by taking polarization measurements with deliberately misaligned beams.}.  

\begin{figure}
\begin{center}
\includegraphics[width=6in]{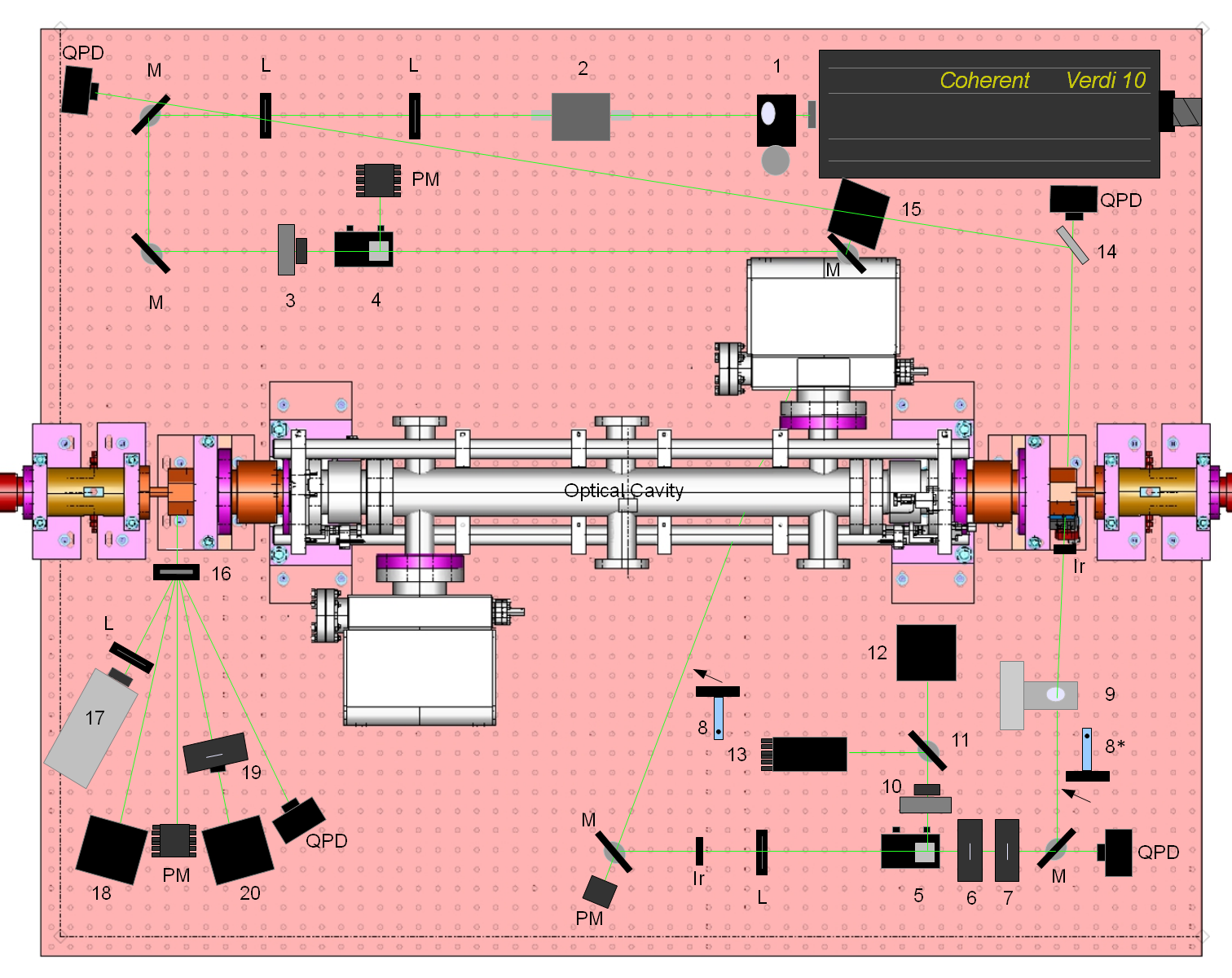}
\caption{Diagram showing a scaled version of the layout of the laser table and optical cavity overlayed by an engineering diagram of the electron beam pipe. Refer to the following list for component labeling.}
\label{fig:tablelayout}
\end{center}
\end{figure}

Figure \ref{fig:tablelayout} shows a detailed diagram of the layout of the laser table with the individual components labeled. The laser emerged from the laser head primarily vertically polarized and was then dropped to 3 inches off the table by a periscope. It was then sent through an electro-optical modulator and two lenses used to shape it as seen in Figure \ref{fig:mode_match}. A half-wave plate followed by a horizontal linear polarizer both acted as a variable attenuator and rotated the beam polarization. Another shaping lens located 2 meters downstream provided the final matching before the cavity. After this lens the beam polarization was changed from linear to circular using a linear polarizer followed by a rotatable quarter-wave and a rotatable half-wave plate. A motorized periscope was then used to elevate and align the beam to enter the cavity inside the beam pipe. The beam arrived at the optical cavity and was partially reflected and partially transmitted. The reflected beam was deflected by the linear polarizer/polarizing beam splitter and was measured by the reflected photodiode attached to an integrating sphere. The transmitted beam was split and analyzed for power, position and polarization.

\newpage
{\footnotesize{\noindent
QPD: quadrant photodiode to track laser position\\
L: lens for shaping beam\\
M: turning mirror\\
PM: power meter for monitor beam power\\
Ir: iris collimator (used only for setup and alignment of laser)\\
1. Manual periscope for dropping laser from 5 inch height at laser opening to 3 inches above table\\
2. EOM (electro-optical modulator) modulates frequency of laser at 6.25~MHz for Pound-Drever-Hall (PDH) cavity locking technique\\
3. Remote-controlled (RC) rotatable half wave plate. In conjunction with the polarizing cube at 4 we could control the fraction of laser power that was dumped into the power meter and the fraction that continued to the cavity.\\
4. Polarizing beam splitter. Here the vertically polarized light from the laser was changed to horizontally polarized.\\
5. Polarizing beam splitter. Creates horizontal linear polarization.\\
6. RC rotatable half-wave plate.\\
7. RC rotatable quarter-wave plate.  Combination of quarter and half wave plates allows one to tune the polarization to create perfect circular polarization at the cavity.\\
8. Flipper mirror inserted to block the beam during laser off times. This is the original position for most of Run 2. On March 16, 2012 the flipper was changed to position ``8*''.\\
9. RC periscope raises laser from 3 inches above the table to 9.5 inches above to the electron beam height.\\
10. Half wave plate used to adjust the fraction of light that goes through the partially reflecting mirror (11) into the Reflection Photodiode (12) and the fraction that is reflected into the beam dump (13).\\
11. Partially reflecting mirror (reflection coefficient sensitive to polarization state).\\
12. Reflection photodiode (RPD) attached to integrating sphere used to maintain cavity lock using PDH method.\\
13. High power beam dump.\\
14. 50\% beam splitter\\
15. Residual reflection photodiode (RRPD) attached to integrating sphere. Note that in this analysis the RRPD is referred to as the ``leakage photodiode''. Monitors component of light that is reflected off the cavity mirror and back through the linear polarizer (5). Adjusting the HWP(6) and QWP(7) to minimize the light in the photodiode (15) maximizes the degree of circular polarization at the optical cavity.\\
16. Holographic beam splitter (HBS) splits the beam into several beam approximately 17 degrees apart. The main beam carries about 98\% of the power with the first order beams on either side having 1\% each and the second order beams a very small fraction.\\
17. LCD camera for viewing transmitted beam. Used for imaging transmitted beam on screen in counting house.\\
18. Transmitted photodiode monitors relative power of transmitted beam and thus the power stored in the cavity.\\
19. RC rotatable Glan Laser linear polarizer.\\
20. Transmitted photodiode. The degree of linear polarization (DOLP) of the transmitted beam was measured by rotating the linear polarizer and measuring the photodiode signal as a function of rotation angle. This signal was normalized to the power in (18) to remove laser and cavity power fluctuations. DOLP=amplitude/offset of sine fit to signal vs angle.\\}
}
\section{Laser Polarization}
This section is devoted to determining the polarization of the photon target and will be divided into two topics. The details of the theory and methods used in determining the laser polarization inside the optical cavity will be first developed including difficulties encountered. Second, these methods will be applied to the Compton polarimeter dataset for Run 2 of \Qs to provide the laser polarization and assign a systematic error.
\subsection{\label{Sctn:Methodology}Methodology and Theory} 
The Compton scattering asymmetry in equations \ref{eq:ameas} and \ref{eq:asymtopol} arises from a dependence on the electron and photon helicities in the scattering cross section. A beam of light with a single definite helicity is said to be circularly polarized. In terms of individual photons, this simply means the photons are either spin left or right. In terms of light as a wave, this means the electric field vector rotates about the axis of propagation. A circularly polarized laser has two perpendicular linearly polarized components of light that are out of phase by $90^{\circ}$ or $\pi/2$ wavelengths. It is this phase difference that causes the electric field to rotate in a continuous circle. By convention, a left(right)-hand spin photon (left(right)-circularly polarized light) is defined as clockwise(counterclockwise) spin when viewed with the light traveling directly toward the viewer.

In 1941 Robert Clark Jones wrote a series of three papers outlining a method for describing polarized light as a complex two-component vector, and optical systems with 2$\times$2 complex matrices \cite{Jones1}\cite{Jones2}\cite{Jones3}. The Jones formulation for light propagation will be followed in the discussion ahead. Light is treated as a plane wave and since only electromagnetic fields transverse to the direction of travel are possible, the plane wave is expressed as a 2-vector:
\begin{equation}
\left[\begin{array}{c}E_xe^{i\phi_x}\\E_ye^{i\phi_y}\end{array}\right]e^{(kz-\omega t)},
\label{eq:jones_vector}
\end{equation}
where $E_{x(y)}$ are the amplitudes of the electromagnetic wave in the x (y) directions, $\phi_{x(y)}$ are the phases of the plane wave in the x (y) directions and $k$ and $\omega$ are the wavenumber and frequency of the light respectively. Since only the relative phase is measurable this equation can be further simplified to
\begin{equation}
\left[\begin{array}{c}E_x\\E_ye^{i\phi}\end{array}\right],
\label{eq:jones_vector_simple}
\end{equation}
where $\phi$ is the relative phase $\phi_y-\phi_x$ and the overall wave propagation term has been removed. Table \ref{tab:jones_vectors} gives examples of some common polarization states using Jones vectors. 
\begin{table}
\caption{\label{tab:jones_vectors}States of light polarization using Jones vectors. All vectors normalized to unity.}
\begin{center}
\begin{tabular}{|l|c|}\hline
Horizontal linear polarization&$\left(\begin{array}{c}1\\0\end{array}\right)$\\\hline
Vertical linear polarization&$\left(\begin{array}{c}0\\1\end{array}\right)$\\\hline
Right circular polarization&$\frac{1}{\sqrt{2}}\left(\begin{array}{c}1\\-i\end{array}\right)$\\\hline
Left circular polarization&$\frac{1}{\sqrt{2}}\left(\begin{array}{c}1\\i\end{array}\right)$\\\hline
\end{tabular}
\end{center}
\end{table}
Jones vectors are manipulated using 2$\times$2 matrices representing optical components. The effect of multiple optical elements is given by matrix multiplication of the respective elements in the order in which they are encountered by the ray of light. The matrix representations of a few common optical elements are shown in Table \ref{tab:jones_matrices}. Some of these elements are special cases of the basic optical elements. For example, the horizontal polarizer is a special case of the partial polarizer with $p_x=1$, $p_y=0$ and circular polarizers are special cases of phase retarders with $\gamma=\pi/4$ and with the birefringent axis rotated $\pm 45^{\circ}$ with respect to the horizontal axis. 
\begin{table}
\caption{\label{tab:jones_matrices}Matrix representations for some common optical components in the Jones formulation. The partial polarizer and phase retarder are shown with their extinction and retardance axes respectively lined up along the vertical and horizontal axes of the coordinate system.}
\begin{center}
\begin{tabular}{|l|c|c|}\hline
Element&Symbol&Matrix Representation\\\hline
Horizontal linear polarizer&$\bf P_x$&$\left(\begin{array}{c c}1&0\\0&0\end{array}\right)$\\\hline
Vertical linear polarizer&$\bf P_Y$&$\left(\begin{array}{c c}0&0\\0&1\end{array}\right)$\\\hline
Right circular polarizer&$\bf P_R$&$\frac{1}{2}\left(\begin{array}{c c}1&i\\-i&1\end{array}\right)$\\\hline
Left circular polarizer&$\bf P_L$&$\frac{1}{2}\left(\begin{array}{c c}1&-i\\i&1\end{array}\right)$\\\hline
Partial polarizer&$\bf P$&$\left(\begin{array}{c c}p_x&0\\0&p_y\end{array}\right),0\ge p_{x,y}\le1$\\\hline
Linear polarization rotator&$\bf S(\omega)$&$\left(\begin{array}{c c}\cos{\omega}&-\sin{\omega}\\\sin{\omega}&\cos{\omega}\end{array}\right)$\\\hline
Phase retarder&$\bf G(\gamma)$&$\left(\begin{array}{c c}e^{i\gamma}&0\\0&e^{-i\gamma}\end{array}\right)$\\\hline
Mirror (normal incidence)&$\bf M$&$\left(\begin{array}{c c}-1&0\\0&1\end{array}\right)$\\\hline
\end{tabular}
\end{center}
\end{table}

In general, perfectly reflecting  mirror at non-normal incidence will have different phase shifts for the polarization states perpendicular and those parallel to the plane of incidence, called the $s$ and $p$ polarizations respectively and can be modeled as an arbitrarily rotated birefringent element (phase retarder) taking the form ${\bf S(\omega)G(\gamma)S(-\omega)}$ \cite{Vansteenkiste}. If one adds to this an arbitrarily-oriented, stress-induced birefringence on the mirror surface, two independent rotation matrices are required as follows: ${\bf S(\omega_1)G(\gamma)S(-\omega_2)}$. This simple form can be extended to include any optical system composed of combinations of polarization rotators and phase retarders (which includes lossless mirrors). These lossless optical systems can be modeled using a single phase retarder with retardance axes aligned along the x and y axes of the coordinate system sandwiched between two rotation matrices as follows (Equation 4 \cite{Jones2}):
\begin{equation}
{\bf S(\omega_2)G(\gamma)S(\omega_1)},
\label{eq:simple_optics}
\end{equation} 
where $\omega_1$ and $\omega_2$ are not in general equal. 

All matrices in this model are unitary. Thus, in this simple model vector length is conserved and no light is lost or absorbed. Polarizing optics do not necessarily conserve light intensity and are, therefore, represented by non-unitary matrices. Optical systems composed of any arrangement of phase retarders, rotators and polarizers can be modeled by a partial polarizer sandwiched between two phase retarders and a rotator added anywhere in the system (see equations 23, 24 in \cite{Jones2}). For the purposes of this study, however, it will be sufficient to include only lossless optical elements (rotators and birefringent plates). This statement will be justified where necessary in the ensuing discussion.  

\begin{figure}[ht]
\centering
\includegraphics[width=5.5in]{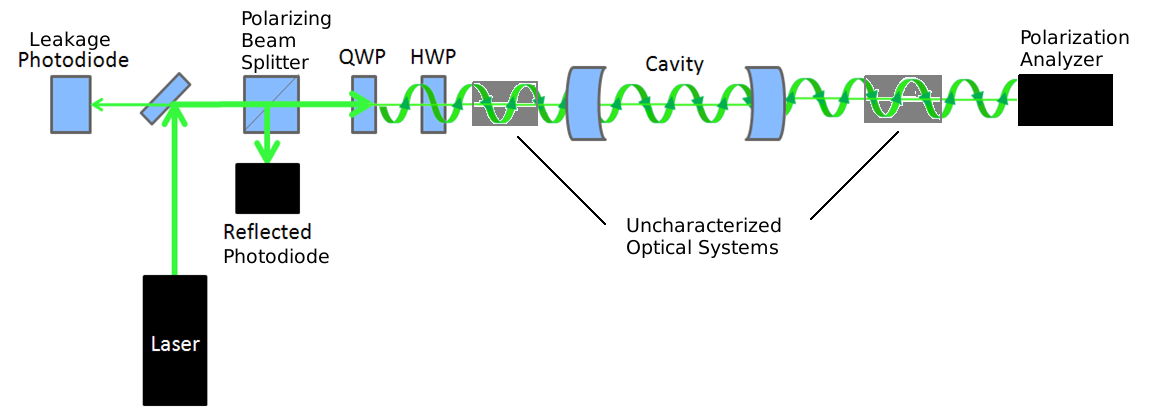}
\caption{\label{fig:optical_schematic}Simplified schematic of optical setup for producing and measuring polarized light inside Fabry-Perot optical cavity.}
\end{figure}

The use of a locked optical cavity as a photon target, together with its placement inside the high-vacuum region of the electron beam enclosure, creates challenges for determination of laser polarization. High losses, steering, and birefringence associated with analyzing optical components make it impossible to directly measure the light inside the cavity. However, in principle, both the reflected and transmitted beams can be analyzed to infer the polarization inside the cavity. Figure \ref{fig:optical_schematic} shows a simple schematic of the optical setup used to produce the photon target. Notice the regions before and after the optical cavity labeled ``uncharacterized''. These regions each have turning mirrors and a vacuum window. The transmitted region also has a holographic beam sampler for splitting the transmitted laser into several beams of different power levels. If the entrance (exit) optical system were fully characterized one could determine the polarization inside the cavity by measuring the polarization characteristics (ellipticity and angle of the polarization ellipse) of the reflected (transmitted) beam. 

Near the beginning of \Q, an attempt was made to characterize the exit region optics to infer laser polarization inside the cavity from exit line measurements of laser polarization. This process involved disassembling the cavity region of the beam pipe and installing optical components inside the optical cavity. A set of polarization states were set up and carefully measured in the optical cavity region. These same states were then measured with the polarization analyzer in the exit line region. Assuming that the optical elements can be modeled as a series of lossless mirrors, arbitrarily rotated birefringent plates and polarization rotators, this optical system can be characterized by three degrees of freedom as given in Equation \ref{eq:simple_optics}. In this setup only beam polarization characteristics are being measured, not absolute beam power. Therefore, this model is also valid for lossy optical components as long as they are not polarization-dependent losses. Fitting for the three parameters in this model, two rotation angles ($\omega_{1}$ and $\omega_{2}$) and one birefringence phase shift ($\gamma$) produces gives an optical system matrix called the transfer function ({\bf TF}) which tells how the polarization evolves from the second mirror of the optical cavity to where it can be measured in the exit line polarization analyzer. This fit involves minimizing the quadrature difference in the degree of circular polarization measured in the exit region and that predicted by the optical transfer matrix. Measurements in the exit line polarization analyzer can be translated into intra-cavity polarizations as follows:
\[
\bf E^{exit}=(TF)E_{cavity}\longrightarrow E_{cavity}=(TF)^{-1}E_{exit},
\]
where ${\bf E_{exit}}$ and ${\bf E_{cavity}}$ are the Jones vectors representing the polarization state of the laser in the cavity and where it is measured in the exit line respectively. 

\begin{figure}[ht]
\centering
\includegraphics[width=4in]{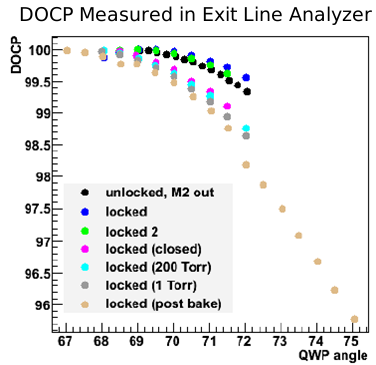}
\caption{\label{fig:TF_change}Degree of circular polarization (DOCP) measured in the exit line as a function of input quarter-wave plate angle. The input state as given by the quarter-wave plate (QWP) angle that produced maximum exit DOCP had to be adjusted by several degrees as modifications (tensioning bolts, pulling vacuum, baking) were made to the cavity. The legend can be interpreted as: 1. ``unlocked, M2 out''--second cavity mirror removed 2. ``locked''--optical cavity locked during measurement 3. ``locked 2''--second measurement with optical cavity locked 4. ``locked (closed)'' cavity region reassembled and bolts tensioned 5. ``locked (200~Torr)''--vacuum pulled to 200~Torr and cavity locked 6. ``locked (1~Torr)''-- vacuum pulled to 1~Torr and cavity locked 7. ``locked (post bake)''-- cavity locked after optical cavity baked producing high vacuum. The QWP setting for maximum intracavity polarization post bake was between 72$^{\circ}$ and 73$^{\circ}$.}
\end{figure}

Although one might be tempted to question the validity of the simple model given the known polarization-dependent reflectivities of typical mirrors, a more outstanding problem with this method arises from stress-dependent birefringence in the vacuum windows. A set of measurements taken during the break between Runs 1 and 2 of the \Qs experiment showed evidence that the transfer matrix depended upon the stress placed on the vacuum window. The main factors that affected the birefringence of the windows were temperature, vacuum pressure and bolt tensioning on the beam pipe components near the windows. The plot in Figure \ref{fig:TF_change} shows how the polarization of the laser measured in the exit line changed as the beam pipe in the optical cavity region was reassembled and vacuum was pulled, implying that the transfer matrix measured with the cavity region disassembled was not the same as the required transfer matrix of the fully reassembled system under high vacuum. 

Analysis of the entrance region optics presents a potential method for circumventing the unknown changes introduced by the vacuum window birefringence. The advantage of characterizing the entrance over the exit line optics is rooted in the fact that the light passes twice through the same system (forward and reverse) and under certain conditions comparison of the changes in the polarization states between the two beams provides sufficient information to characterize the intermediate optical system. This means that the optical transfer matrix for the entrance can be found without disassembling the cavity or breaking vacuum. Although it is satisfying to be able to properly model the optics of the system by building a transfer matrix, it will be shown that this is not strictly necessary and that maximum circular polarization at the cavity can be achieved without knowledge of the intermediate optical system. It is instructive, however, to demonstrate the possibility of creating an optical model by analysis of the reflected beam only.

Once again consider an optical model of birefringent plates, mirrors and rotators, all of which are either lossless or have polarization-independent losses, as a model for the uncharacterized optics at the entrance region of the cavity in Figure \ref{fig:optical_schematic}. The uncharacterized optics can be modeled by two rotators and a single birefringent plate as in Equation \ref{eq:simple_optics}. The QWP and HWP shown are rotatable and are directly introduced into the model but are not assumed to be perfect. Unknown birefringence errors on these plates are allowed in the model as are arbitrary rotation offsets of the optical axes. If the coordinate system is assumed constant no matter which direction the light propagates (forward or reverse), the optical system matrix for the returning light is simply the transpose of the forward system matrix. Thus the optics model has six fit parameters: two offsets for the waveplates, two errors for the waveplate birefringences and an arbitrary birefringence at an unknown rotation angle. Notice (Figure \ref{fig:optical_schematic}) that the incoming light passes through a linear polarizer (polarizing beam splitter) and the reflected light is analyzed by the same polarizer into its two linear states. The ``reflected photodiode''(labeled 12 in Figure \ref{fig:tablelayout}) and ``leakage photodiode'' (labeled 15 in Figure \ref{fig:tablelayout}) measure the intensity in the vertical and horizontal polarization states respectively. These photodiodes produce the signal measured to determine the output polarization state, that is, the Jones vector of the reflected light. The entrance function model matrix ${\bf M}$ is then
\begin{equation}
\begin{array}{lcl}
{\bf M}&=&\left[M_{element}(\alpha_0,\alpha_1)\right]\left[M_{HWP}(\alpha_2,\alpha_3)\right]\left[M_{QWP}(\alpha_4,\alpha_5)\right],\\
\text{with}&~&~\\
M_{element}&=&{\bf S}(\alpha_0){\bf G}(\alpha_1){\bf S}(-\alpha_0)\\
M_{HWP}&=&{\bf S}(\theta_{\frac{\lambda}{2}}\times\alpha_2){\bf G}(\pi/2 +\alpha_3){\bf S}(-\theta_{\frac{\lambda}{2}}\times\alpha_2)\\
M_{QWP}&=&{\bf S}(\theta_{\frac{\lambda}{4}}\times\alpha_4){\bf G}(\pi/4 +\alpha_5){\bf S}(-\theta_{\frac{\lambda}{4}}\times\alpha_4),\end{array}
\end{equation}
where the $\alpha_i's$ are fit parameters, $\theta_{\frac{\lambda}{2}(\frac{\lambda}{4})}$ are the rotation angles of the half(quarter)-waveplates and the matrices are defined in Table \ref{tab:jones_matrices}. If one chooses to use the same coordinate system for the optical system regardless of the direction of the light, the optical matrix for the returning beam is simply the transpose of ${\bf M}$ (Equation 11 \cite{Vansteenkiste}). Analysis of the reflected beam requires no additional degrees of freedom. The output measured in the two photodiodes is then given as
\begin{equation}
\left[\begin{array}{c}a_0(PD-p_0)_{Leakage}\\a_1(PD-p1)_{Reflected}\end{array}\right]={\bf M^T M}\left[\begin{array}{c}1\\0\end{array}\right],
\label{eq:entrance_model}
\end{equation}
where $(PD-p)$ are the pedestal subtracted signals in the two photodiodes and $a_{0,1}$ are the normalization constants. The initial state $\left[\begin{array}{c}1\\0\end{array}\right]$ shows that the initial state of the laser before entering the optical setup is horizontally polarized. The leakage photodiode measures the returning horizontally polarized component of the reflected beam and the reflected photodiode measures the vertically polarized component. Both the pedestals $p_{0,1}$ and the normalization constants $a_{0,1}$ are included as fit parameters since the photodiode signals are not absolute power measurements, bringing the total number of fit parameters to 8.

To determine the parameters of the model a full scan of the half-wave and quarter-wave plates was done and the signals in the photodiodes measured as a function of the rotation angle. Although data from either of the photodiodes is sufficient to obtain the parameters of the model, scans were taken of both the leakage and reflected photodiodes. Fits were performed to find the best parameters of the model using MINUIT to minimize the square of the difference between the signals measured in the photodiodes and that predicted by the model\footnote{More accurately the square of the difference normalized to the estimated variance of the difference was minimized. This is just a $\chi^2$ minimization.}.  The model parameters found separately using the data from the two photodiodes are given in Table \ref{tab:minuit_params} and show excellent agreement. 
\begin{table}[ht]
\caption{\label{tab:minuit_params}Model parameters from fit to signals measured in the reflected and leakage photodiodes during a scan of quarter-wave and half-wave plates.}
\begin{tabular}{l|c|c|c|c}
Description&Name&Leakage PD& Reflected PD&Units\\\hline
Pedestal&$p_{0,1}$&$32831.8\pm0.3$&$32749.8\pm0.3$& arb\\
Power normalization&$a_{0,1}$&$2486.4\pm0.6$&$894.4\pm0.5$& arb\\
Birefringence angle&$\alpha_0$&$-2.2980\pm0.0004$&$-2.29\pm0.01$& degrees\\
Birefringence phase&$\alpha_1$&$-0.1051\pm0.0001$&$-0.1048\pm0.0002$& radians\\
$\lambda/2$ plate rotation offset&$\alpha_2$&$78.784\pm0.002$ &$79.2\pm0.3$& degrees\\
$\lambda/2$ plate phase error&$\alpha_3$&$0.9671\pm0.0002$&$0.9652\pm0.0004$&fractional\\
$\lambda/4$ plate rotation offset&$\alpha_4$&$43.516\pm0.003$&$44.1\pm0.6$&degrees\\
$\lambda/4$ plate phase error&$\alpha_5$&$1.0141\pm0.0002$&$1.0117\pm0.0006$&fractional\\
\end{tabular}
\end{table}

\begin{figure}
\centering
\includegraphics[width=5.6in]{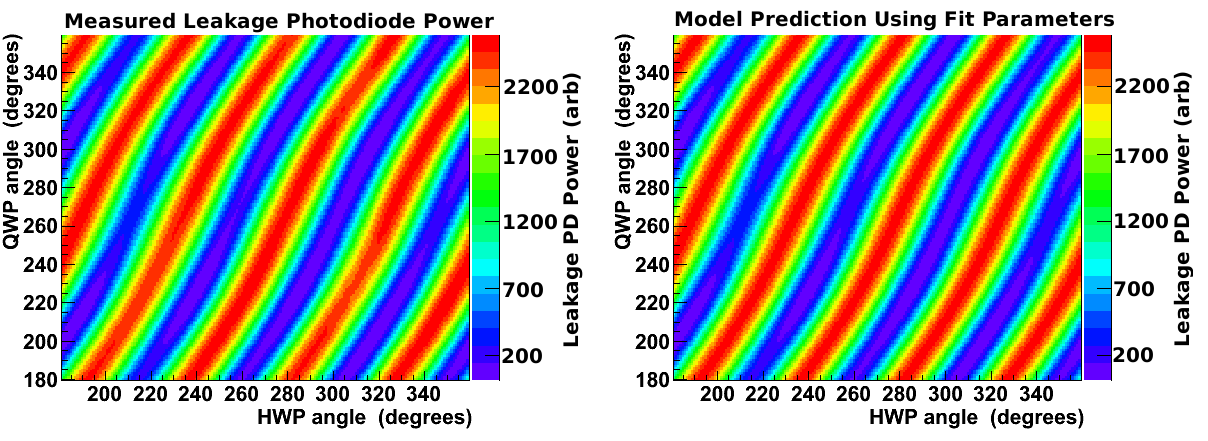}
\caption{\label{fig:RRPD_scan}Leakage photodiode signal as a function of half-wave and quarter-wave plate rotation angle. Measured signal shown on the left and prediction from model (Equation \ref{eq:entrance_model}) using fit parameters in Table \ref{tab:minuit_params}.}
\end{figure}

\begin{figure}
\centering
\includegraphics[width=5.6in]{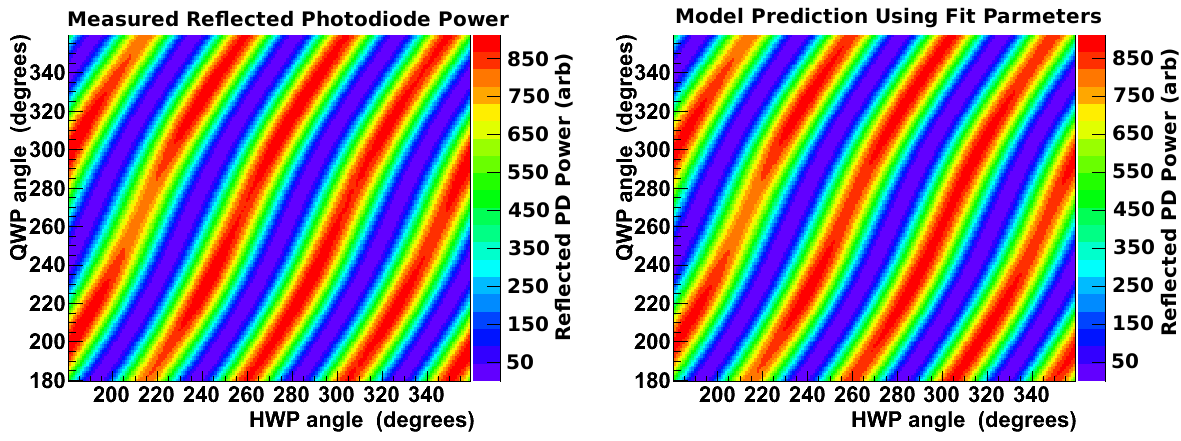}
\caption{\label{fig:RPD_scan}Reflected photodiode signal as a function of half-wave and quarter-wave plate rotation angle. Measured signal shown on the left and prediction from model (Equation \ref{eq:entrance_model}) using fit parameters in Table \ref{tab:minuit_params}.}
\end{figure}

Figures \ref{fig:RRPD_scan} shows the measured leakage photodiode signal along with the model predictions (model parameters taken from MINUIT fit) as function of quarter and half-wave plate positions. The model appears to accurately reproduce the measured power in the photodiode. A similar pair of plots is provided for the reflected photodiode in Figure \ref{fig:RPD_scan}. The reflected and leakage photodiodes show similar patterns with the the minimum signal in the reflected occurring at the maximum of the leakage and vice versa. 

\begin{figure}[ht]
\centering
\includegraphics[width=5.6in]{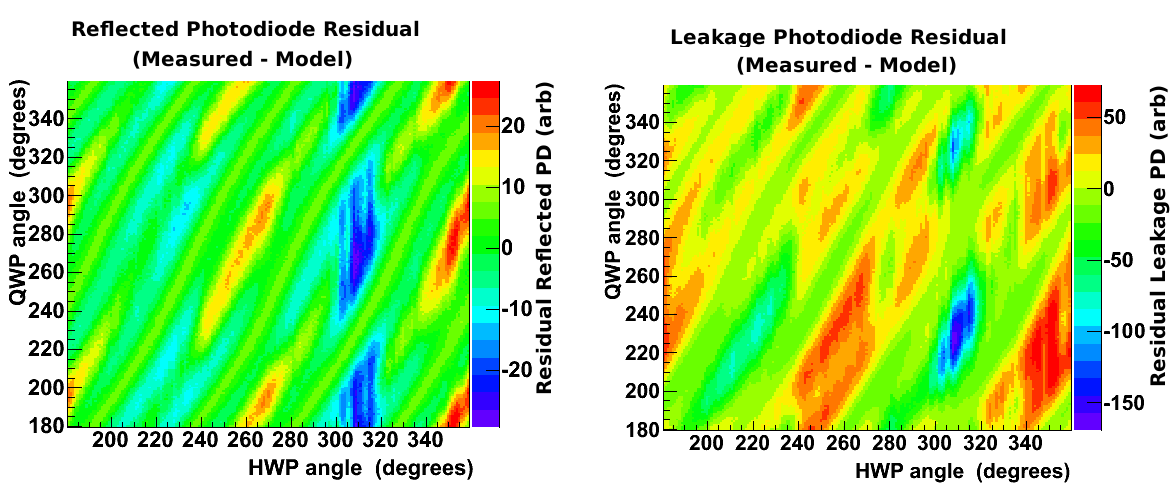}
\caption{\label{fig:PD_residuals}Residuals (Measured Photodiode $-$ Model Prediction) for reflected (left) and leakage (right) photodiodes. The $3\times$ larger range of residuals in the leakage photodiode are mainly due to a $3\times$ larger overall signal size (compare maximum 900 channels in Figure \ref{fig:RPD_scan} and 2500 channels in Figure \ref{fig:RRPD_scan}).}
\end{figure}

Figure \ref{fig:PD_residuals} shows the residuals, that is, the difference between the measured values and those predicted in the model, as a function of wave plate rotation angles. An ideal model has residuals that vary randomly around 0. The obvious structure in these residual plots show that there is something missing from the simple model. First, back-reflections off optical elements such as the quarter-wave and half-wave plates end up in the photodiodes. A better fit could be obtained if the beam were blocked after these elements and a full scan done to simply measure and subtract off the background (ambient light plus back-reflections) as a function of wave-plate angles. This is especially obvious in the vertical blue stripe apparent in the reflected photodiode residual and less so in the leakage photodiode residual. This stripe between (HWP angles 300 and 320 degrees) appears to be associated with the half-wave plate rotation angle and could be something as simple as a speck of dust on the wave plate surface. Structure that follows the striped features of the measured photodiode power plots is evidence of an imperfect model as opposed to a poorly measured background. In any case, if one neglects this systematically different region of the model, the range of residuals is only 4-6\% of the range of the photodiode powers. If the determination of polarization were dependent upon the absolute accuracy of the model, these details would be important and to ``nail down''. However, it can be shown that the circular polarization at the optical cavity can be accurately determined without a precise determination of the model. 

\begin{figure}[ht]
\centering
\includegraphics[width=5.6in]{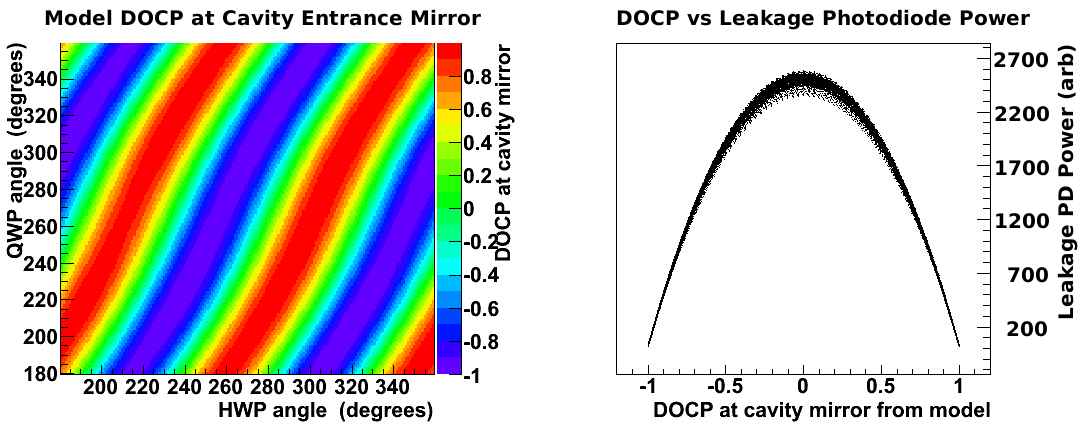}
\caption{\label{fig:DOCPvsRRPD}(Left) Degree of circular polarization  at entrance cavity mirror obtained from model as a function of quarter-wave and half-wave plate rotation angles. (Right)Model prediction of DOCP versus power measured in the leakage photodiode.}
\end{figure}

Figure \ref{fig:DOCPvsRRPD} shows the DOCP obtained from the model at the entrance to the optical cavity as a function of quarter-wave and half-wave plate angles clearly demonstrating the ability of the setup to create any circular polarization state. Left and right helicity states are shown as -1 and +1 DOCP respectively. The strength of this model is not its ability to accurately predict DOCP, but the correlation that it shows between power measured in the photodiodes and the DOCP at the entrance to the optical cavity. The right plot in Figure \ref{fig:DOCPvsRRPD} shows the tight correlation predicted by the model for the leakage photodiode. A similar (but inverted parabola) correlation exits for the reflected photodiode but its usefulness as a measurement of polarization is reduced by the fact that its power is maximized at maximum DOCP and by its sensitivity to cavity power. The leakage photodiode has a minimum signal when the DOCP at the cavity entrance is maximized and near maximum DOCP its signal changes little with cavity lock quality making it a very sensitive parameter for adjusting DOCP. Of course, this correlation between cavity DOCP and photodiode power comes from a model that has already been shown to have 5\% residuals; however, a simple verification of the correlations between DOCP and the two photodiode powers is sufficient since one can then simply minimize leakage photodiode power to achieve maximum DOCP. Furthermore, if polarizations near the maximum 100\% DOCP are found to be achievable, the effect of small changes in linear polarization on the DOCP will be greatly suppressed.

\begin{figure}[ht]
\centering
\includegraphics[width=5.6in]{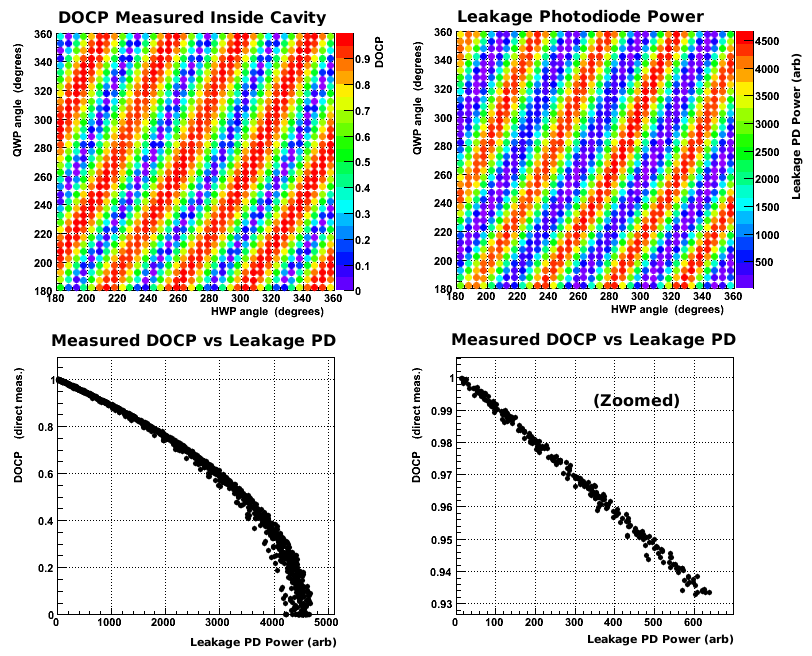}
\caption{\label{fig:DOCP_measured_correl}(Top Left) Degree of circular polarization measured inside the optical cavity as a function of wave plate rotation angles. (Top Right)Leakage photodiode power measured over same scan of wave plates. (Bottom Left)Measured correlation of DOCP inside optical cavity and leakage photodiode power. (Bottom Right)Measured correlation of DOCP inside optical cavity and leakage photodiode power zoomed in to the region near 100\% DOCP.}

\end{figure}

\begin{figure}[ht]
\begin{center}
\includegraphics[width=5.9in]{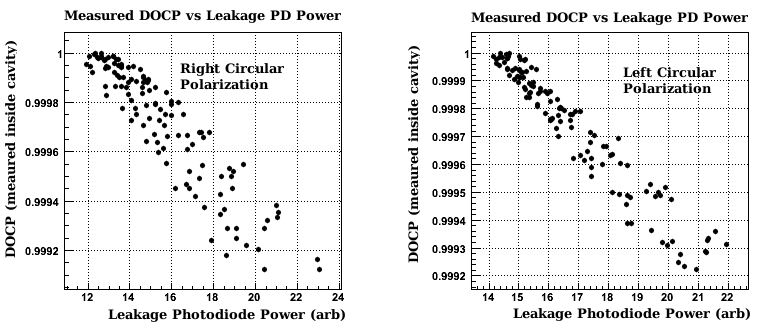}
\caption{\label{fig:landr_docp_correl}Correlation of left and right circular polarization states directly measured inside the unlocked optical cavity versus power in the leakage photodiode. Statistical uncertainty near 100\% DOCP is shown in the worst case (right circular polarization) to be 0.02\% full width.}
\end{center}
\end{figure}

A series of measurements were taken with the cavity disassembled so that states inside the cavity could be directly measured albeit with the cavity unlocked. Figure \ref{fig:DOCP_measured_correl} shows the results of one such scan. The correlation of the cavity DOCP with the measured leakage photodiode power is confirmed and is shown to be approximately linear in the region near 100\% circular polarization. Figure \ref{fig:landr_docp_correl} shows a zoomed in region of the plot of DOCP versus leakage photodiode power clearly demonstrating the ability to reach 100.00\% circular polarization in both left and right circular states. Notice that the leakage photodiode signal is not 0 at 100\% circular polarization since the signal is also composed of back-reflections from optical components as well as ambient light present on the laser table.

 
In conclusion, the issue of inaccurate exit and entrance line transfer matrices can thus be bypassed by analyzing the reflected beam. The laser polarization state was set up using a combination of a linear polarizer (polarizing beam splitter), a QWP and a HWP (see Figure \ref{fig:optical_schematic}). This combination allows any arbitrary state of elliptical polarization to be created. The optimal rotation of the two wave plates is found by minimizing the power in the leakage photodiode. One can think of this process as adjusting the waveplates to effectively cancel the effects of residual birefringence in the remaining optics producing 100\% circularly polarized light at the entrance to the cavity.

\begin{figure}[ht]
\centering
\includegraphics[width=4in]{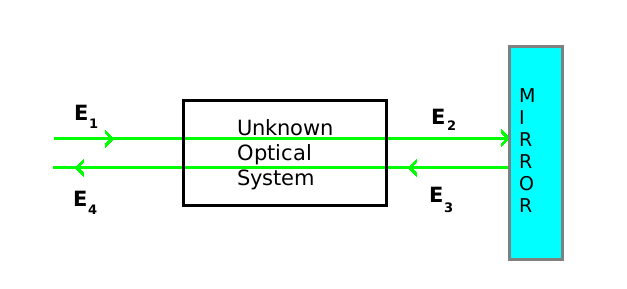}
\caption{\label{fig:reversibility}Simple optical setup to illustrate the polarization optical reversibility theorems as seen in \cite{Vansteenkiste}. Polarized light ${\bf E_1}$ goes through an unknown optical system is reflected by a mirror and goes back through the same optical system in reverse emerging as ${\bf E_4}$.}
\end{figure}

The justification of this method is rooted in reversibility theorems for polarization states of light and can be found in \cite{Vansteenkiste} as well as the original papers by R. C. Jones \cite{Jones1}\cite{Jones2}. The most relevant theorem proved in this paper\cite{Vansteenkiste} states that linearly polarized light entering an unknown optical system will emerge as circularly polarized if and only if the same beam retroreflected directly back along the same path in the reverse direction through the optical system emerges in a linearly polarized state orthogonal to the original input linear polarization. The unknown optical system is once again restricted to polarization rotators and birefringent elements represented by unitary matrices, although the arguments hold for elements with polarization independent losses. Imagine an optical setup similar to the one in Figure \ref{fig:reversibility} with incoming polarized light beam ${\bf E_1}$ going through an unknown optical system before emerging as ${\bf E_2}$. Polarization state ${\bf E_2}$ is reflected directly backward as ${\bf E_3}$, traverses the unknown optics in the reverse direction and emerges as vector ${\bf E_4}$. This theorem states that for ${\bf E_1}$ linearly polarized, ${\bf E_2}$ will be circularly polarized if and only if ${\bf E_4}$ is linearly polarized in the direction orthogonal to ${\bf E_1}$. In terms of the optical setup for the Compton photon target, the linear polarizer produces a (horizontal) linearly polarized beam which then goes through a series of optical components before reaching the optical cavity. The fraction of the beam that is circularly polarized at the entrance to the optical cavity will be retroreflected back reaching the linear polarizer as vertically polarized light and be deflected into the reflected photodiode. Linearly polarized light at the cavity will arrive at the linear polarizer as horizontal linear polarization and be measured in the leakage photodiode . Thus, minimizing the leakage photodiode signal will maximize circular polarization at the optical cavity.

\subsection{Determination of Polarization and Error}
Having demonstrated the possibility of achieving 100.00\% circular polarization it is now essential to determine what actually was achieved during Run 2 of the \Qs experiment. Since the leakage photodiode is most sensitive to the circular polarization of the laser in the optical cavity this analysis will focus solely on the signals seen by this photodiode. The discussion will first outline the method for determining the laser polarization. This will be followed by a discussion of possible systematic errors introduced by assumptions or requirements of the method.

\subsubsection{Methodology}
The lower left plot of \ref{fig:DOCP_measured_correl} shows that maximum DOCP does not coincide with zero in the leakage photodiode, indicating that the polarization signal is being added to a background that may or may not be constant. To find the true polarization signal, that is, the portion of the signal in the photodiode correlated with polarization, one must find a way to determine this background. The most obvious source of this background is back-reflections from optical elements downstream of the leakage photodiode. Given the configuration shown in Figure \ref{fig:tablelayout} one can see the linear polarizer, a lens, the waveplates and the vacuum window could easily provide such a background. It is important to take note of the position of the ``flipper mirror'' labeled as 8 in the figure. This mirror was used to deflect the laser into a beam dump periodically to measure backgrounds in the photon and electron detectors. The same flipper mirror is shown also at 8* in diagram (\ref{fig:tablelayout}) reflecting the final configuration of the optics at the end of Run 2 where the flipper mirror was moved just upstream of the periscope. At this position the back-reflections from optical elements installed between it and the leakage photodiode which are all potential sources of back-reflection background in the leakage photodiode remain even when the flipper mirror is inserted to block the beam. Thus the flipper-mirror-inserted signal in the leakage photodiode is a good measure of the background. For the period before the flipper mirror position was optimized for background subtraction, the background must be estimated by other means.

Consider a theoretically optimized optical cavity with a perfectly mode-matched laser. When this cavity is stably locked nearly 100\% of the light is coupled in, leaving no reflected light. Thus, the locked cavity signal in the leakage photodiode becomes a good measure of the background, and the difference between unlocked and locked is the polarization signal. For the optical cavity used in the Compton polarimeter, the coupling was typically 60-80\%. A sophisticated analysis could be attempted to multiply the difference between unlocked and locked by the reciprocal of the coupling to obtain a more precise polarization signal. Instead, what is done in this analysis is to simply assume a conservative 50\% coupling and estimate the polarization signal as 2$\times$ the difference between locked and unlocked as follows:
\begin{equation}
Polarization~Signal = \frac{\left[PD_{leakage}(unlocked)-PD_{leakage}(locked)\right]}{0.5}.
\label{eq:pol_sig}
\end{equation}

The reason this estimate is valid is that the polarization signal is used, not to measure the laser polarization, but to set an upper bound on how far it could have strayed from the assumed 100.00\% circular polarization. Even with conservative estimates of the error, the polarization will be shown to be constrained at the $< 0.2\%$ level.

\begin{landscape}
\begin{figure}
\centering
\includegraphics[width=9in]{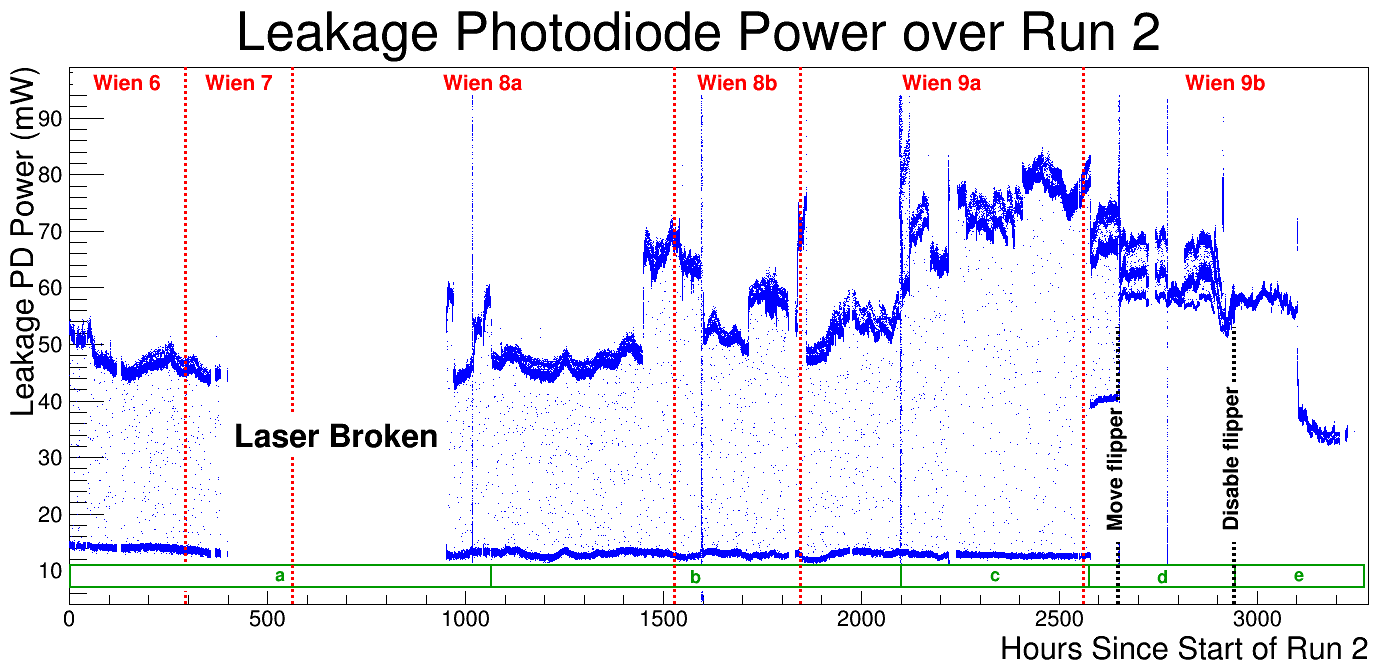}
\caption{\label{fig:rrpd_vs_time}Leakage photodiode versus hours from beginning of Run 2 of \Qs experiment. The triple valued signal comes from flipper mirror inserted (lowest signal), flipper out and cavity locked (intermediate signal) and flipper out and cavity unlocked (largest signal). Twice the difference between the top two values is used as an upper bound for the polarization signal (see Equation \ref{eq:pol_sig}).  Red vertical lines show Wien state divisions used in analysis of the \Qs dataset. Notice the background change after flipper mirror moved. This allowed back-reflections from downstream optical elements to remain when the flipper was inserted providing a measure of the background seen by the photodiode. Near the end of the run the flipper was disabled to mitigate heat load changes on optical elements that were affecting cavity lock stability and mode-matching. Periods of similar estimated polarization signal are labeled a to e.}
\end{figure}
\end{landscape}

Figure \ref{fig:rrpd_vs_time} shows the signal of the leakage photodiode over Run 2\footnote{Note that the leakage photodiode sits behind a turning mirror and is positioned to catch only backward traveling light. It is calibrated to milliwatts of light incident on the mirror.  The actual power entering the integrating sphere is down from this by about 2 orders of magnitude, thus this measurement is potentially sensitive to ambient light sources. The integrating sphere is close to the mirror but not completely shielded to prevent light from sources other than the reflected beam from entering. Light that enters from other sources such as glow in the mirror produced by the forward going beam are not subject to the same attenuation and will likely be overestimated by a factor of something like 100$\times$. These are all part of the background that must be subtracted.}. As explained in the caption, the signal has three levels, the lowest when the flipper mirror is inserted, the intermediate level when the flipper is retracted and the optical cavity locked and the highest level when the optical cavity is unlocked. The difference between the two upper bands in the diagram (unlocked-locked) is the polarization signal estimate as given in Equation \ref{eq:pol_sig}. The dataset has been subdivided into periods where a similar polarization signal is observed. These periods are labeled ``a'' to ``e'' in the Figure \ref{fig:rrpd_vs_time}. For periods ``a'' through  ``c'' the flipper mirror is positioned directly downstream of the leakage photodiode and the polarization signal must come from the estimated value in Equation \ref{eq:pol_sig}. After the flipper is moved downstream in period ``d'', the back-reflections from optical elements downstream of the leakage photodiode are still present when the flipper is inserted allowing the polarization signal to be obtained from the straight difference between flipper in and cavity unlocked, that is, the difference between the bottom and top bands\footnote{Back-reflections from the vacuum window will still be blocked. It is likely then that the polarization signal found by subtracting the flipper-in signal from the unlocked and flipper-out signal will be over-estimated. Thus, the error can only be overestimated.}. This is cleaner and avoids the overestimate of multiplying the unlocked-locked by a factor of 2. Table \ref{tab:pol_sig} gives the polarization signal estimates for the 5 periods of interest.

\begin{table}[!!h]
\begin{center}
\caption{\label{tab:pol_sig}Polarization signal estimates for periods labeled ``a'' to ``e'' in Figure \ref{fig:rrpd_vs_time}. Implied polarization shifts are calculated using Equation \ref{eq:leakage_to_DOCP} with maximum leakage signal of 6300~mW as explained in the text. Method column shows how the polarization signal was calculated for a given period. Period ``d'' was the only time when the difference between unlocked and blocked (shutter inserted) was a valid measure of the polarization signal.}
\begin{tabular}{l|c|c|c|c}\hline
Period&Hour Range&Polarization Signal&Method&Implied $\Delta$DOCP\\ \hline
a&0 - 1065&4-6 mW&Eq. \ref{eq:pol_sig}&0.03\% -- 0.05\%\\
b&1065 - 2100&6-8 mW&Eq. \ref{eq:pol_sig}&0.05\% -- 0.06\%\\
c&2100 - 2575&8-10 mW&Eq. \ref{eq:pol_sig}&0.06\% -- 0.08\%\\
d&2575 - 2943&10-14 mW&$Unlocked-Blocked$&0.08\% -- 0.10\%\\
e&2943 - 3100&4-6 mW&Eq. \ref{eq:pol_sig}&0.03\% -- 0.05\%\\\hline
\end{tabular}
\end{center}
\end{table}  

The zoomed (lower right) plot of measured degree of circular polarization (DOCP) versus leakage photodiode power in Figure \ref {fig:DOCP_measured_correl} provides a linear correlation which can be used to translate measured polarization signal into actual shifts in circular polarization. The slope of the linear relationship is $-0.5\%$ change in DOCP per 45 channels increase in leakage photodiode. The lower left plot in Figure \ref {fig:DOCP_measured_correl} gives the normalization of 4500 channels at maximum leakage photodiode power. This maximum happens when the laser is 100\% linearly polarized at the entrance to the optical cavity. Every 1\% shift in leakage photodiode power relative to the maximum gives a 0.5\% shift in DOCP yielding the following empirical relationship:
\begin{equation}
Cavity~DOCP = 1-0.5\times\frac{Polarization~Signal}{Maximum~Leakage~PD~Signal}.
\label{eq:leakage_to_DOCP}
\end{equation}

There is no direct measurement of the maximum value of the leakage photodiode. Sending the full power of the beam back to the laser head damages the laser and makes it unstable. This practice was avoided, although a couple of mistakes were made, where, for short periods, this happened. One such time when it happened, the leakage photodiode reached the upper limit of the ADC voltage range with a reading of 5.4~W. With $\sim$10.5~W coming from the laser, it seems safe to assume that more than 7~W makes it back to the mirror in front of the leakage photodiode when the waveplates are set to maximize leakage photodiode signal. Allowing a 10\% error in the calibration of the photodiode gives a conservative effective maximum of 6.3~W. Even with this conservative estimate, for the largest polarization signal of 14~mW in period ``d'', this implies a shift in DOCP of only 0.11\%. The total polarization shifts implied by the polarization signals are given in Table \ref{tab:pol_sig}.

To summarize, in the setup used for the Compton polarimeter, the leakage photodiode is sensitive to the degree of circular polarization at the entrance to the optical cavity.  When the laser polarization at the entrance to the optical cavity is close to 100\% circular, the polarization signal (the part of the signal in the leakage photodiode that is sensitive to polarization) is proportional to the DOCP. Direct measurements determined the relationship between polarization signal and changes in DOCP at the cavity entrance to be 0.5\% change in DOCP for every 1\% change in the leakage photodiode power (relative to what its reading would be at maximum returning laser power). This relationship, along with methods outlined for conservative estimates of the polarization signal, is used to bound shifts in DOCP to $<0.11$\%. Thus the polarization is reported to be 100.00\% with possible fluctuations as large as 0.11\%. Several potential sources of systematic error introduced by assumptions of this method are discussed next. 

\subsubsection{Potential Systematic Errors}

The method outlined in the previous section overlooks a few small potential sources of systematic error. These errors naturally fall into two general categories. The first category of systematic error arises from the assumption of a 100\% polarized laser beam incident on the cavity. This assumption is implicit in the use of the Jones calculus, which is only applicable to fully polarized light. An upper bound on the amount of unpolarized light entering the optical cavity will be provided in this section. The second category of systematic errors arises from differences between the optical cavity locked and unlocked states. Although the method outlined in the previous paragraphs is guaranteed to maximize circular polarization at the entrance to the optical cavity, an uncertainty will be assigned to account for any changes in the polarization of the light in the locked cavity after 200+ mirror bounces. The polarization could change by either birefringence of the cavity mirrors creating linear polarization or by incoherent scattering producing an unpolarized component. Furthermore, ambient light levels on the table may contribute marginally to the reading in the leakage photodiode. In particular, shifts of the ambient light between locked and unlocked cavity states and between flipper in and out states may change the ambient light seen by the leakage photodiode, and either accentuate or cancel the true polarization signal, depending upon the sign of the shift.\\
{\bf 1. Unpolarized Light}\\
An unpolarized beam of light has randomly oriented electromagnetic fields as opposed to a polarized beam with well-defined orientations. Since the laser in this system passes through a linear polarizer, it begins in a well-defined and fully polarized state. Depolarization of a fully polarized beam can happen when the beam encounters random scattering centers such as imperfections/residue on the surface of mirrors. The issue of unpolarized light is two-fold. First, is the issue of unpolarized laser light incident on the optical cavity (depolarization occurring before the optical cavity) and second is the issue of depolarization that occurs inside the optical cavity. 

The issue of unpolarized light incident on the optical cavity was addressed by direct measurement inside the optical cavity region (with the cavity unlocked). An upper bound of 0.04\% unpolarized light was determined. This means that light that arrives at the optical cavity is at least 99.96\% polarized. This measurement did not bound the depolarization inside the locked cavity. Depolarization that happens before the optical cavity will be measured equally on both the locked and unlocked signals in the leakage photodiode and will thus be absent from their difference whenever the polarization signal is computed using Equation \ref{eq:pol_sig}. 

Depolarization that happens inside the cavity will only show up on the locked signal and will inflate the locked signal in the leakage photodiode falsely diminishing the polarization signal computed using Equation \ref{eq:pol_sig}. This issue of depolarization inside the locked cavity will be discussed in the next section on errors arising from differences between locked and unlocked states. It will be shown that data taken during period ``d''  can be used to bound the effects of both incident unpolarized light and intracavity depolarization. Although this data provides an even tighter bound on incident unpolarized light, the more conservative (and directly measured) 0.04\% bound was included in the error in Table \ref{tab:pol_error} to account for depolarization.

{\bf 2. Differences between Locked and Unlocked Cavity}\\
The remaining sources of error concern unknown differences between locked and unlocked optical cavity states on the laser table. Three sources of potential differences between the two states will be considered and arguments made to bound each. The three sources are as follows: increases in linear polarization that occur inside the locked optical cavity due to residual birefringence of the reflective coating on the cavity mirrors; ambient light shifts (or changes in background) between locked and unlocked states; and depolarization that occurs inside the optical cavity. 

The degree to which shifts in ambient light affect the polarization signal would be best found by direct measurement on the laser table by isolating the leakage photodiode from all ambient light except that which comes through the mirror. If the ambient light seen by the leakage photodiode increased when the cavity locked, this effect would inflate the locked signal and partially cancel the polarization signal which was measured as unlocked minus locked. The author believes this is a negligible effect but it would be best to directly measure or eliminate it in the future. Furthermore, if the polarization state changes when the cavity locks, either by depolarization inside the cavity or an increase in linear polarization from birefringence on the cavity mirrors, this too would, at least partially, cancel the polarization signal by adding signal in the leakage photodiode only in the locked state. It is a difficult task to disentangle these three effects (linear polarization inside the cavity, depolarization inside the cavity and ambient shifts) without direct measurement. A measurement was made to bound the change in polarization of the laser between the locked and unlocked cavity states. Evidence is provided in the following paragraphs that this measurement along with data taken during \Qs is sufficient to bound errors from each of these effects on the circular polarization inside the locked cavity.

\begin{figure}
\centering
\includegraphics[width=5in]{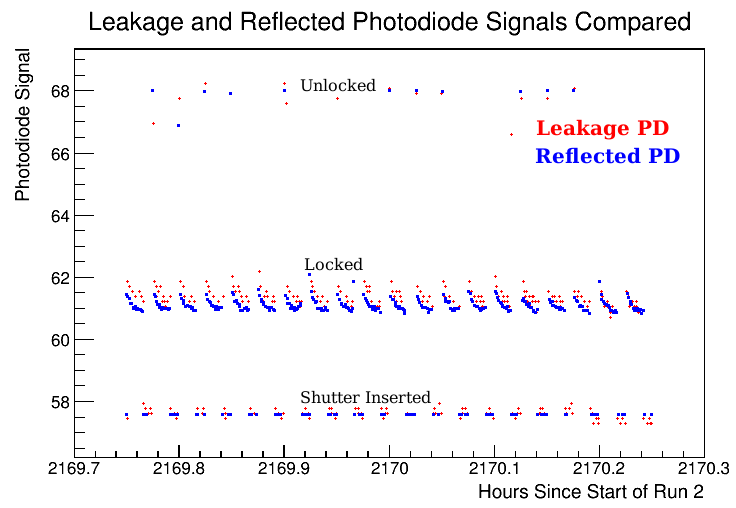}
\caption{\label{fig:rpd_rrpd_overlay}Zoomed in section of Figure \ref{fig:rrpd_vs_time} showing the leakage photodiode signal levels in milliwatts for unlocked, locked and shutter inserted. Overlayed is a scaled and offset reflection photodiode signal to demonstrate the expected signal change from power only. A small 0.5~mW excess signal can be seen in the leakage photodiode. }
\end{figure}

Evidence for polarization changes with cavity lock may be present in deviations from the expected signal level seen in the leakage photodiode. The polarization signal estimated as the difference between unlocked and locked divided by coupling (Equation \ref{eq:pol_sig}) assumes that the only change visible to the leakage photodiode when the cavity locks is a change in reflected {\bf power}. Under this assumption, the polarization signal, that is, the signal after the polarization-independent background has been subtracted, is simply the difference between unlocked and locked scaled by 1/coupling to estimate what would be measured with 100\% of the power coupled into the cavity. However, the leakage photodiode is sensitive to both polarization and power changes. On the other hand, the reflection photodiode is sensitive to power but insensitive to small changes in polarization and ambient background light. Thus, the fractional drop in power in the reflected photodiode is used as a measure of the coupling into the cavity. Comparing the changes in both of these photodiodes allows one to test the assumption of a ``power-only'' change in the leakage photodiode between locked and unlocked. If the laser polarization does not change when the cavity locks\footnote{This means that the intracavity polarization is the same as it was measured to be at the entrance to the cavity and multiple bounce reflections from cavity mirrors do not change polarization.}, then one might expect an identical fractional shift in both the leakage and reflection photodiodes. An identical fractional shift would provide evidence (although not conclusive) that the shift is solely due to the power of the reflected beam. Figure \ref{fig:rpd_rrpd_overlay} shows an example of the leakage photodiode signal over a half hour period during Run 2. A scaled reflection photodiode signal has been overlayed to demonstrate the expected locked signal level due to power shifts in the reflected beam. The small excess of $\sim0.5$~mW in the locked leakage photodiode above what is expected from the change in reflected power may be attributed to a change in either polarization or ambient background light or to a combination of both. With this hint of a polarization/ambient light shift with cavity lock, it becomes necessary to bound these effects.
 
\begin{figure}[ht]
\begin{center}
\includegraphics[width=5.0in]{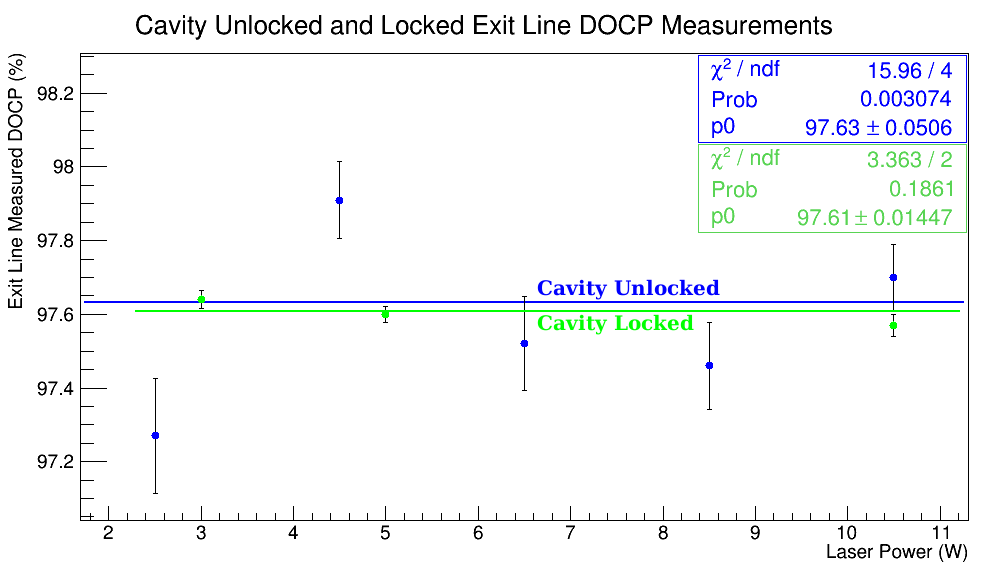}
\caption{\label{fig:pol_change}Laser polarization measured in the exit line for both cavity locked and cavity unlocked. Cavity mirrors with reflection coefficients R=99.5\% allow enough light through even in the unlocked state to measure polarization although the accuracy is reduced. The difference in polarization between locked and unlocked is statistically consistent with zero.}
\end{center}
\end{figure}

A dedicated measurement was made after Run 2 of \Qs was completed to bound the difference between locked and unlocked polarization. In this test, the laser transmitted through the cavity in both locked and unlocked states was polarization analyzed. This measurement was possible when the cavity was unlocked due to the relatively low reflectivity of the mirrors ($R=99.5\%$). The results, shown as a function of laser power in Figure \ref{fig:pol_change}, are consistent with no difference in circular polarization between locked and unlocked cavity states as measured on the transmitted beam in the exit line. For these measurements the linear polarization of the transmitted beam was measured in the exit line by rotating a linear polarizer and measuring the power transmitted through the polarizer as a function of rotation angle. A sinusoidal fit to the data of transmitted intensity versus rotation angle was then used to calculate the degree of linear polarization as DOLP = amplitude/constant offset. Circular polarization is then calculated as
\begin{equation}
DOCP =\sqrt{1-DOLP^2}
\label{eq:DOCP}
\end{equation}
 under the assumption of 100\% polarization, that is, that no light is unpolarized. Even if one were to argue that a small difference in DOCP is observed, it is important to keep in mind that the measurements in this figure are taken in the exit line where the DOCP is around 97.6\%, the value that roughly translates into 100\% polarization in the optical cavity \footnote{The polarization state of the light is changed by the turning mirrors and vacuum window encountered between the optical cavity and the exit line. Thus, 100\% circularly polarized light in the optical cavity is measured as approximately 97.6\% circularly polarized in the exit line.}. Suppose one were to conservatively estimate a circular polarization change as high as 0.2\% between locked and unlocked states in the exit line from Figure \ref{fig:pol_change}. The sensitivity of DOCP to changes in linear polarization is given by
\begin{equation}
\frac{d DOCP}{d DOLP} =\frac{-DOLP}{\sqrt{1-DOLP^2}}.
\label{eq:DOCP_sens}
\end{equation}
Thus the sensitivity in the exit line to shifts in linear polarization is 
\[\frac{dDOCP}{dDOLP}\rvert_{(DOCP=97.6\%)}=-0.223,\] whereas the sensitivity in the cavity is closer to \[\frac{d DOCP}{d DOLP}\rvert_{(DOCP=99.9\%)}=-0.0474\] giving almost a factor of 5 suppression\footnote{This assumes an identical shift in linear polarization both in the exit line and inside the cavity. A more sophisticated model would utilize a measured optical transfer matrix to translate the change measured in the exit line to that inside the cavity. However, this estimate of suppression is valid to first order under the assumption that the exit line optics create small changes in laser polarization.}. Even a shift as large as 0.2\% in DOCP in the exit line would translate to only a 0.04\% shift of DOCP inside the cavity.

The previous method utilizes a rotating linear polarizer to measure laser linear polarization which then implies a circular polarization under the assumption of no depolarization (Equation \ref{eq:DOCP}). One could argue that the above measurement provides a bound on shifts in linear polarization not circular polarization since linear polarization is what is actually measured. Circularly polarized light and unpolarized light will be indistinguishable when analyzed by a rotating linear polarizer. The conservative 0.2\% shift in DOCP measured in the exit line implies a 0.9\% change in linear polarization in the exit line. It is safe to assume to first order that this also implies a shift of approximately 0.9\% inside the cavity as well, setting an upper limit of $\Delta$DOLP$<0.9\%$ between locked and unlocked. Table \ref{tab:pol_error} includes a 0.04\% error associated with the implied shift in intracavity DOCP from a 0.9\% increase in linear polarization.

\begin{figure}[!!!!ht]
\begin{center}
\includegraphics[width=5.5in]{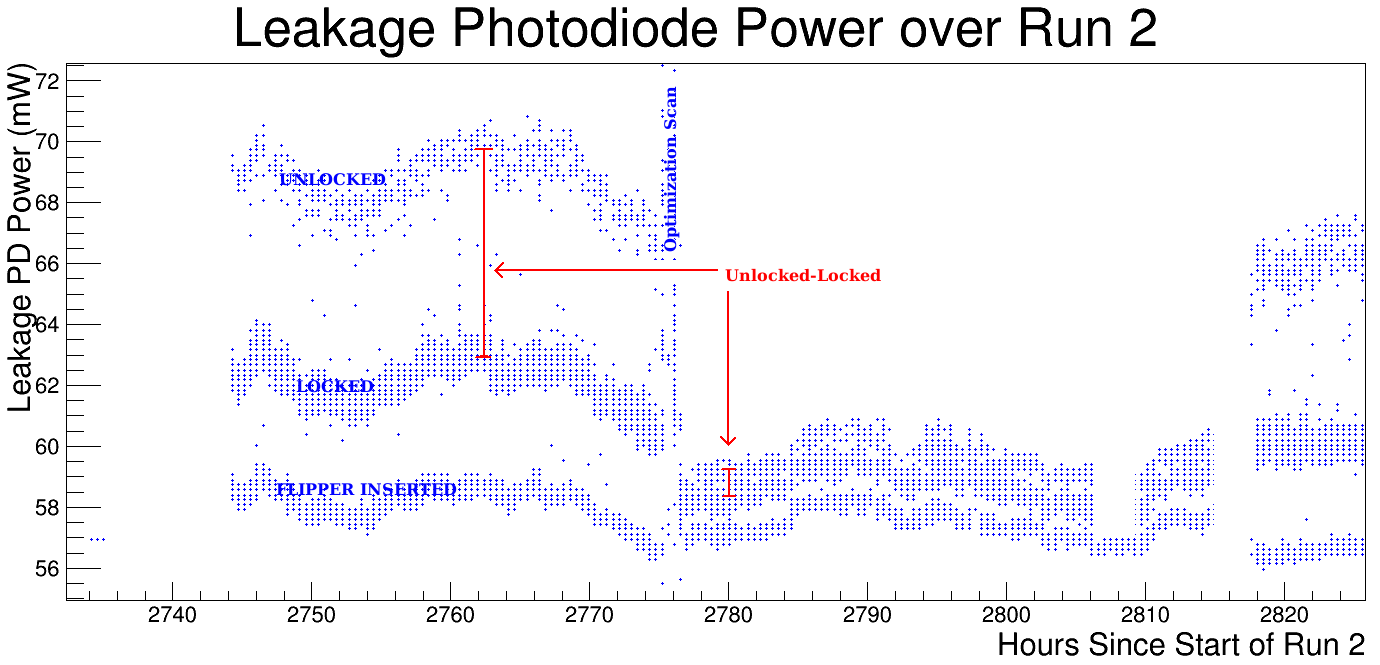}
\caption{\label{fig:LCPandRCP}Example of optimization scan of half-wave and quarter-wave plates to minimize polarization signal in leakage photodiode. This is a zoomed-in section of Figure \ref{fig:rrpd_vs_time} where the laser helicity state was flipped and an optimization scan was done to find the best position. This was during period ``d''. The polarization signal was about 14~mW before the optimization and less than 2~mW after the optimization.}
\end{center}
\end{figure}

The two remaining potential sources of error, depolarization inside the locked cavity and ambient light shifts on the laser table between locked and unlocked states, can be bounded indirectly by a combination of plausibility arguments and data. Direct measurement inside the optical cavity confirmed that 100.00\% DOCP was achievable for both left and right circular polarization states at a statistical uncertainty level of $\pm 0.02\%$ (see Figure \ref{fig:landr_docp_correl})\footnote{The upper limit of $\pm 0.04\%$  for unpolarized light actually limits the accuracy of the measured correlation in Figure \ref{fig:landr_docp_correl}. The slope remains the same but the maximum shifts down when there is an unpolarized component. The method used to measure circular polarization in the correlation plot also assumes 100\% polarized light and cannot differentiate between circularly polarized and unpolarized light. }. When the system is optimized, that is, the minimum signal in the leakage photodiode is achieved by adjusting the rotation angles of the waveplates, this minimum signal corresponds to 100.00\% DOCP (see Figure \ref{fig:landr_docp_correl}). One such waveplate optimization scan was executed during period ``d''. The results of this optimization can be seen in Figure \ref{fig:LCPandRCP}. As illustrated in the figure, the polarization signal was greatly reduced by the scan. The total difference in leakage photodiode signal between unlocked and blocked (flipper inserted) can be used to place a tighter bound on depolarization at the entrance to the cavity than even the direct measurement of $<0.04\%$. Using Equation \ref{eq:leakage_to_DOCP} with a 2~mW polarization signal from the unlocked minus blocked in Figure \ref{fig:LCPandRCP} gives an upper bound on this source of depolarization of 0.02\%. Neither ambient light nor depolarization are waveplate dependent, thus the fact that the optimization reduced both locked and unlocked leakage signals shows that both had a valid polarization signal that could not be mocked by ambient light or depolarization. At the end of the optimization, the locked signal was about 1~mW less than the unlocked and the blocked (flipper inserted) signal was about 1~mW below the blocked. If depolarization inside the cavity were a large effect, one might expect the order to be inverted and the unlocked signal to be smaller than the locked after optimization. It is the author's opinion that the ambient light seen by the leakage photodiode is the smallest when the shutter is inserted since the beam is dumped preventing both reflection and transmission of light. In that case, the blocked beam signal is actually an underestimate of the ambient light background and the leakage photodiode signal when the cavity is locked will be composed of any shifts from depolarized light (always a positive addition) and ambient light (also positive by this argument). Thus the measured difference in the leakage photodiode between locked and blocked directly after optimization can be used to bound depolarization giving 
\[
Unpolarized \leq0.5\times\frac{Polarization~Signal}{Maximum~Leakage~PD}=0.5\times\frac{6~mW}{6300~mW}=0.05\%.
\]
The polarization signal was calculated as follows. From Figure \ref{fig:LCPandRCP}, the difference between locked and blocked is about 1.5~mW. Assuming this is totally from unpolarized light, this difference was multiplied by a factor of 2 to allow for the fact that the linear polarizer only transmits half the total unpolarized power. Multiplying by another factor of 2 allows for only 50\% coupling of power into the cavity which implies that 50\% of the reflected beam comes from inside the cavity giving a total of $1.5\times2\times2=6$~mW. This translates into a 0.05\% shift in DOCP. Similarly, if one assumes the total difference between locked and blocked to be from ambient light shifts, the polarization signal is $1.5\times2=3$~mW, which translates into upper bound on error from ambient light shifts of 0.02\%. Table \ref{tab:pol_error} includes a 0.02\% error for possible ambient light shifts and 0.05\% error for depolarization inside the cavity based upon the arguments presented here. 
 
As already alluded to, the reflected beam returning from the locked cavity is composed of both single bounce light from the entrance cavity mirror as well as multiple bounce beam from inside the cavity. A 2011 publication by Peter Asenbaum and Mark Arndt in {\it Optics Letters} describes a setup where the sensitivity of the reflected beam to birefringence of the cavity mirrors is used to create an error signal for maintaining cavity lock \cite{Asenbaum}. It would be useful for future experiments to utilize a modified version of the setup described in this paper to directly measure the effect of cavity mirror birefringence. It is expected to be a tiny effect for \Qs but a direct measurement is more satisfying than a bound.  

To summarize the discussion of systematic error, the polarization signal is used to measure the shifts in DOCP with time. The polarization signal is the part of the reflected laser that is sensitive to changes in polarization of light at the optical cavity. The polarization signal is estimated by taking the difference of unlocked and locked signals in the leakage photodiode and multiplying by 2 to allow for imperfect coupling. The largest polarization signal is seen in period ``d'' where 10-14~mW was recorded. Using the correlation between leakage photodiode signal and DOCP measured directly inside the optical cavity provides a means of translating polarization signals into shifts in DOCP and yields 0.5\% percent change in DOCP per percent change in polarization signal. An error of of 0.02\% associated with the statistical uncertainty of the correlation of DOCP and leakage photodiode signal must also be included \ref{fig:landr_docp_correl}. The largest measured polarization signal of 14~mW implies a laser DOCP change of $\leq 0.11\%$. Since this method assumes that only the power of the reflected beam changes when the cavity locks (as opposed to the polarization of the beam or the ambient background light), upper bounds were established for real changes in intracavity circular polarization due to depolarization ($\leq 0.05\%$) and cavity mirror birefringence ($\leq 0.05\%$) as well as false shifts from ambient backgrounds ($\leq 0.02\%$). Finally, the method used for direct measurement of DOCP in the cavity region versus leakage photodiode power (see Figure \ref{fig:landr_docp_correl}) cannot distinguish between unpolarized and circularly polarized light. If the beam at the entrance to the optical cavity were to have a component that was unpolarized, this would shift the correlation plot downward by the percentage of unpolarized light. The unpolarized component was bounded by direct measurement in the cavity region to be $\leq 0.04\%$. The laser polarization then becomes 100.00\% with a conservatively estimated error of 0.14\%. Table \ref{tab:pol_error} lists the sources and values for each of the error contributions to the laser polarization.

\begin{table}
\begin{center}
\caption{\label{tab:pol_error}Table of errors.}
\begin{tabular}{|l|c|p{7cm}|}\hline
Category&Error&Explanation\\\hline
Polarization Tracking&0.11\%&Error derived from period ``d'' with a maximum polarization signal of 14~mW (see Table \ref{tab:pol_sig}).\\
Model&0.02\%&Statistical variation around linear correlation of DOCP vs. Leakage PD signal (see Figure \ref{fig:DOCP_measured_correl}).\\
Unpolarized Light (unlocked)&0.04\%&Unpolarized light at entrance to cavity. Conservative upper bound from direct measurement. This produces an offset in the correlation plot \ref{fig:DOCP_measured_correl}. Changes in polarization are relative to the maximum which can shift from 100.00\% to as low as 99.96\%.\\
Intracavity $\Delta$DOCP&0.04\%&Implied DOCP shift from change in linear polarization inside locked cavity. Upper bound of 0.9\% shift in DOLP between locked and unlocked. A 0.9\% shift in DOLP would shift DOCP from 99.90\% to 99.86\%\\
Intracavity Depolarization&0.05\%&Depolarization inside locked cavity.\\
Ambient Light Shifts&0.02\%&Difference in ambient light on the laser table between locked and unlocked that mimics polarization signal.\\\hline
{\bf Total}&{\bf 0.14\%}&Quadrature sum of errors.\\\hline

\end{tabular}

\end{center}
\end{table}

\subsubsection{Lessons from Periods of High Polarization Signal}
Before leaving the discussion of laser polarization it is useful to mention three short periods that are outliers with much larger polarization signals. Although, these should be removed from the dataset for calculation of electron beam polarization for \Q, their inclusion will have a negligible effect. 

\begin{figure}[ht]
\begin{center}
\includegraphics[width=4.0in]{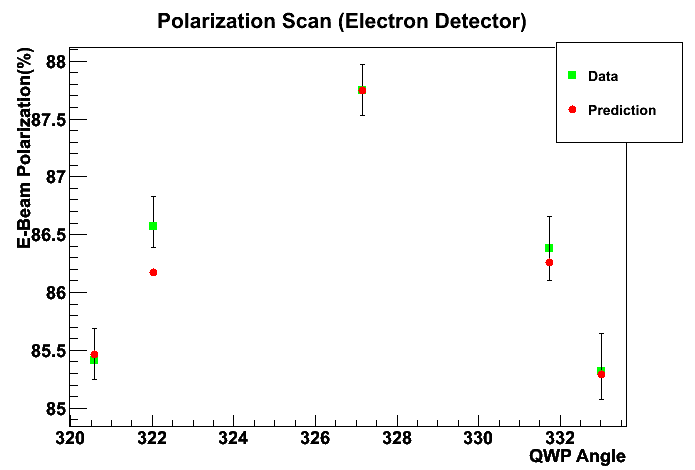}
\caption{\label{fig:las_pol_scan}Plot of measured electron beam polarizations reported by electron detector while laser circular polarization was scanned by rotating the QWP. The optics model predicted a shift from 100\% to 97.4\% in laser polarization, in good agreement with what was reported by the electron detector. Predicted values normalized to be equal with electron beam value at position of highest laser polarization. The error bars on the electron detector polarizations are only statistical. No model errors are shown for the predicted values.}
\end{center}
\end{figure}

The first period of non-optimal laser polarization occurred on Feb 25,2012 between 12:00 and 19:00. This period corresponds to hours 2169--2175 on Figure \ref{fig:rrpd_vs_time} but has leakage photodiode signals that are too large to be included in the range of this plot. The polarization signal was as large as 400 channels which implies that the laser polarization was near 97.3\%, polarization--a change easily observed by the electron detector. This period corresponds to a laser polarization scan where the laser polarization was deliberately changed during a period of stable electron beam conditions to verify that the electron detector measurements tracked the predicted laser polarization change from the model. The results of the electron-detector-measured electron beam polarizations while the laser polarization was being scanned can be seen in Figure \ref{fig:las_pol_scan}.

\begin{figure}[ht]
\begin{center}
\includegraphics[width=5.5in]{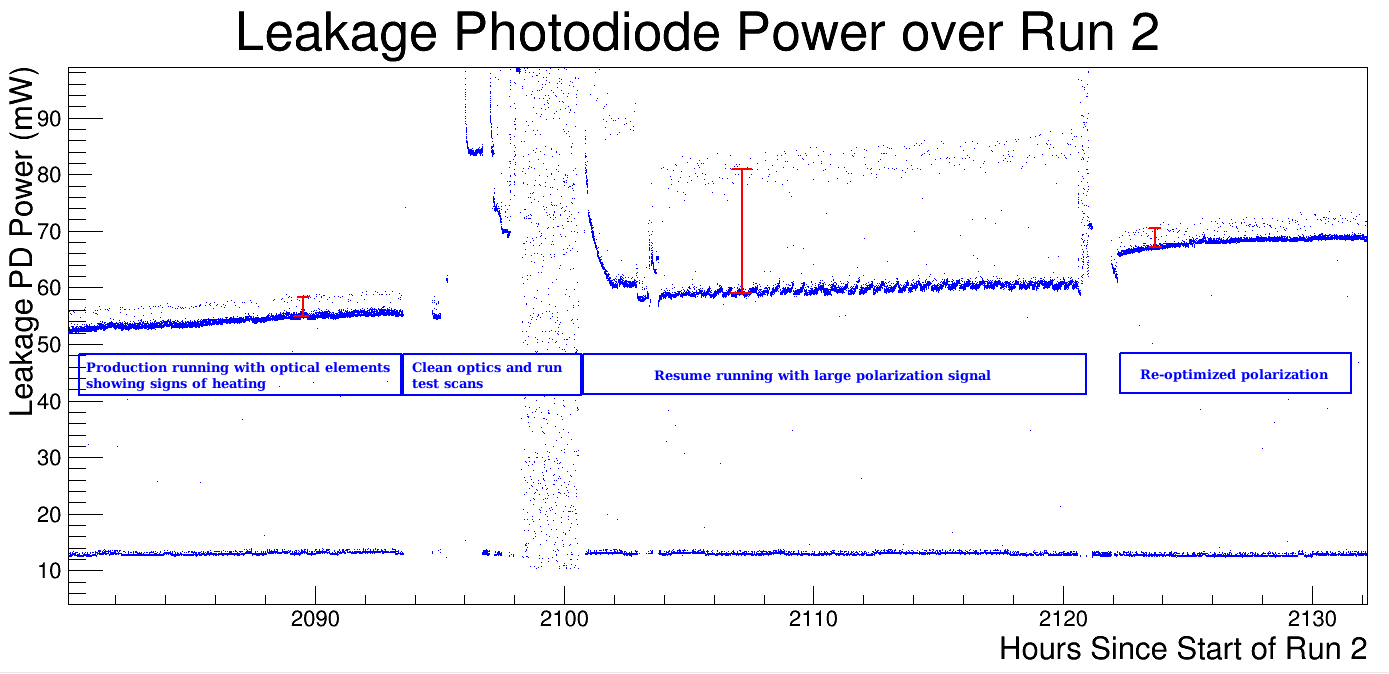}
\caption{\label{fig:optics_cleaning}Plot showing leakage photodiode response versus time. A large change in laser polarization is evidenced by the change in polarization signal ($2\times$ red bars) after the optics are cleaned and a turning mirror replaced to mitigate evidence of heating effects. Notice the rapid change in the leakage photodiode for a two hour period after the optical cleaning likely due to residual film on the optics from cleaning.}
\end{center}
\end{figure}

The second period occurred on Feb 22,2012 15:50 - Feb 23, 2012 11:50 which corresponds to  2100.8 -- 2120.8 hours in Figure \ref{fig:rrpd_vs_time} and  has a difference between locked and unlocked of about 25 mW. This corresponds to a polarization signal of 50 mW and a 0.4\% polarization change. The story of this period according to the electronic logbook is interesting and shows just how sensitive the leakage photodiode is to changing birefringence -- changes that would not be accounted for in the previous method of measuring the exit line transfer matrix. Heating effects had been observed in the optics, which were creating issues with optical cavity mode-matching. When the laser flipper was removed and when the cavity locked, changes in laser power on optical elements were creating shifts in the mode-matching quality of the beam. It was decided to try cleaning optical elements on the laser table. One lens was cleaned and the bottom turning mirror on the periscope replaced. This particular turning mirror was particularly subject to contamination from dust landing on its surface and burning due to its upward facing position. When the laser was once again turned on and locked the polarization signal had increased from $\sim 7$~mW to $45$~mW. Hours later a new scan of half-wave and quarter-wave plates was done to re-optimize the system and the polarization signal once again was reduced to $\sim 7$~mW. This is very interesting because simply cleaning optics and replacing a turning mirror changed the circular polarization inside the optical cavity by 0.3\% near the 100\% DOCP position where sensitivity to changes in linear polarization are smallest. Figure \ref{fig:optics_cleaning} shows the response of the leakage photodiode over this period.

The final period of large polarization signal is seen on March 27, 2012 between 11:00 and 15:00, which corresponds to hours 2912 to 2916 in Figure \ref{fig:rrpd_vs_time}. This period has a $\sim50$~mW polarization signal. The only evidence of an issue during this time is a laser chiller failure which affects the temperature stabilization of the laser head. Although the temperature inside the laser table enclosure is not recorded, perhaps it increased enough to create small birefringence changes in elements such as the vacuum window. 

In conclusion, the setup on the laser table for the Compton polarimeter worked very well for the creating highly polarized right and left circular states inside the optical cavity. Although laser polarization is a function of time, due to largely unmonitored  processes such as temperature changes, damage to optical components and laser position drift, the leakage photodiode provides a very sensitive measurement of those drifts and provides a bound on polarization drifts at the 0.14\% level. 

Past parity-violation experiments in Hall A at Jefferson lab utilizing Compton polarimetry have relied on exit line polarization measurements and a measured transfer matrix to translate these measurements into intracavity polarizations.  Due to the difficulties inherent in this method as outlined in Section \ref{Sctn:Methodology}, laser polarization has been a significant source systematic error on electron beam polarization measurements for these experiments. The HAPPEX-III experiment quoted a total uncertainty on electron beam polarization of $\pm$0.94\% with the laser polarization uncertainty contributing most of that at $\pm$0.80\% \cite{Friend}. The error budget of \Qs for electron beam polarization was $\pm 1\%$. Although uncertainty in laser polarization of $\pm$0.80\% would have been sufficient for the \Qs experiment, the reduction of this systematic error is a key victory for future parity-violation experiments at Jefferson Lab, such as MOLLER and SOLID, with more stringent requirements for electron beam polarimetry at the $\pm 0.4\%$ level\cite{MOLLER}\cite{SOLID}.


\chapter{Extracting the Weak Charge of the Proton}
\captionsetup{justification=justified,singlelinecheck=false}

\label{Ch:extracting_qwp}

\lhead{Chapter 7. \emph{Extracting \qwp}} 

The weak charge of the proton is determined by extrapolating the $Q^2$ dependence of the  parity-violating $ep$ scattering asymmetry to $Q^2=0$. This chapter serves to illustrate the process of taking the measured asymmetry and translating it into a value for the proton's weak charge. Extracting the weak charge requires a determination of the parity-violating asymmetry $A_{PV}$ from the measured asymmetry. Determining $A_{PV}$ requires many separate pieces of information, many of which are analysis topic in themselves, and which at the time of this writing are not yet mature. Furthermore, the final dataset may be expanded to include more data than were used in this analysis. To prevent unintentional bias toward the Standard Model prediction, the asymmetry dataset for the \Qs experiment is blinded, meaning that an undisclosed constant has been added to all asymmetry measurements. Until all pieces of information required to extract the weak charge from the measured asymmetry are finalized, the size of this blinding term will not be revealed. For these reasons what is presented in this chapter is intended for the purposes of illustration and will not represent the final result of \Q. This analysis will focus on the subset of Run 2 data for which a modulation correction exists.  

Determining the weak charge of the proton from the measured asymmetry can be divided into two analysis processes: the parity-violating asymmetry must be extracted from the measured asymmetry and some method must be employed for translating the parity-violating asymmetry measured at the kinematics of the \Qs experiment to $Q^2=0$. The parity-violating asymmetry evaluated at $Q^2=0$ is proportional to the weak charge of the proton (see Equation \ref{eq:reduced_asym}). The next two sections separately work through both of these processes to determine a final weak charge of the proton. For the measured asymmetry a straight average of the 16 PMT asymmetries will be used to eliminate bias introduced by unequal weighting. This average will be referred to as ``PMT Average''.

\section{\label{Sctn:qwpExtract}Extracting the Parity-Violating Asymmetry}
In Chapter \ref{Ch:instrumentation} the measured asymmetry was expressed as a sum of constituent parts, one of which was the parity-violating asymmetry of interest (see Equation \ref{eq:raw_asymmetry}). Using this equation to solve for the parity-violating asymmetry $A_{PV}$ gives
\begin{equation}
A_{PV}=\frac{R}{1-\sum_bf_b}\left[\frac{A_{meas}}{P}-\sum_bf_bA_b\right],
\label{eq:pv_asymmetry}
\end{equation} 
where the asymmetry $A_{meas}$ is the raw asymmetry corrected for helicity-correlated false asymmetries
\begin{equation}
A_{meas}=A_{raw}-A_{beam}-A_{bb}-A_T-A_{\epsilon}.
\label{eq:a_meas}
\end{equation} 

With a total of three backgrounds, there are 13 values that must be determined before the parity-violating asymmetry can be evaluated. Each of these is dealt with below.
\subsection{Bias Corrections}
$R$ is a multiplicative factor to account for changes from the measured to the reported $Q^2$ of the experiment and is the product of several factors which bias the asymmetry: $R=R_{RC}R_{Det}R_{Bin}R_{Q^2}$.

The experimental asymmetry will be reported at a given effective angle, beam energy and $Q^2$, whereas the detector accepted a range of angles and energies. Multiple scattering and {\it bremsstrahlung} processes, including higher order vacuum loops, alter the angle, energy and polarization of the electron both before and after scattering. The aim is to report the ``tree level'' parity-violating asymmetry rather than the measured asymmetry which includes radiative effects. $R_{RC}=\frac{A_{tree}}{A_{RC}}$ comes from the GEANT3 simulation for \Qs and is the ratio of the simulated asymmetry without radiative effects to the asymmetry with radiative effects. For Run 2, this factor is estimated to be $R_{RC}=1.0101\pm 0.0007$.

A detector light-weighting correction is required due to the non-uniformity of the distribution of $Q^2$ across the detector bars, coupled with position/angle dependent production and collection of \v{C}erenkov light. GEANT4 simulations comparing light-weighted $Q^2$ distributions to those without light weighting gave a correction factor to the measured asymmetry of $R_{Det}=0.9921\pm 0.0044$ . 

$R_{Bin}$ is a factor which corrects the measured asymmetry which is averaged over the accepted $Q^2$ distribution, $\left<A(Q^2)\right>$, to the reported asymmetry at a single $Q^2$ value, $A(\left< Q^2\right>)$. This factor was determined by simulation to be $R_{Bin}=0.98\pm 0.005$. Although this is a model-dependent correction, its sensitivity to the parametrization used to extract the $Q^2$-dependence of the asymmetry is negligible.

With an additional systematic error of 0.016 to account for the uncertainty in determining the central $Q^2$ upon which all the other factors depend, the total correction factor $R$ becomes
\[
 R=R_{RC}R_{Det}R_{Bin}R_{Q^2}=0.982\pm 0.019.
\]

\subsection{Background Corrections}
\begin{figure}[ht]
\begin{center}
\includegraphics[width=4in]{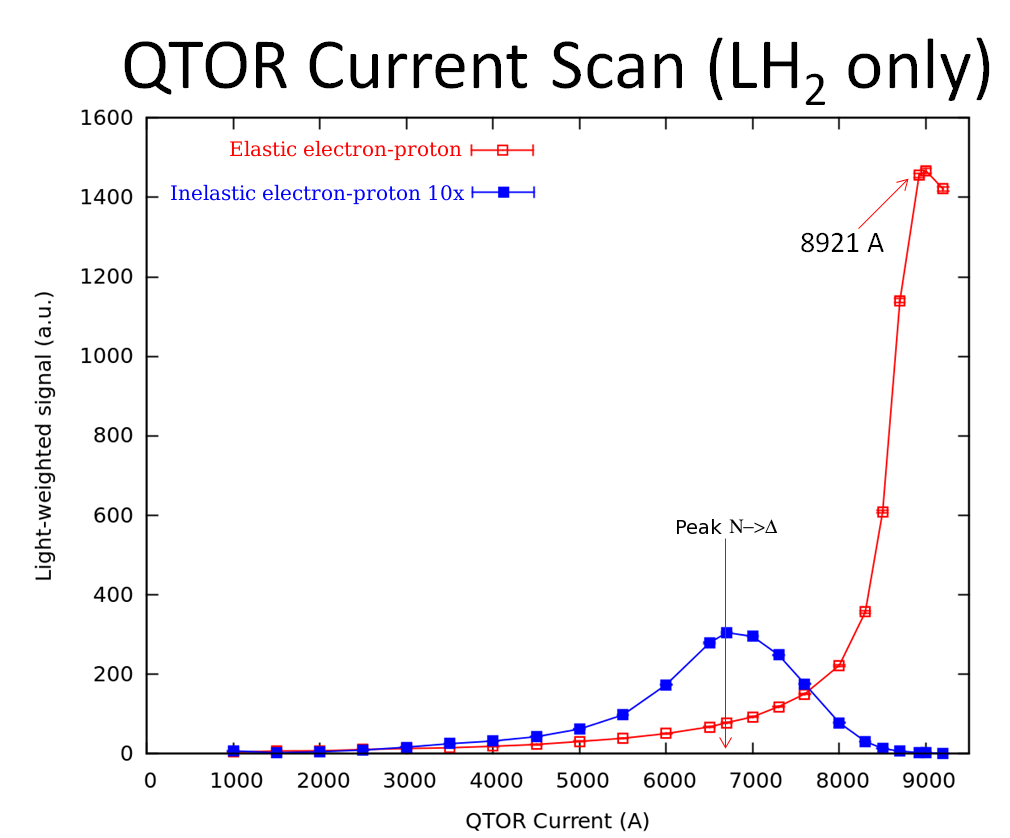}
\caption{\label{fig:sim_current_scan}Simulation of liquid hydrogen (LH2) elastic and inelastic scattering events versus \qtor spectrometer current clearly showing the peak $N\rightarrow\Delta$ production at 6700~A. The tail of the inelastic distribution under the elastic peak creates the fourth background asymmetry $A_3$.}
\end{center}
\end{figure}

There are three sources considered as background contributions to the measured asymmetry in \Q. For each of these, the best estimates are given at the time of writing for both the size of the background $A_b$ and the fraction to which it contributes to the observed detector yield $f_b$.  
\begin{itemize}
\item{The largest background comes from the aluminum entrance and exit windows on the target. A thick 4\% (0.04 radiation lengths) aluminum target made from the same material as the target windows and located near the position of the downstream aluminum window was used to measure the size of the asymmetry. The measured asymmetry was scaled individually for the upstream and downstream windows to account for differences in asymmetry between the thick target and the windows as follows:
\[
A_{u,d}=A_{4\%}\frac{A_{u,d}^{sim}}{A_{4\%}^{sim}},
\]
where $A_{u(d)}$ are the asymmetries for the upstream (downstream) windows and $A_{4\%}$ is the asymmetry measured on the thick aluminum target. The simulation corrects for changes due to $z$-location, energy loss in the hydrogen target, as well as radiative effects associated with a thicker target material. Included in the simulation are generators for elastics, quasi-elastics, inelastics, inelastic single particle states and giant dipole resonances. The total asymmetry is calculated as \
\[
A_{windows}=\frac{R_uA_u+R_dA_d}{R_u+R_d},
\]
where $R_{u(d)}$ is the detector rate coming from the upstream (downstream) windows. The total polarization-corrected asymmetry is $A_1=1506\pm 72$~ppb. 

The dilution fraction from the aluminum windows is measured to first order as $f_{win}=\frac{Y_{empty}}{Y_{full}}$, where $Y_{empty}$ is the current-normalized detector yield with an evacuated target and $Y_{full}$ the current-normalized detector yield with the target full of liquid hydrogen. This ratio needs to be corrected for energy losses associated with the presence of liquid hydrogen. Without the hydrogen in the target the image of the upstream aluminum window mostly falls off the bar. A light-weighting correction needs to be applied to translate the measured rates to light yields in the detector bars. The corrected dilution fraction for the aluminum windows is $f_1=0.02590\pm 0.00011$.}
\item{A second background is a soft neutral background arising from secondary electron scattering from collimator edges, the shielding wall and the \qtor spectrometer along the scattered electron transport channel. For this reason it is often referred to as the ``QTOR transport channel'' background. The dilution fraction for this background was measured using the Region 3 trigger scintillator detectors. The two scintillator detectors, used during tracking mode, were situated in opposite octants and could be rotated to cover the acceptance of any two opposite detector bars (see Figure \ref{fig:vdcs}). Since the scintillators were insensitive to neutral particles, they were used as a vetoes for the main detector, accepting only events that fired the main detector but not the trigger scintillator. The dilution fraction was measured to be $f_2=0.0013\pm 0.0014$. The asymmetry associated with this background is taken from simulation and the best estimate as of this writing is $A_2=-283\pm 57$~ppb }
\item{A third background comes from inelastic scattering events associated with $N\rightarrow \Delta(1232)$ production. Figure \ref{fig:sim_current_scan} illustrates the presence of the inelastic distribution tail under the elastic peak at a spectrometer current of 8921~A where production data was taken. The dilution fraction of the inelastic distribution was determined using simulation to be $f_3=0.0002\pm 0.0002$. The asymmetry associated with this inelastic process was measured at the inelastic peak by reducing the spectrometer current to 6700~A where the inelastic peak was focused on the detector bars. This asymmetry was corrected for an elastic radiative tail, the aluminum window, and neutral backgrounds. The corrected inelastic asymmetry under the production elastic peak at 8921~A was determined to be $A_3=-3020\pm970$~ppb.}
\end{itemize}
Therefore the total background asymmetry correction is 
\[
A_{Bkgd}=\sum_{b=1}^3f_bA_b=38.0\pm2.0~\text{ppb}
\] 
and the fraction of the measured yield coming from elastic electron-proton scattering is 
\[
f_{ep}=1-\sum_{b=1}^3f_b=0.9726\pm 0.0014.
\]
\subsection{Measured Asymmetry}
The measured asymmetry $A_{meas}$ given by Equation \ref{eq:a_meas} is composed of the raw asymmetry with helicity-correlated false asymmetries subtracted off. Of these false asymmetries, only two have signficant corrections to apply:  $A_{beam}$, a false asymmetry associated with with helicity-correlated beam motion, and $A_{bb}$, a false asymmetry associated with a neutral background seen in the main detector.

The error-weighted raw asymmetry $A_{raw}$ for the main detector PMT average over the Run 2 data set was $-159.4\pm 8.2$(stat)~ppb. Runlet-level (4-5 minute) averages of the main detector asymmetry distribution were first calculated. The error for each runlet-level asymmetry was calculated as the RMS width of the distribution divided by the square root of the number of quartets in the runlet. The total asymmetry was then calculated as the error-weighted average of runlet-level asymmetries \footnote{The runlet-level error weights were determined as the inverse variance of the modulation-corrected quartet MDallbars asymmetry distributions divided by the number of quartets in the runlet ($\frac{1}{\sigma^2_{quartet}/N_{runlet}}$)}. This total asymmetry measurement includes a total of 760542988 quartets\footnote{Recall that 1 quartet is approximately 4~ms of data.} and has a PMT-average quartet asymmetry RMS width of 225.4~ppm.  

Detector non-linearity, bounded by bench tests of the PMT + readout electronics, was estimated to be less than 1\%. An uncertainty of 2~ppb is assigned to account for this potential source of systematic error.   

\Qs operated as a ``blinded'' experiment, meaning that the reported asymmetries were offset from their true value by an additive constant. As previously mentioned, this blinding was done as a barrier to prevent unintentional biasing of the data toward the Standard Model expectation during the analysis process. The blinding term, selected from a uniform distribution over the range $\pm 60$~ppb, was added to each quartet-level asymmetry. Blinding was implemented in the software analyzer such that the raw data remained untouched but all processed files had the blinding term added to main detector calculated asymmetries. This blinding term remains a secret at the time of writing and will only be unveiled after all the data analysis is complete. As a result, the asymmetry quoted in this analysis is blinded.

The correction $A_{beam}$ is associated with helicity-correlated beam parameters position, angle, and energy. The details of this correction were given in Chapter \ref{Ch:BMod_correction}. For Run 2 this correction is $A_{beam}=-2.1\pm 0.8$~ppb.

$A_{bb}$ is a false asymmetry associated with helicity-correlated scattering in the beamline downstream of the target. This background was measured to be independent of spectrometer current implying that it is neutral particles and is believed to be primarily created in the tungsten collimator just downstream of the target. This background asymmetry was studied by blocking octants 1 and 5 of the primary collimator with 5~cm of tungsten. Asymmetries measured in main detector bars 1 and 5 during blocked octant running were seen to be highly correlated with the three background detectors (PMTonly, PMTltg and MD9) as well as the upstream luminosity monitors, giving evidence that they were all observing the same background. Since the luminosity monitors provide the most statistically accurate measure of the background asymmetry, they were chosen as the natural candidate for monitoring the background asymmetry during production running. The correlation between slug-level ($\sim$8~hr blocks of data) averages of the main detector and the upstream luminosity monitors over Run 2 provides a scale factor $\alpha$ for correcting the main detector asymmetry as follows:
\[
A_{bb}=\alpha A_{bb}^{uslumi},
\]
  
where $A_{bb}^{uslumi}$ is the beamline background asymmetry seen in the upstream luminosity monitor. This scale factor $\alpha=3.9\pm 1.5$~ppb/ppm, was determined from this correlation over Run 2. The asymmetry of the upstream luminosity monitors averaged over Run 2 is  $A_{bb}^{uslumi}=0.99$~ppm (sign-corrected for slow helicity reversals), giving a total correction $A_{bb}=3.9\pm2.2$~ppb.

\begin{figure}[ht]
\begin{center}
\includegraphics[width=4in]{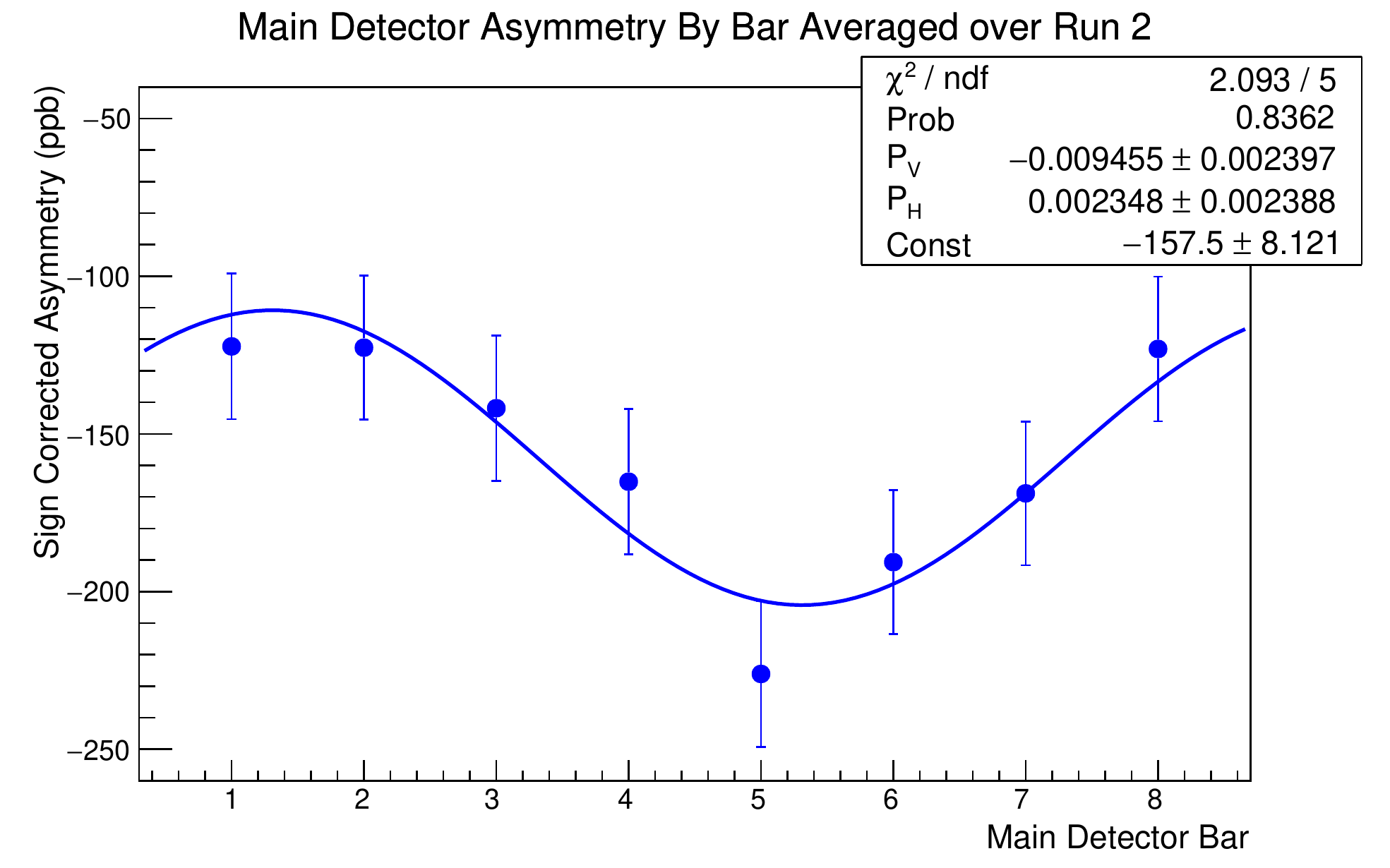}
\caption{\label{fig:transverse_fit}Modulation-corrected (and sign-corrected for slow helicity reversals) PMT average  asymmetry by bar showing manifest azimuthal dependence. The values shown are averages over Run 2 and the following function is fit to the data: $-4800\times\left[P_V\cos{\left(\frac{Bar\#-1}{4\pi}\right)-P_H\sin{\left(\frac{Bar\#-1}{4\pi}\right)}}\right]+Const$. $P_{V(H)}$ is the vertical (horizontal) transverse polarization.}
\end{center}
\end{figure}

$A_T$ is the false asymmetry from residual transverse polarization on the electron beam. The parity-conserving scattering asymmetry associated with a fully transverse-polarized beam was measured to be $-4.8\pm 0.6$~ppm \cite{Waidyawansa}. Although the \Qs experiment nominally ran with longitudinally polarized beam, evidence of a residual transverse polarization remained. Transverse electron polarization manifests itself as an azimuthal dependence of the scattering asymmetry in the main detector. Figure \ref{fig:transverse_fit} shows the azimuthal dependence of the average asymmetry over Run 2 with a sinusoidal fit to find the residual of the transverse polarization projected into its horizontal and vertical components. For Run 2, the residual vertical transverse polarization was $P_V=-0.0095$ and the residual horizontal transverse polarization was $P_H=0.0023$, which corresponds to a rotation of the ``polarization vector'' by $0.5^{\circ}$ vertically and $0.1^{\circ}$ horizontally from perfect longitudinal polarization.

To first order the transverse asymmetry cancels around the azimuth but broken symmetries in the main detector slightly degrade this cancellation, creating a small leakage term. Details on the method of extracting the leakage are provided in \cite{Waidyawansa}. The azimuthal dependence of the detectors was fit during separate running periods with maximal horizontal and vertical transverse polarizations. The fit similar to that shown in Figure \ref{fig:transverse_fit} was performed. The constant term of the fit measures the lack of cancellation. The constant leakage terms for the horizontal and vertical transverse polarizations were measured to be $C_H=11\pm 61$~ppb and $C_V=12\pm 55$~ppb respectively. Both are consistent with zero, but the upper bound is given by the large uncertainty. The transverse leakage is given as  
\[
A_T=\left|C_V\times P_V\right|+\left|C_H\times P_H\right|=0.14\pm0.55~\text{ppb.}
\]
Because the correction is negligible relative to the its error, it will be assumed to be zero and only the uncertainty of $\pm0.6$~ppb will be applied. 

\begin{figure}[ht]
\begin{center}
\includegraphics[width=3.5in]{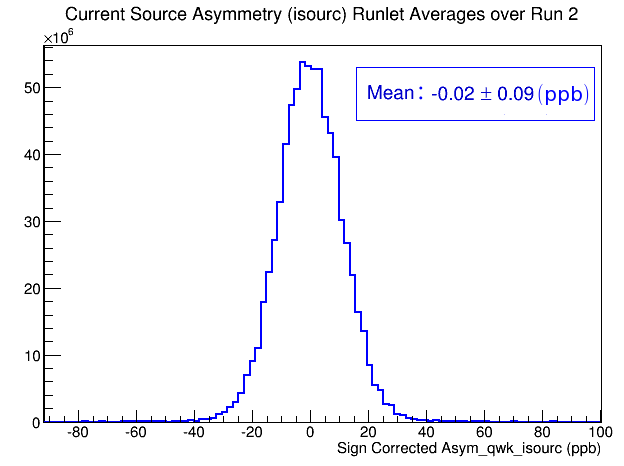}
\caption{\label{fig:isourc_asym}Distribution of asymmetry measurements of a battery fed channel in the \Qs data acquisition. Data has been corrected to reverse sign with slow helicity reversals so that it can be directly compared with the parity-violating physics asymmetry. The distribution is formed of runlet level averages (4-5 minute averages) weighted by the number of quartets in the runlet. The error on the mean is calculated by the average quartet RMS width divided by the square root of the total number of quartets (7.57e8 quartets)\protect\footnotemark.}
\end{center}
\end{figure}

 The false asymmetry associated with electronics pickup correlated with the helicity reversal was measured to be consistent with zero over Run 2 with high precision using a null channel. Figure \ref{fig:isourc_asym} shows the measured distribution of a current source (battery) plugged into one of the \Qs data acquisition channels. The signal in this channel was adjusted to approximately the same level as the detector channels so that the effect of helicity pickup for the battery channel would be similar to that of the detectors. The mean asymmetry of -0.02$\pm$0.09~ppb is consistent with zero and completely negligible.

Putting all this together gives 
\[
A_{meas}=A_{raw}-A_{beam}-A_{bb}=-161.2\pm 8.2\text{(stat)}\pm3.1\text{(sys)~ppb},
\]
where errors are included from statistics, the modulation correction, the beamline background correction, transverse leakage and detector non-linearity.
\footnotetext{The cut applied to the dataset including isourc was different from that applied to the main detector dataset in the analysis of the past chapter. This resulted in 0.5\% fewer quartets in for the isourc asymmetry. This difference does not alter the conclusions.} 

\subsection{Polarization}
\begin{figure}[!h]
\begin{center}
\includegraphics[width=5.9in]{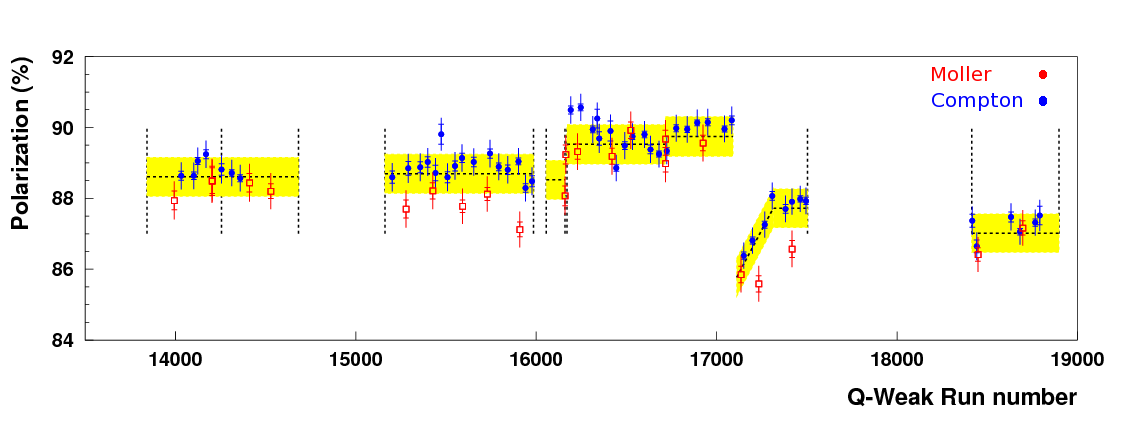}
\caption{\label{fig:run2_epol}The electron beam polarization values for the Compton electron detector and the M\o ller polarimeter over Run 2. The yellow bands represent the central value and full error (systematic and statistical) for the given period.}
\end{center}
\end{figure}

The electron polarization was determined using a combination of the Compton electron detector and M\o ller polarimeters. The electron detector delivered continuous polarizations at nominal running conditions and its values were chosen to give the time-dependent polarization. A small systematic difference of 0.7\% was observed between Compton and M\o ller values averaged over Run 2. Given the systematic error of 0.83\% reported for the M\o ller and 0.58\% for the Compton (see Chapter \ref{Ch:instrumentation}), this difference is not alarming. The reported polarization was weighted to include both. The total relative error for the polarization over Run 2 including both systematic and statistical errors is 0.62\%. The yellow bands in Figure \ref{fig:run2_epol} show the scaled polarization values and errors over Run 2. The polarization for the data set included in this analysis, averaged with the same weights as are applied to the main detector asymmetry, is $P=0.889\pm0.006$.

\subsection{The Parity-Violating Asymmetry}
Re-expressing Equation \ref{eq:pv_asymmetry} in simpler terms, the parity-violating asymmetry is then given as

\begin{equation}
A_{PV}=\frac{R}{f_{ep}}\left(\frac{A_{meas}}{P}-A_{Bkgd}\right)= -221.5\pm 9.3\text{(stat)}\pm 6.1\text{(syst)~ppb}.
\label{eq:apv_exp}
\end{equation}
The total error added in quadrature is 11.1~ppb. Values and errors for the individual terms in the expression are summarized in Table \ref{tab:asym_err}.

\begin{table}[h]
\begin{center}
\caption{\label{tab:asym_err}Summary of values and errors for various terms in the calculation of the parity-violating asymmetry $A_{PV}$. }
\begin{tabular}{|l|c|p{5cm}|}\hline
Term&Value&Comment\\\hline\hline
R&0.982$\pm$0.019&$R=R_{RC}R_{Det}R_{Bin}R_{Q^2}$\\\hline
P&0.889$\pm$0.006&Electron beam polarization\\\hline
$A_{meas}$&$-161.2\pm$8.2(stat)$\pm$3.1(sys)~ppb&Error includes statistics, non-linearity, transverse leakage, and modulation correction\\\hline
$f_{ep}$&0.9726$\pm$0.0014&$f_{ep}=1-\sum_{b=1}^3f_b$\\\hline
$A_{Bkgd}$&38.0$\pm$2.0~ppb&$A_{Bkgd}=\sum_{b=1}^3f_bA_b$\\\hline
\end{tabular}
\end{center}
\end{table}

\begin{figure}
\centering
\includegraphics[width=5.8in]{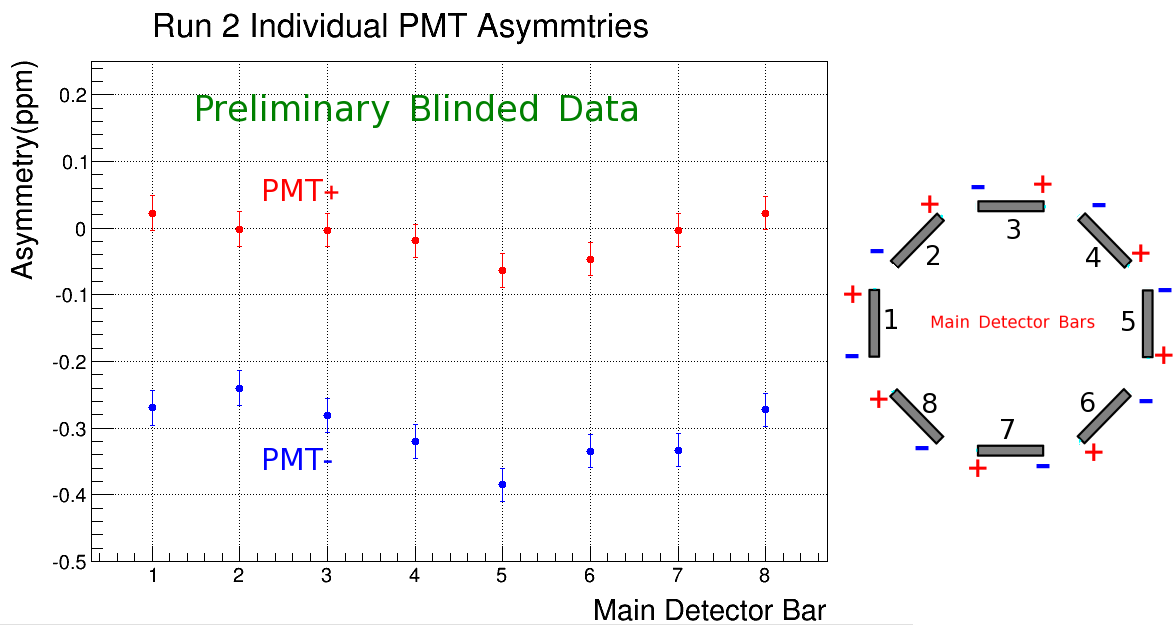}
\caption{\label{fig:pmt_dd}Main detector asymmetries averaged over Run 2 showing the systematic difference between positive and negative PMT's. This ``PMT double-difference'' effect is uniform in all octants. The sinusoidal variation around the azimuth arises from residual transverse polarization on the electron beam and is a distinct from the double-difference believed to arise from transverse polarization of the scattered electrons produced by g-2 precession in the spectrometer. }
\end{figure}

As detailed in Chapter 3, each of the eight main detector bars are read out by two PMT's, one at each end. The PMTavg Asymmetry (straight average of the 16 PMT asymmetries) gives the physics asymmetry for \Q. An interesting/troubling phenomenon has been found in the \Qs production data. A well-determined discrepancy exists between the asymmetries measured by the two PMT's on each detector bar, although both are collecting light from the same quartz bar, albeit from opposite ends. Figure \ref{fig:pmt_dd} shows this discrepancy between the asymmetry measured on the opposite ends of each bar. This difference in asymmetry, measured to be uniform in all octants of the main detector and stable over time, has been termed the ``PMT double difference''. This effect is believed to originate from transverse polarization of the scattered electrons arising from g-2 precession as the electrons travel through the \qtor spectrometer which rotates the electron spin $\sim 35^{\circ}$ relative to its trajectory. These electrons then scatter and shower in the lead pre-radiators on the front of each main detector bar. Non-uniformity of light collection as a function of angle and position along the detector bar, coupled with a small polarization-dependent shift in the event distribution, is believed to cause this discrepancy. Two models currently being simulated show promise of contributing to this effect. GEANT4 simulations of spin-dependent Mott scattering of transversely polarized electrons which have radiated to low energies in the pre-radiators, appear to show a small polarization-dependent distribution shift. Another model being simulated is the higher energy parity-conserving transverse scattering asymmetry from the analyzing power in lead, with an effect similar to that measured by \Qs for scattering from the proton \cite{Waidyawansa}. 

Analysis of this effect is ongoing. In either of these models, the double difference effect will cancel in the average of all PMT's to first order. Broken symmetries in the main detector ensure this cancellation will not be perfect. However, light sensitivity distributions measured using the tracking system will allow a correction for this leakage of the double difference into the PMT average asymmetry. Conservatively estimating the leakage due to broken detector symmetry at $\pm$10\% and a correction factor calculated from tracking data known to the $\pm$10\% level would give a systematic uncertainty of 1\% to the correction of the 300~ppb double-difference. Although at this point it is premature to assign an uncertainty for this effect and it will not be considered in this analysis, the reader should keep in mind that this effect will contribute to the final uncertainty of the unblinded result.

\section{\label{Sctn:qwpExtract}From Parity-Violating Asymmetry to Weak Charge}
To obtain the weak charge of the proton, \qwp, from the measured parity-violating asymmetry, recall the following equation introduced in Chapter 2.
\begin{equation}
A_{PV}/A_0=\left[\frac{{\epsilon}G^{\gamma,p}_EG^{Z,p}_E+{\tau}G^{\gamma,p}_MG^{Z,p}_M-\left(1-4\sin^2{\theta}_W\right){\epsilon}^{\prime}G^{\gamma,p}_MG^{Z,p}_A}{{\epsilon}(G^{\gamma,p}_E)^2+{\tau}(G^{\gamma,p}_M)^2} \right],
\label{eq:qw_asymmetry2}
\end{equation} 
with $A_0=\frac{-G_FQ^2}{4\pi \alpha\sqrt{2}}$.
The electromagnetic Sachs form factors of the proton, $G^{\gamma,p}_E$ and $G^{\gamma,p}_M$, are well-determined from parity-conserving electron scattering. Values for the weak neutral current (WNC) vector form factors can be derived from measured electromagnetic form factors of the proton and neutron under the assumption of charge symmetry which asserts the invariance under the following interchanges: $p\leftrightarrow n$, $u\leftrightarrow d$ and $s\leftrightarrow s$. Here $p~(n)$ are proton (neutron), $u~(d)$ are up (down) quarks and $s$ is the strange quark\footnote{Contributions from heavier quarks are neglected. In fact, for the purposes of this analysis it is appropriate to drop the strange contribution as well}. Thus charge symmetry implies that the form factors of the up (down) quark for the proton are identical to the form factors for the down (up) quark of the neutron and that strangeness contributes equally to both. The effect of charge symmetry breaking is expected to alter the electromagnetic form factors less $1\%$ at the kinematics of \Q\cite{Miller2014}. Expressing the form factor of the proton and neutron in terms of the quark distributions $G_{E,M}^u$, $G_{E,M}^d$ and $G_{E,M}^s$ weighted by their electric charges gives 
\begin{equation}
G_{E,M}^{\gamma p}=\frac{2}{3}G_{E,M}^{u}-\frac{1}{3}(G_{E,M}^{d}+G_{E,M}^{s}).
\label{eq:gemp}
\end{equation}
Using the same distributions under the assumption of charge symmetry yield the following for the neutron:
\begin{equation}
G_{E,M}^{\gamma n}=\frac{2}{3}G_{E,M}^{d}-\frac{1}{3}(G_{E,M}^{u}+G_{E,M}^{s})
\label{eq:gemn}
\end{equation}
Rearranging these equations allows the quark form factors to be expressed in terms of the measured proton and neutron form factors plus an addition from strangenes:
\begin{align}
G_{E,M}^u=2G_{E,M}^{\gamma p}+G_{E,M}^{\gamma n}+G_{E,M}^{s}\\
G_{E,M}^d=G_{E,M}^{\gamma p}+2G_{E,M}^{\gamma n}+G_{E,M}^{s}
\label{eq:qFFs}
\end{align}
The WNC vector form factor $G_{E,M}^{Z p}$ can also be expressed in terms of the quark form factors weighted with the appropriate weak charges
\begin{equation}
\begin{array}{lcc}
G_{E,M}^{Zp}&=&\left(1-\frac{8}{3}\sin^2\theta_W\right)G_{E,M}^u+\left(-1+\frac{4}{3}\sin^2\theta_W\right)(G_{E,M}^d-G_{E,M}^s)\\
~&=&\left(1-4\sin^2\theta_W\right)G_{E,M}^p+(-1)(G_{E,M}^n-G_{E,M}^s)\\
~&=&Q_W^pG_{E,M}^p+Q_W^nG_{E,M}^n-G_{E,M}^s\end{array},
\label{eq:gemzp}
\end{equation}
where the proton and neutron measured form factors have been substituted using Equations \ref{eq:qFFs}. In the final line the form factor weights have been labeled to clearly show that they are the weak charges of the proton and neutron. At $Q^2=0$, the WNC electric form factor $G_{E}^{Zp}$ gives the weak charge of the proton. 

In the following analysis, strange quark content will be assumed to be zero. Strange content of the proton form factors measured by the HAPPEX-III experiment was found to be consistent with zero\cite{HAPPEX3}. Recently-published lattice QCD results provide even tighter constraints on form factor strange quark content, with evidence of non-zero strangeness but at levels that are negligible for this analysis\cite{Green2015}. Estimates based on figures in \cite{Green2015} suggest the following contributions to the form factors:
\begin{align}
G_E^s\left(Q^2=0.025(Gev/c)^2\right)=0.0006\pm0.0002\\
G_M^s\left(Q^2=0.025(Gev/c)^2\right)=0.017\pm0.004.\\
\end{align}
These enter the nucleon form factors (Equations \ref{eq:gemp} and \ref{eq:gemn}) weighted by the strange quark charge (-1/3). With estimates of the proton electric and magnetic form factors at the \Qs kinematics of the following size,
\begin{align*}
G_E^p\left(Q^2=0.025(Gev/c)^2\right)\approx 0.9\\
G_M^p\left(Q^2=0.025(Gev/c)^2\right)\approx 2.6,
\end{align*}
strange quark contributions to the form factors are expected to be less than 0.3\%. A further suppression by a factor of 3 comes from the fact that the form factors only contribute to the hadronic correction terms, which are about 33\% of $A_{PV}$ at the kinematics of \Q.

For the analysis ahead, the formalism outlined in \cite{Lui2007} is followed. The parity-violating asymmetry is re-expressed in terms of proton and neutron form factors giving
\begin{equation}
\begin{array}{lcl}
A_{PV}/A_0&=&(1-4\sin^2\theta_W)(1+R_V^p)\\~&~&-\left[\left(\epsilon G_E^{\gamma p}G_E^{\gamma n}+\tau G_M^{\gamma p}G_M^{\gamma n}\right)(1+R_V^n)+\epsilon^{\prime}(1-4\sin^2\theta_W)G_M^pG_A^e\right]/\sigma_{red},\end{array}
\label{eq:apv_pnFFs}
\end{equation} 
where 
\[
\tau=Q^2/(4M_p^2),~~~~\epsilon=(1+2(1+\tau)\tan^2\theta_e)^{-1},~~~~\sigma_{red}=\epsilon (G_E^{\gamma p})^2+\tau (G_M^{\gamma p})^2,
\]
and $\theta_e$ is the electron scattering angle. The factors $R_{V}^p$ and $R_V^n$ are radiative corrections that are both $Q^2$ and process-dependent and are required to correct for higher order lepton-quark scattering diagrams. The values for $R_V^{p,n}$ are given in Table \ref{tab:Apv_parameters}. 

Fits to world electron scattering data have yielded precise functional parametrizations of the form factors \cite{Galster}\cite{Walcher}\cite{Kelly}\cite{ArringtonSick}\cite{Venkat}. Of these schemes, the form factor parametrization by Arrington and Sick was most closely designed for low $Q^2$ parity-violation experiments and is the parametrization utilized in this analysis for the extraction of \qwp. The Arrington-Sick parametrization utilizes a continued-fraction expansion\cite{ArringtonSick}:
\begin{equation}
G_{CF}(Q)=\frac{1}{1+\frac{b_1Q^2}{1+\frac{b_2Q^2}{1+\cdots}}},
\label{eq:Gcf}
\end{equation} 
suitable for lower momentum transfers $Q<0.8$~GeV/c. The values of $b_i$ that parametrize the proton and neutron form factors appropriate for \Qs kinematics are given in Table \ref{tab:AS_parameters}. Although parameters have been provided in \cite{ArringtonSick} that correct the form factors for two-photon exchange, these are said to be valid in the range $Q=0.3-1.0$~GeV/c which clearly does not include the momentum transfer of \Qs ($Q\approx 0.16$~GeV/c). Instead the parameter values corrected only for Coulomb distortion are used. 

The form factor for $G_E^n$ uses a slightly modified continued fraction expansion as follows:
\[
G_E^n = 0.484\times Q^2\times G_{CF},
\]
where the factor of 0.484 ensures the slope at $Q^2=0$ matches the measured neutron mean-square charge radius\cite{ArringtonSick}\cite{Koester}.
\begin{table}
\centering
\caption{\label{tab:AS_parameters}Fit parameters for Arrington-Sick continued fraction form factor parametrizations\cite{ArringtonSick}. Proton data include corrections for Coulomb distortion. Parameters assume $Q^2$ is in units of (Gev/c)$^2$.}
\begin{tabular}{lccccc}\hline
~$b_1$&$b_2$&$b_3$&$b_4$&$b_5$\\\hline
$G_E^p$&3.440&$-$0.178&$-$1.212&1.176&$-$0.284\\
$G_M^p/\mu_p$&3.173&$-$0.314&$-$1.165&5.619&$-$1.087\\
$G_E^n$&0.977&$-$20.82&22.02&$-$&$-$\\
$G_M^n/\mu_n$&3.297&$-$0.258&0.001&$-$&$-$\\\hline
\end{tabular}
\end{table}

The form factors in the denominator are treated differently since the denominator includes the full electron scattering cross section. Instead of correcting for two-photon exchange, the form factors in the denominator of Equation \ref{eq:apv_pnFFs} are empirically fit to include these effects. The denominator is parametrized using form factors $F_E^p$ and $F_M^p$ yielding the following expression:
\[
\sigma_{red}=\epsilon(F_E^p)^2+\tau (F_M^p)^2.
\] 
The parameter values for $F_E^p$ and $F_M^p$ are given in Table \ref{tab:AS_denom_parameters}.
\begin{table}
\centering
\caption{\label{tab:AS_denom_parameters}Fit parameters for Arrington-Sick continued fraction form factor parametrizations for the denominator $\sigma_{red}$\cite{ArringtonSick}. Parameters assume $Q^2$ is in units of (Gev/c)$^2$.}
\begin{tabular}{lccccc}\hline
~$b_1$&$b_2$&$b_3$&$b_4$&$b_5$\\\hline
$F_E^p$&3.366&$-$0.189&$-$1.263&1.351&$-$0.301\\
$F_M^p/\mu_p$&3.205&$-$0.318&$-$1.228&5.619&$-$1.116\\\hline
\end{tabular}
\end{table}

The weak neutral current (WNC) axial form factor $G_A^Z$, with both isoscalar and isovector components, is parametrized using a modified dipole form factor with axial dipole mass $\Lambda_A$ as follows\cite{Lui2007}:
\begin{equation}
G_A^e(Q^2)=G_D(q^2)\left[\frac{g_A}{g_V}(1+R_A^{T=1})+\frac{3F-D}{2}R_A^{T=0}\right],
\label{eq:G_A}
\end{equation}
 where the strange quark contribution to spin is not considered and where the dipole form factor is given by
\[
G_D(Q^2)=\frac{1}{(1+q^2/\Lambda_A^2)^2}.
\]
The ratio $-g_A/g_V$ is the isovector axial form factor of the proton at $Q^2=0$, $F$ and $D$ are SU(3) reduced matrix elements\cite{Filippone} and $R_A^{T=0}$, $R_A^{T=1}$ and $R_A^{(0)}$ are radiative corrections to the isovector and isoscalar axial vector amplitudes. Values for $\Lambda^2$, $-g_A/g_V$, $(3F-D)$ and $R_A^{T=0,1}$ can be found in Table \ref{tab:Apv_parameters}. 

The electron beam energy during Run 2, acceptance-averaged and corrected for ionization energy loss in the target, was $1.153\pm0.003$~GeV. The four-momentum transfer squared averaged over the acceptance was $\langle Q^2\rangle=0.02455\pm0.00032$ which gives an average scattering angle of 7.84$^{\circ}$ using Equation \ref{eq:Qsquared}.

\begin{table}
\caption{\label{tab:Apv_parameters}Values (and errors in parentheses where appropriate) for parameters in Equations \ref{eq:apv_pnFFs} and \ref{eq:G_A}. Values taken from Tables I and II in \cite{Lui2007} and \cite{PDG2014}.}
\begin{center}
\begin{tabular}{l|l}\hline
Parameter&Value(Error)\\\hline\hline
$\alpha$&$7.29735257\times10^{-3}$\\
$M_p$&$0.938272$~GeV\\
$G_F$&$1.1663787\times10^{-5}$\\
$\sin^2\theta_W(M_Z)$&$0.23126(5)$\\
$\Lambda_A^2$&$1.00(0.04)$(GeV/c)$^2$\\
$g_A/g_V$&$-1.2695$\\
$3F-D$&$0.58(0.12$)\\
$R_V^p$&$-0.0520$\\
$R_V^n$&$-0.0123$\\
$R_V^{(0)}$&$-0.0123$\\
$R_A^{T=0}$&$-0.239(0.20)$\\
$R_A^{T=1}$&$-0.258(0.34)$\\\hline
\end{tabular}
\end{center}
\end{table}

Substituting in the required parameters to Equations \ref{eq:Gcf} and \ref{eq:G_A} gives the following values for the form factors at the kinematics of \Q:
\begin{align*}
G_E^p = +0.922\pm 0.004\\
G_M^p = +2.587\pm 0.015\\
G_E^n = +0.0115\pm 0.0006\\
G_M^n = -1.766\pm 0.014\\
G_A^e = -0.9649\pm 0.0019\\
\end{align*}

Rearranging Equation \ref{eq:apv_pnFFs} and allows one to solve for the measured weak charge of the proton:
\begin{equation}
Q_W^p=(1-4\sin^2\theta_W)(1+R_V^p),=\left(A_{PV}/A_0+H(Q^2,\theta_e)\right)\\
\end{equation}
where the term $H(Q^2,\theta_e)$ containing hadronic corrections is given by
\begin{equation}
H(Q^2,\theta_e)=\left[\left(\epsilon G_E^{\gamma p}G_E^{\gamma n}+\tau G_M^{\gamma p}G_M^{\gamma n}\right)(1+R_V^n)+\epsilon^{\prime}(1-4\sin^2\theta_W)G_M^pG_A^e\right]/\sigma_{red}.
\end{equation} 
Substituting the experimentally measured asymmetry $A_{PV}=-221.5$~ppb (Equation \ref{eq:apv_exp}) gives
\[
Q_W^p=0.0741.
\]
Applying a correction to $Q_W^p$ of $0.00560\pm0.00036$ for the $\Box_{\gamma Z}$ diagram as outlined in Section \ref{Sctn:EWcorr} gives the (blinded) weak charge of the proton as

\begin{equation}
Q_W^p(\text{Blinded})=0.0685\pm0.0042\text{(stat)}\pm0.0027\text{(syst)}\pm0.0022\text{(theory)}.
\end{equation}
The quadrature sum of the statistical and systematic errors gives a total uncertainty of $\pm 0.0055$ which is $\pm 7.8\%$ of the Standard Model value of $Q_W^p(SM)=0.0708\pm0.0003$ \cite{PDG2014}. The range in $Q_W^p$ associated with asymmetry blinding box ($\pm60$~ppb) extends from $Q_W^p=0.045$ to $Q_W^p=0.096$. The systematic errors used in the calculation of the total systematic error are summarized in Table \ref{tab:Qw_syst_err}. 

\begin{table}[h]
\centering
\caption{\label{tab:Qw_syst_err}Summary of systematic errors contributing to the determination of the proton weak charge $Q_W^p$.}
\begin{tabular}{l|l|l|l}\hline
Term&Value&Systematic Error&Reference\\\hline\hline
$A_{PV}$&$-221.5$~ppb&6.1~ppb&Equation \ref{eq:apv_exp}\\
$Q^2$&0.02455~(GeV/c)$^2$&0.00036~(GeV/c)$^2$&Section \ref{Sctn:qwpExtract}\\
$E$&1.153~GeV&0.003~GeV&Section \ref{Sctn:qwpExtract}\\
$G_E^p$& +0.922 & 0.004&\cite{ArringtonSick}\\
$G_M^p$& +2.587 & 0.015&\cite{ArringtonSick}\\
$G_E^n$& +0.0115& 0.0006&\cite{ArringtonSick}\\
$G_M^n$& $-1.766$ & 0.014&\cite{ArringtonSick}\\
$\sigma_{red}$&0.891149&0.0062&\cite{ArringtonSick}\\
$R_A^{T=0}$&$-0.239$&0.20&\cite{Lui2007}\\
$R_A^{T=1}$&$-0.258$&0.34&\cite{Lui2007}\\
$\Lambda_A^2$&1.00~(GeV/c)$^2$&0.04~(GeV/c)$^2$&\cite{Lui2007}\\
$3F-D$&0.58&0.12&\cite{Lui2007}\\
$\sin^2\theta_W(M_Z)$&0.23126&0.00005&\cite{PDG2014}\\\hline
\end{tabular}
\end{table}
 

\chapter{Concluding Discussion} 
\captionsetup{justification=raggedright,singlelinecheck=false}

\label{Conclusion}  

\lhead{Concluding Discussion. \emph{Conclusion}} 

A parity-violating scattering asymmetry of longitudinally polarized electrons from unpolarized protons was determined using approximately 2/3 of the \Qs dataset. The asymmetry, measured to be  $A_{PV}=-221.5\pm9.3\text{(stat)}\pm3.1\text{(sys)~ppb}$, remains blinded due to ongoing analysis. This represents a 5\% measurement of the parity-violating asymmetry. Assuming that the final result includes a dataset 1.5 times larger, one could expect the full result to have a statistical error on the asymmetry of 7.6~ppb. A similar increase in systematic error in the added dataset gives a total error of $\delta A/A\approx$4.1\%. Although this falls short of the original proposal, it represents the most precise measurement of the parity-violating electron scattering asymmetry ($A_{PV}$) ever made. When unblinded, this result will test the Standard Model prediction of the weak charge of the proton. This measured parity-violating asymmetry using 2/3 of the full dataset translates into a $\pm$7.8\% measure of the proton's weak charge. The blinding term on the asymmetry is somewhere in the $\pm$60~ppb range, translating into a range for the proton weak charge from 0.045 to 0.096. This range allows for as large as a 4.7$\sigma$ deviation from the Standard Model.

Insights gained during the \Qs experiment will impact the future parity program going forward. Reduction of key systematic errors in Compton polarimetry achieved during \Q, make the stringent polarimetry requirements of future experiments such as SOLID and MOLLER appear to be achievable. Efforts underway to understand the processes giving rise to the observed ``PMT double-difference'' will influence the design of future parity experiments. 

Experimental physics at the precision frontier is expected to be challenging and, in this regard, it never seems to disappoint; however, the potential for discovery makes it worth the effort.


\addtocontents{toc}{\vspace{2em}} 

\appendix 



\chapter{Maximum Likelihood and the $\chi^2$ Statistic} 
\captionsetup{justification=justified,singlelinecheck=false}

\label{AppendixA} 

\lhead{Appendix A. \emph{Chi-Squared Statistic}} 

The parent distribution from which a given observation or measurement has been extracted determines the probability of that observation. Often in physics, a specific distribution is first assumed and then the mean and error of a given set of observations is calculated from the known characteristics of that distribution. For example, starting from a Gaussian parent distribution with mean $\mu$ and variance $\sigma^2$, the probability distribution is given as 
\begin{equation}
P(x)=\frac{1}{\sigma\sqrt{2\pi}}\int e^{-\frac{(x-\mu)^2}{2\sigma^2}}dx.
\label{eq:gaussian_pdf}
\end{equation}
Therefore, the differential probability which gives the probability of observing a single value $x_i$ in an interval $dx_i$ is given as 
\begin{equation}
dP_i=\frac{1}{\sigma\sqrt{2\pi}}e^{-\frac{(x-\mu)^2}{2\sigma^2}}dx_i.
\end{equation}

Suppose instead that we have a functional form $y=f(x)$ for a given data set and the data are expected to be normally distributed about this function with variance $\sigma^2$. The assumption of a normal distribution allows us to calculate the probability of a given observation $y_i$ as
\begin{equation}
dP_i=\frac{1}{\sigma\sqrt{2\pi}}e^{-\frac{(y_i-y)^2}{2\sigma^2}}dy_i.
\label{eq:differential_probability}
\end{equation}
Extending this to N independent observations gives a total probability of 
\begin{equation}
dP=\prod_{i=1}^NdP_i=\prod_{i=1}^N\frac{1}{\sigma\sqrt{2\pi}}e^{-\frac{(y_i-y)^2}{2\sigma^2}}dy_i.
\label{eq:probability_product}
\end{equation}
If the functional form $y$ is parameterized by n coefficients $\alpha_n$, the best values for these coefficients can be found by maximizing the probability P. This is the essence of the \emph{method of maximum likelihood} by which best estimates of the parameters of the parent distribution are considered to be those which maximize the probability of the observed values. Maximizing \ref{eq:probability_product} is the same thing as minimizing the sum of the exponentials:
\[
\frac{dP}{d\alpha_n}=0\longrightarrow\frac{d}{d\alpha_n}\left(\sum_{i=1}^N(\Delta y_i/\sigma)^2\right)=0.
\]

Therefore, maximizing the probability for a Gaussian distribution leads to minimizing the $\chi^2$ statistic which is given as $\sum_i(\Delta y_i/\sigma)^2$.

A detailed treatment of the $\chi^2$ statistic as a tool for finding the best fit and a test of the goodness of fit can be found in the well known text\emph{ Data Reduction and Error Analysis for the Physical Sciences}~\cite{Bevington} by Bevington and Robinson.


\chapter{Monitor Differences for \Qs} 
\captionsetup{justification=justified,singlelinecheck=false}

\label{AppendixB} 

\begin{figure}
\centering
\includegraphics[width=5.9in]{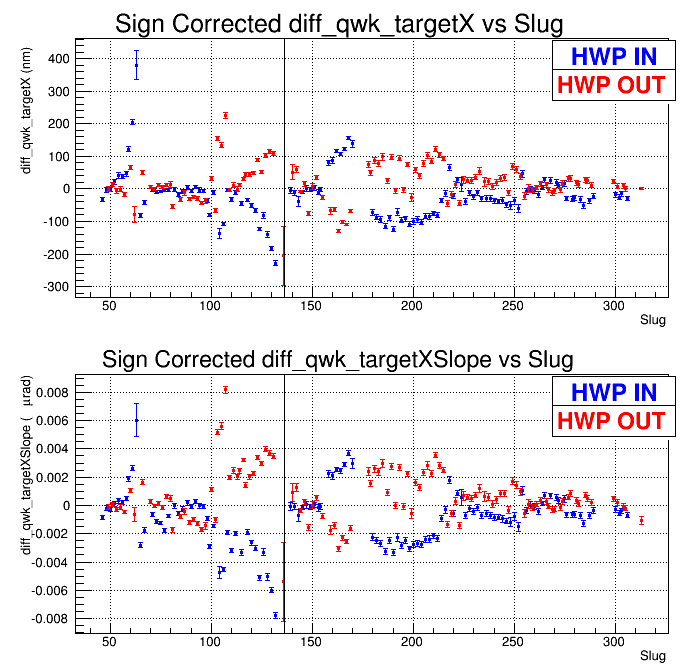}
\caption{\label{fig:Xdiff_by_slug}Slug-averaged helicity-correlated differences in horizontal position (top) and angle (bottom) at the target color-labeled by half-wave plate state. The errors shown are derived from monitor difference RMS widths and reflect beam motion not measurement precision.}
\end{figure}

\begin{figure}
\centering
\includegraphics[width=5.9in]{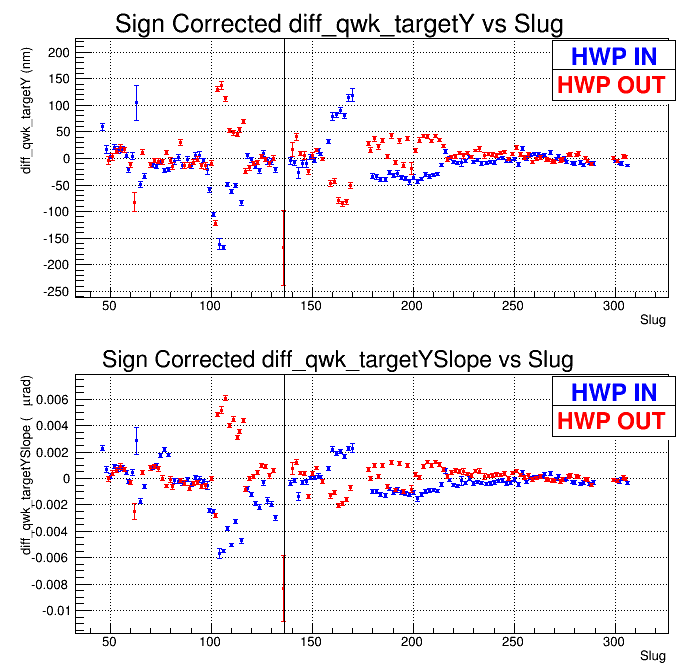}
\caption{\label{fig:Ydiff_by_slug}Slug-averaged helicity-correlated differences in vertical position (top) and angle (bottom) at the target color-labeled by half-wave plate state. The errors shown are derived from monitor difference RMS widths and reflect beam motion not measurement precision. }
\end{figure}

\begin{figure}
\centering
\includegraphics[width=5.9in]{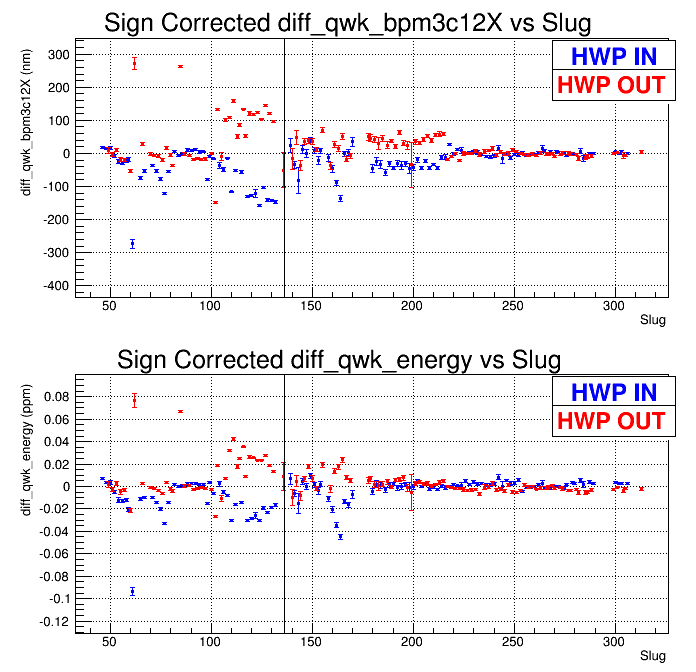}
\caption{\label{fig:Ediff_by_slug}Slug-averaged helicity-correlated differences in horizontal position at BPM3c12X (top) and \Qs energy variable (bottom) color-labeled by half-wave plate state. The errors shown are derived from monitor difference RMS widths and reflect beam motion not measurement precision.}
\end{figure}

\begin{figure}
\centering
\includegraphics[width=5.9in]{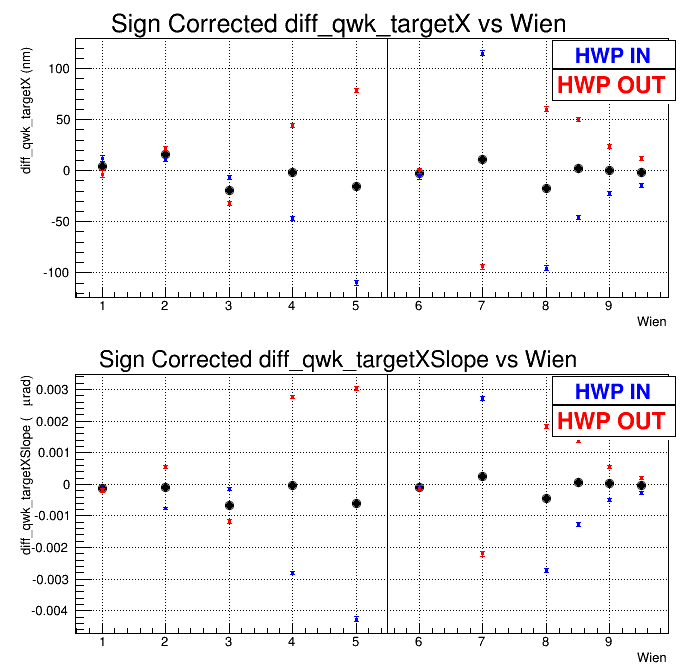}
\caption{\label{fig:Xdiff_by_wien}Wien-averaged helicity-correlated differences in horizontal position (top) and angle (bottom) at the target color-labeled by half-wave plate state. The average of the In and Out IWHP states is shown in black. The errors shown are derived from monitor difference RMS widths and reflect beam motion not measurement precision. For measurement precision estimates see Table \ref{tab:monitor_diff_by_wien}.}
\end{figure}

\begin{figure}
\centering
\includegraphics[width=5.9in]{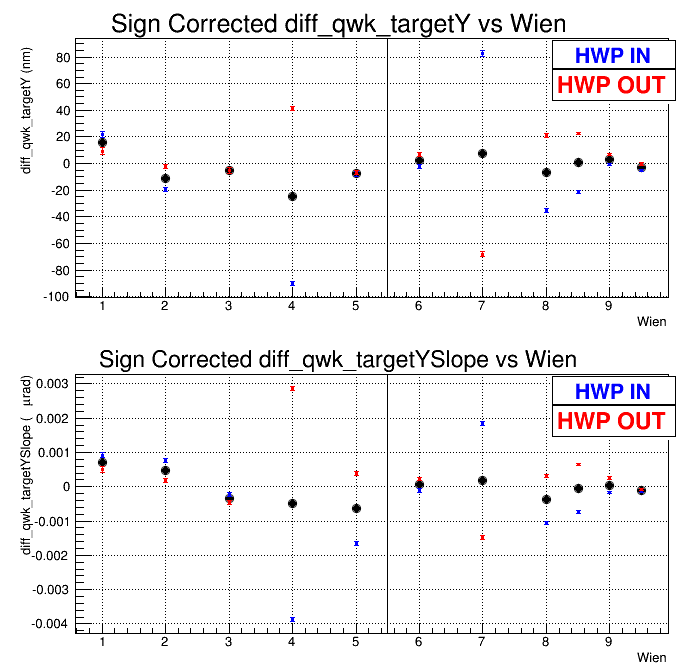}
\caption{\label{fig:Ydiff_by_wien}Wien-averaged helicity-correlated differences in vertical position (top) and angle (bottom) at the target color-labeled by half-wave plate state. The average of the In and Out IWHP states is shown in black. The errors shown are derived from monitor difference RMS widths and reflect beam motion not measurement precision. For measurement precision estimates see Table \ref{tab:monitor_diff_by_wien}. }
\end{figure}

\begin{figure}
\centering
\includegraphics[width=5.9in]{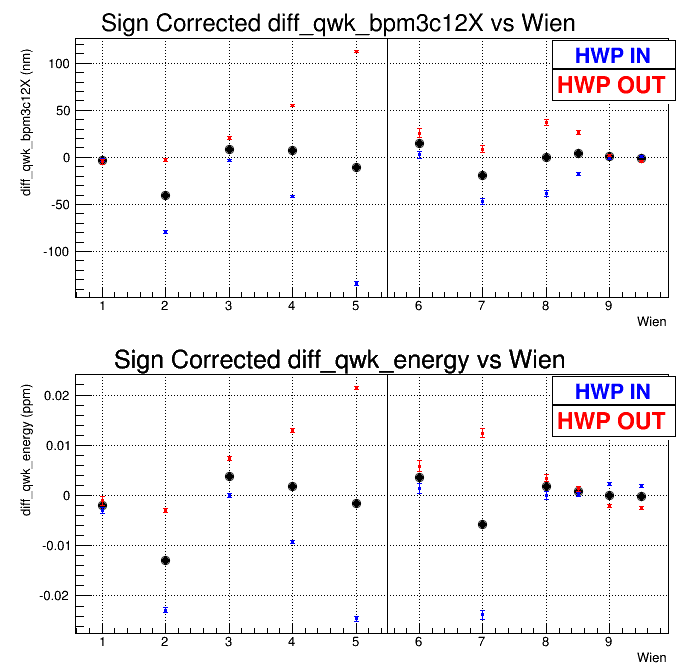}
\caption{\label{fig:Ediff_by_wien}Wien-averaged helicity-correlated differences in horizontal position at BPM3c12X (top) and \Qs energy variable (bottom) color-labeled by half-wave plate state. The average of the In and Out IWHP states is shown in black. The errors shown are derived from monitor difference RMS widths and reflect beam motion not measurement precision. For measurement precision estimates see Table \ref{tab:monitor_diff_by_wien}.}
\end{figure}

\lhead{Appendix B. \emph{Monitor Differences}} 
This gives a summary of the monitor differences over Runs 1 and 2 (Wiens 1-9b) of \Q. The monitors shown are those used in the beam modulation analysis. The \Qs energy variable is also included since it is used in the Standard Regression scheme (see \ref{Chapter4}). Figures \ref{fig:Xdiff_by_slug} to \ref{fig:Ediff_by_slug} show the monitor differences averaged over slugs ($\sim$8~hr periods of the same IHWP slow-reversal state). Figures \ref{fig:Xdiff_by_wien} to \ref{fig:Ediff_by_wien} show the monitor differences averaged over Wiens ($\sim$1~month periods of the same Wien slow-reversal state). All averages are error-weighted with the same weights as the production dataset (MDallbars errors). The error bars shown come from the RMS width of the measured beam position and are thus mostly an indication of the beam properties as opposed to how accurately the differences in the beam parameters were measured. The precision of the measurements are given in Table \ref{tab:monitor_diff_by_wien} and are found by dividing the monitor resolution by the square root of the number of samples. A single BPM such as BPM3c12X has a resolution of approximately 1~$\mu$m. Studies were performed during \Qs to determine the resolution of the target variables which are composed of several BPM's extrapolated to the target position. The results of this study, found in the \Qs Analysis and Simulation Elog 788, are summarized in Table \ref{tab:res_mon}.
\begin{table}[h]
\begin{center}
\caption{\label{tab:res_mon}Resolution of target variables at the MPS ($\sim$1~ms) timescale. The larger resolutions in Run 2 are due to the failure of one of the BPM's used in the target variable composition. The resolution of helicity-correlated differences at quartet level are given by dividing these values by $\sqrt{2}$. The average values do not depend on the choice of errors (that is, errors derived from monitor resolution versus errors derived from RMS width of the differnce distribution) since the averaging weights come from the main detector error. Thus, these values correspond to the black dots in Figures \ref{fig:Xdiff_by_wien} to \ref{fig:Ediff_by_wien} even though the error bars are different.}
\begin{tabular}{lcc}\\\hline
~&Run 1&Run 2\\\hline\hline
targetX(Y)&0.98~$\mu$m&1.72~$\mu$m\\
targetXSlope(YSlope)&0.13~$\mu$rad&0.21~$\mu$rad\\
BPM3c12X(Y)&1.0~$\mu$m&1.0~$\mu$m\\\hline
\end{tabular}
\end{center}
\end{table}

\begin{landscape}
\begin{table}[h]
\begin{center}
\caption{\label{tab:monitor_diff_by_wien}Beam monitor helicity-correlated differences averaged over Wien states. The monitor differences are calculated at the runlet level and then weighted with the same weights as the production data to give the effective monitor differences. The errors are first calculated at the runlet level as the resolution of the monitors divided by the square root of the number of samples and then propagated using the same weights as the main detector production data. Resolutions used are the values in Table \ref{tab:res_mon} divided by $\sqrt{2}$ to account for quartet-level averaging.}
\begin{tabular}{|l|c|c|c|c|c|}\hline
Wien&targetX(nm)&targetXSlope(nrad)&targetY(nm)&targetYSlope(nrad)&bpm3c12X(nm)\\\hline\hline
1 &$4.32\pm 0.09$&$-0.15\pm 0.01$&$15.74\pm 0.09$&$0.71\pm 0.01$&$-3.71\pm 0.09$\\\hline
2 &$16.43\pm 0.07$&$-0.10\pm 0.01$&$-11.05\pm 0.07$&$0.48\pm 0.01$&$-40.95\pm 0.07$\\\hline
3 &$-19.56\pm 0.07$&$-0.66\pm 0.01$&$-5.21\pm 0.07$&$-0.34\pm 0.01$&$8.69\pm 0.08$\\\hline
4 &$-1.43\pm 0.08$&$-0.02\pm 0.01$&$-24.40\pm 0.08$&$-0.50\pm 0.01$&$6.77\pm 0.08$\\\hline
5 &$-15.42\pm 0.08$&$-0.62\pm 0.01$&$-7.23\pm 0.08$&$-0.63\pm 0.01$&$-10.97\pm 0.09$\\\hline
6 &$-2.96\pm 0.13$&$-0.11\pm 0.02$&$2.29\pm 0.13$&$0.05\pm 0.02$&$14.25\pm 0.08$\\\hline
7 &$10.58\pm 0.16$&$0.26\pm 0.02$&$7.31\pm 0.16$&$0.17\pm 0.02$&$-18.95\pm 0.09$\\\hline
8a&$-16.51\pm 0.11$&$-0.42\pm 0.01$&$-6.77\pm 0.11$&$-0.37\pm 0.01$&$-0.23\pm 0.07$\\\hline
8b&$2.19\pm 0.11$&$0.05\pm 0.01$&$0.60\pm 0.11$&$-0.04\pm 0.01$&$4.41\pm 0.06$\\\hline
9a&$0.23\pm 0.09$&$0.02\pm 0.01$&$3.22\pm 0.09$&$0.05\pm 0.01$&$-0.02\pm 0.05$\\\hline
9b&$-0.88\pm 0.09$&$-0.02\pm 0.01$&$-2.59\pm 0.09$&$-0.10\pm 0.01$&$-1.09\pm 0.05$\\\hline
\end{tabular}
\end{center}
\end{table}
\end{landscape}

\chapter{Effect of FFB Stability on Modulation Analysis} 
\captionsetup{justification=justified,singlelinecheck=false}

\label{AppendixC}

\lhead{Appendix C. \emph{Stability of FFB}}

The stability of the FFB response was investigated by dividing up the modulation periods into smaller equal-time slices. The idea was to see if the response when modulation first turned was different than its steady-state response, if such a thing as a steady-state response even existed. The 4-second modulation periods were divided into thirds and a complete analysis done on each to find correction slopes. The first 1.33 seconds of data are termed ``tertile 1'', the second ``tertile 2'' and so on. 

\begin{figure}[ht]

\centering
\framebox{\includegraphics[width=5.6in]{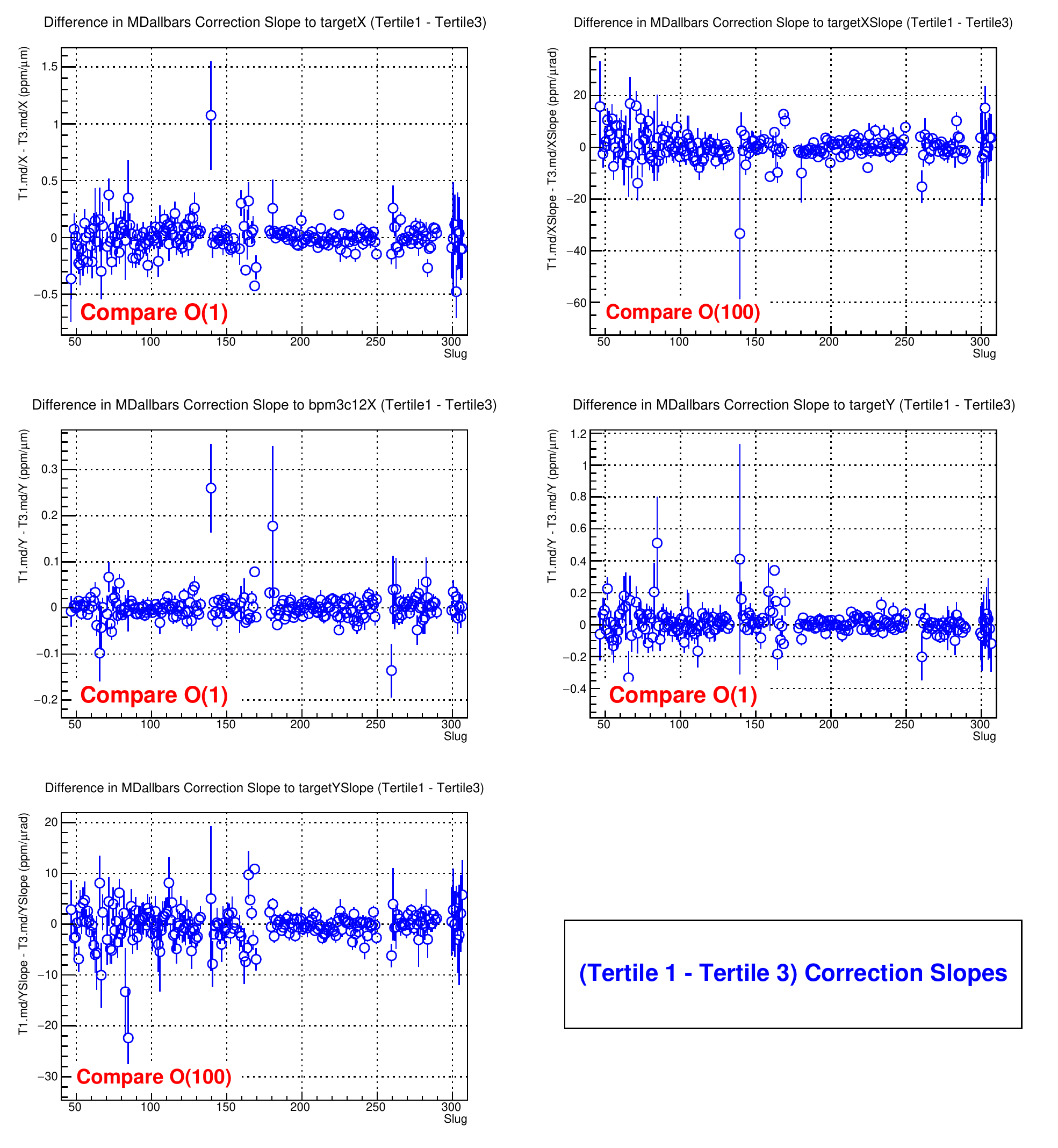}}
\caption{Change in MDallbars correction slopes between tertiles 1 and 3. Each point represents the average for a given slug of data and the error bar is $\sigma/\sqrt{N}$. The ``pull plot'' distributions shown in Figure \ref{fig:tert_mdall_pull_plots} show that these differences are consistent with what is expected from statistics.}
\label{fig:tert_mdall_slope change}
\end{figure}

\begin{figure}[ht]

\centering
\framebox{\includegraphics[width=5.6in]{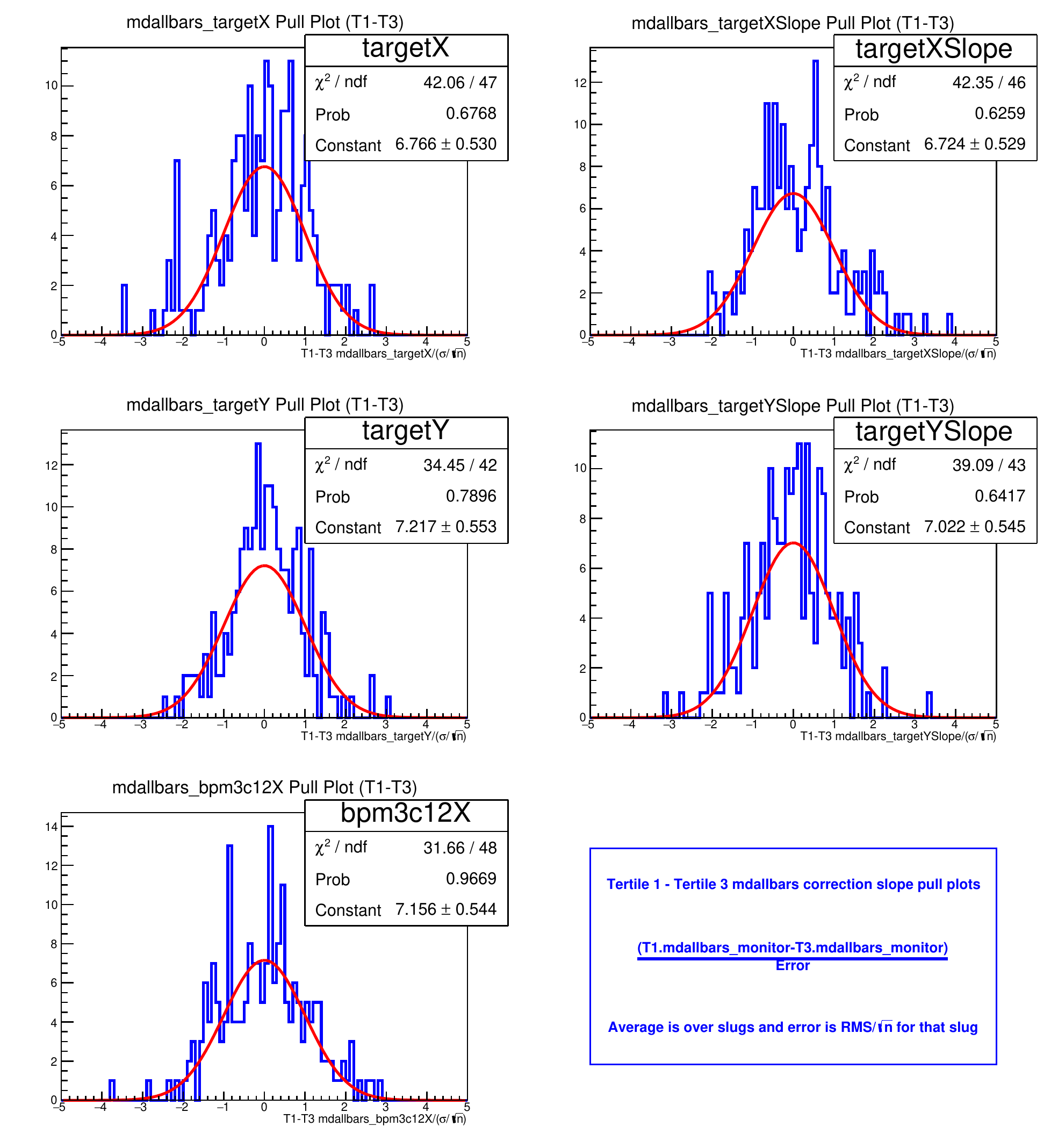}}
\caption{Pull distribution plots for difference in MDallbars to monitor correction slopes between tertiles 1 and 3. The histograms are the slug averages divided by their standard error. Pure statistical distributions (shown in red) are expected to have zero mean and unit standard deviation. The p-value of the fit of the distributions to the data indicate that the data are consistent with purely statistical deviations.}
\label{fig:tert_mdall_pull_plots}
\end{figure}

First and most important, is the stability of the final detector to monitor correction slopes. That is, are variations in the slopes between tertiles that are obviously non-statistical? Figure \ref{fig:tert_mdall_slope change} shows the difference in the MDallbars correction slopes to the five monitors chosen for the correction. These plots show the average difference for each slug with error bars formed from the standard deviation of the slopes in each slug divided by the square root of the number of slopes in the average. Shown on the inset in each plot is an estimate the size of the slopes for that variable. It would appear from these plots that there is no significant difference in the slopes from the two tertiles. The pull distributions shown in Figure \ref{fig:tert_mdall_pull_plots} show the same data where each slug data point is normalized to its error bar. If the differences between the tertiles is completely statistical the ``pull'' distributions should have a mean of 0 and a standard deviation of 1. The Gaussian functions shown in red overlaying the data are the standard normal distribution multiplied by a single scale parameter which is allowed to vary in the fit. The probability for each of these plots being explained by purely statistical fluctuations is high. In fact, perhaps the probability is too high, indicating that RMS/$\sqrt{n}$ overestimates the error.  

Figures \ref{fig:tert_coeffX} to  \ref{fig:tert_coeff3c12X} show the differences in the fit coefficients (sine and cosine fit amplitudes) between various tertiles, scaled by the total amplitude of the response for scale. ``Sine'' refers to the responses that are in phase with the modulation driving signal and ``Cosine'' to the responses that are $\pi/2$ out of phase with the driving signal.  The figure show a percent change in monitor and detector responses between tertiles. For example, the percent change in the in-phase response of ``targetX'' between tertiles 1 and 3 is given as
\begin{equation}
\frac{\Delta X_{sin}}{|X|}(\%)=100\%\times\frac{(X_{sin}^{(T1)}-X_{sin}^{(T3)})}{\sqrt{ \left(\frac{X_{sin}^{(T1)}+X_{sin}^{(T3)}}{2}\right)^2+\left(\frac{X_{cos}^{(T1)}+X_{cos}^{(T3)}}{2}\right)^2}},
\label{eq:fractional_tertile_change}
\end{equation}
where, for example $X_{sin}^{(T1)}$ is the sine response amplitude of targetX to a given type of modulation.

The results shown in figures \ref{fig:tert_coeffX} to  \ref{fig:tert_coeff3c12X}
suggest that for the most part, the FFB response as seen by the monitors is fairly stable over the modulation cycle time. When examining these plots it is helpful to remember that monitors less sensitive to a given modulation type may see larger percent changes but much smaller absolute changes. For example, we would expect targetX, targetXSlope and bpm3c12X to be most sensitive to X1 modulation with targetY and targetYSlope only marginally sensitive. Looking at the plots we can see percent level changes in  targetX, targetXSlope and bpm3c12X during X1 modulation. While targetY and targetYSlope see changes at the 10's of percent level during X1 modulation, the means of the distributions appear to be consistent with 0. Although one can point to seemingly large outliers where shifts of order 100\% occur, these outliers never occur in a monitor expected to have a large response to a given modulation type. These observations lead to the following conclusions:
\begin{itemize}
\item{The monitors that are expected to be sensitive to a given type of modulation appear to have a relatively stable response at the few percent level when different ``slices'' of the modulation cycles are compared.}
\item{Monitors not expected to be sensitive to a given type of modulation do not show large (greater than a few percent) systematic shifts in response when different ``slices'' of the modulation cycles are compared.}
\item{The largest shifts occur during energy modulation when the FFB system is paused. Shifts during this time must be attributed to instability in the energy response, not to FFB. It is conceivable that some unknown feedback system such as MOMOD remains active during energy modulation and creates the energy instability.} 
\end{itemize}

\begin{figure}[h]

\centering
\framebox{\includegraphics[width=5.6in]{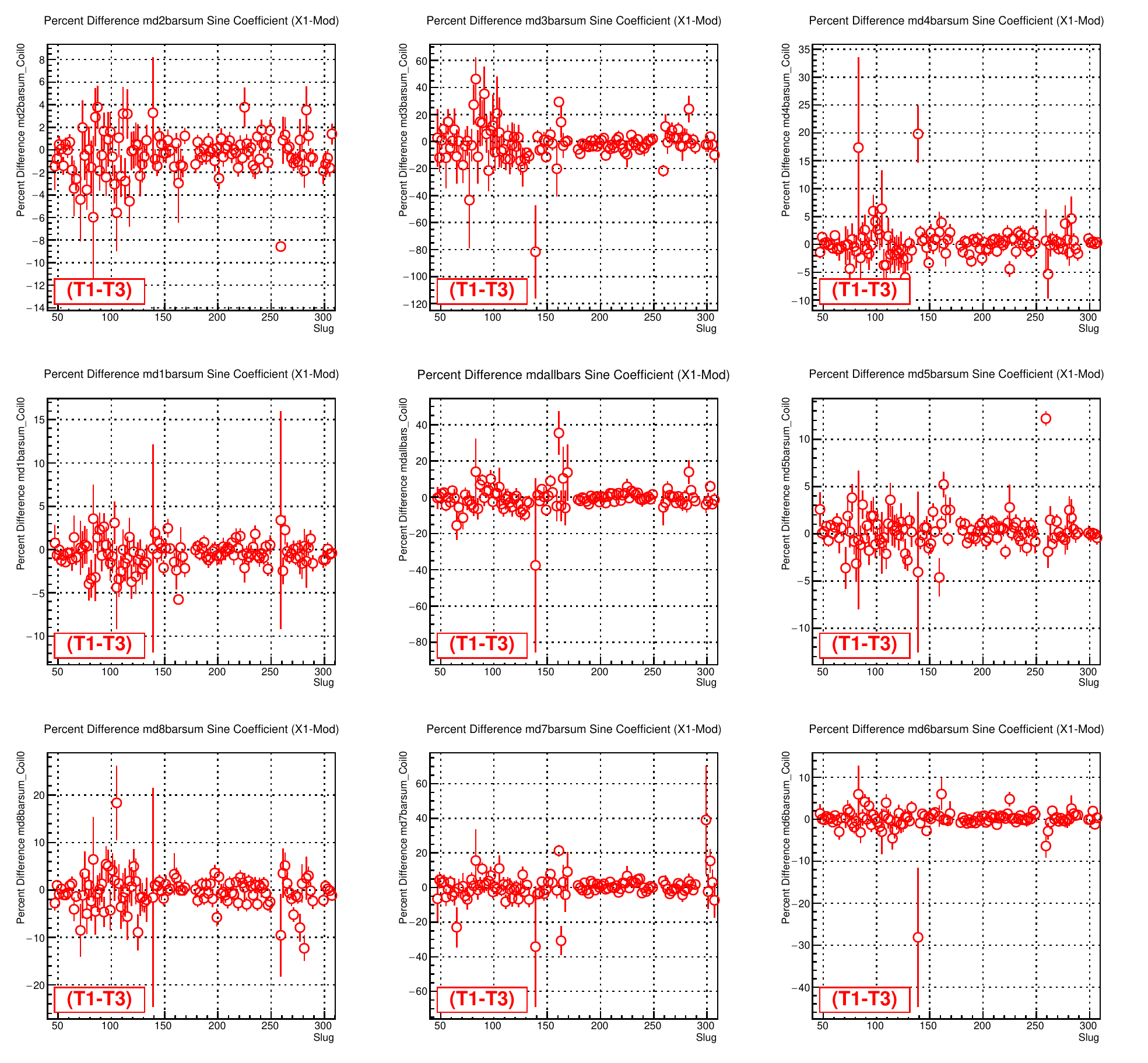}}
\caption{Percent change in the in-phase (sine) main detector response between data tertiles 1 and 3 during X1-modulation (coil 0). See Equation \ref{eq:fractional_tertile_change} for an explanation of the percent change shown.}
\label{fig:tert_md_coeff_coil0}
\end{figure}

\begin{figure}[h]

\centering
\framebox{\includegraphics[width=5.6in]{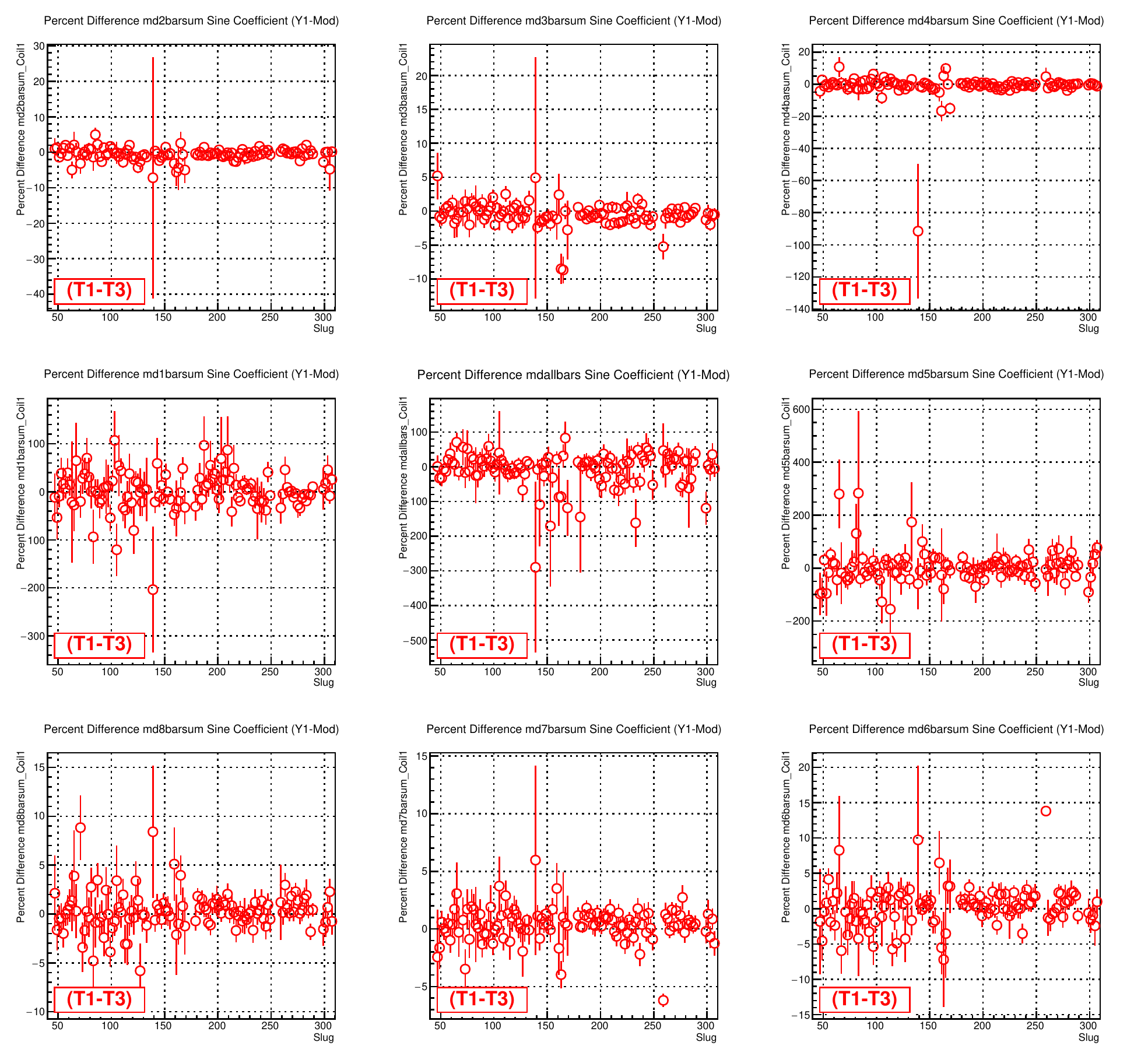}}
\caption{Percent change in the in-phase (sine) main detector response between data tertiles 1 and 3 during Y1-modulation (coil 1). See Equation \ref{eq:fractional_tertile_change} for an explanation of the percent change shown.}
\label{fig:tert_md_coeff_coil1}
\end{figure}

\begin{figure}[h]

\centering
\framebox{\includegraphics[width=5.6in]{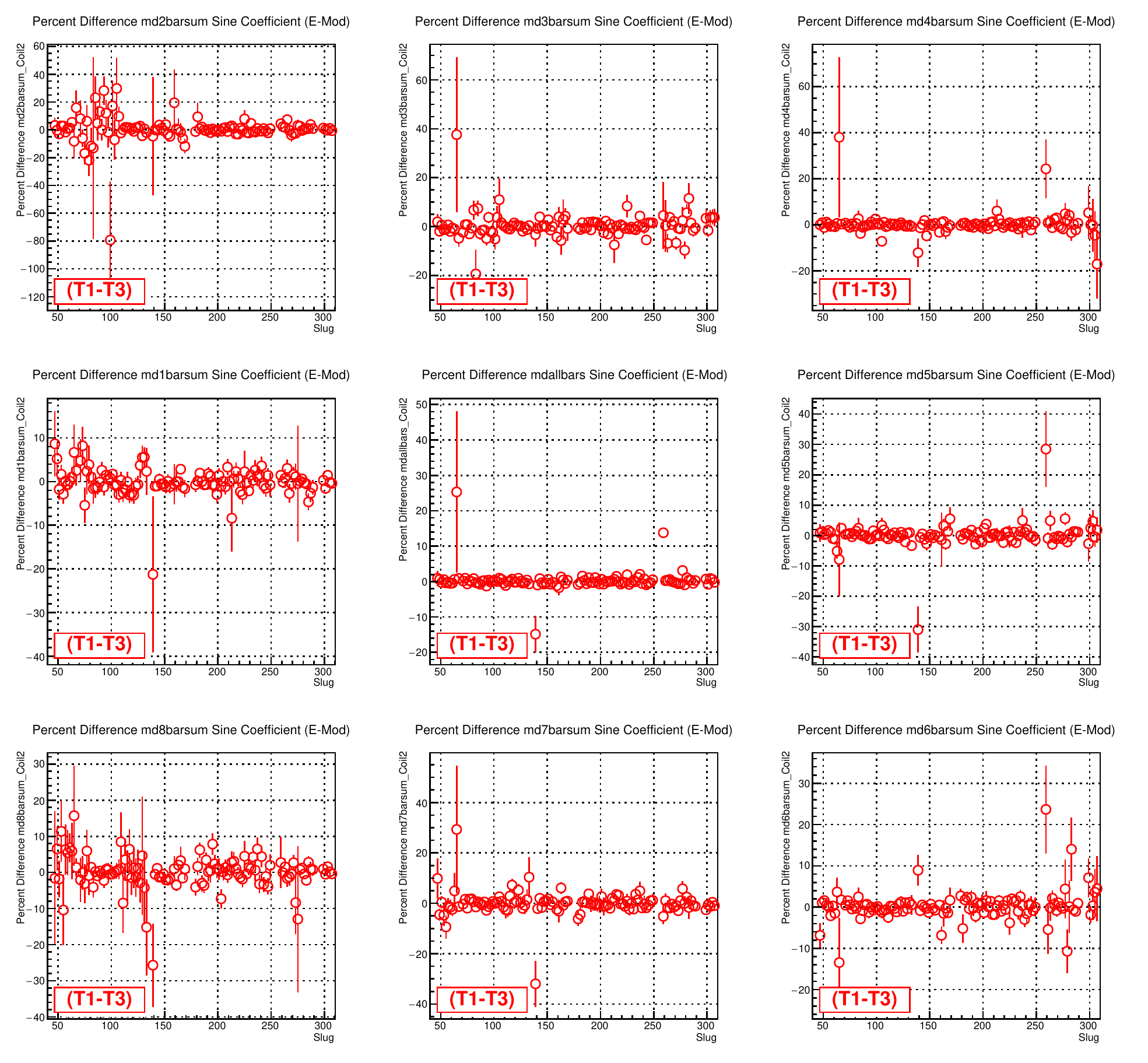}}
\caption{Percent change in the in-phase (sine) main detector response between data tertiles 1 and 3 during energy modulation(coil 2) . See Equation \ref{eq:fractional_tertile_change} for an explanation of the percent change shown.}
\label{fig:tert_md_coeff_coil2}
\end{figure}

\begin{figure}[h]

\centering
\framebox{\includegraphics[width=5.6in]{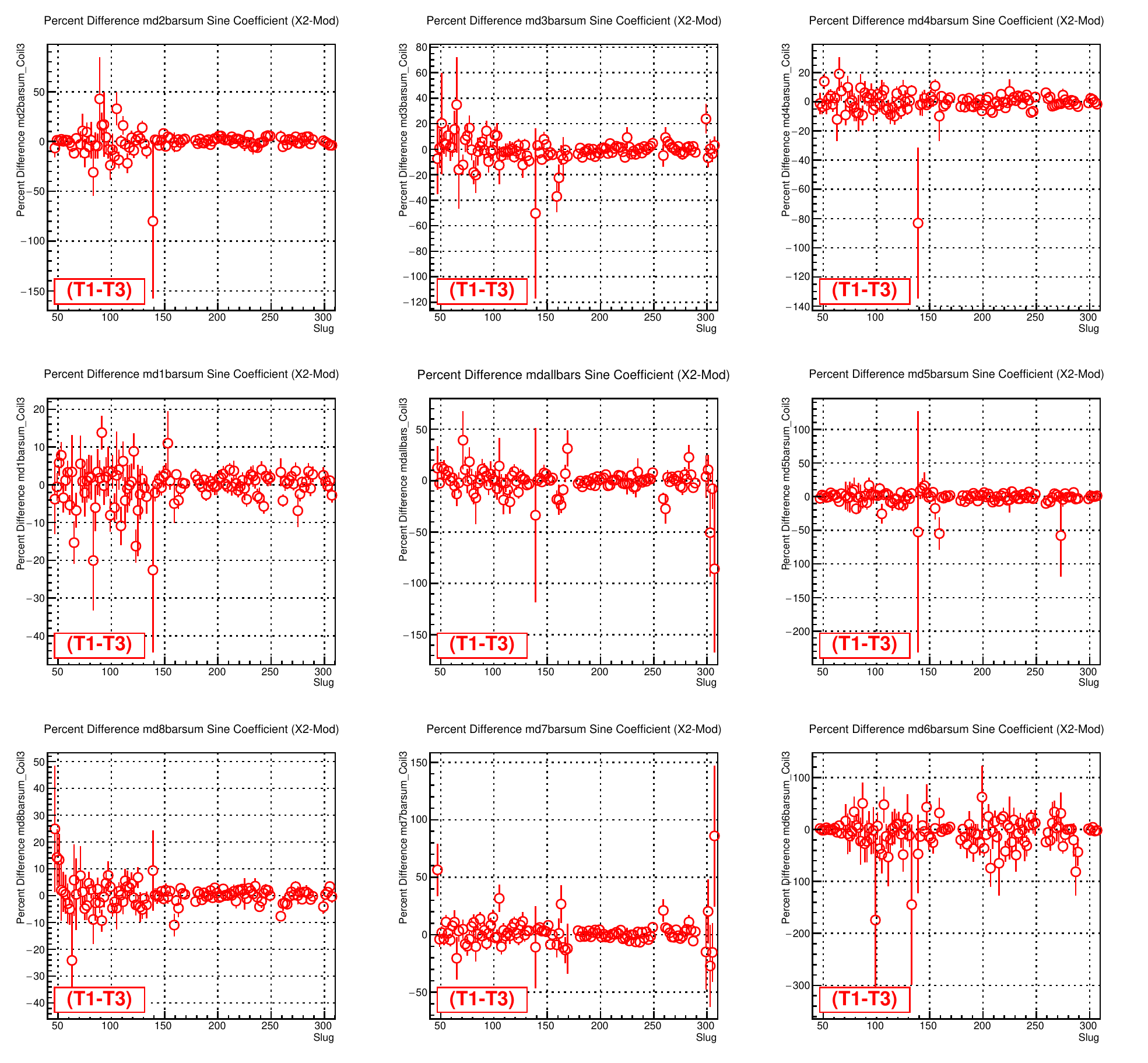}}
\caption{Percent change in the in-phase (sine) main detector response between data tertiles 1 and 3 during X2-modulation (coil 3). See Equation \ref{eq:fractional_tertile_change} for an explanation of the percent change shown.}
\label{fig:tert_md_coeff_coil3}
\end{figure}

\begin{figure}[h]

\centering
\framebox{\includegraphics[width=5.6in]{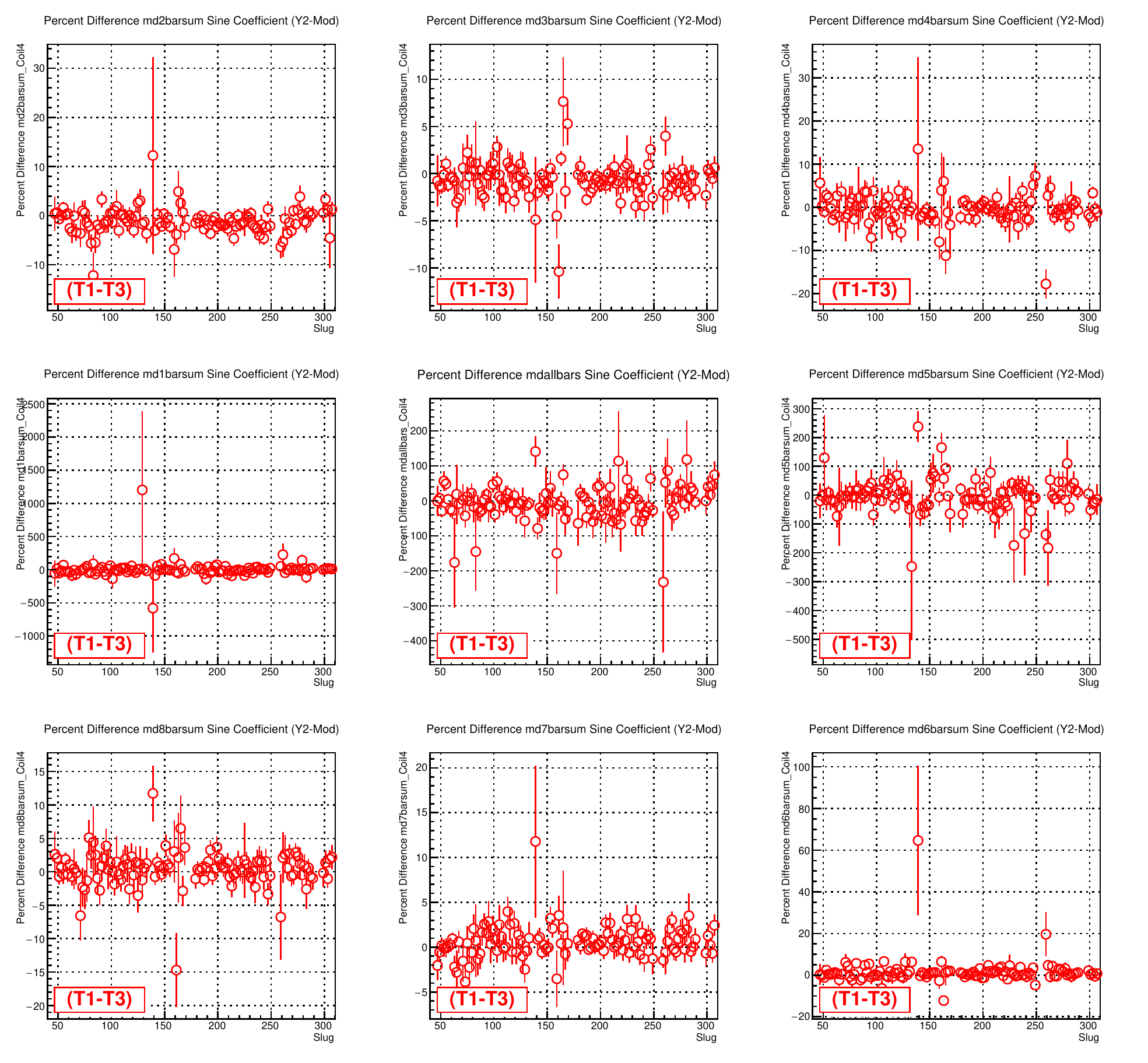}}
\caption{Percent change in the in-phase (sine) main detector response between data tertiles 1 and 3 during Y2-modulation (coil 4). See Equation \ref{eq:fractional_tertile_change} for an explanation of the percent change shown.}
\label{fig:tert_md_coeff_coil4}
\end{figure}

\begin{figure}[h]

\centering
\framebox{\includegraphics[width=5.6in]{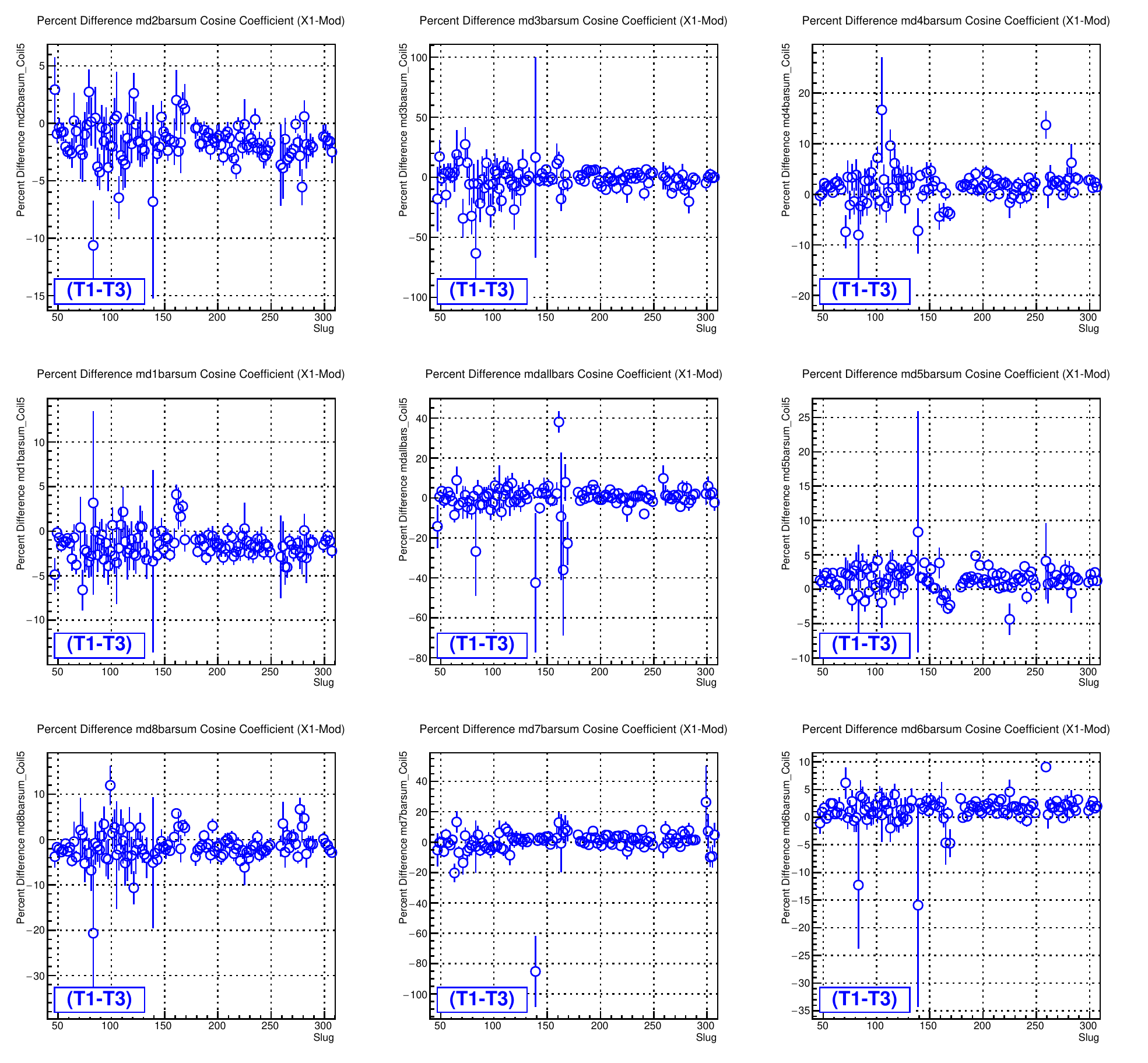}}
\caption{Percent change in the out-of-phase (cosine) main detector response between data tertiles 1 and 3 during X1-modulation (coil 5). See Equation \ref{eq:fractional_tertile_change} for an explanation of the percent change shown.}
\label{fig:tert_md_coeff_coil5}
\end{figure}

\begin{figure}[h]

\centering
\framebox{\includegraphics[width=5.6in]{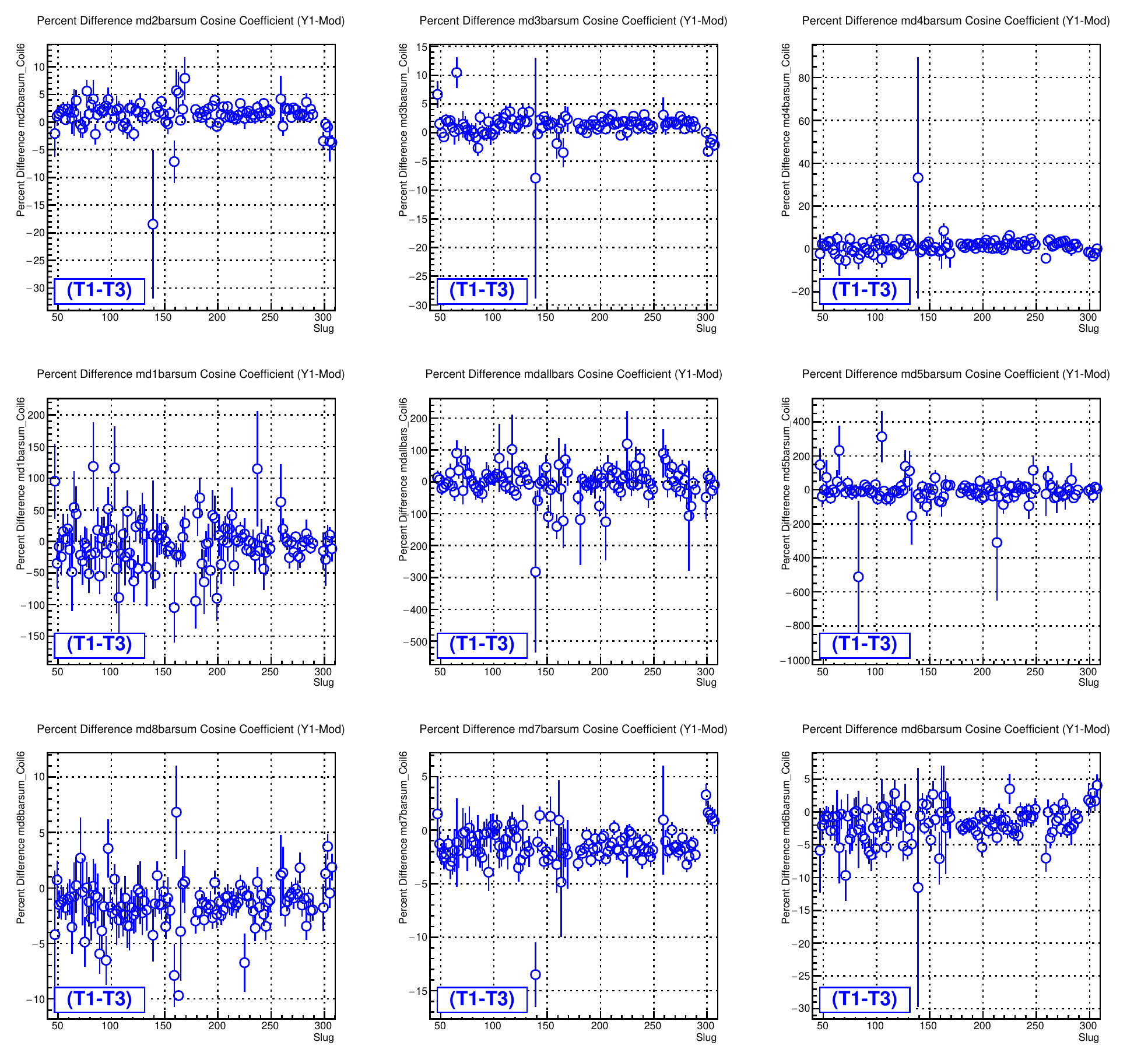}}
\caption{Percent change in the out-of-phase (cosine) main detector response between data tertiles 1 and 3 during Y1-modulation (coil 6). See Equation \ref{eq:fractional_tertile_change} for an explanation of the percent change shown.}
\label{fig:tert_md_coeff_coil6}
\end{figure}

\begin{figure}[h]

\centering
\framebox{\includegraphics[width=5.6in]{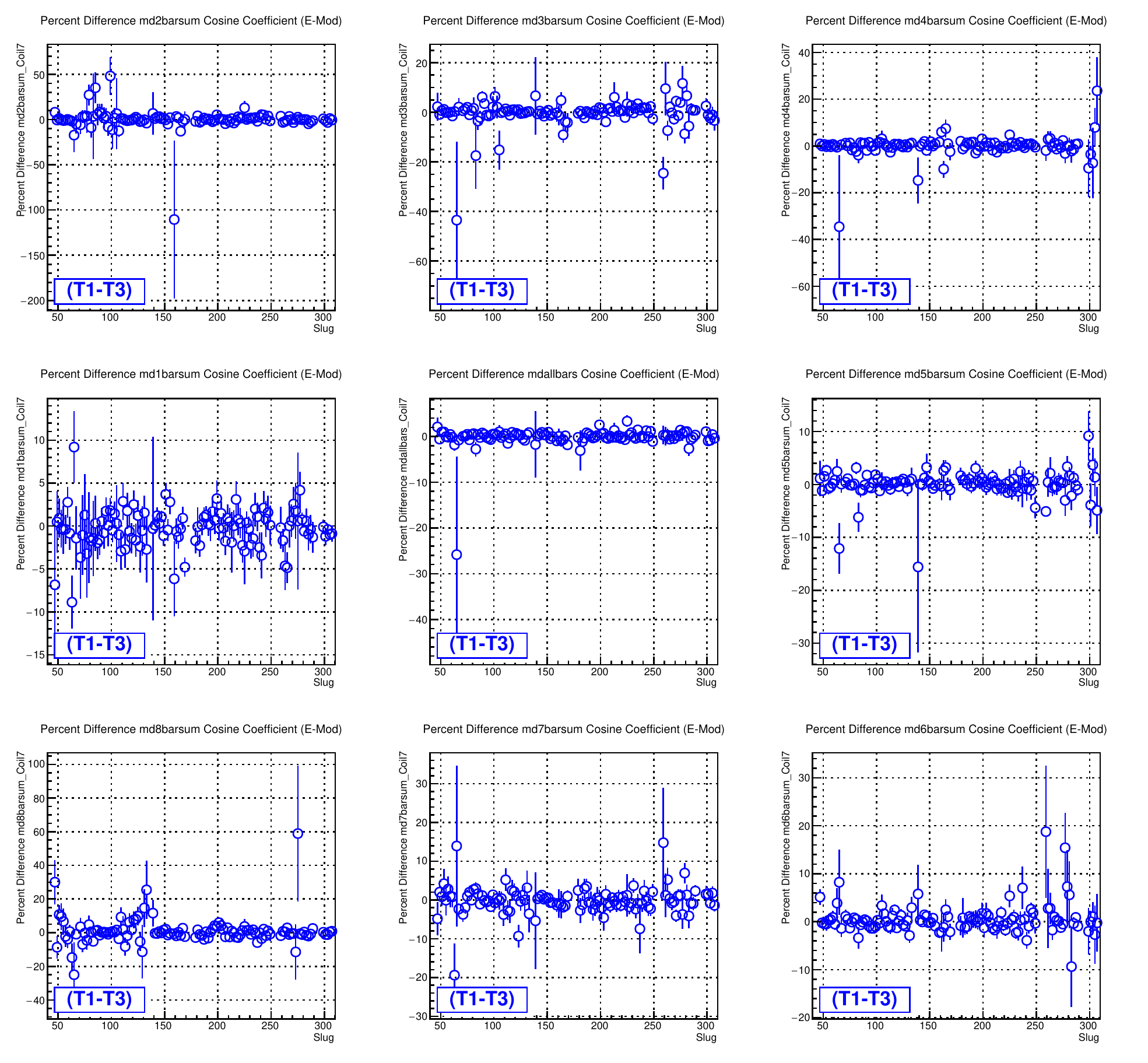}}
\caption{Percent change in the out-of-phase (cosine) main detector response between data tertiles 1 and 3 during energy modulation(coil 7) . See Equation \ref{eq:fractional_tertile_change} for an explanation of the percent change shown.}
\label{fig:tert_md_coeff_coil7}
\end{figure}

\begin{figure}[h]

\centering
\framebox{\includegraphics[width=5.6in]{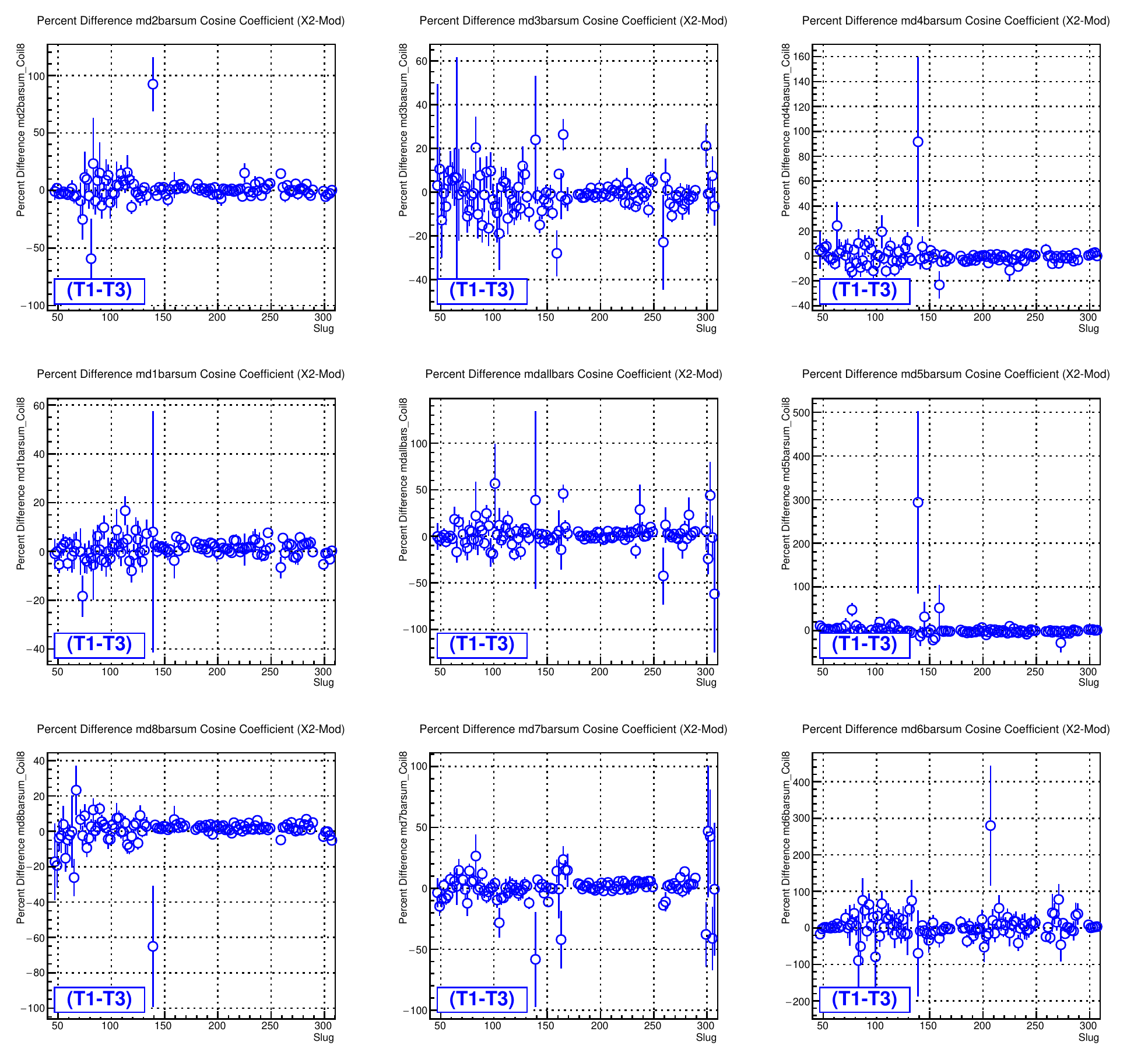}}
\caption{Percent change in the out-of-phase (cosine) main detector response between data tertiles 1 and 3 during X2-modulation (coil 8). See Equation \ref{eq:fractional_tertile_change} for an explanation of the percent change shown.}
\label{fig:tert_md_coeff_coil8}
\end{figure}

\begin{figure}[h]

\centering
\framebox{\includegraphics[width=5.6in]{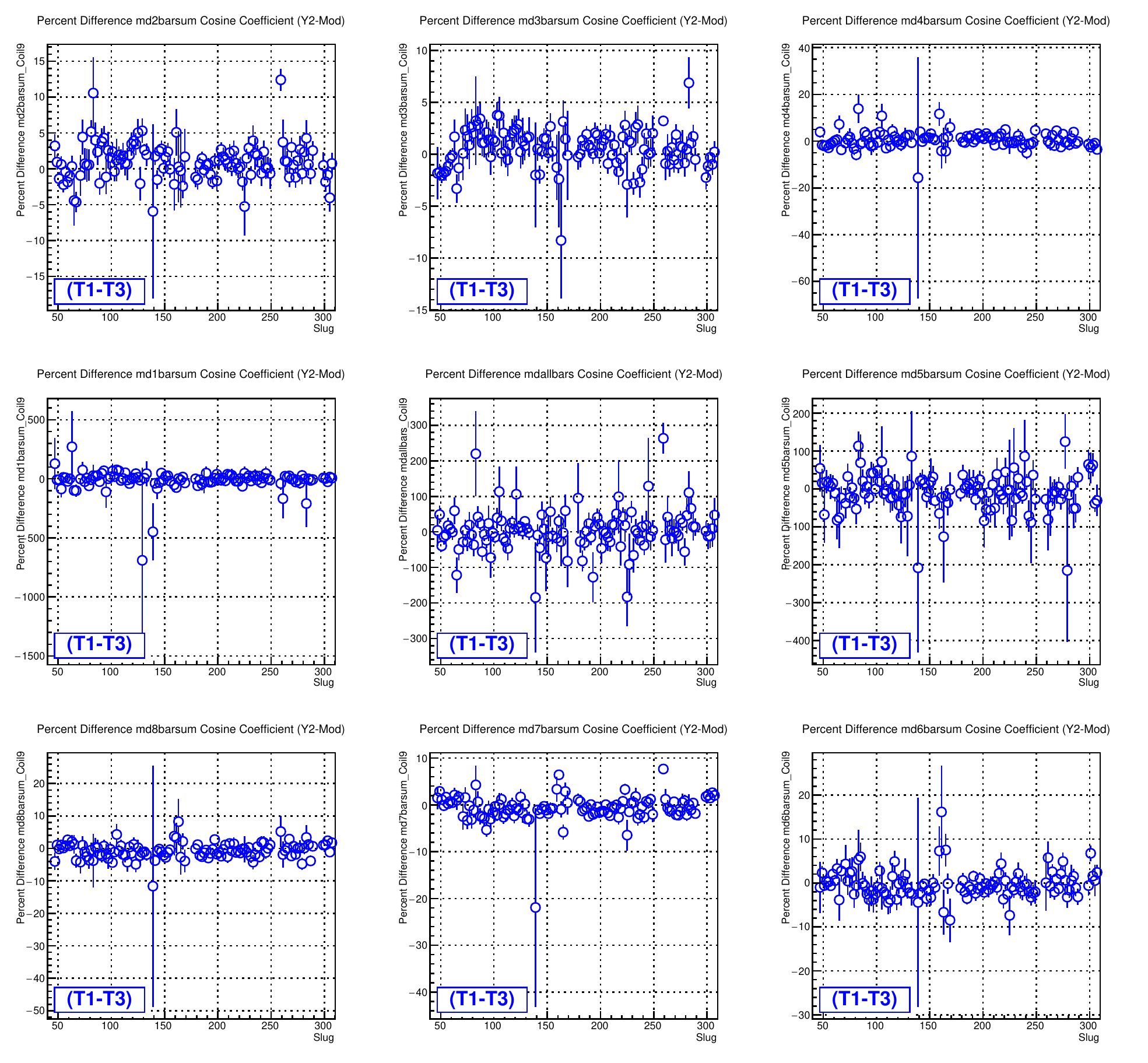}}
\caption{Percent change in the out-of-phase (cosine) main detector response between data tertiles 1 and 3 during Y2-modulation (coil 9). See Equation \ref{eq:fractional_tertile_change} for an explanation of the percent change shown.}
\label{fig:tert_md_coeff_coil9}
\end{figure}

Figures \ref{fig:tert_md_coeff_coil0} to \ref{fig:tert_md_coeff_coil9} show similar small variations between tertile 1 and tertile 3 in the main detector responses. 

It may be observed, however, that the stability observed in the slopes appears to be greater than that in the monitor and detector coefficients. In fact, they appear to be consistent with statistics. How is the stability of the detector to monitor correction slopes to be understood in the light of the larger instability of the monitor and detector coefficients? Although stability is desired because any noise will increase the uncertainty, perfect stability is not critical to the modulation analysis procedure. In the end, the periodic modulation is only used to determine a relationship between the detectors and monitors and all that is necessary is that there be a well-determined average response of both monitors and detectors over the cycles. It is useful to remember that this whole analysis assumes that the detectors and monitors are linear and only removes the linear response. Beyond that there is no condition on the stability of the driving system built into Equations \ref{eq:det_to_coil} or \ref{eq:chisquaresolution_no_variance}. The basic conditions for success of the modulation analysis method are:
\begin{itemize}
\item{Modulation must span space of beam distortion modes: \\The 5 main ``distortion'' modes are thought to be horizontal and vertical motion, horizontal and vertical angle and energy. The beam must be driven in such a way that all 5 parameters are modulated.}
\item{Monitor set must also span this beam distortion space:\\ The combined set of monitors must have a collective sensitivity to the full phase space in which the beam is driven by the modulation coils. We must have monitors that are sensitive to the five main beam parameters.}
\item{A calibration signal precisely proportional to and in phase with the coil driving signal must be available: \\
Unlike linear regression, monitor and detector responses are not directly compared. Instead the response of each is found relative to the calibration signal. For the Qweak modulation system, a sinusoidal signal sent to the coils (both air coil magnets and the RF cavity) was used to drive beam motion. The physical calibration signal came in the form of a sawtooth wave synchronous with the drive signal to the coils. The sawtooth wave was then translated into a modulation phase and our actual calibration signal was the sine and cosine of this phase. Without this information about the drive signal, the only possible correction for trajectory related false asymmetries would be linear regression.}
\item{The response to the motion must be well-determined in both the monitors and detectors. This is not the same as saying it must non-zero.}
\end{itemize}

An unstable response amplitude will create greater uncertainty in the determination of correction slopes, but it will not in general bias the results unless there exists some instability that is somehow coherent with the driving signal. Since the detectors and the monitors used to correct them both respond to the same unstable behavior, the slopes should be the same within error for both a stable response and a slowly changing response amplitude and this is consistent with our observations.

\begin{landscape}
\begin{figure}[t]

\centering
\framebox{\includegraphics[width=8.9in]{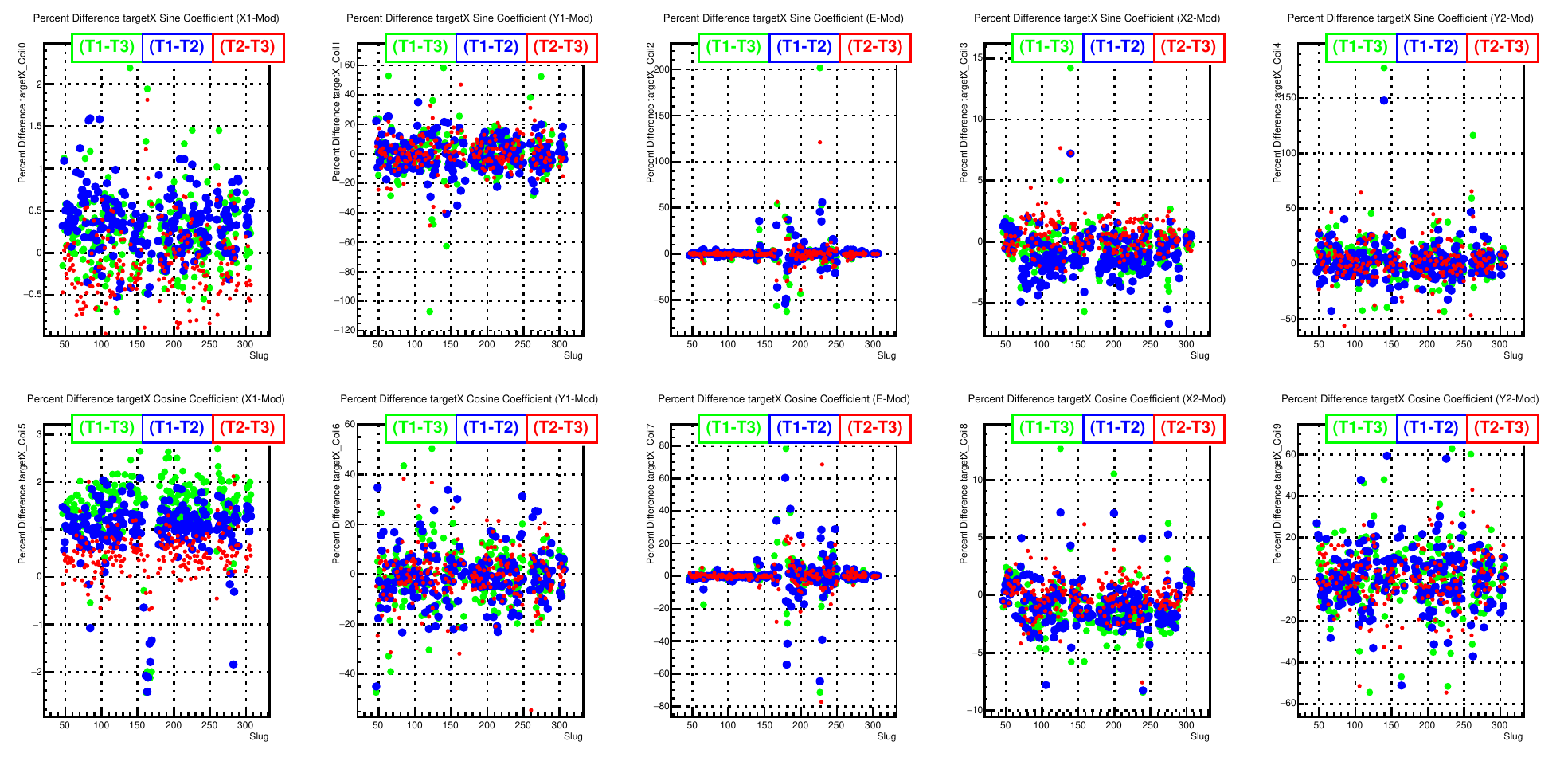}}
\caption{Percent change in targetX response between data tertiles. See Equation \ref{eq:fractional_tertile_change} for an explanation of the percent change shown.}

\label{fig:tert_coeffX}
\end{figure}

\begin{figure}[t]

\centering
\framebox{\includegraphics[width=8.9in]{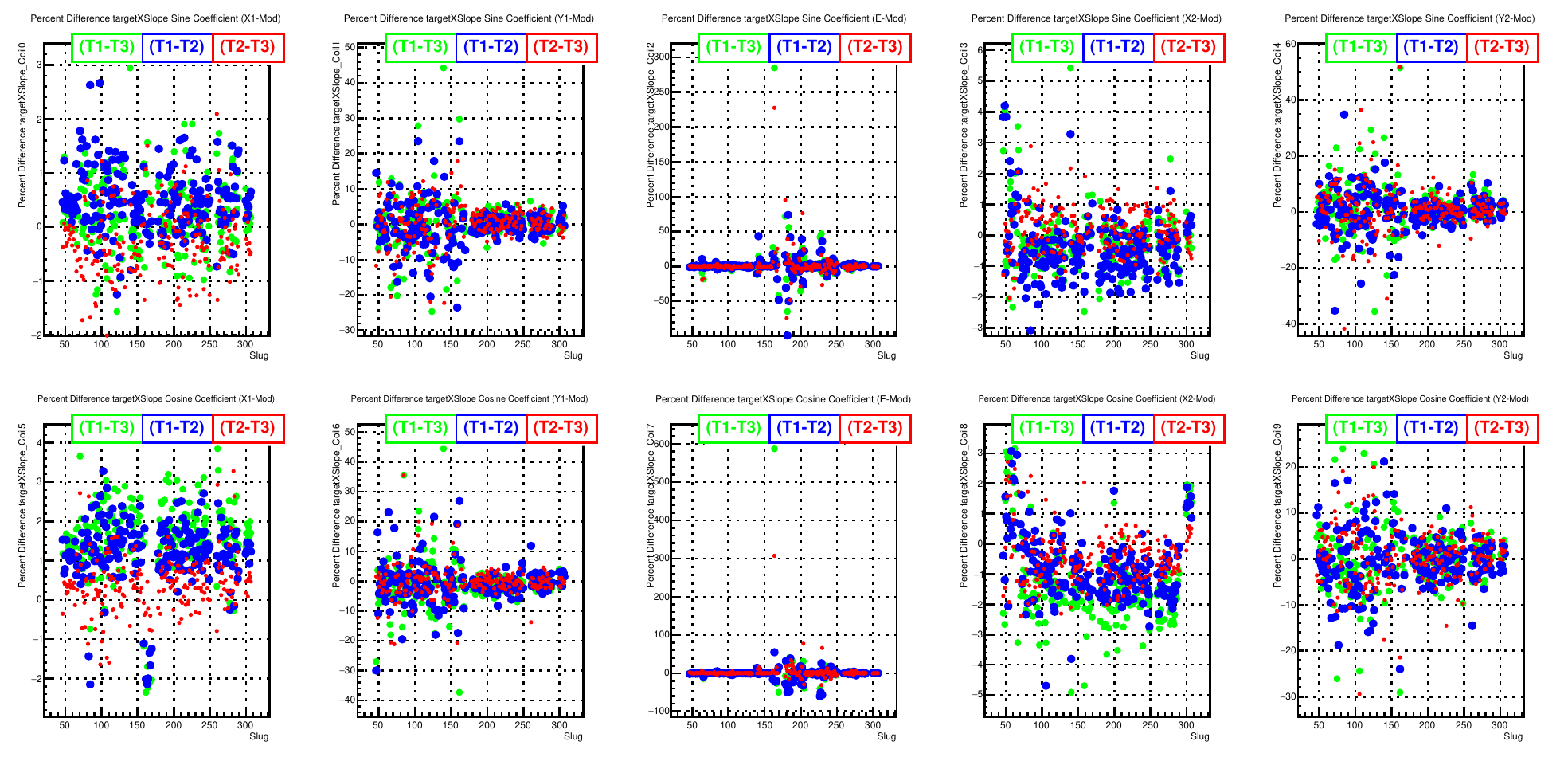}}
\caption{Percent change in targetXSlope response between data tertiles. See Equation \ref{eq:fractional_tertile_change} for an explanation of the percent change shown.}

\label{fig:tert_coeffXSlope}
\end{figure}
\begin{figure}[t]

\centering
\framebox{\includegraphics[width=8.9in]{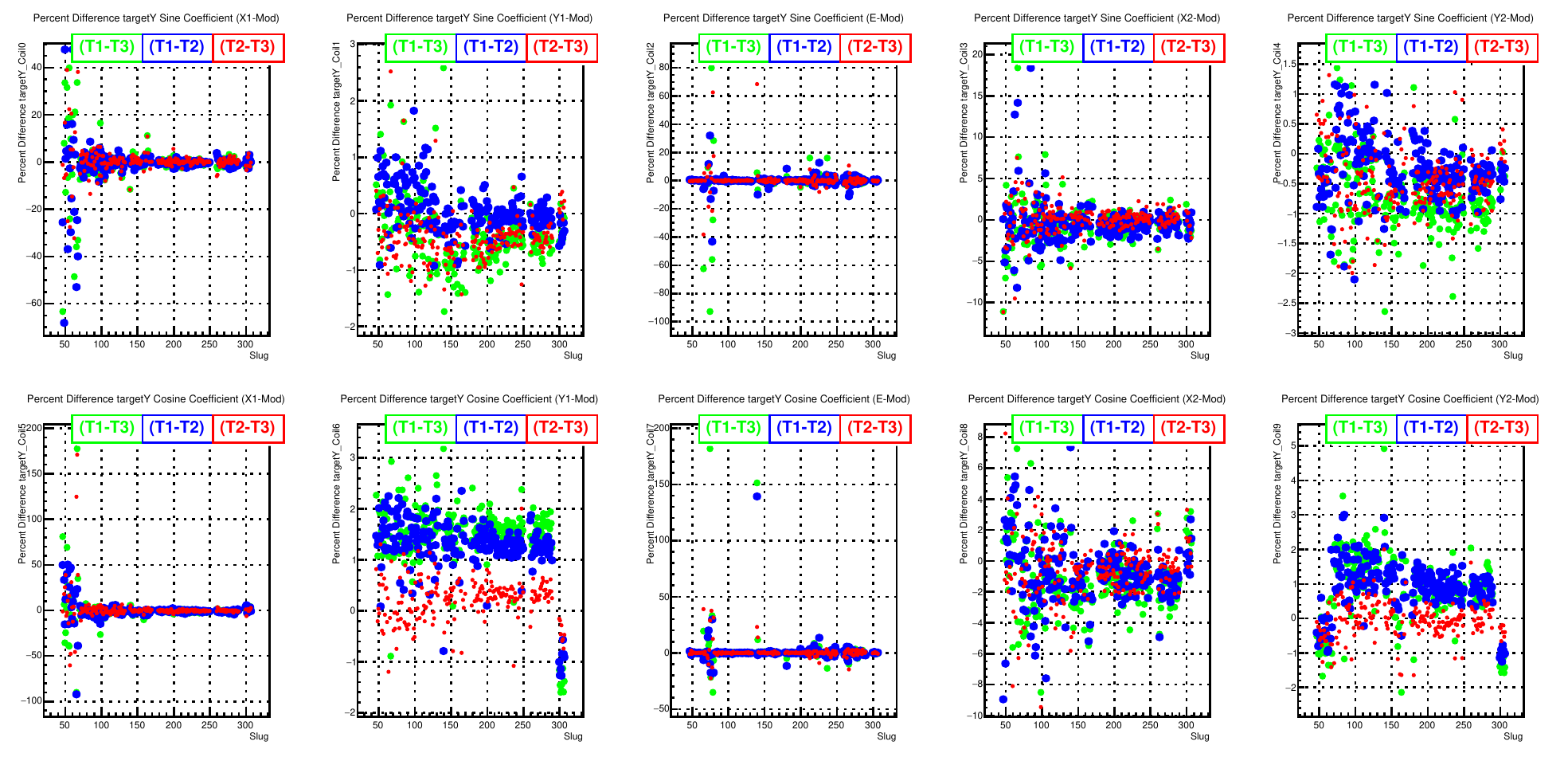}}
\caption{Percent change in targetY response between data tertiles. See Equation \ref{eq:fractional_tertile_change} for an explanation of the percent change shown.}

\label{fig:tert_coeffY}
\end{figure}

\begin{figure}[t]

\centering
\framebox{\includegraphics[width=8.9in]{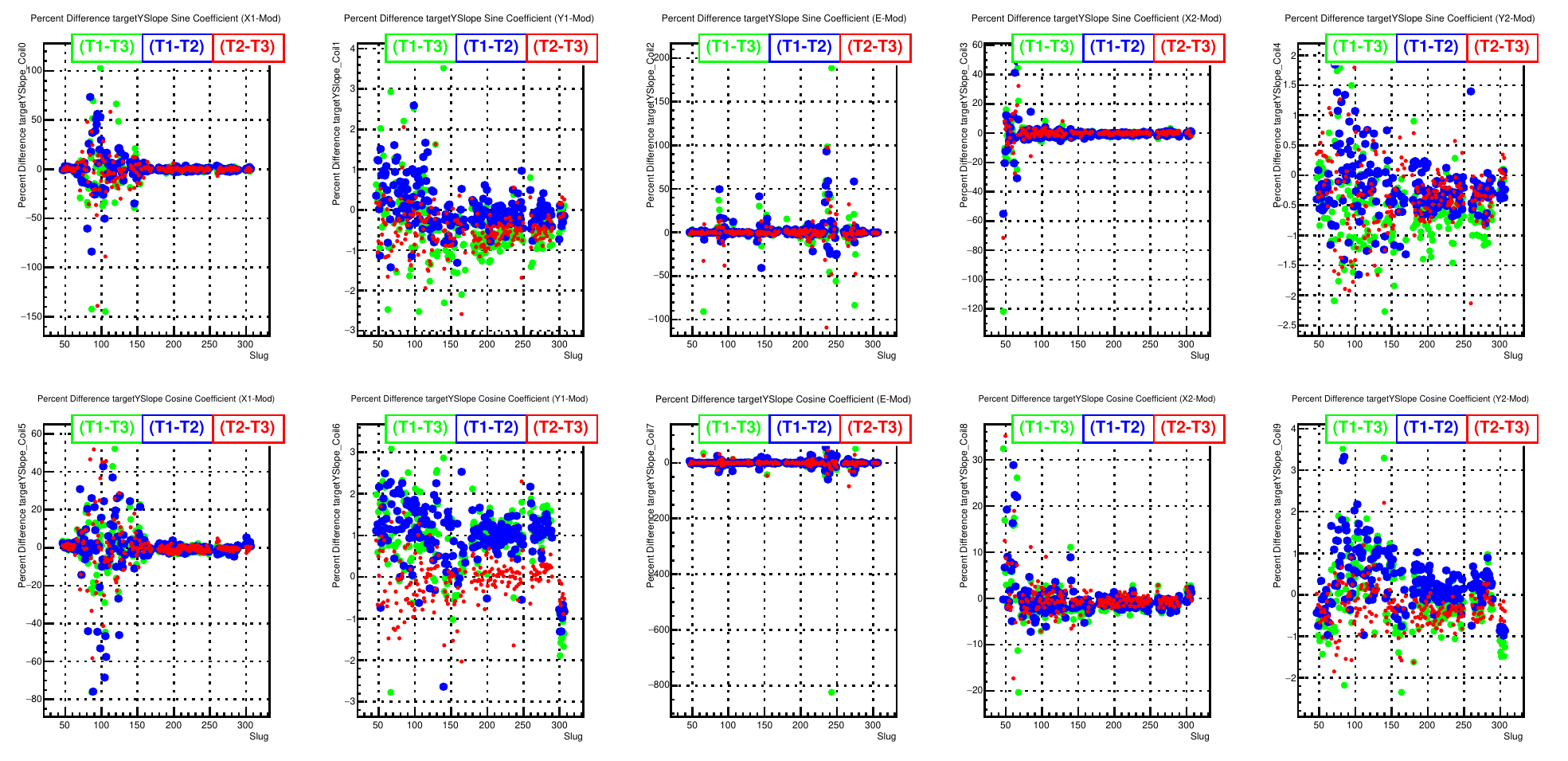}}
\caption{Percent change in targetYSlope response between data tertiles. See Equation \ref{eq:fractional_tertile_change} for an explanation of the percent change shown.}

\label{fig:tert_coeffYSlope}
\end{figure}
\begin{figure}[t]

\centering
\framebox{\includegraphics[width=8.9in]{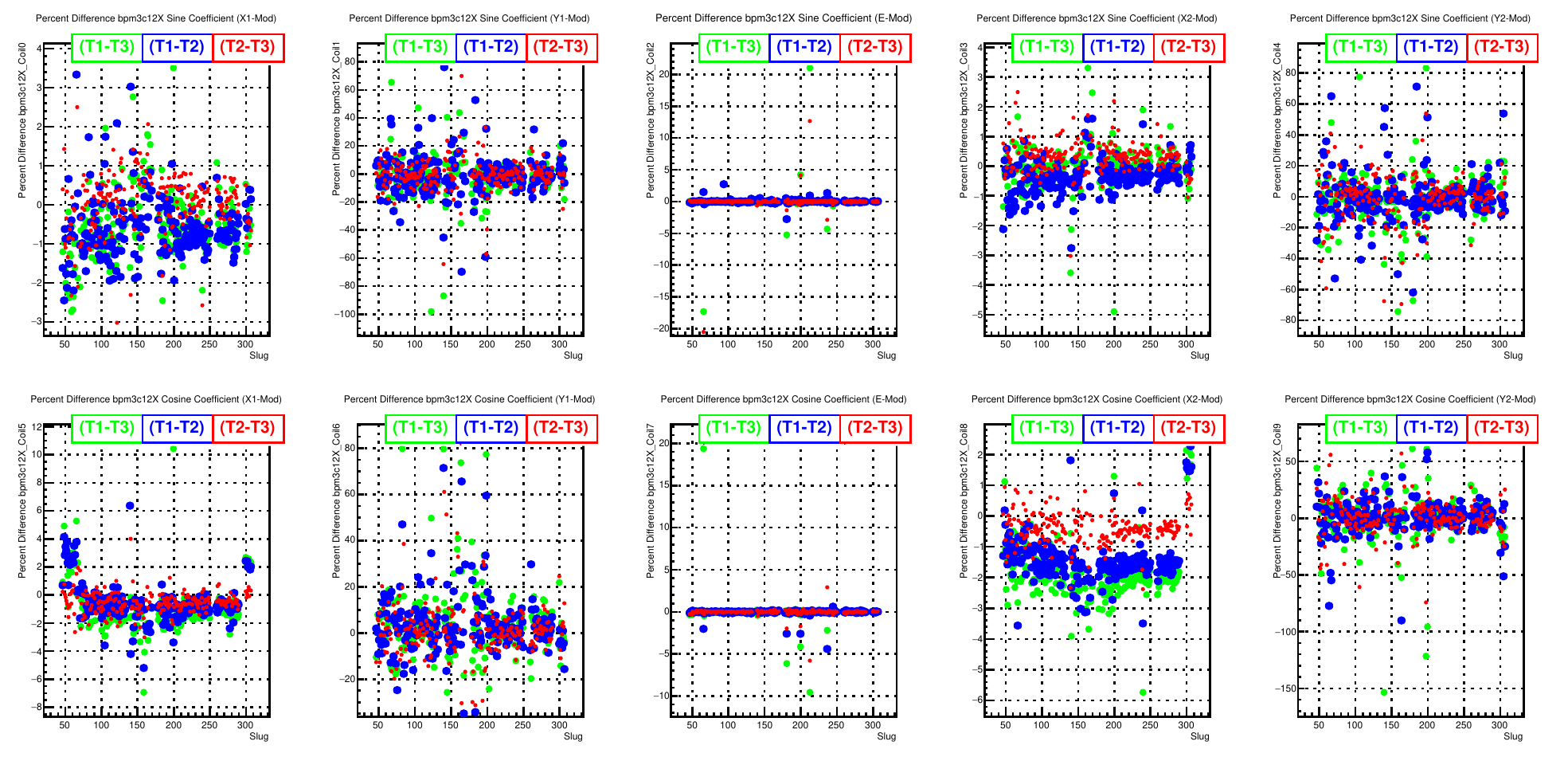}}
\caption{Percent change in bpm3c12X response between data tertiles. See Equation \ref{eq:fractional_tertile_change} for an explanation of the percent change shown.}

\label{fig:tert_coeff3c12X}
\end{figure}
\end{landscape}


\chapter{Main Detector Monopole and Dipole Responses} 
\captionsetup{justification=justified,singlelinecheck=false}

\label{AppendixD} 

\lhead{Appendix D. \emph{Detector Dipole and Monopole}} 

\begin{figure}[ht]
\centering
\includegraphics[width=0.7\textwidth]{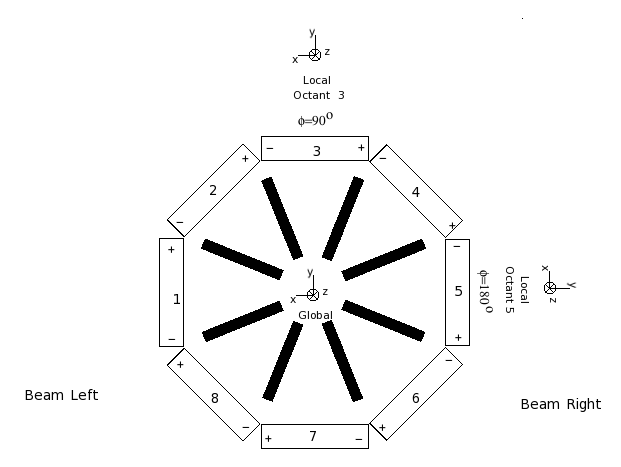}
\caption{Diagram of \Qs detector system looking downstream showing octant labeling. Numbered rectangles represent detector bars while solid spokes represent spectrometer coils.}
\label{fig:octant_coords_D}
\end{figure}
Consider the azimuthal arrangement of the main detector bars in Figure \ref{fig:octant_coords_D}. Response patterns on the azimuth, that is, responses as a function of main detector bar azimuthal location can offer clues as to the source or origin of the signal creating the response. For example, the response of the main detector to position or angle modulation as a function of the octant angle around the azimuth should have an approximately sinusoidal pattern. These sinusoidal patterns are referred to as ``dipoles'' and in the jargon of \Qs when the maximum response is in the vertical direction (maximum response on bars 3 and 7) it is referred to  as a ``vertical dipole''. Likewise when the maximum response is in the horizontal direction (maximum response on bars 1 and 5), it is called a ``horizontal dipole''. Thus, X-type modulations will create horizontal dipole responses and Y-type modulations are expected to produce vertical dipole responses. Any effect that creates equal responses in all detector bars is called a ``monopole response''. Shifts in energy and charge are candidates for monopole responses.

 Examples of dipole and monopole responses to the various types of X and Y modulation before and after correction using the full 10-Coil analysis can be seen in figures \ref{fig:Run2_10coil_Xdipoles} and \ref{fig:Run2_10coil_Ydipoles}. In these plots the dipoles are given as the amplitude of the sinusoid and the monopole as the center or offset of the sinusoid. A few observations can be made from the plots.

1. As expected, obvious horizontal and vertical dipole responses can be observed for X-type and Y-type modulations respectively. In both cases, these large dipoles are removed after correction suggesting that the modulation analysis is very good at removing the geometric responses. \\
2. The failure mode of the beam modulation analysis affects all the main detectors equally i.e. the residuals are monopole. For some coils there are significant monopole residual sensitivities whereas the dipole sensitivities are consistent with zero.\\
3. With only a couple of exceptions -- Coil 3 (in-phase X-2 modulation) during Run 1  and Coil 4 (in-phase Y-2 modulation) -- the monopole responses are smaller after correction than before the correction. \\
4. The residual monopole responses are largest for the X-type coils and appear to be insignificant for Y-type coils averaged over Runs 1 and 2.\\ 

Figures \ref{fig:Run1_10coil_Edipoles_app} and \ref{fig:Run2_10coil_Edipoles_app} show the responses of the main detector bars to energy modulation. The relatively large horizontal dipole response of energy can be partially explained by the mixing of horizontal position and angle with energy. Although the optics of the beamline are supposed to minimize dispersion at the target, a predominantly horizontal sensitivity to energy shifts remains. The modulation correction does an especially good job of removing the dipole and monopole in-phase responses of the main detector bars to energy.

A similar set of plots are shown for the ``Omit 0,5'' modulation analysis in figures \ref{fig:Run1_Omit05_Xdipoles} -- \ref{fig:Run2_Omit05_Edipoles_app}. Although the previous observations largely remain true for this analysis one significant difference is apparent. Omitting coils 0 and 5 shrinks the monopole responses in the other two X-coils (3 and 8) to be consistent with 0 while enlarging the monopole response in the omitted coils 0 and 5. However, even for coils omitted from the analysis, the residual dipole response is consistent with zero.\\ 

A final set of plots in Figure \ref{fig:Run2_Omit38_Edipoles} illustrates the characteristic of the failure for the modulation scheme ``Omit 3,8'' considered unreliable due to the large residual correlations it produces between the main detector and monitors over long timescales (see table \ref{tab:run2_residual_correlations_table}). There is a very large monopole detector sensitivity to coil 3 after correction. Even though coils 3 and 8 are not included in this scheme, dipole responses to these coils are still consistent with zero.

\begin{landscape}
\begin{figure}[!ht]
\begin{center}
\includegraphics[width=9in]{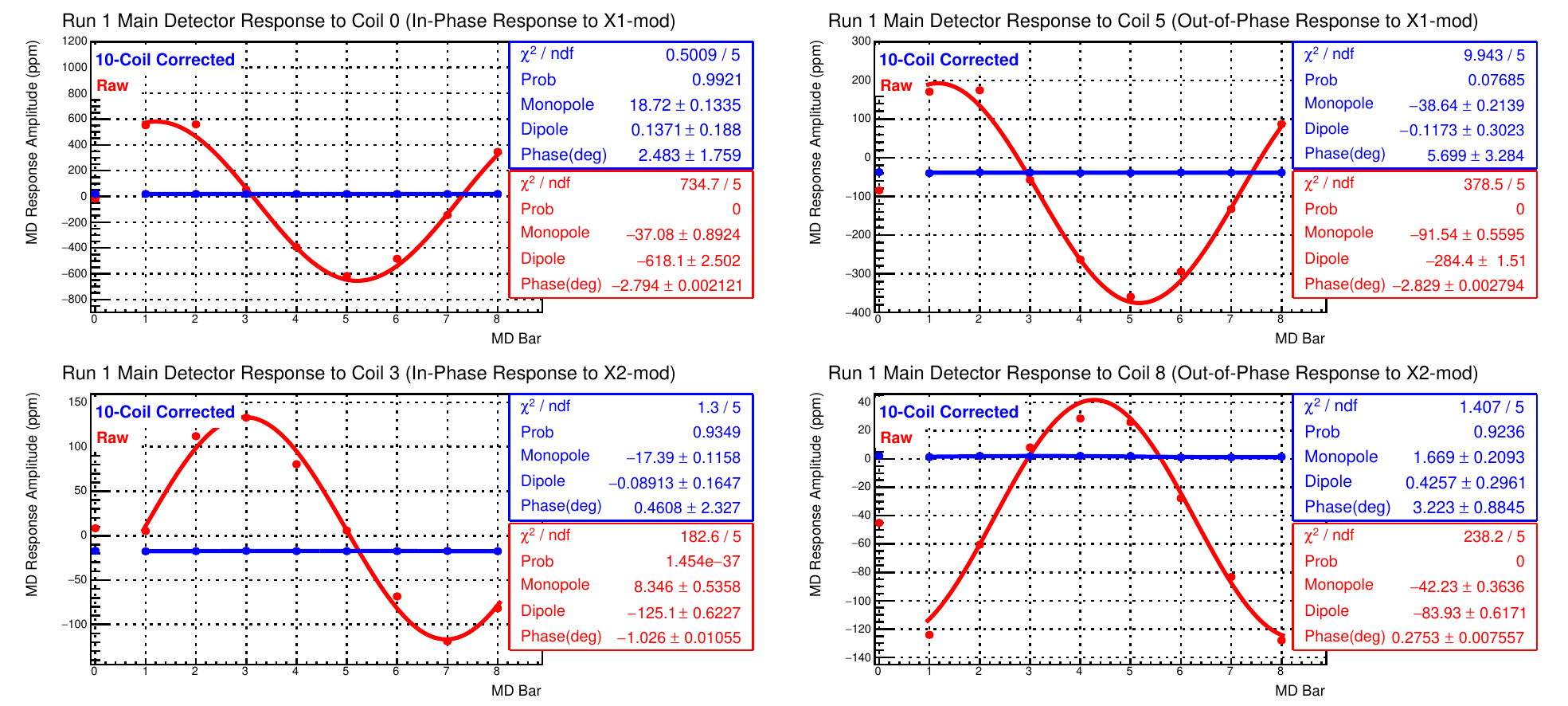}
\caption{\label{fig:Run1_10coil_Xdipoles}Responses of individual main detector bars averaged over Run 1 to X-type modulation coils before correction and after correction using a full 10-Coil analysis. The dipole response is given by the amplitude of the sinusoid while the monopole is the offset.}
\end{center}
\end{figure}
\begin{figure}[!ht]
\begin{center}
\includegraphics[width=9in]{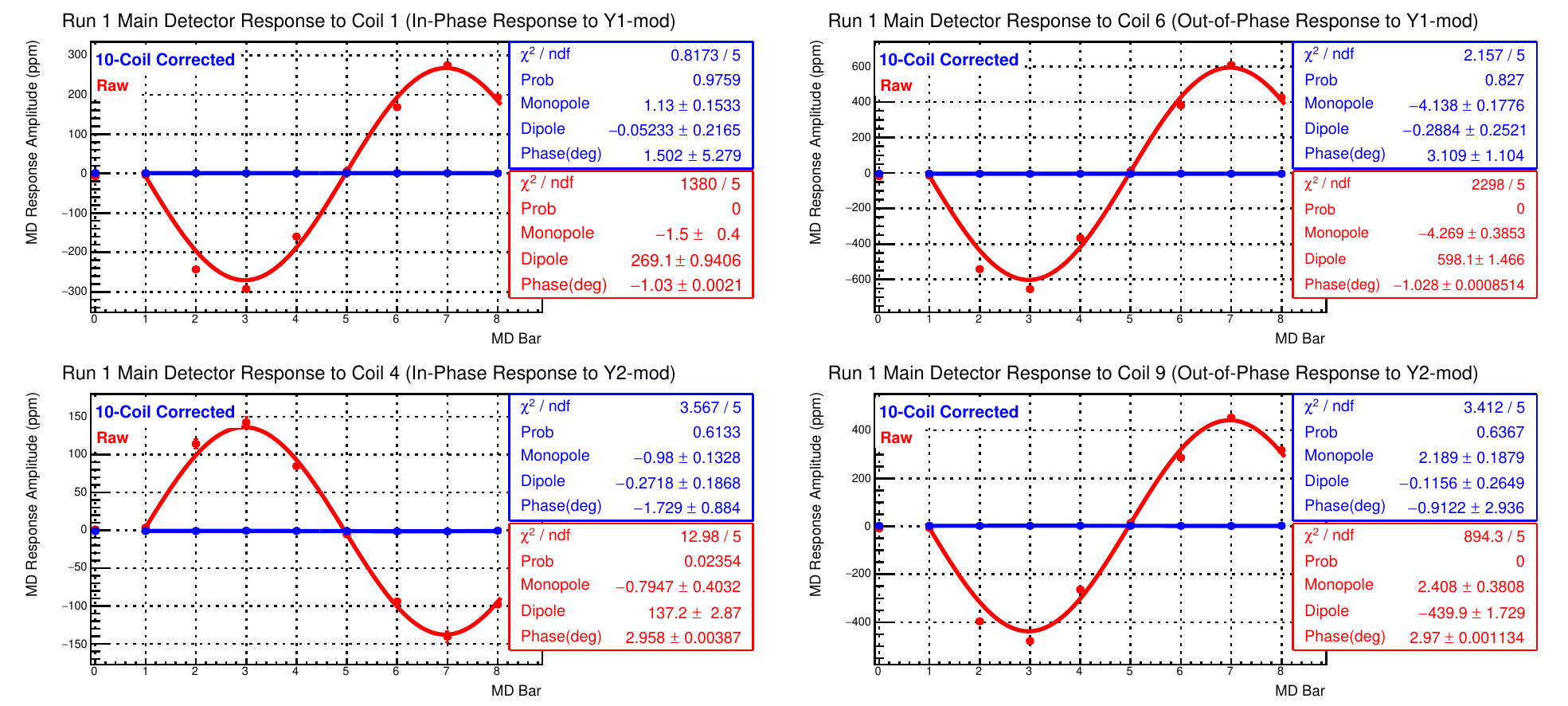}
\caption{\label{fig:Run1_10coil_Ydipoles}Responses of individual main detector bars averaged over Run 1 to Y-type modulation coils before correction and after correction using a full 10-Coil analysis. The dipole response is given by the amplitude of the sinusoid while the monopole is the offset.}
\end{center}
\end{figure}

\begin{figure}[!ht]
\begin{center}
\includegraphics[width=9in]{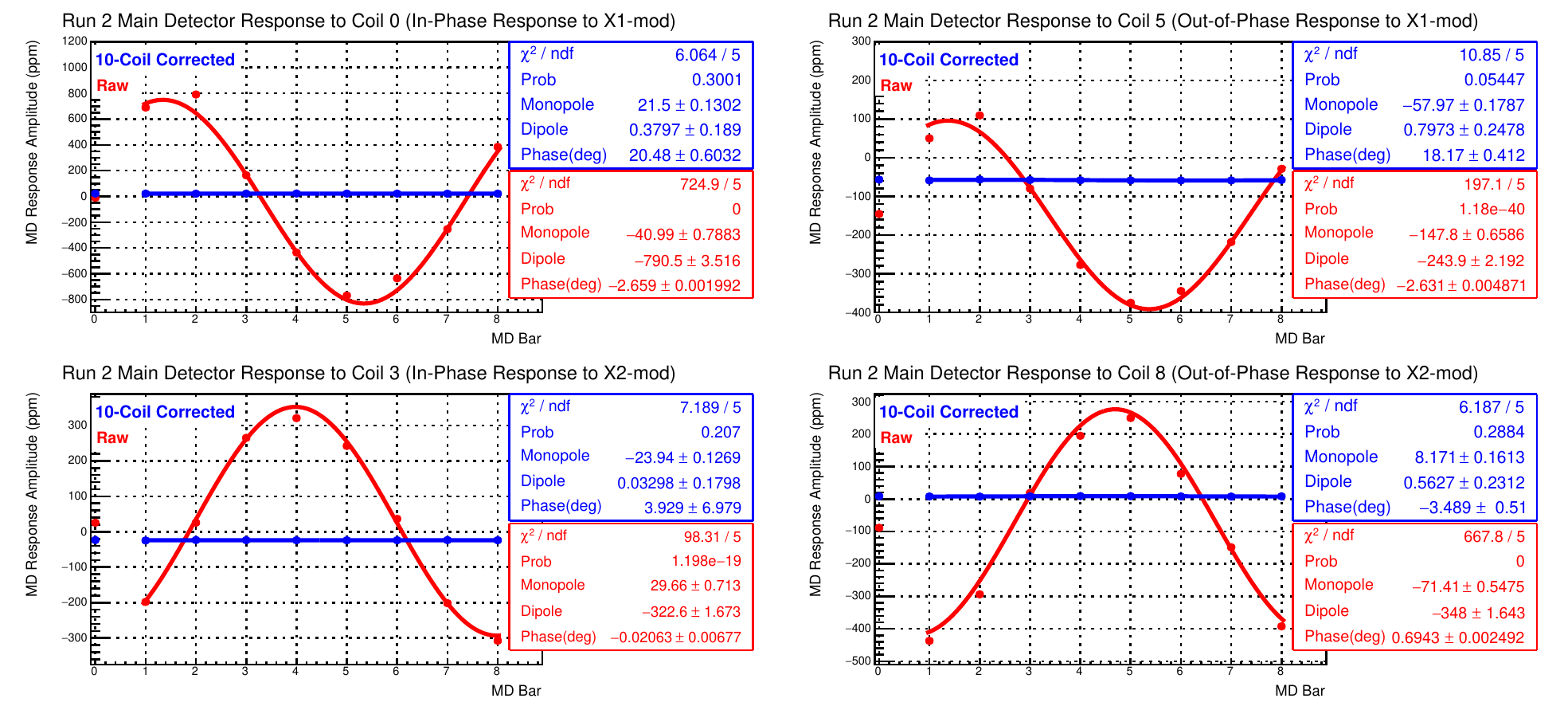}
\caption{\label{fig:Run2_10coil_Xdipoles_app}Responses of individual main detector bars averaged over Run 2 to X-type modulation coils before correction and after correction using a full 10-Coil analysis. The dipole response is given by the amplitude of the sinusoid while the monopole is the offset.}
\end{center}
\end{figure}

\begin{figure}[!ht]
\begin{center}
\includegraphics[width=9in]{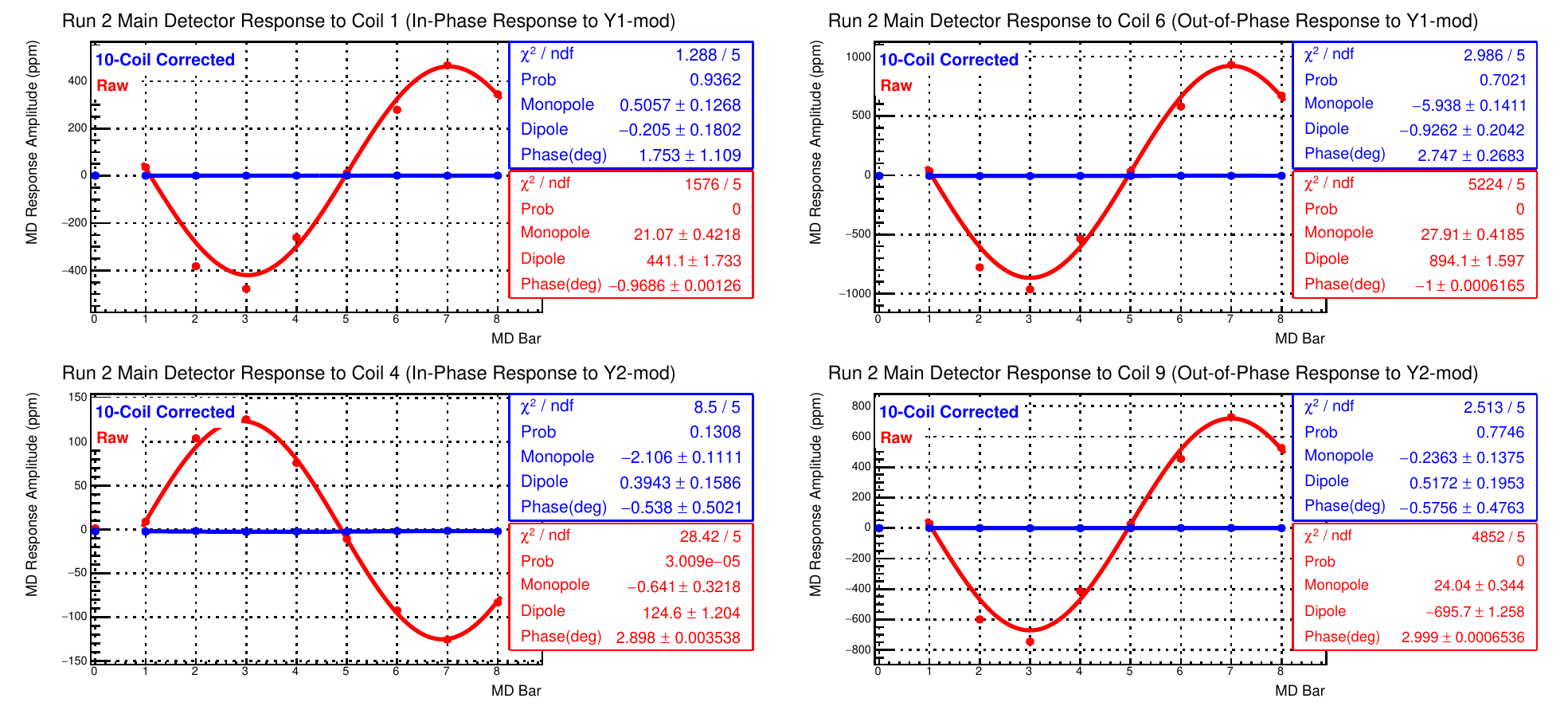}
\caption{\label{fig:Run2_10coil_Ydipoles_app}Responses of individual main detector bars averaged over Run 2 to Y-type modulation coils before correction and after correction using a full 10-Coil analysis. The dipole response is given by the amplitude of the sinusoid while the monopole is the offset.}
\end{center}
\end{figure}

\begin{figure}[!ht]
\begin{center}
\includegraphics[width=9in]{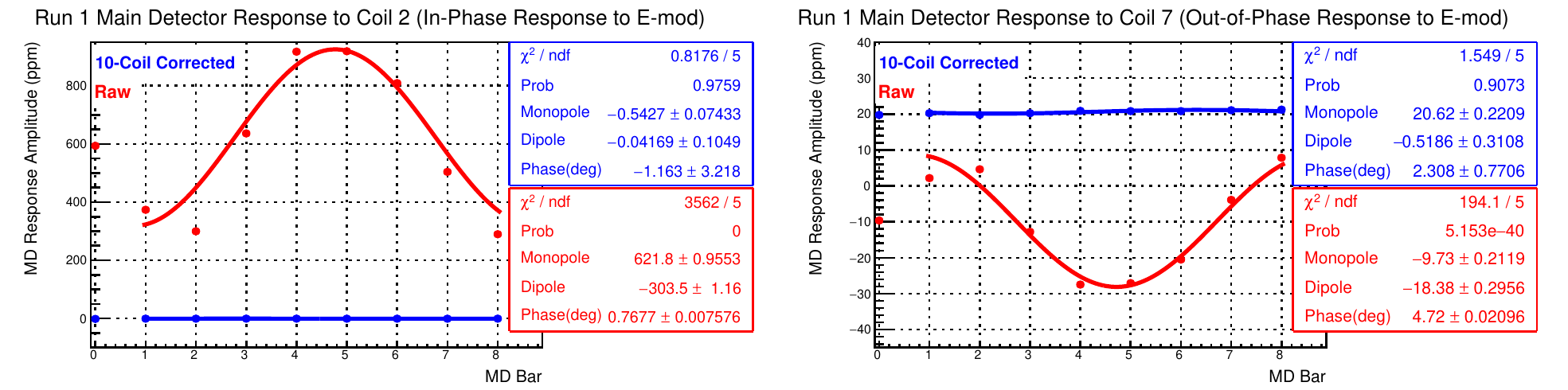}
\caption{\label{fig:Run1_10coil_Edipoles_app}Responses of individual main detector bars averaged over Run 1 to energy modulation coils before correction and after correction using a full 10-Coil analysis. The dipole response is given by the amplitude of the sinusoid while the monopole is the offset.}
\end{center}
\end{figure}

\begin{figure}[!ht]
\begin{center}
\includegraphics[width=9in]{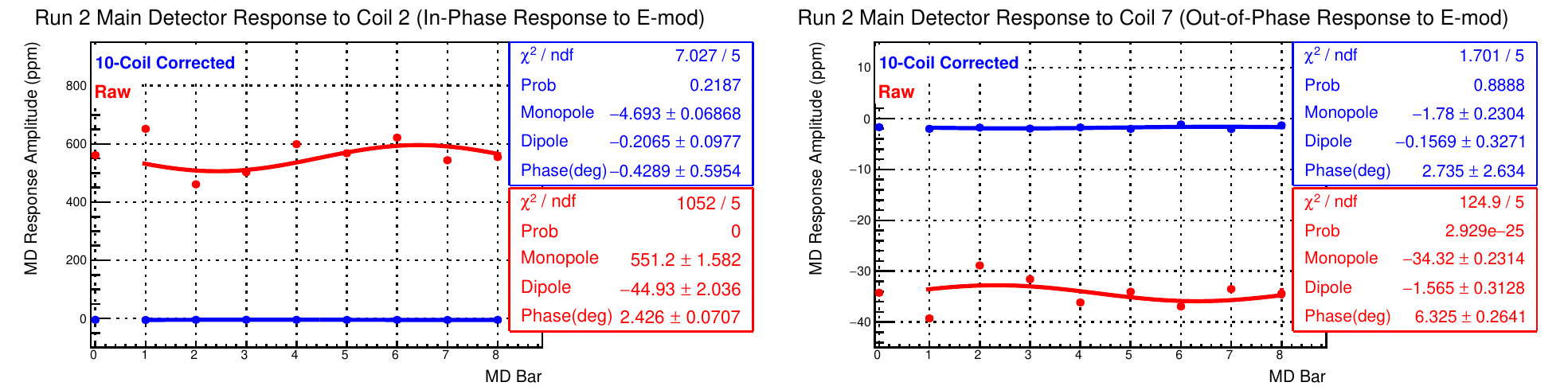}
\caption{\label{fig:Run2_10coil_Edipoles_app}Responses of individual main detector bars averaged over Run 2 to energy modulation coils before correction and after correction using a full 10-Coil analysis. The dipole response is given by the amplitude of the sinusoid while the monopole is the offset.}
\end{center}
\end{figure}

\begin{figure}[!ht]
\begin{center}
\includegraphics[width=9in]{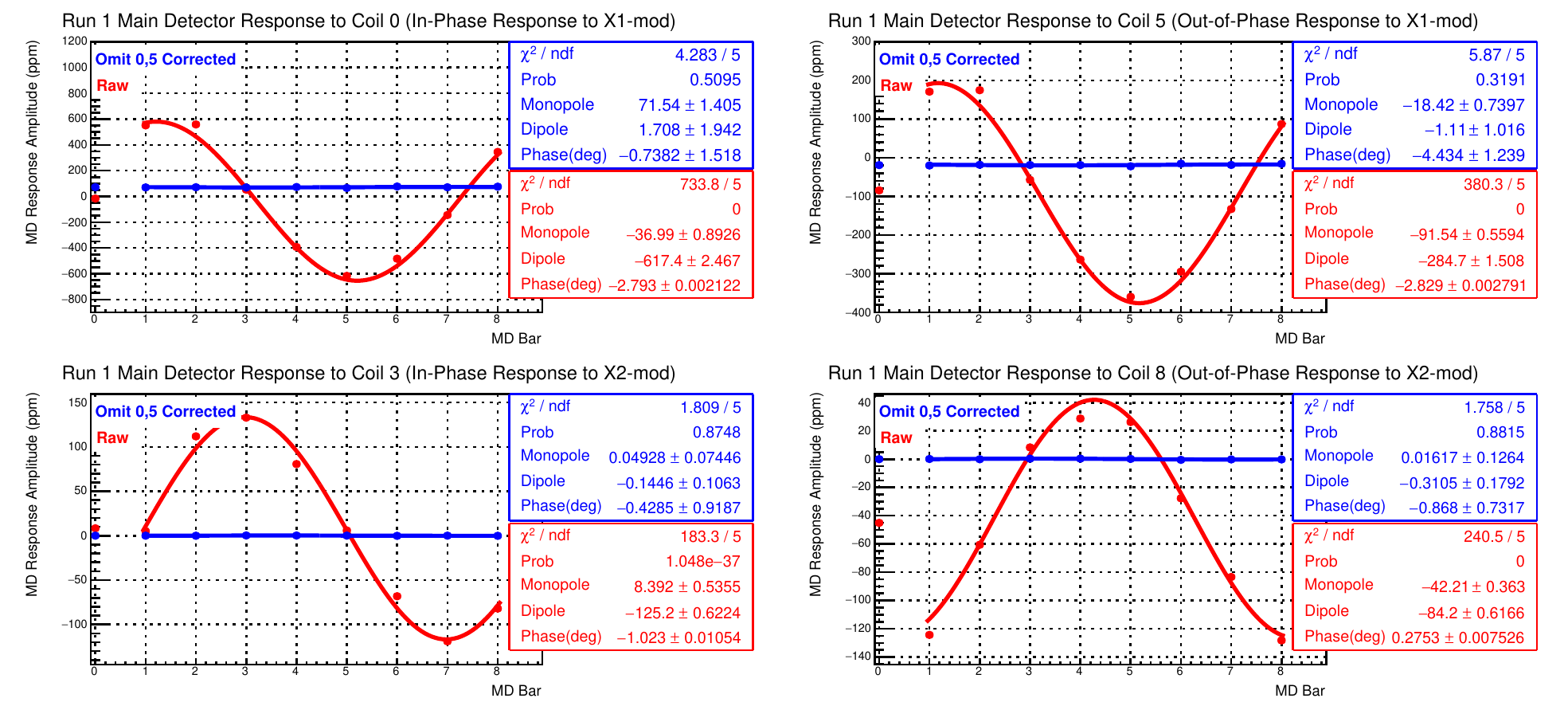}
\caption{\label{fig:Run1_Omit05_Xdipoles}Responses of individual main detector bars averaged over Run 1 to X-type modulation coils before correction and after correction using the ``Omit 0,5'' analysis. The dipole response is given by the amplitude of the sinusoid while the monopole is the offset.}
\end{center}
\end{figure}
\begin{figure}[!ht]
\begin{center}
\includegraphics[width=9in]{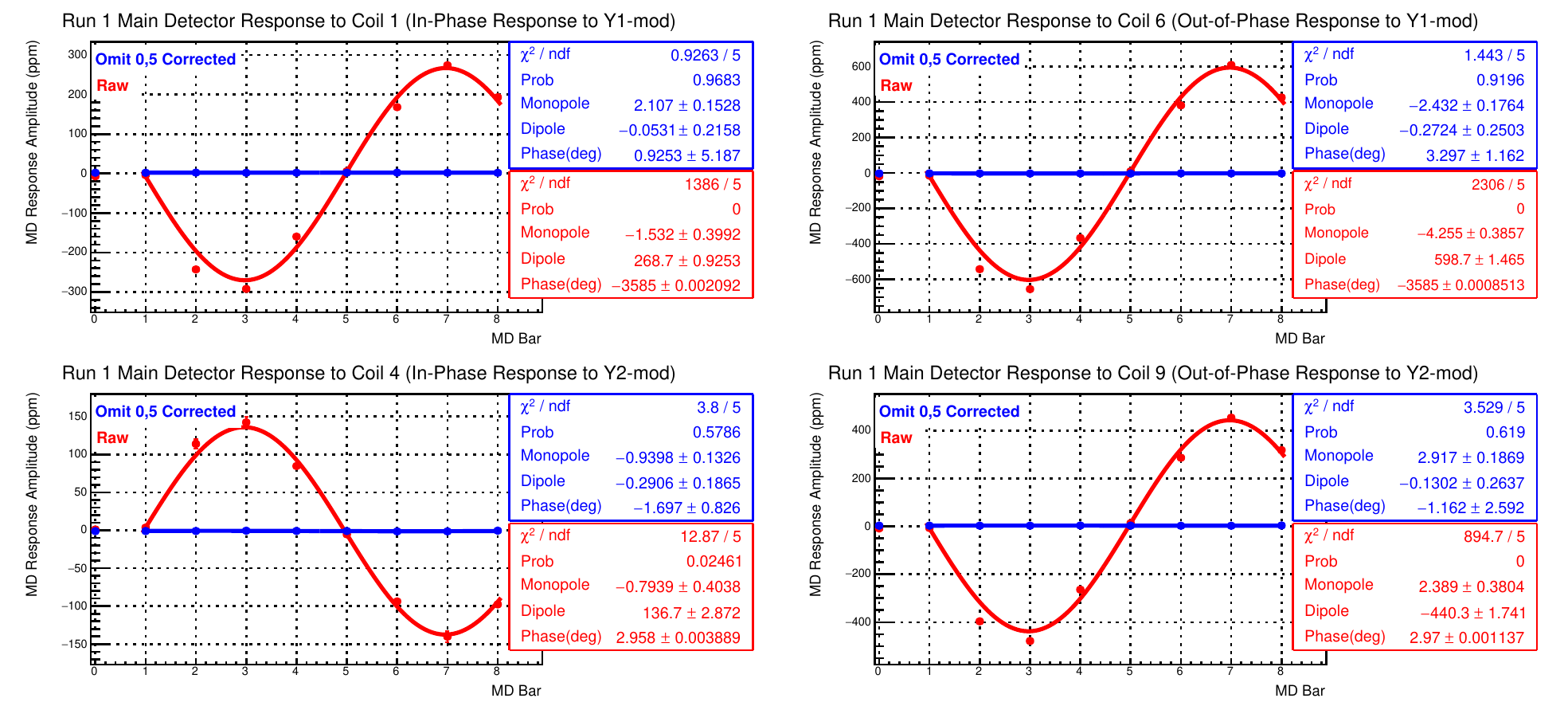}
\caption{\label{fig:Run1_Omit05_Ydipoles}Responses of individual main detector bars averaged over Run 1 to Y-type modulation coils before correction and after correction using the ``Omit 0,5'' analysis. The dipole response is given by the amplitude of the sinusoid while the monopole is the offset.}
\end{center}
\end{figure}

\begin{figure}[!ht]
\begin{center}
\includegraphics[width=9in]{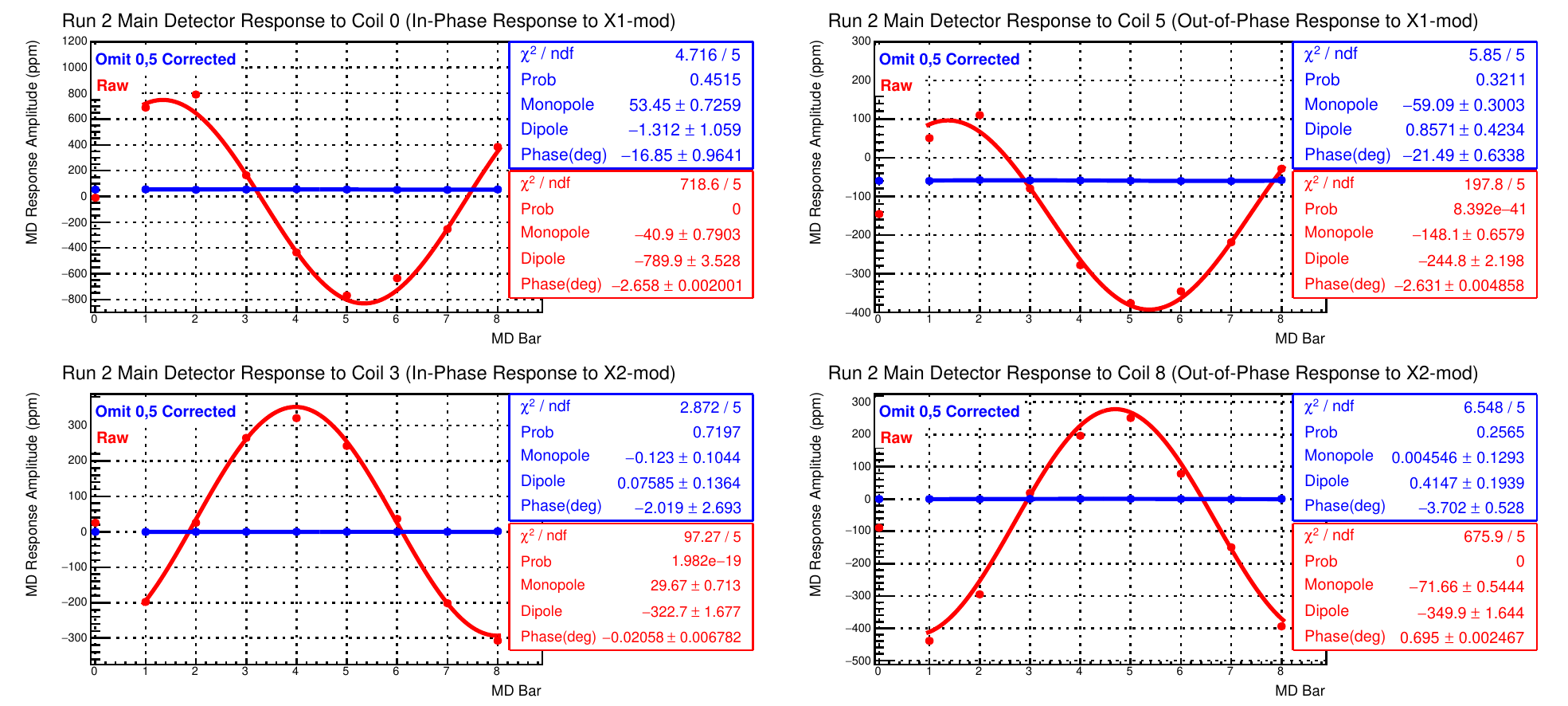}
\caption{\label{fig:Run2_Omit05_Xdipoles_app}Responses of individual main detector bars averaged over Run 2 to X-type modulation coils before correction and after correction using the ``Omit 0,5'' analysis. The dipole response is given by the amplitude of the sinusoid while the monopole is the offset.}
\end{center}
\end{figure}

\begin{figure}[!ht]
\begin{center}
\includegraphics[width=9in]{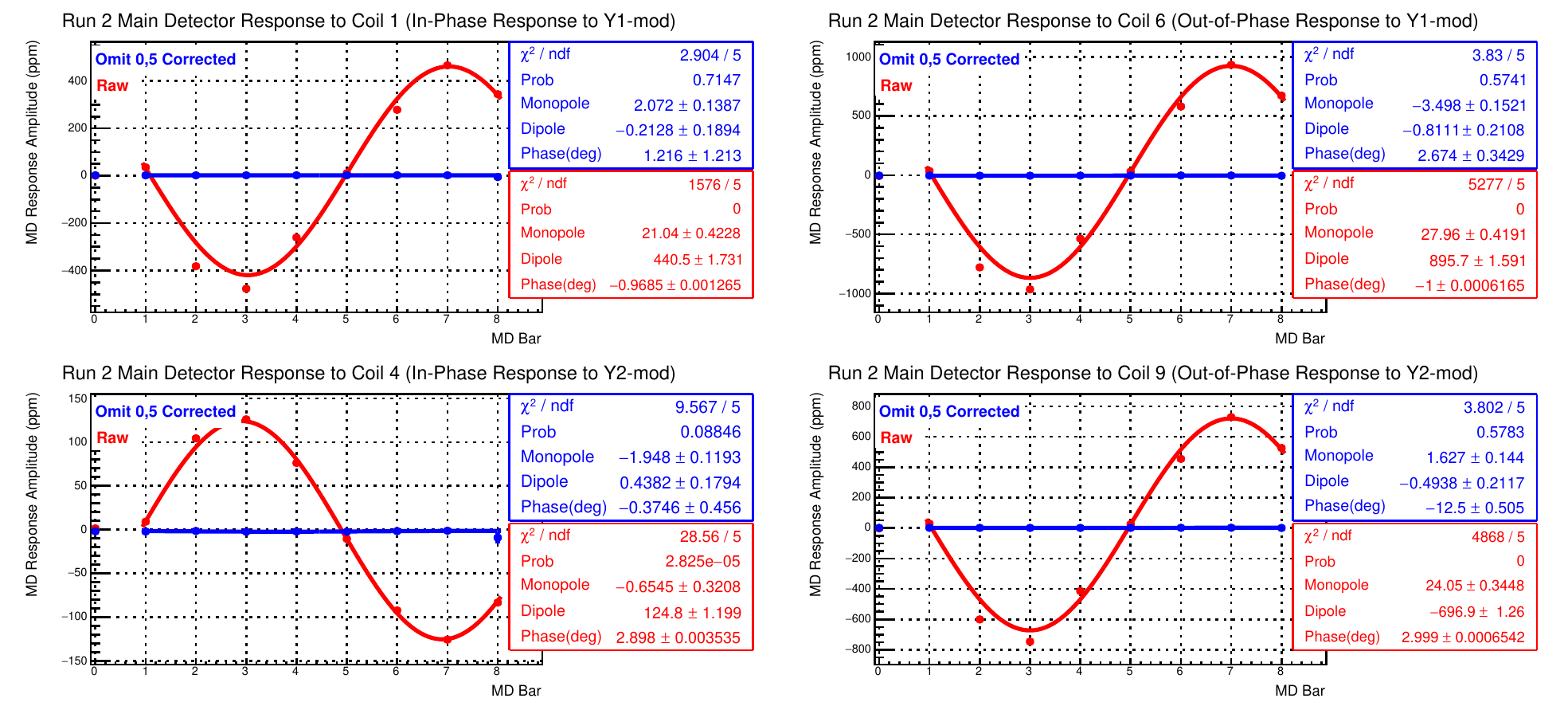}
\caption{\label{fig:Run2_Omit05_Ydipoles_app}Responses of individual main detector bars averaged over Run 2 to Y-type modulation coils before correction and after correction using the ``Omit 0,5'' analysis. The dipole response is given by the amplitude of the sinusoid while the monopole is the offset.}
\end{center}
\end{figure}

\begin{figure}[!ht]
\begin{center}
\includegraphics[width=9in]{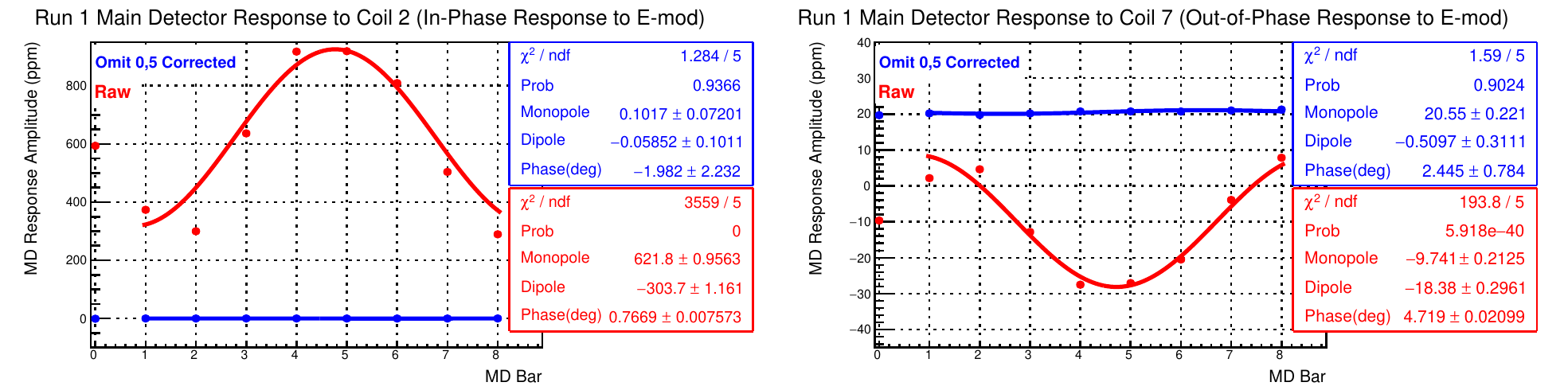}
\caption{\label{fig:Run1_Omit05_Ydipoles_app}Responses of individual main detector bars averaged over Run 1 to energy modulation coils before correction and after correction using the ``Omit 0,5'' analysis. The dipole response is given by the amplitude of the sinusoid while the monopole is the offset.}
\end{center}
\end{figure}

\begin{figure}[!ht]
\begin{center}
\includegraphics[width=9in]{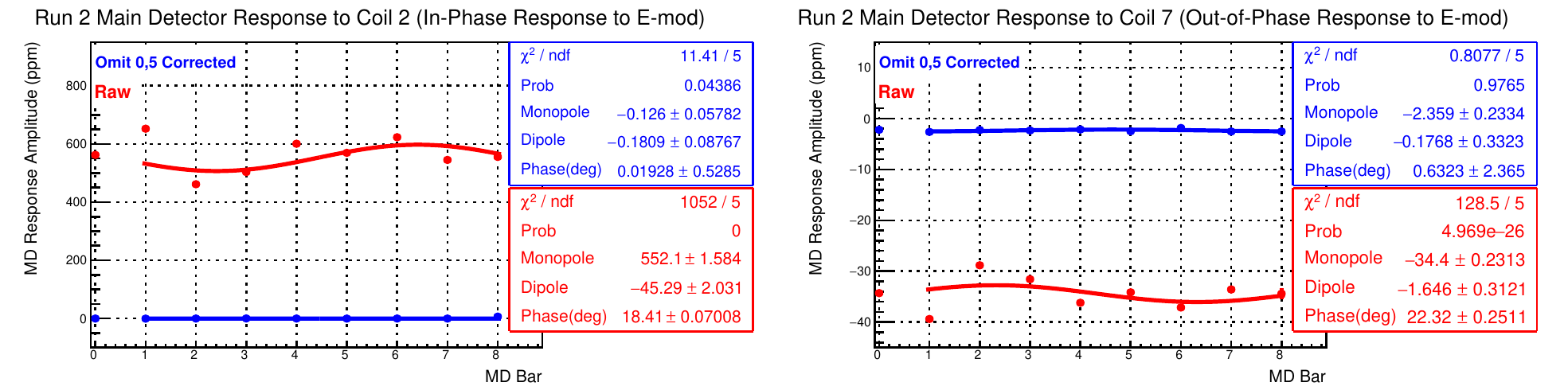}
\caption{\label{fig:Run2_Omit05_Edipoles_app}Responses of individual main detector bars averaged over Run 2 to energy modulation coils before correction and after correction using the ``Omit 0,5'' analysis. The dipole response is given by the amplitude of the sinusoid while the monopole is the offset.}
\end{center}
\end{figure}

\begin{figure}[!ht]
\begin{center}
\includegraphics[width=9in]{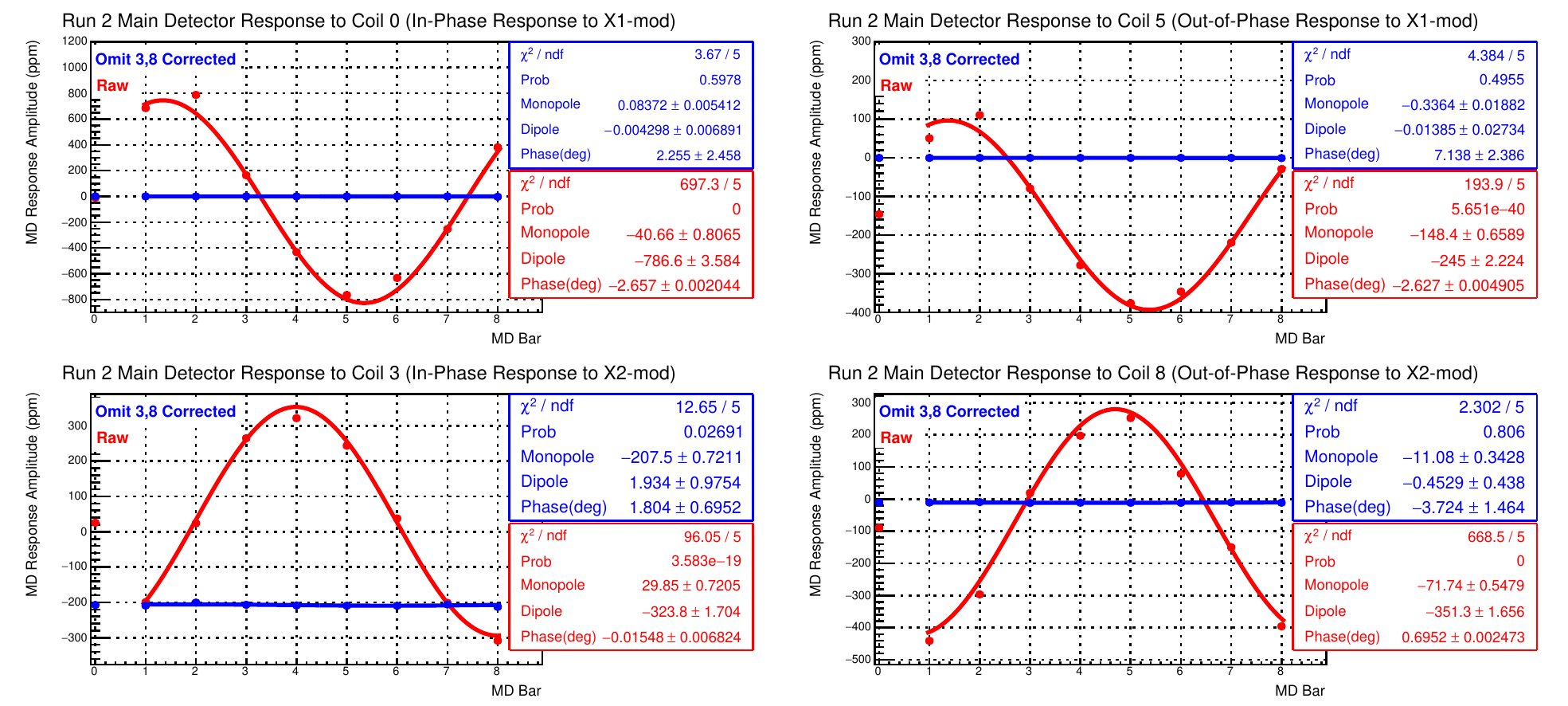}
\caption{\label{fig:Run2_Omit38_Edipoles}Responses of individual main detector bars averaged over Run 2 to X-type modulation coils before correction and after correction using the ``Omit 3,8'' analysis considered unreliable due to large residual correlations it produces between the main detector and the monitors (see table \ref{tab:run2_residual_correlations_table}). Notice the large monopole residual in the dipole response is given by the amplitude of the sinusoid while the monopole is the offset.}
\end{center}
\end{figure}
\end{landscape}

\chapter{Qweak Terminology} 
\captionsetup{justification=justified,singlelinecheck=false}

\label{AppendixE} 

\lhead{Appendix E. \emph{Terminology}}
This Appendix serves as a reference for the terminology used in the thesis which may be isolated to the \Qs experiment or to parity-violation experiments.

\begin{itemize}
\item {targetX(Y): electron beam horizontal(vertical) position measured by extrapolating positions from beam position monitors (BPM's) in the drift region before the target downstream to the target position.}
\item {targetXSlope(YSlope): electron beam horizontal angle(vertical angle) relative to the ideal beam axis and measured by finding differences between BPM's in the drift region before the target.}
\item{BPM3c12X: the X or horizontal measurement of the BPM located in the region of highest dispersion of the electron beam in the arc leading into Hall C. BPM3c12X is highly sensitive to energy shifts and is often referred to as our ``energy monitor''.}
\item { qwk\_energy: a calculated variable designed to represent true energy shifts in the electron beam. It mainly utilizes BPM3c12X, the BPM most sensitive to energy, but subtracts measured position and angle sensitivities.}
\item{ qwk\_charge: a variable representing the measurement of electron beam current used to normalize the main detector. For Run 1 qwk\_charge was an average of BCM1 and BCM2. For most of Run 2 it was BCM8.}
\item{BCM\#: beam current monitors used to measure beam current in the experimental hall and numbered according to their position on the beam line with BCM1 being the most upstream.}
\item{BPMwxyz: these are beam position monitors which read out electron beam position transverse to the ideal beam trajectory. ``w'' is a number given to the experimental hall. For BPM's reading the Hall C beam position w=3.  ``x'' gives a clue as to the location of the monitor. For example ``x=c'' means the BPM is located in the beam tunnel leading to Hall C whereas ``x=h'' means it is located in the experimental hall. ``y''gives the girder number on which the BPM is located. ``z'' takes values either ``X'' or ``Y'' referring to measurements of horizontal and vertical displacement respectively.}
\item{MPS: stands for ``Macro-Pulse Synchronization'' and is the timing signal that controls the timing of helicity state changes on the electron beam. In common usage in \Qs it is used to refer to a time of approximately 1~ms in which data for a specific helicity state was taken continuously.}
\item{Quartet: all measurements for \Qs were taken in groups of four distinct consecutive MPS windows called ``quartets''. The helicity pattern over a quartet was either +--+ or -++- with the sign of the first chosen pseudo-randomly. This pattern was chosen to cancel slow linear drifts.}
\item{Yield, Difference, Asymmetry: these are the three measurements recorded for all detectors and monitors used in the \Qs experiment. ``Yield'' refers to the absolute signal integrated over the gated MPS window for any given detector. Yield is normalized to integration time and beam current and has units of V/($\mu$A$\cdot$s). ``Difference'' refers to the average difference between yields of opposite helicity states measured over a quartet. Specifically it refers to $\pm(Y_1+Y_4-Y_2-Y_3)/4$, where $Y_i$ refers to the number of the MPS in the quartet and where ``+'' refers to the $+--+$ pattern and ``$-$'' refers to the $-++-$ pattern. ``Asymmetry'' is the difference normalized to the average yield.}
\item{PMT Average Asymmetry: called ``AsymPMTavg'' or ``AsymMD\textunderscore PMTavg'' where ``MD'' acknowledges that the average is over the ``main detector''. The main detector has eight separated bars and each bar is readout with two PMT's, one on each end for a total of 16 signals. These signals can be combined different ways. One choice is the equal-weighted average of the asymmetries to obtain an average asymmetry. ``PMT Average Asymmetry'' is this straight average. Since there is no {\it a priori} reason to weight the results of one PMT over others any average yield must apply a weighting factor to equalize signal strengths that naturally exist in the PMT's before averaging. Thus, there is no such thing as a PMT Average yield or difference since this average would involve weighting the signals.}
\item{MDallbars: \Qs jargon for a particular weighting used to produce an average yield for the main detector PMT's. The weighting approximately equalizes the natural variation in signal from the PMT's. This weighted average of yields can then be used to produce MDallbars differences and asymmetries.}
\end{itemize}

\addtocontents{toc}{\vspace{2em}} 

\backmatter


\label{Bibliography}

\lhead{\emph{Bibliography}} 

\bibliographystyle{unsrtnat} 

\bibliography{Bibliography} 

\end{document}